\documentclass[12pt,twoside]{report}
\usepackage[a4paper,portrait,top=3cm,bottom=3cm,left=3.5cm,right=3cm]{geometry}
\usepackage{hyperref}
\usepackage{setspace}
\usepackage{titlesec}
\usepackage{amsmath}
\usepackage{amsfonts}
\usepackage{rotating} 
\usepackage{float} 
\usepackage[font=normalsize,margin=1cm]{subfig,caption} 
\usepackage{graphicx}
\usepackage[british]{babel}
\usepackage{mdwlist}
\usepackage{booktabs} 
\usepackage{tablefootnote} 
\usepackage{emptypage} 
\usepackage[bottom]{footmisc}
\usepackage[backend=biber,
            style=numeric,
            sorting=none,
            maxnames=3,
            firstinits=true,
            isbn=false,
            url=false,
            doi=false,
            eprint=false
           ]{biblatex}

\newbibmacro{string+doiurlisbnissn}[1]{%
  \iffieldundef{doi}{%
    \iffieldundef{url}{%
      \iffieldundef{isbn}{%
        \iffieldundef{issn}{%
          #1%
        }{%
          \href{http://books.google.com/books?vid=ISSN\thefield{issn}}{#1}%
        }%
      }{%
        \href{http://books.google.com/books?vid=ISBN\thefield{isbn}}{#1}%
      }%
    }{%
      \href{\thefield{url}}{#1}%
    }%
  }{%
    \href{http://dx.doi.org/\thefield{doi}}{#1}%
  }%
}
\DeclareFieldFormat{title}{\usebibmacro{string+doiurlisbnissn}{\mkbibemph{#1}}}
\DeclareFieldFormat[article,incollection]{title}{\usebibmacro{string+doiurlisbnissn}{\mkbibquote{#1}}}

\renewbibmacro{in:}{%
  \ifentrytype{article}{}{%
    \printtext{\bibstring{in}\intitlepunct}%
  }%
}

\AtEveryBibitem{\clearlist{language}}
\AtEveryBibitem{\clearfield{note}}

\hypersetup{pdfauthor={Wojciech J. Szlachta},
            pdftitle={First principles interatomic potential for tungsten based on Gaussian process regression.},
            linktocpage=true,
            bookmarks=true,
            bookmarksnumbered=true,
            breaklinks=true
           }

\titleformat{\chapter}[hang]{\LARGE\bfseries}{\thechapter}{1cm}{\LARGE\bfseries}
\numberwithin{equation}{chapter}
\numberwithin{figure}{chapter}

\makeatletter
\def\input@path{{.}{chapter_1_introduction/}{chapter_2_classical_and_quantum_simulation_of_solids/}{chapter_3_simulation_techniques/}{chapter_4_gaussian_approximation_potential/}{chapter_5_bulk_properties_and_lattice_defects_in_tungsten/}{chapter_6_bispectrum-gap_potential_for_tungsten/}{chapter_7_soap-gap_potential_for_tungsten/}{chapter_8_bond-based_soap-gap_potential/}{chapter_9_conclusions_and_further_work/}{appendices/}}
\makeatother
\graphicspath{{.}{chapter_1_introduction/}{chapter_2_classical_and_quantum_simulation_of_solids/}{chapter_3_simulation_techniques/}{chapter_4_gaussian_approximation_potential/}{chapter_5_bulk_properties_and_lattice_defects_in_tungsten/}{chapter_6_bispectrum-gap_potential_for_tungsten/}{chapter_7_soap-gap_potential_for_tungsten/}{chapter_8_bond-based_soap-gap_potential/}{chapter_9_conclusions_and_further_work/}{appendices/}}

\begin{document}

\begin{titlepage}
\begin{center}
\vspace*{1cm}
\includegraphics[height=4cm]{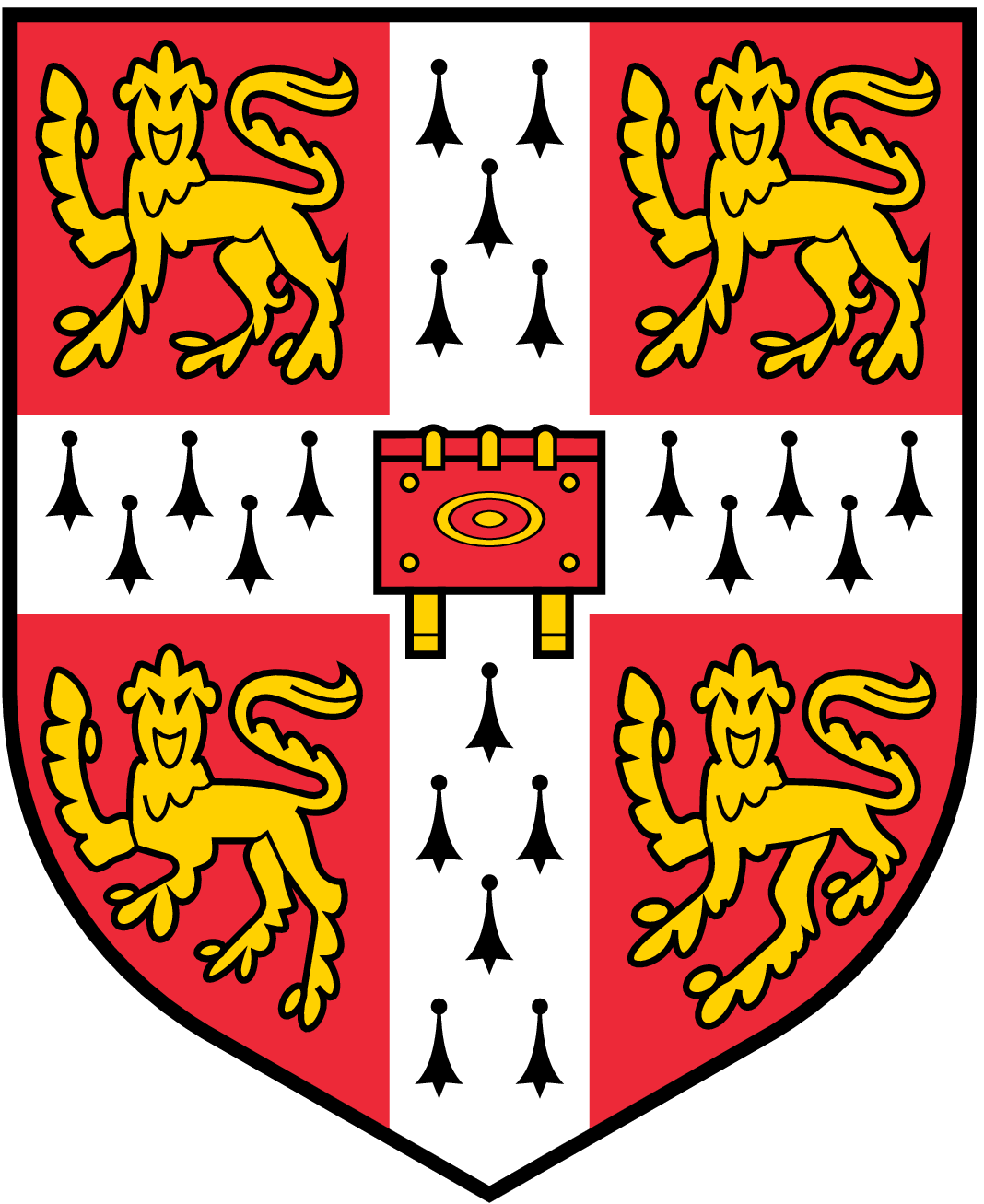} \\
\vspace*{1cm}
\LARGE{\bfseries{First principles interatomic potential \\ for tungsten based on \\ Gaussian process regression.}} \\[2.5cm]
\Large{Wojciech Jerzy Szlachta \\[0.5cm] Pembroke College} \\
\vfill
\normalsize{This dissertation is submitted for the degree of Doctor of Philosophy.} \\[1.5cm]
\normalsize{September 2013}
\end{center}
\end{titlepage}

\cleardoublepage

\pagenumbering{roman}

%
%

\phantomsection
\chapter*{Declaration}
\label{chapter:declaration}

The content of this dissertation describes the work I carried out between October 2009 and September 2013 in the Cavendish Laboratory and the Department of Engineering at the University of Cambridge.

\vspace{1.5cm}

This dissertation is the result of my own work and includes nothing which is the outcome of work done in collaboration except where specifically indicated in the text. It has not been submitted in whole or in part for any degree or diploma at this or any other university, and does not exceed 60,000 words, including tables, footnotes, bibliography and appendices.\\
\ 

Wojciech Jerzy Szlachta

\cleardoublepage

%
%

\thispagestyle{empty}

\begin{center}
\large{\bfseries{First principles interatomic potential for tungsten \\ based on Gaussian process regression.}} \\[0.5cm]
\normalsize{Wojciech Jerzy Szlachta \\ Pembroke College}
\end{center}

\vspace{1.5cm}

\noindent\LARGE{\bfseries{Summary}}

\vspace{1.5cm}


\normalsize{An accurate description of atomic interactions, such as that provided by first principles quantum mechanics, is fundamental to realistic prediction of the properties that govern plasticity, fracture or crack propagation in metals. However, the computational complexity associated with modern schemes explicitly based on quantum mechanics limits their applications to systems of a few hundreds of atoms at most.

\begin{sloppypar} 
This thesis investigates the application of the Gaussian Approximation Potential (GAP) scheme to atomistic modelling of tungsten --- a bcc transition metal which exhibits a brittle-to-ductile transition and whose plasticity behaviour is controlled by the properties of $\frac{1}{2} \langle 111 \rangle$ screw dislocations. We apply Gaussian process regression to interpolate the quantum-mechanical (QM) potential energy surface from a set of points in atomic configuration space. Our training data is based on QM information that is computed directly using density functional theory (DFT). To perform the fitting, we represent atomic environments using a set of rotationally, permutationally and reflection invariant parameters which act as the independent variables in our equations of non-parametric, non-linear regression.
\end{sloppypar} 

We develop a protocol for generating GAP models capable of describing lattice defects in metals by building a series of interatomic potentials for tungsten. We then demonstrate that a GAP potential based on a Smooth Overlap of Atomic Positions (SOAP) covariance function provides a description of the $\frac{1}{2} \langle 111 \rangle$ screw dislocation that is in agreement with the DFT model. We use this potential to simulate the mobility of $\frac{1}{2} \langle 111 \rangle$ screw dislocations by computing the Peierls barrier and model dislocation-vacancy interactions to QM accuracy in a system containing more than 100,000 atoms.}

\cleardoublepage

%
%

\phantomsection
\chapter*{Acknowledgements}
\label{chapter:acknowledgements}

I would like to thank my supervisor G\'{a}bor Cs\'{a}nyi for all his guidance and support, and for the influence his encouragement and enthusiasm for the subject had on my doctoral studies. I am also most grateful to Mike Payne for providing me with an opportunity to carry out my research in the Theory of Condensed Matter Group in the Cavendish Laboratory.

\begin{sloppypar} 
I am especially thankful to Albert Bart\'{o}k-P\'{a}rtay who was always happy to share his knowledge and ideas with me on a daily basis. I am indebted to all (both present and former) members of G\'{a}bor Cs\'{a}nyi's research group: Letif Mones, Noam Bernstein, Alan Nichol, Livia Bart\'{o}k-P\'{a}rtay, Thomas Stecher, Csilla V\'{a}rnai, Silvia Cereda, James Kermode, Sebastian John and Robert Baldock, and to all members of the Theory of Condensed Matter Group for creating a great environment to work in. I would also like to thank the staff of the Cavendish Laboratory and the Department of Engineering, and Michael Rutter for his advice on the computer matters.
\end{sloppypar} 

Finally, I would like to acknowledge the financial support I received from EPSRC during the course of my PhD and for several conferences and workshops.

\cleardoublepage

%
%

\tableofcontents

\listoffigures

\listoftables

\cleardoublepage

\pagenumbering{arabic}

%
%

\chapter{Introduction}
\label{chapter:introduction}

A detailed knowledge of material properties is crucial to understanding and exploiting their characteristics. As manufacturing and experimental methods shrink in the length scales they can access, this allows new materials with the desired properties to be investigated, driving the creation of new products, new industries, or even opening up of new areas of engineering and science. However, as experimental characterisation of many materials on the atomic scale is either impossible or impractical due to technological or economical barriers, atomistic computational modelling and simulation of materials has became an important method in the fields of physics, materials science and engineering.

The modern computational schemes based on quantum mechanics (\emph{ab initio} methods) are very effective in predicting the structure, properties and behaviour of a wide range of materials. However, in spite of very rapid advances in computational power over the last decades, these schemes still remain computationally very demanding, limiting the size of systems studied to just a few hundreds of atoms (more details in \cite{Goringe19971}). As a result, only properties that depend on short length and time scales can be well described. Although empirical methods can be used to model larger systems over longer periods of time, they often cannot predict the physical properties to the required level of accuracy. Consequently, the modelling of plasticity, fracture, crack propagation or any other phenomena involving long range interactions, requires computational schemes that are both accurate and efficient, and also scale favourably with the system size.

One possible way of creating a new empirical model, or improving an existing one, is through the addition of information based on the results of quantum-mechanical calculations in an indirect way, usually through parameterisation of the underlying Hamiltonian (tight binding methods) or parameterisation of the functional form of the atomic energy function (interatomic potentials). However, none of these methods are generally capable of providing a sufficient compromise between accuracy, efficiency and scalability, thus failing to explain many of the complicated phenomena arising in the field of condensed matter physics.

This thesis presents my research that explores an alternative way of improving current models, or creating new empirical models altogether, by augmenting them with information directly computed using the quantum-mechanical methods and applying probabilistic inference in order to correct the discrepancies between empirical and quantum-mechanical predictions.

The existing first principles methods, although restricted by the system size, are nevertheless capable of producing huge amounts of reliable and consistent data such as total energies, stresses and individual forces acting on the atoms. Unlike the experimental results, this data can be easily and cheaply obtained over a broad range of conditions such as different temperatures and pressures. Consequently, the problem of transferability can be directly addressed with a potential specifically tailored to the problem at hand. Furthermore, potentials for classes of materials so far unexplored by experimient can be generated.

At the same time using state of the art non-linear, non-parametric regression methods we can construct models that contain an arbitrary amount of information derived from the quantum-mechanical calculations. This allows then to predict physical properties of various materials to the required degree of accuracy while also providing fine and systematic control of the computational cost, which can be tuned at will for a given application.

The modus operandi is that a variety of configurations are sampled in small unit cells. This in turn enables simulations of large systems in which the individual atomic environments are nevertheless familiar. In this process we rely on the ability of the existing quantum-mechanical methods to reproduce the experimental results accurately. Consequently, we are aware that any discrepancies between quantum mechanics and experiment will be reflected in the resulting potential as our potential is ``trained'' from the quantum-mechanical data exclusively.

In this work we compute our reference training data using density functional theory which is the currently widely used band theory. The existing results of DFT calculations for solid-state systems have been found to be in excellent agreement with experimental data. However, as higher accuracy techniques are developed and computational power increases our potentials can be fit to more and more accurate calculations which should get closer and closer to experimental results.

\clearpage 

This thesis is organised as follows: in chapter \ref{chapter:classical_and_quantum_simulation_of_solids} I introduce the relevant theoretical background behind the methods available for computational modelling of solids, and in chapter \ref{chapter:simulation_techniques} I outline the simulation techniques that can be used with these methods to investigate the character of atomic interactions, enabling prediction of a wide range of both microscopic and macroscopic properties. In chapter \ref{chapter:gaussian_approximation_potential} I describe the methodology of Gaussian process regression, and how it can can be applied for the purpose of Gaussian Approximation Potential.

I present the outcome of my own work in chapters \ref{chapter:bulk_properties_and_lattice_defects_in_tungsten}, \ref{chapter:bispectrum-gap_potential_for_tungsten}, \ref{chapter:soap-gap_potential_for_tungsten} and \ref{chapter:bond-based_soap-gap_potential} --- the methods involved in quantum-mechanical and classical simulations of tungsten (the transition metal that was selected as a ``testing ground'' for our Gaussian Approximation Potential for metals), convergence testing and preliminary results of these simulations are given in chapter \ref{chapter:bulk_properties_and_lattice_defects_in_tungsten}. Finally, I present the results on generating Gaussian Approximation Potentials for tungsten using the bispectrum atomic descriptor in chapter \ref{chapter:bispectrum-gap_potential_for_tungsten}, and using the Smooth Overlap of Atomic Positions kernel in chapter \ref{chapter:soap-gap_potential_for_tungsten}. I conclude this thesis with a brief theoretical investigation of how some of the limitations of the above potentials can be overcome by applying the existing methodology of Smooth Overlap of Atomic Positions kernel to a new bond-based, rather than an atom-centred formulation of Gaussian Approximation Potential in chapter \ref{chapter:bond-based_soap-gap_potential}.

\cleardoublepage

\chapter[Classical and Quantum Simulation of Solids]{\texorpdfstring{Classical and Quantum \\ Simulation of Solids}{Classical and Quantum Simulation of Solids}}
\label{chapter:classical_and_quantum_simulation_of_solids}

\section{Introduction}
\label{chapter:classical_and_quantum_simulation_of_solids:section:introduction}

In this chapter I introduce the relevant theoretical background behind the computational modelling of materials, with a particular emphasis on solids. While most of these methods can be applied to other phases, the work contained in this thesis is focussed on crystalline solids. Hence, I will only attempt to outline the most relevant topics and reference the reader to the sources of further information when appropriate.

I start with a brief discussion of the most common methods for simulating material properties, using both quantum-mechanical and classical approaches. In this work we are purely concerned with the description of the physical phenomena on the atomic level. Thus although we employ quantum-mechanical methods in our calculations, properties of materials that are solely determined by the electron behaviour (such as electronic band structure) are beyond the scope of this work --- we are primarily interested in the total energy of the system and its derivatives (forces, and at times Hessian matrix as well).

In this chapter I review the application of different simulation methods across a range of length and time scales and their accuracy and associated computational cost in section \ref{chapter:classical_and_quantum_simulation_of_solids:section:computational_modelling_of_solids}. We follow by outlining the basic theory behind these methods in sections \ref{chapter:classical_and_quantum_simulation_of_solids:section:ab_initio_methods}, \ref{chapter:classical_and_quantum_simulation_of_solids:section:tight_binding} and \ref{chapter:classical_and_quantum_simulation_of_solids:section:interatomic_potentials}, where I discuss the quantum-mechanical \emph{ab initio} approach to electronic structure calculations, their semi-empirical approximations, as well as fully classical interatomic potential models.

\section{Computational Modelling of Solids}
\label{chapter:classical_and_quantum_simulation_of_solids:section:computational_modelling_of_solids}

The concept of computational modelling of materials, and in particular solids, spans a multitude of fields, ranging from quantum chemistry, across solid state physics, and extending to materials engineering (to name a few). Each approach has its advantages and disadvantages, making it appropriate for applications involving a range of time and length scales.

Starting from the smallest scale, the most accurate description of a small molecule (or indeed an isolated atom) comes from the laws of quantum mechanics, where properties are dictated by the electronic configuration. In its most basic form, molecular orbital (MO) theories use a linear combination of atomic orbitals to represent molecular orbitals of the entire molecule, where the motion of all electrons is correlated and electronic configurations correspond to a set of discrete energy levels.

In solids the electron states are very numerous, effectively blending into a continuous range of configurations, thus making the notion of individual electron configurations of lower relevance. The currently widely used band theory --- density functional theory (DFT) --- is discussed in more detail in the following section. However, at present it suffices to say that DFT can be used to compute structure and underlying quantum-mechanical properties of solids, especially when periodic boundary conditions dictate the character of the electronic structure (although DFT has also been successful in predicting molecular properties due to its relative accuracy and increased speed over other quantum chemical methods).

While all of the modern computational schemes based on quantum mechanics are very effective in predicting the structure, properties and behaviour of a wide range of materials, in spite of very rapid advances in computational power over the past few decades, the computational cost of these methods still remains very large. Therefore, the size of systems studied is limited to a few thousand atoms at best, and in many cases to equilibrium configurations (more details in \cite{Goringe19971}). Consequently, to describe larger systems one needs to resort to semi-empirical methods where the granularity (the extent to which a system is broken down into small parts) is dictated by the size of the system, as well as the time scale of investigated phenomena.

In practice we find that commonly used modelling techniques that are applied on the nanometre scale employ interatomic potentials (as outlined in section \ref{chapter:classical_and_quantum_simulation_of_solids:section:interatomic_potentials}), while modelling on the micrometer scale usually requires coarse-graining schemes that increase the granularity beyond the atomic level, as summarised in figure \ref{figure:simulation_scales} below.

\begin{figure}[H]
\begin{center}
\vspace{0.5cm}
\resizebox{12cm}{!}{\footnotesize{}\input{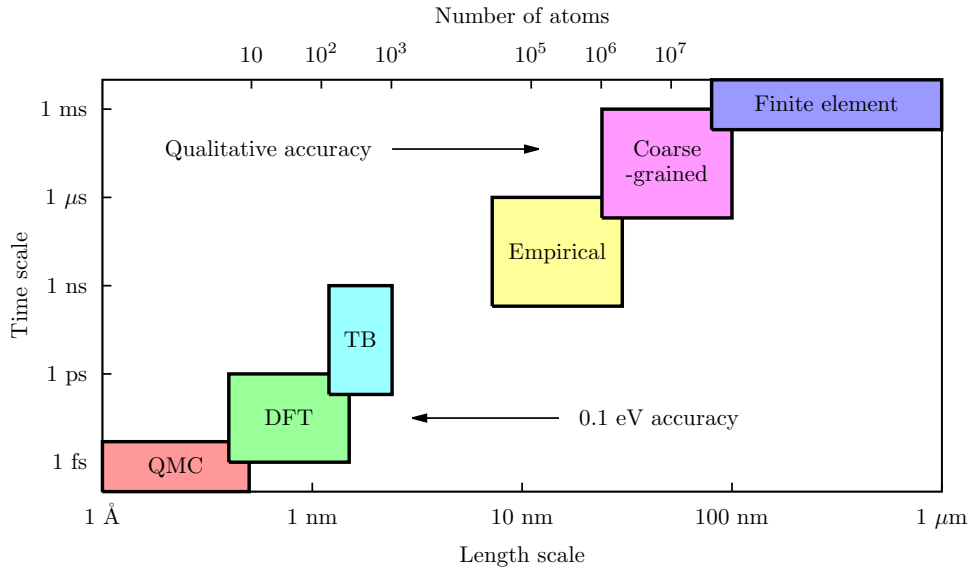}\normalsize{}}
\vspace{0.75cm}
\caption[Modelling techniques for a range of length and time scales.]{Representation of the range of length and time scales that are accessible for modelling techniques ranging from \emph{ab initio} methods, through empirical models, and extending to coarse-grained and finite-element schemes.}
\label{figure:simulation_scales}
\end{center}
\end{figure}

\noindent Although tight binding (TB) schemes (outlined briefly in section \ref{chapter:classical_and_quantum_simulation_of_solids:section:tight_binding}) attempt to fill the gap between explicitly quantum-mechanical and fully classical (empirical) methods, they are not completely successful in that task. While they often provide sufficient accuracy, they have an unfavourable scaling, and their computational cost remains prohibitive. Hence, there is clearly a need for methods that can approach accuracy of quantum-mechanical schemes, but with a more favourable scaling and smaller computational cost --- namely quantum-mechanical accuracy, but without explicit treatment of electrons.

\section{\emph{Ab Initio} Methods}
\label{chapter:classical_and_quantum_simulation_of_solids:section:ab_initio_methods}

\subsection{Born-Oppenheimer Approximation}

In order to calculate the electronic structure of materials the theory of quantum mechanics needs to be employed. However, in general, it is not possible to solve the many-body Schr\"{o}dinger equation directly (except for very simple, highly symmetrical systems), so a number of approximations need to be applied (more details can be found in \cite{RevModPhys.64.1045}).

Firstly, the Born-Oppenheimer approximation decouples the electronic structure from the nuclear motion --- electronic structure calculations are performed for fixed nuclear configurations:

\begin{equation}
\hat{H} \Psi = \left[ \sum_i^{\text{elec.}} - \frac{\hbar^2}{2 m_e} \boldsymbol{\nabla}_i^2 + \sum_{i, j}^{\substack{\text{elec.} \\ \text{nucl.}}} \frac{Z_j e^2}{4 \pi \epsilon_0 |\mathbf{r}_i - \mathbf{R}_j|} + \sum_{i < k}^{\text{elec.}} \frac{e^2}{4 \pi \epsilon_0 |\mathbf{r}_i - \mathbf{r}_k|} \right] \Psi.
\end{equation}

\noindent Secondly, even the electronic Schr\"{o}dinger equation in general cannot be solved exactly (again, except for very simple, highly symmetrical systems).

An in-depth description of quantum chemistry Hartree-Fock, post-Hartree-Fock or Quantum Monte Carlo methods is beyond the scope of this thesis. I will only mention here that Hartree-Fock methods usually rely on a basis set built from the linear combination of atomic orbitals (LCAO) ansatz. While the Hartree-Fock solution is usually a good starting point for an accurate description of small many-electron systems, its computational cost and nominal scaling of \(\mathcal{O}(N^4)\) is often prohibitive for condensed matter systems. On the other hand, even though Quantum Monte Carlo has been successfully applied for the calculation of bulk and surface energetics of small crystalline systems (more details can be found in \cite{PhysRevB.82.165431}), its computational cost limits this method to a few dozens of atoms at most.

Electron-electron interactions can also be approximated using an alternative approach, whereby electron exchange and correlation is modelled using general functionals of electron density (and its derivatives) --- density functional theory (DFT). DFT has been extremely popular in solid-state physics and in many cases the results of DFT calculations for solid-state systems have been found to be in excellent agreement with experimental data (in particular for condensed matter systems). Their computational cost is also significantly lower compared to Hartree-Fock based methods and their descendants that involve complex, many-electron wavefunctions (more details in \cite{RevModPhys.61.689}).

\subsection{Density Functional Theory}

Density functional theory relies on the fact that there is a one-to-one correspondence between the ground-state many-electron density and the external potential acting on it. The ground-state energy is then a functional of the ground-state density (which uniquely determines the ground-state properties of a many-electron system), and the external potential acting on the system (more details in \cite{PhysRev.136.B864}). Within this framework, the total energy $E$ of the system in an external potential $V_{\text{ext.}}(\mathbf{r})$ is given by the Hohenberg-Kohn functional:

\begin{equation}
E[n(\mathbf{r})] = F_{\text{H-K}}[n(\mathbf{r})] + \int V_{\text{ext.}}(\mathbf{r}) n(\mathbf{r}) \mathbf{d}^3\mathbf{r},
\end{equation}

\noindent where the ground-state density $n_{0}$ and ground-state energy $E_{0}$ can be obtained through variational minimisation.

The unknown functional $F_{\text{H-K}}$ can be rewritten in terms of the kinetic energy of non-interacting electrons, the Hartree electron-electron interaction energy and an unknown electron exchange and correlation term, resulting in the Kohn-Sham energy functional, which for a set of doubly occupied electronic states is given by:

\begin{align}
E &= 2 \sum_i^{\text{elec.}} \int \psi_i \left[ - \frac{\hbar^2}{2 m_e} \boldsymbol{\nabla}_i^2 \right] \psi_i \mathbf{d}^3\mathbf{r} + \frac{1}{2} \frac{e^2}{4 \pi \epsilon_0} \int \frac{n(\mathbf{r}) n(\mathbf{r'})}{|\mathbf{r} - \mathbf{r'}|} \mathbf{d}^3\mathbf{r} \mathbf{d}^3\mathbf{r'} \nonumber \\
&+ E_{\text{XC}}[n(\mathbf{r})] + \int V_{\text{ext.}}(\mathbf{r}) n(\mathbf{r}) \mathbf{d}^3\mathbf{r},
\end{align}

\noindent and where the electronic density is given by:

\begin{equation}
n(\mathbf{r}) = \sum_i^N |\psi_i(\mathbf{r})|^2.
\end{equation}

\noindent This, together with Kohn-Sham equations (more details in \cite{PhysRev.140.A1133}), provides a way to systematically map the problem of a strongly interacting electron gas onto a system of non-interacting electrons moving in an effective potential due to all the other electrons:

\begin{equation}
\left[ - \frac{\hbar^2}{2 m_e} \boldsymbol{\nabla}_i^2 + \frac{e^2}{4 \pi \epsilon_0} \int \frac{n(\mathbf{r'})}{|\mathbf{r} - \mathbf{r'}|} \mathbf{d}^3\mathbf{r'} + V_{\text{XC}}(\mathbf{r}) + V_{\text{ext.}}(\mathbf{r}) \right] \psi_i = \epsilon_i \psi_i,
\label{equation:kohn-sham}
\end{equation}

\noindent where the exchange-correlation potential is given by:

\begin{equation}
V_{\text{XC}}(\mathbf{r}) = \frac{\partial E_{\text{XC}}[n(\mathbf{r})]}{\partial n(\mathbf{r})}.
\end{equation}

\noindent The Kohn-Sham equations need to be solved self-consistently --- once the exchange-correlation energy and the solutions to the set of eigenequations are known, the occupied electronic states need to generate the charge density that corresponds to the electronic potential that was used to construct the original equations.

\begin{sloppypar} 
Within the density functional theory framework the exact value of the exchange-correlation energy remains unknown and it needs to be approximated. The local-density approximation (LDA; more details in \cite{PhysRev.140.A1133}) assumes that the exchange-correlation energy per electron at any given point is equal to that of a homogeneous electron gas of the same density and that the exchange-correlation energy functional is purely local. A more accurate approximation including the next term in a derivative expansion of the charge density is provided by the generalised gradient approximations (GGAs; more details in \cite{PhysRevB.45.13244}, \cite{PhysRevLett.77.3865} and \cite{PhysRevB.59.7413}) --- this approach is significantly more accurate in many systems.
\end{sloppypar} 

Pseudopotential theory replaces the strong electron-ion potential with a much weaker interaction between pseudo-valence electrons and pseudo-ion cores, which encompasses the features of the valence electron moving through the solid (more details in \cite{PhysRev.112.685}). This allows the wave function to be expanded in a relatively small set of plane waves thus making the solution of Schr\"{o}dinger equation more tractable computationally.

Finally, aperiodic geometries can be approximated using supercells (more details in \cite{ashcroft1976solid}) and iterative minimisation techniques can be used to minimise the total energy functional.

While density functional theory has established itself as a means of performing quantum-mechanical calculations in many fields of physics and chemistry these days, the scope of such calculations is limited by the scaling of the computational cost which increases asymptotically as the cube of the system size (\(\mathcal{O}(N^3)\)) for plane-wave methods that are applicable to metals (while there exist DFT schemes approaching \(\mathcal{O}(N)\) scaling for insulators, these are usually not suitable for simulation of metallic systems).

\section{Tight Binding}
\label{chapter:classical_and_quantum_simulation_of_solids:section:tight_binding}

\subsection{Empirical Tight Binding}

While the need for an accurate description of physical phenomena usually implies that a quantum-mechanical model of the system is necessary, computationally efficient alternatives that can handle larger systems are equally important. The tight binding method is one possible compromise --- this approach calculates the electronic band structure using an approximate set of wave functions for isolated atoms based on each lattice site and replaces the Hamiltonian operator with a parameterised Hamiltonian matrix (more details in \cite{0034-4885-60-12-001}):

\begin{equation}
H_{i \alpha j \beta} = \sum_{\mathbf{R}_j} \exp[i \mathbf{k} \cdot (\mathbf{R}_j - \mathbf{R}_i)] \times \int \psi_{i \alpha}^\ast(\mathbf{r} - \mathbf{R}_i) H \psi_{j \beta}(\mathbf{r} - \mathbf{R}_j) \mathbf{dr}.
\end{equation}

\noindent The key idea underlying all tight binding calculations, introduced originally by Slater and Koster (more details in \cite{PhysRev.94.1498}), is to replace the integral in the above equation with a parameter depending on the internuclear distance alone. In the original formulation, when the basis functions consist of Bloch sums formed from L\"{o}dwin functions:

\begin{equation}
\psi_{i \alpha} = \sum_{i' \alpha'} S_{i \alpha i' \alpha'}^{- \frac{1}{2}} \phi_{i' \alpha'},
\end{equation}

\noindent where \(\phi_{i \alpha}\) are the original atomic orbitals, and \(S_{i \alpha i' \alpha'}\) are the overlap matrix elements, it can be shown that the Hamiltonian matrix elements can be written as:

\begin{equation}
H_{i \alpha j \beta} = \sum_{\mathbf{R}_j J} \exp[i \mathbf{k} \cdot (\mathbf{R}_j - \mathbf{R}_i)] \times h_{\alpha \beta J}(|\mathbf{R}_j - \mathbf{R}_i|) G_{\alpha \beta J}(k, l, m),
\end{equation}

\noindent where \(J\) represents the angular momentum of the bond, \(h_{\alpha \beta J}\) is the constant for a given \(|\mathbf{R}_j - \mathbf{R}_i|\) and \(G_{\alpha \beta J}\) is the angular dependence (as given in \cite{PhysRev.94.1498}).

Since the basis functions do not need to be evaluated in the tight binding approach, the only information required to compute the electronic structure of the system are the Hamiltonian matrix elements, which are written in the parameterised form. Hence, the system is described by the parameterisation scheme alone, and the quality of the tight binding calculation is only as good as the parameters used --- band structures of different polymorphs are frequently used as part of the data set to be fit during construction of tight binding models (more details in \cite{0034-4885-60-12-001}).

\begin{sloppypar} 
The tight binding approximation provides a methodology in which the quantum mechanics of the system is directly included, but which also (in spite of its \(\mathcal{O}(N^3)\) scaling due to matrix diagonalisation) is computationally far less demanding than \emph{ab initio} methods. It has been extended more recently, improving the accuracy and transferability of the tight binding method (tight binding like \emph{ab initio} methods --- more details in \cite{PhysRevB.40.3979}, \cite{PhysRevB.51.12947} and in section \ref{chapter:classical_and_quantum_simulation_of_solids:section:tight_binding:density_functional_tight_binding}), or alternatively improving the scalability and thus increasing the size of the systems that can be investigated (through linear-scaling, low-order approximations to tight binding, such as BOP potential --- more details in \cite{0965-0393-5-3-002} and in section \ref{chapter:classical_and_quantum_simulation_of_solids:section:interatomic_potentials:bond_order_potential}).\footnote{While this appears to be an appealing way of closing the gap between \emph{ab initio} and empirical methods, these developments have been hindered by the extreme complexity of the functional forms that result (more details in \cite{0034-4885-60-12-001}).}
\end{sloppypar} 

\subsection{Density Functional Tight Binding}
\label{chapter:classical_and_quantum_simulation_of_solids:section:tight_binding:density_functional_tight_binding}

Density functional tight binding (DFTB) tries to avoid the difficulties of empirical tight binding, where the procedure of how to determine the desired matrix elements is arbitrary. Instead, within the DFTB formalism the elements of the Hamiltonian and overlap matrices are calculated with the help of density functional theory using integral approximations. 

In a similar manner to the Kohn-Sham equations of density functional theory (equation \ref{equation:kohn-sham}), a basis set $\{\psi_i\}$ of pseudoatomic wave functions can be used to solve a modified Kohn-Sham equation that consists of a kinetic energy term, Hartree term, exchange correlation potential, nuclear potential, as well as an additional term $(r/r_{0})^{N}$. This term is introduced to concentrate charge density closer to the nucleus and improve band-structure calculations within the LCAO formalism because the wave function is forced to avoid areas away from the nucleus:

\begin{equation}
\left[ - \frac{\hbar^2}{2 m_e} \boldsymbol{\nabla}_i^2 + \frac{e^2}{4 \pi \epsilon_0} \int \frac{n(\mathbf{r'})}{|\mathbf{r} - \mathbf{r'}|} \mathbf{d}^3\mathbf{r'} + V_{\text{XC}}(\mathbf{r}) + V_{\text{nuc.}}(\mathbf{r}) + \left(\frac{r}{r_{0}}\right)^{N} \right] \psi_i = \epsilon_i \psi_i.
\end{equation}

\noindent Finally, the solutions for $\{\psi_i\}$ are used to tabulate the Hamiltonian matrix elements. The model also needs to be supplemented by a completely empirical pair repulsion, which is calculated from the energy difference between the DFTB band energy and that of self-consistent solution to the modified Kohn-Sham equation.

DFTB retains many aspects of the traditional tight binding formalism, and for that reason it can be seen as an approximate LCAO-DFT scheme. It yields exactly the same energy expression as common non-orthogonal tight binding schemes but with a well-defined procedure for determining the desired matrix elements (more details in \cite{PhysRevB.51.12947}).

\section{Interatomic Potentials}
\label{chapter:classical_and_quantum_simulation_of_solids:section:interatomic_potentials}

\subsection{Linear Scaling}

In quantum-mechanical methods the potential energy of the system is a complicated many-body function, incorporating the explicit treatment of electron interactions and encompassing the non-local character of quantum mechanics. On the other hand, in classical simulations relying on interatomic potentials, the atoms are treated as elementary particles (usually represented as point-like masses) and they interact through a many-body interaction potential, which can be approximated as:

\begin{equation}
E = \sum_{i = 1}^N \epsilon(\mathbf{x}^{(1)} - \mathbf{x}^{(i)}, ..., \mathbf{x}^{(N)} - \mathbf{x}^{(i)}) = \sum_{i = 1}^N \epsilon(\{\mathbf{x}^{(j)} - \mathbf{x}^{(i)}\}_{j = 1}^N) = \sum_{i = 1}^N \epsilon_i .
\end{equation}

\noindent The sum is over all \(N\) atoms in the system and the atomic energy function \(\epsilon\) represents local energies of atoms (and therefore embodies the local character of the classical approximation). Atomic forces on the atoms are simply computed by differentiating the potential energy function \(E\):

\begin{equation}
\mathbf{f}^{(i)} = - \boldsymbol{\nabla}^{(i)} E(\{\mathbf{x}^{(1)}, ..., \mathbf{x}^{(N)}\}) = - \boldsymbol{\nabla}^{(i)} E(\{\mathbf{x}^{(j)}\}_{j = 1}^N).
\end{equation}

Another approximation, which is present in essentially all interatomic potentials (if electrostatics effects are removed), is the limited range of the atomic energy function \(\epsilon\):

\begin{equation}
\lim_{|\mathbf{x}^{(i)} - \mathbf{x}^{(j)}| \to \infty} \frac{\partial \epsilon_i}{\partial \mathbf{x}^{(j)}} = 0,
\end{equation}

\noindent which is implemented using a finite range cutoff.

These two approximations, namely the decomposability of the potential energy into the sum of atomic energy functions and the limited range of the atomic energy functions, result in a small computational cost (since the number of required computations scales linearly with the number of atoms in a system) and the ease of parallelisation of the interatomic potentials.

While there have been many developments in the area of interatomic potentials over the last few decades involving the formulation of a wide variety of many-body potentials, the most relevant in the study of bulk, surface and cluster properties of metallic compounds appears to be a group of potentials based on the second moment approximation to the tight binding method. These include the embedded atom potential (EAM; originally formulated in \cite{PhysRevB.29.6443}), and Finnis-Sinclair potential (FS; originally formulated in \cite{doi:10.1080/01418618408244210}), amongst others. Although the embedded atom potential and Finnis-Sinclair potential have different functional forms, the underlying formulation is similar, representing the total energy of the system as a sum of pairwise interactions and an \(n\)-body term.

At the same time, over the last decade significant developments have been made in using machine learning techniques to determine potential energy surfaces of various systems using artificial neural networks, or more precisely, multilayer perceptrons (MLP; more details in \cite{doi:10.1021/jp9105585}). This, together with the work drawing connections between the energy of the system expressed as a functional of the atomic density distribution function and the corresponding interaction potentials (more details in \cite{PhysRevB.80.024104}), points towards new directions for further development of the fundamental framework of interaction potentials for materials modelling.

Consequently, the problem of approximating quantum mechanics using interatomic potentials can be related to the problem of fitting an atomic energy function to the quantifiable properties of the real material. These are governed by the equations of quantum mechanics, and are calculated using one of the quantum-mechanical methods. Although the quantum-mechanical potential energy, in general, cannot be decomposed into separate atomic energy functions, one relies on the fact that even quantum-mechanical properties are usually local, which can be verified by investigation of the decay of Hessian matrix elements.

\subsection{Atomic Environments}
\label{chapter:classical_and_quantum_simulation_of_solids:section:interatomic_potentials:subsection:atomic_environments}

In the fitting problem, where one needs to create a mapping from the atomic environment of the system to the total energy of the system, it is crucial that one can describe the atomic environment in a quantitative way. Although an array of three-dimensional vectors provides a simple and complete representation of an atomic system, distinguishing between structures using this as a sole input is difficult. This is because the description of the system is affected by re-ordering of the vectors, as well as simple symmetry transformations such as rotations, translations, reflections or inversions. Consequently, identical, or very similar (related by these simple symmetry transformations) structures can have drastically different representations, even though they often correspond to an exactly the same value of the energy.

There are multiple methods of constructing atomic representation invariants. In most existing interatomic potentials bond lengths and bond angles are commonly used as function arguments and they are rotationally invariant by definition. However, they are not, in general, invariant with respect to permutation of the neighbouring atoms, and furthermore, if the accuracy of such representation is to be systematically improved (by including higher order, many-body terms), the size of a complete set of such parameters grows exponentially (as \(\mathcal{O}(N^m)\), with $m$ being the highest order term, and $N$ the number of neighbours). There also exist no systematic way of reducing it.

Hence, an atomic representation that is useful in condensed matter physics and computational chemistry should remain invariant under rotations, translations and permutations of the identical atoms at the same time, and also remain accurate. The most well known and universally established set of such invariants are the bond-order parameters (originally introduced in \cite{PhysRevB.28.784}) which have been used extensively to analyse the atomic structure of solids in the field of computational chemistry (more details in \cite{duijneveldt:4655}, \cite{cape:2366}, \cite{nose:1803}). Although the bond-order parameters do not provide a complete representation of the system (the mapping between bond-order parameters and the atomic structures is not one-to-one), they have proved successful in numerous studies of nucleation and phase transitions.

However, the set of bond-order parameters, in fact, forms a subset of a more general set of invariants called the bispectrum --- an infinite array of rotational and permutational invariants which provides an almost complete representation of atomic configurations. The bispectrum parameters, originally introduced by the signal processing community, have been recently adapted for the purpose of representing crystal structures (more details in \cite{2010PhDT}, \cite{PhysRevLett.104.136403}) and they provide a systematic way of obtaining atomic environment representations, with a sensitivity that can be systematically tuned at will. We will explore the theoretical background behind this approach, and show how the bispectrum can be used within the Gaussian Approximation Potential formalism in section \ref{chapter:gaussian_approximation_potential:section:description_of_atomic_environments}.

\subsection{Lennard-Jones Potential}

One of the simplest interatomic potentials still in use today was originally proposed by John Lennard-Jones in 1924 (more details in \cite{1924}). It has the simple mathematical form of a negative order polynomial which approximates the interaction between a pair of neutral noble gas atoms. The most common form of the Lennard-Jones potential is given by:

\begin{equation}
\epsilon(\{\mathbf{x}^{(j)} - \mathbf{x}^{(i)}\}_{j = 1}^N) = \sum_{j = 1}^N \left( \frac{A}{r_{ij}^{12}} - \frac{B}{r_{ij}^{6}} \right),
\end{equation}

\noindent where the $r^{-12}$ term approximates Pauli repulsion at short range due to overlapping electron orbitals, and the $r^{-6}$ term approximates the long-range attraction due to the van der Waals force. While the attractive term has a clear physical justification (it is van der Waals force between two spheres of constant radii), the repulsion term has been selected primarily due to its computational efficiency (it can be written as square of the attractive term) and the fact that it is a reasonable approximation for Pauli repulsion.

Although we are not going to use the Lennard-Jones potential in this work, it serves as an excellent demonstration of the concept of a pair-potential --- a potential where the atomic energy can be decomposed into a sum of energies associated with bonds, $V_2$:

\begin{equation}
\epsilon_{i} = \sum_{j = 1}^N \frac{1}{2} V_{2}(r_{ij}),
\end{equation}

\noindent and where the total energy of the system is given by:

\begin{align}
E = \sum_{i = 1}^N \epsilon_{i} &= \sum_{i = 1}^N \sum_{j = 1}^N \frac{1}{2} V_{2}(r_{ij}) \nonumber \\
&= \sum_{i}^N \sum_{\substack{j \\ j<i}}^N V_{2}(r_{ij}),
\end{align}

\noindent where the conditional sum avoids double-counting of bonds.

This leads us to a more general concept, that any many-body interatomic potential can be decomposed into a sum of one-body, two-body, three-body, etc. contributions (the one-body term $V_{1}$ describes an external force applied to the system, so is usually either assumed to be a constant, or is absent altogether):

\begin{align}
E = \sum_{i}^N V_{1}(\mathbf{x}^{(i)}) &+ \underbrace{\sum_{i}^N \sum_{\substack{j \\ j<i}}^N V_{2}(\mathbf{x}^{(i)}, \mathbf{x}^{(j)})}_\text{bonds} \nonumber \\
&+ \underbrace{\sum_{i}^N \sum_{\substack{j \\ j<i}}^N \sum_{\substack{k \\ k<j}}^N V_{3}(\mathbf{x}^{(i)}, \mathbf{x}^{(j)}, \mathbf{x}^{(k)})}_\text{angles} \nonumber \\
&+ \dots
\end{align}

\noindent For this expression to be useful, we need fast convergence of the total energy $E$ and the decrease in the value of functions $V_n$ as $n$ increases. Unfortunately, this is not always the case, and even three-body interatomic potentials are usually not sufficient to describe simple atomic systems to the required degree of accuracy. Consequently, to achieve a sufficient level of precision, resembling that of explicitly quantum-mechanical methods, a truly many-body approach is usually necessary.

\subsection{Finnis-Sinclair Potential}

A large variety of empirical potentials have been introduced since the 1980s, but among these schemes the one introduced by Finnis and Sinclair (more details in \cite{doi:10.1080/01418618408244210}) has been particularly successful in the description of body-centred cubic metals. It is based on the second moment approximation to tight binding theory --- it incorporates the band character of metallic cohesion and it has been extensively used to model lattice point defects and grain boundaries (more details in \cite{doi:10.1080/01418618608245292}, \cite{Yan19964351}, \cite{Landa19992477}, \cite{PhysRevB.58.14020}).

Although simple pair potentials have their merits, one of the drawbacks is that they cannot account for the Cauchy discrepancy. Unless the elastic constants of a cubic crystal satisfy $C_{12} = C_{44}$ (which is usually not the case for metallic systems), they cannot be reproduced by a pair potential. The solution proposed by Finnis and Sinclair is to include in the potential a term which provides the simplest expression of band character, namely the second moment approximation to the tight binding model, so the cohesive energy per atom varies with a square root of atomic coordination. This is achieved by adding an $n$-body term to the total energy of the system:

\begin{equation}
E = \sum_{i}^N \sum_{\substack{j \\ j<i}}^N V_{2}(r_{ij}) - A \sum_{i}^N \sqrt{\rho_{i}},
\end{equation}

\noindent where:

\begin{equation}
\rho_{i} = \sum_{\substack{j \\ j \neq i}} \phi(r_{ij}),
\end{equation}

\noindent and $\phi(r_{ij})$ can be interpreted as a sum of squares of overlap integrals.

The functional forms $\phi$ and $V_{2}$ are fitted to experimental data using a small number of adjustable parameters:

\begin{equation}
\phi(r_{ij}) = \left\{ \begin{array}{l l} (r_{ij} - d)^2 & r_{ij} \leq d \\ 0 & r_{ij} > d \end{array} \right. ,
\end{equation}

\noindent and:

\begin{equation}
V_{2}(r_{ij}) = \left\{ \begin{array}{l l} (r_{ij} - c)^2 (c_0 + c_1 r_{ij} + c_2 r_{ij}^2) & r_{ij} \leq c \\ 0 & r_{ij} > c \end{array} \right. ,
\end{equation}

\noindent where the range cutoff parameters $c$ and $d$ usually are assumed to lie between second and third nearest neighbour.

\subsection{Embedded Atom Model}

The embedded atom model (EAM), as originally formulated by Daw and Baskes (more details in \cite{PhysRevB.29.6443}), shares many ideas with the Finnis-Sinclair potential. The two were developed independently, but they share the common idea that the strength of a chemical bond depends on the bonding environment. The EAM potential is based on the concept that the energy required to place a small impurity atom in a lattice is solely a function of electron density at that particular site. Consequently, each atomic species has a unique energy function that depends on electron density alone (more details in \cite{PhysRevB.37.3924}).

The basic equations of the embedded atom model are:

\begin{equation}
E = \sum_{i}^N \sum_{\substack{j \\ j<i}}^N V_{2, \alpha \beta}(r_{ij}) + \sum_{i}^N F_{\alpha}(\rho_{i}),
\end{equation}

\noindent and:

\begin{equation}
\rho_{i} = \sum_{\substack{j \\ j \neq i}} f_{\beta}(r_{ij}),
\end{equation}

\noindent where the model presented here takes into account multiple species, which are designated by $\alpha$ for atom $i$, and $\beta$ for atom $j$. Consequently, $V_{2, \alpha \beta}$ is simply a pair-wise potential function for species $\alpha$ and $\beta$, $F_{\alpha}$ is an embedding function that represents the energy required to place atom of type $\alpha$ in the electron cloud and $f_{\beta}$ is the contribution to the electron charge density from an atom of type $\beta$. To use the embedded atom model these three functions must be specified for each atomic species combination, for example giving three functions for a monoatomic metal, seven functions for a binary alloy, etc. They are usually given in a tabularised form obtained through cubic spline interpolation (example in \cite{0965-0393-12-4-007}).

It is easy to see that although physical interpretation is quite different, the equations for EAM and Finnis-Sinclair potentials for a monoatomic metal are identical when the embedding function is proportional to a square-root of electron density, which in turn is taken to correspond to linear superposition of squares of overlap integrals. This concept has been taken further by Brenner, who also demonstrated (more details in \cite{PhysRevLett.63.1022}) that Tersoff and Brenner potentials for covalently bonded solids (more details in \cite{PhysRevB.37.6991}, \cite{PhysRevB.42.9458}) can be expressed using similar equations as EAM and Finnis-Sinclair potentials although with slightly different functional forms.

\subsection{Bond Order Potential}
\label{chapter:classical_and_quantum_simulation_of_solids:section:interatomic_potentials:bond_order_potential}

Analytic bond order potentials (BOP), formulated by Pettifor and Oleinik (more details in \cite{PhysRevB.59.8487}, \cite{PhysRevLett.84.4124}, \cite{PhysRevB.65.172103}, \cite{Pettifor20042}) are a further extension of the bond order ideas used in the Finnis-Sinclair, EAM, Tersoff and Brenner potentials, where the analytic form of the $\sigma$ and $\pi$ bond orders are derived as an approximation to the exact many-atom expansion of bond energy within the two-centre, orthogonal tight binding representation of the electronic structure.

Without going into too much mathematical detail (which is beyond the scope of this chapter), the total energy of the system can be expressed as:

\begin{equation}
E = \sum_{i}^N \sum_{\substack{j \\ j<i}}^N V_{2, \alpha \beta}(r_{ij}) + E_{prom} + E_{bond},
\label{equation:bop}
\end{equation}

\noindent where the first term contains the overlap repulsion interaction between atom $\alpha$ at site $i$ and atom $\beta$ at site $j$. The second term represents the promotion energy of bringing the $sp$-valent atoms together from infinity where the $(E_{p}^{\alpha} - E_{s}^{\alpha})$ term is the splitting of valence $s$ and $p$ energy levels:

\begin{equation}
E_{prom} = \sum_{i}^N (E_{p}^{\alpha} - E_{s}^{\alpha})(\Delta N_p)_{i}^{\alpha},
\end{equation}

\noindent and $(\Delta N_p)_{i}^{\alpha}$ represents the change in the number of p electrons of atom $\alpha$ at site $i$. Finally, the bonding energy term is given by:

\begin{equation}
E_{bond} = \sum_{i}^N \sum_{\substack{j \\ j<i}}^N 2 \sum_{L, L'} H_{iL, jL'}^{\alpha \beta} \Theta_{jL', iL}^{\beta \alpha},
\end{equation}

\noindent where $H_{iL, jL'}^{\alpha \beta}$ and $\Theta_{jL', iL}^{\beta \alpha}$ represent the Hamiltonian and the bond-order matrix elements on sites $i$ and $j$, respectively, $L = (l, m)$ and $L' = (l', m')$ are the appropriate orbital and magnetic quantum numbers. The factor of $2$ accounts for the spin degeneracy.

Analytic BOP potentials represent the best potentials to date in terms of accuracy, as far as conventional interatomic potentials are concerned. They are successful in representing the different properties of the $\sigma$ and $\pi$ bonds correctly, and provide an efficient $\mathcal{O}(N)$ method for performing large scale simulations, although computationally they are significantly more complex and expensive than FS or EAM methods.

The earlier, non-analytic formulation of the BOP potentials suffer from the fact that the Hellmann-Feynman forces only become exact as the bond orders converge to the exact values, which usually cannot be achieved at reasonable computational expense. Consequently, application of these potentials to large scale molecular dynamics simulations of transition metals has been limited (more details in \cite{PhysRevB.53.1656}, \cite{PhysRevB.53.12694}, \cite{PhysRevB.69.094115} and \cite{PhysRevB.75.104119}). Although one could obtain the forces through direct differentiation of equation \ref{equation:bop}, evaluating derivatives of the bond-order matrix elements is extremely difficult in practice, and consequently the computed forces are usually not consistent with the total energy of the system. This situation has been remedied recently with the development of the valence-dependent analytic BOP potential for transition metals by Drautz and Pettifor (more details in \cite{PhysRevB.74.174117}), where the true forces can be computed analytically.

\cleardoublepage

\chapter{Simulation Techniques}
\label{chapter:simulation_techniques}

\section{Introduction}

In the last chapter I outlined the basic classical and quantum-mechanical approaches to computational simulation of solids. While the calculation of the total energy of the system and its derivatives (forces) is critical in describing the instantaneous state of the system, once combined with a number of well established techniques one can use them to investigate how the atoms interact over time. This in turn enables prediction of a wide range of both microscopic and macroscopic properties.

I start this chapter with a review of the molecular dynamics techniques as applied to the most commonly used thermodynamic ensembles (i.e. microcanonical and canonical ensembles) in section \ref{chapter:simulation_techniques:section:molecular_dynamics}. I follow with section \ref{chapter:simulation_techniques:section:geometry_optimisation} where I describe the commonly used minimisation techniques, as used in geometry optimisation problems. These concepts are taken further in section \ref{chapter:simulation_techniques:section:transition_state_search}, where I briefly outline the most commonly used methods of transition state searching. Finally, I finish this chapter with section \ref{chapter:simulation_techniques:section:monte_carlo_methods}, which outlines the Monte Carlo approach to the problem of predicting material properties.

\section{Molecular Dynamics}
\label{chapter:simulation_techniques:section:molecular_dynamics}

\subsection{Microcanonical Ensemble}

Developed in the 1950s and 1960s, molecular dynamics (MD) is a method of numerically solving Newton's equations of motion for a system of interacting particles. It employs numerical techniques to perform computer ``experiments'' that allows one to evaluate the dynamics of the system and therefore compute structural and thermodynamic properties of complex systems that would otherwise be impossible to study analytically. 

In its simplest form, a molecular dynamics simulation of an isolated system of $N$ particles, with masses $\{m_i\}_{i=1}^N$, where the volume and total energy of the system are conserved (i.e. the microcanonical ensemble, with $N$, $V$ and $E_{total}$ all fixed) is carried by integrating Newton's equation of motion:

\begin{equation}
\mathbf{f}_{i} = - \boldsymbol{\nabla}_{i} U = m_{i} \mathbf{v}_i
\end{equation}

\noindent and

\begin{equation}
\mathbf{v}_{i} = \frac{\partial \mathbf{x}_i}{\partial t},
\end{equation}

\noindent where $U$ is the potential energy of the system, $T = \sum_{i=1}^N \frac{1}{2} m_{i} v_{i}^2$ is the kinetic energy and the total energy of the system is given by $E_{total} = U + T$.

Since molecular dynamics relies on Newtonian equations of motion alone, it is worth noting here that it is completely independent of how the potential energy $U$ is computed. As long as one can evaluate its derivatives in order to obtain atomic forces this method can be used with interatomic potentials or with more complex quantum-mechanical \emph{ab initio} methods using exactly the same principles.

We begin the integration of Newton's equations of motion by computing the Taylor expansion of the position vector $\mathbf{x}_{i}$:

\begin{equation}
\mathbf{x}_{i}(t + \Delta t) = \mathbf{x}_{i}(t) + \left. \frac{\partial \mathbf{x}_{i}}{\partial t} \right|_t \Delta t + \left. \frac{1}{2!} \frac{\partial^2 \mathbf{x}_{i}}{\partial t^2} \right|_t \Delta t^2 + \left. \frac{1}{3!} \frac{\partial^3 \mathbf{x}_{i}}{\partial t^3} \right|_t \Delta t^3 + \mathcal{O}(\Delta t^4).
\label{equation:taylor-plus}
\end{equation}

\noindent The Euler method is the simplest, first-order method for integrating an ordinary differential equation. It is implemented through:

\begin{align}
&\mathbf{x}_{i}(t + \Delta t) = \mathbf{x}_{i}(t) + \mathbf{v}_{i}(t) \Delta t + \mathcal{O}(\Delta t^2) \nonumber \\
&\mathbf{v}_{i}(t + \Delta t) = \mathbf{v}_{i}(t) + \frac{\mathbf{f}_{i}(t)}{m} \Delta t + \mathcal{O}(\Delta t^2).
\end{align}

\noindent The Euler method often serves as a basis for more complicated methods, but it is very rarely used in practice as it suffers from numerical stability problems due to its low accuracy. The local error (error per step) is proportional to the square of the step size, $\mathcal{O}(\Delta t^2)$, and the global error (error at any given time) is proportional to the step size,  $\mathcal{O}(\Delta t)$.

The Euler method relies on a forward difference approximation to the first derivative. A much more accurate method which is no more computationally intensive relies on a central difference approximation to the second derivative. This is usually referred to as the St\"{o}rmer method, or more recently as the Verlet method since being rediscovered by Verlet in 1967 (more details in \cite{PhysRev.159.98}, \cite{ANU:165633}). One proceeds again by computing the Taylor expansion of the position vector $\mathbf{x}_{i}$:

\begin{equation}
\mathbf{x}_{i}(t - \Delta t) = \mathbf{x}_{i}(t) - \left. \frac{\partial \mathbf{x}_{i}}{\partial t} \right|_t \Delta t + \left. \frac{1}{2!} \frac{\partial^2 \mathbf{x}_{i}}{\partial t^2} \right|_t \Delta t^2 - \left. \frac{1}{3!} \frac{\partial^3 \mathbf{x}_{i}}{\partial t^3} \right|_t \Delta t^3 + \mathcal{O}(\Delta t^4),
\label{equation:taylor-minus}
\end{equation}

\noindent and by adding and subtracting equations \ref{equation:taylor-plus} and \ref{equation:taylor-minus}, we obtain:

\begin{align}
&\mathbf{x}_{i}(t + \Delta t) = 2 \mathbf{x}_{i}(t) - \mathbf{x}_{i}(t - \Delta t) + \frac{\mathbf{f}_{i}(t)}{m} \Delta t^2 + \mathcal{O}(\Delta t^4) \nonumber \\
&\mathbf{v}_{i}(t) = \frac{\mathbf{x}_{i}(t + \Delta t) - \mathbf{x}_{i}(t - \Delta t)}{2 \Delta t} + \mathcal{O}(\Delta t^2).
\end{align}

\noindent Due to a cancellation of errors, the Verlet integration is significantly more accurate than the Euler method, with no need to evaluate third-order derivatives. The local error in the position in the Verlet method is of the order $\mathcal{O}(\Delta t^4)$, while the global error is of the order $\mathcal{O}(\Delta t^2)$, which can be demonstrated by showing that:

\begin{equation}
\text{error}(\mathbf{x}_{i}(t + n \Delta t)) = \frac{n (n + 1)}{2} \mathcal{O}(\Delta t^4).
\end{equation}

\noindent While the Verlet integration gives good numerical stability and has time-reversal symmetry its disadvantages include its treatment of velocities, which always lag behind the positions. They are recalculated at each time step from atomic positions using a central difference approximation to the first derivative.

A related method to the Verlet scheme, called the Velocity-Verlet algorithm (more details in \cite{swope:637}), is more appropriate when a more accurate treatment of velocities is necessary (for example when calculating kinetic energies):

\begin{align}
&\mathbf{x}_{i}(t + \Delta t) = \mathbf{x}_{i}(t) + \mathbf{v}_{i}(t) \Delta t + \frac{1}{2} \frac{\mathbf{f}_{i}(t)}{m} \Delta t^2 + \mathcal{O}(\Delta t^4) \nonumber \\
&\mathbf{v}_{i}(t + \Delta t) = \mathbf{v}_{i}(t) + \frac{1}{2} \frac{\mathbf{f}_{i}(t) + \mathbf{f}_{i}(t + \Delta t)}{m} \Delta t + \mathcal{O}(\Delta t^4).
\end{align}

\noindent Velocity-Verlet integration is again more accurate than the Euler method, and as in the case of the standard Verlet method error, it has a local error in position of the order $\mathcal{O}(\Delta t^4)$ while the global error is of the order $\mathcal{O}(\Delta t^2)$.

\subsection{Canonical Ensemble}

In the canonical ensemble the number of particles $N$, volume $V$ and temperature $T$ of the system are conserved, and consequently the total energy of the system is allowed to change. This is often necessary as in a molecular dynamics simulation of a relatively small system (a few tens or hundreds atoms) a localised excitation, caused for example by a process involving annihilation of a lattice defect, could contribute to an appreciable change of system temperature. One would never observe this in a macroscopic solid, where any excess energy would be transported through and shared among an extremely large number of atoms (of the order $\gg 10^{20}$).

In a fixed temperature molecular dynamics simulation (popularly referred to as $NVT$ MD), the energy of exothermic and endothermic processes is exchanged with a thermostat which adds or removes heat from the system simulating coupling of the system to a heat bath. This allows the temperature of the system to remain constant, thus better replicating experimental conditions in a spatially restricted simulation cell (as the system size approaches infinity, the $NVT$ and $NVE$ ensembles become equivalent, with the system itself acting as its own heat bath).

There have been a number of schemes proposed to generate constant temperature MD simulations. The first method proposed by Andersen (more details in \cite{andersen:2384}) relied on picking particles at random and allocating them new velocities chosen from the appropriate Maxwell distribution. While this method generates the correct thermodynamical ensemble it also has a significant effect on the particle dynamics because the impulses applied to random particles can cause problematic behaviour.

Another approach was proposed by Berendsen (more details in \cite{berendsen:3684}), where the kinetic energy is smoothly rescaled towards the target value and the temperature of the system is corrected such that the deviation of the temperature from its required value decays exponentially with a time constant $\tau$. This gives a stable and easy to implement method, but for a small system the Berendsen thermostat does not generates the correct thermodynamic ensemble. For large systems approaching the size of a few hundreds or thousands of atoms the scheme usually approximates the correct results for most thermodynamic properties reasonably well though. Consequently, the Berendsen thermostat is sometimes used to equilibrate the temperature of a system before another thermostat that does generate a canonical ensemble is used to calculate the thermodynamic properties.

Another popular scheme was proposed by Nos\'{e} and Hoover (more details in \cite{nose:511}, \cite{PhysRevA.31.1695}), and it is usually referred to as the Nos\'{e}-Hoover thermostat. It relies on the concept of introducing a heat bath with an extra degree of freedom $s$, which has an artificial mass and velocity associated with it. The kinetic energy is then included explicitly into the Hamiltonian, and the potential energy term (of the form $\propto \ln(s)$) can be adjusted, so that the algorithm reproduces a canonical ensemble for the correct number of independent momentum degrees of freedom for the system. This leads to the modified Nos\'{e} equations of motion and it can be demonstrated that sampling a microcanonical ensemble of this extended system (which allows for fluctuations in $s$, corresponding to heat transfer between system and the heat bath) is equivalent to sampling the canonical ensemble in the real system. In the limit of Nos\'{e}-Hoover thermostat with ``mass'' that approaches infinity, one recovers the result of a microcanonical ensemble in the real system.

While the Nos\'{e}-Hoover thermostat offers many advantages such as existence of a conserved quantity in the dynamics of the extended system with heat bath, and it guarantees sampling from a canonical ensemble, it can behave non-ergodically due to the lack of a stochastic component.

A different approach altogether that guarantees ergodic sampling is given by Langevin dynamics (more details in \cite{adelman:2375}), where the equations of motion are modified to include a dissipative term due to viscous damping caused by fictitious heat bath particles. The advantages of the Langevin thermostat include straightforward implementation (it can be easily integrated with Velocity-Verlet method; more details in \cite{quigley:11432}), and in addition to maintaining ergodicity it also guarantees sampling that is thermodynamically consistent with a canonical ensemble.

\section{Geometry Optimisation}
\label{chapter:simulation_techniques:section:geometry_optimisation}

\subsection{Steepest Descent and Conjugate Gradients}

Given the potential energy $U(\{\mathbf{x}_{i}\}_{i=1}^N)$ of the system of $N$ particles, one can formulate a problem of finding positions $\{\mathbf{x}_{i}\}_{i=1}^N$ such that the potential energy is minimised. This is what we refer to as a geometry optimisation problem, and local and global minima of the potential energy surface correspond to the stable and metastable states of the system. Starting from some non-equilibrium configuration the usual method of tackling this problem involves iterative movement of the atoms to reduce the net forces on them (gradients of the potential energy surface).

Locating minimum energy states can be performed using a damped MD method (more details in \cite{Probert2003130}, \cite{sheppard:134106}), but a more common approach involves iterative minimisation following the downhill gradient of the potential energy surface. In contrast to MD simulations, which calculate atomic trajectories with kinetic energy, particle velocities and therefore any effects of temperature are not included, and hence the minimisation trajectories have no physical sense. Thus only the final state of the system is of relevance, as it corresponds to the local minimum energy state of the system when the temperature approaches zero.

In the most general terms, we can define the iteration step of geometry optimisation problem as follows:

\begin{equation}
\mathbf{x}_{n+1} = \mathbf{x}_{n} + \alpha_n \mathbf{p}_n,
\end{equation}

\noindent where $\alpha_n$ is a positive scalar corresponding to the step length, and $\mathbf{p}_n$ is the search direction.

The simplest gradient-based geometry optimisation method, often referred to as the ``steepest descent'' method, is given by:

\begin{equation}
\mathbf{x}_{n+1} = \mathbf{x}_{n} - \alpha_n \boldsymbol{\nabla} U,
\end{equation}

\noindent where the step size $\alpha_n$ is chosen by the means of line search so that it satisfies the Wolfe conditions (more details in \cite{nocedal1999numerical}) and the search direction is simply given by:

\begin{equation}
\mathbf{p}_n = - \boldsymbol{\nabla} U.
\label{equation:search-direction}
\end{equation}

A more sophisticated method that usually has a much higher convergence rate and which is guaranteed to converge in at most $n$ steps for a system consisting of $n$ degrees of freedom whose energy is quadratic is the ``conjugate gradients'' method (more details in \cite{nocedal1999numerical}). It uses conjugate directions as the search directions instead of the local gradient of the ``steepest descent'' method, and consequently it ensures that each successive step minimises $U$ over the hyperplane that contains all of the previous search directions. Without going into too much mathematical detail, in the conjugate gradient algorithm $\alpha_n$ is given by:

\begin{equation}
\alpha_n = - \frac{\mathbf{r}_n^T \mathbf{p}_n}{\mathbf{p}_n^T \mathbf{A} \mathbf{p}_n},
\end{equation}

\noindent where $\mathbf{A}$ is a symmetric and positive definite matrix that is used to approximate the underlaying potential surface $U = \frac{1}{2} \mathbf{x}^T \mathbf{A} \mathbf{x} - \mathbf{b}^T \mathbf{x}$, and $\mathbf{r} = \boldsymbol{\nabla} U = \mathbf{A} \mathbf{x} - \mathbf{b}$. Unlike steepest descents, where each successive search direction is orthogonal to the previous one, the conjugate gradient method starts by searching along the steepest descent direction, but each successive $\mathbf{p}_n$ is a combination of the steepest direction, and the previous direction $\mathbf{p}_{n - 1}$:

\begin{equation}
\mathbf{p}_n = - \mathbf{r}_n + \beta_n \mathbf{p}_{n - 1},
\end{equation}

\noindent where:

\begin{equation}
\beta_n = \frac{\mathbf{r}_n^T \mathbf{A} \mathbf{p}_{n - 1}}{\mathbf{p}_{n - 1}^T \mathbf{A} \mathbf{p}_{n - 1}}.
\end{equation}

A comparison of two iterations of steepest-descent and conjugate-gradients methods for a two-dimensional, quadratic potential well is shown in figure \ref{figure:conjugate_gradients} below.

\begin{figure}[H]
\begin{center}
\resizebox{12cm}{!}{\footnotesize{}\input{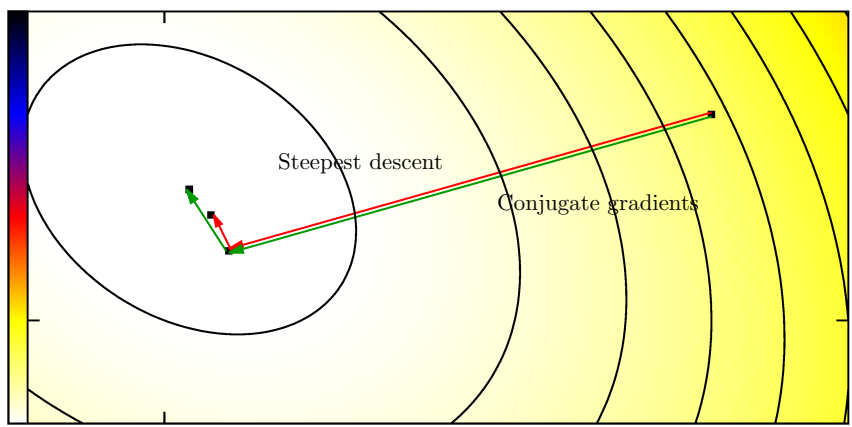}\normalsize{}}
\caption[Comparison of steepest-descent and conjugate-gradient methods.]{Comparison of the first two iterations using steepest-descent and conjugate-gradient methods in a quadratic potential well.}
\label{figure:conjugate_gradients}
\end{center}
\end{figure}

\subsection{Newton and Quasi-Newton Methods}

Building on the formalism presented in the previous section, extending the equation \ref{equation:search-direction} of the steepest-descent algorithm, the search direction adopted in the Newton's method is given by:

\begin{equation}
\mathbf{p}_n = - \mathbf{B}_n^{-1} \boldsymbol{\nabla} U,
\end{equation}

\noindent where $\mathbf{B}_n$ is the exact Hessian $\boldsymbol{\nabla}^2 U(\mathbf{x}_n)$ of the potential energy surface $U$ (one can think of the steepest descent method as one corresponding to an identity Hessian matrix). However, the major disadvantage of the Newton's method is that inversion of the Hessian matrix is usually quite costly and the method becomes impractical unless the Hessian matrix can be evaluated easily.

In contrast, quasi-Newton methods such as the Broyden-Fletcher-Goldfarb-Shanno (BFGS) method (more details in \cite{BROYDEN01031970}) do not compute the Hessian matrix directly, and instead they attempt to estimate its inverse $\mathbf{B}_n^{-1}$ from successive gradient vectors. Consequently, these methods often provide a convergence rate which approaches that of Newton's method but at a significantly reduced computational cost. The BFGS method has been shown to provide good performance, even when dealing with non-smooth potential energy surfaces and consequently they are commonly used in the context of atomistic simulations.

\section{Transition State Search}
\label{chapter:simulation_techniques:section:transition_state_search}

\subsection{Nudged Elastic Band}

For most systems the dynamics is usually characterised by the property that some regions of phase space, those of lower energy, are occupied for significant portion of time, but the system occasionally finds its way through the bottleneck to another region of phase space. This process can continue with a transition to new region of phase space or return to a region visited previously. In the language of reaction dynamics we would call them ``reactants'' and ``products'' respectively although in our situation initial and final state are perhaps more appropriate. The bottleneck, or transition state separating the two lower energy regions of phase space in turn corresponds to a saddle point of the potential energy surface which corresponds to a particular arrangement of the constituent atoms.

Finding transition state structures and their corresponding energies reduces to the problem of finding first-order saddle points of the potential energy surface --- i.e. an atomic configuration that is equivalent to a point in phase space where the potential energy surface has a minimum in all but one dimension, in which it has a maximum. It suffices to say that, as was the case with molecular dynamics or geometry optimisation techniques, essentially any classical or quantum-mechanical method of evaluating the potential energy surface, be it an interatomic potential or a DFT method, can be used to find transition states, although at radically different computational costs. It is also worth mentioning that one of the byproducts of locating a transition state is the minimum energy pathway (MEP). In fact the transition state is usually found by guessing the initial MEP, usually by the means of linear approximation, and iteratively optimising it.

The nudged elastic band (NEB; more details in \cite{henkelman:9978}) is one of the most commonly used methods for finding reaction pathways when both the initial and final states are known. The algorithm works by linearly interpolating a set of intermediate images between the known initial and final states, each image being a snapshot of the system along the reaction path. The position of the images is then iteratively adjusted according to the true force acting perpendicular to the reaction path and the force that results from an artificial spring connecting the neighbouring images that keeps them spaced along the transition path. In the original implementation, the total force acting on the image $i$ is given by:

\begin{equation}
\mathbf{f}_{i} = - \boldsymbol{\nabla} U |_{\perp} + \mathbf{f}_{i}^{S} |_{\parallel},
\end{equation}

\noindent where the true force acting perpendicular to the path is given by:

\begin{equation}
- \boldsymbol{\nabla} U |_{\perp} = - \boldsymbol{\nabla} U + (\boldsymbol{\nabla} U \cdot \boldsymbol{\hat{\tau}}_i) \boldsymbol{\hat{\tau}}_i,
\label{equation:potential-perp}
\end{equation}

\noindent and the force due to the artificial spring is given by:

\begin{equation}
\mathbf{f}_{i}^{S} |_{\parallel} = k \left( (\mathbf{R}_{i+1} - \mathbf{R}_{i}) - (\mathbf{R}_{i} - \mathbf{R}_{i-1}) \right) \cdot \boldsymbol{\hat{\tau}}_i,
\end{equation}

\noindent where $k$ is an arbitrarily chosen spring constant, and normalised tangent $\boldsymbol{\hat{\tau}}$ is computed by bisecting two unit vectors:

\begin{equation}
\boldsymbol{\tau}_{i} = \frac{\mathbf{R}_{i+1} - \mathbf{R}_{i}}{| \mathbf{R}_{i+1} - \mathbf{R}_{i} |} + \frac{\mathbf{R}_{i} - \mathbf{R}_{i-1}}{| \mathbf{R}_{i} - \mathbf{R}_{i-1} |}.
\end{equation}

Within the NEB formalism, the image positions can be evolved using any optimisation method, such as damped MD or the conjugate gradients minimiser (as outlined in the previous section \ref{chapter:simulation_techniques:section:geometry_optimisation}; more details in \cite{sheppard:134106}). Convergence of the transition path to the minimum energy pathway can be recognised once the magnitude of force on the images does not fall any further. However, if the parallel component of the force is large compared to the perpendicular one (such as near the inflection points of the MEP), formation of kinks that fluctuate forwards and backwards can be observed. Finally, particular attention has to be given to an appropriate selection of the spring constants to avoid ``cutting'' corners of the potential energy surface and at the same time to maintain the spacing of the images. This is critical for obtaining an accurate estimate of the saddle point energy.

Recent studies of dislocations with the NEB method also suggest that while NEB ensures equal spacing of the system images, the dislocation positions are not distributed uniformly along the MEP. Consequently, clustering of dislocation positions near potential minima can be observed which results in an error in the predicted slope of the Peierls barrier and the Peierls stress (more details in \cite{0965-0393-20-3-035019}).

\subsection{String Method}

The string method (more details in \cite{PhysRevB.66.052301}, \cite{e:164103}) is similar to the nudged elastic band method in that it also involves a series of images generated along a guessed transition path that are iteratively moved towards the MEP. However, unlike the NEB, the optimisation procedure involves two separate steps: firstly the images are moved according to the force perpendicular to the transition path. This is then followed by a reparameterisation step, which in turn ensures that the images are evenly spaced along the new path.

In the original string method formulation, the transition path, initially obtained by the means of linear interpolation or otherwise ``guessed'', is given by $\boldsymbol{\gamma}$, and for the MEP it satisfies the equation:

\begin{equation}
\boldsymbol{\nabla} U(\boldsymbol{\gamma}) |_{\perp} = 0,
\label{equation:stationary-states}
\end{equation}

\noindent i.e. the force acting perpendicular to the path approaches zero, where $\boldsymbol{\nabla} U |_{\perp}$ is the component of $\boldsymbol{\nabla} U$ perpendicular to $\boldsymbol{\gamma}$ as in equation \ref{equation:potential-perp}.

The idea behind the string method relies on evolving the path $\boldsymbol{\gamma}$ under the potential force field. The simplest dynamics for the evolution of such a path is given by:

\begin{equation}
\mathbf{v}_n = - \boldsymbol{\nabla} U(\boldsymbol{\gamma}) |_{\perp},
\end{equation}

\noindent where $\mathbf{v}_n$ is the normal velocity of the path (only the normal component is of relevance, as the tangential component redistributes the images along the path). In order to use this equation numerically, we parameterise path $\boldsymbol{\gamma} = \{\varphi(\alpha)\}_{\alpha}$ (the simplest parameterisation being that of a constant arc length $|\varphi(\alpha)| = \text{const.}$, although other parameterisations are possible) to obtain:

\begin{equation}
\boldsymbol{\dot{\varphi}} = - \boldsymbol{\nabla} U |_{\perp} + \lambda \boldsymbol{\hat{\tau}},
\end{equation}

\noindent where $\boldsymbol{\hat{\tau}}(\alpha) = \boldsymbol{\varphi}_{\alpha} / | \boldsymbol{\varphi}_{\alpha} |$, $\boldsymbol{\varphi}_{\alpha}$ denotes the derivative of $\boldsymbol{\varphi}$ with respect to $\alpha$, and the term $\lambda \boldsymbol{\hat{\tau}}$ is a Lagrange multiplier term added to enforce our parameterisation of a constant arc length so that $|\varphi(\alpha)|$ is a constant.

Since the term $\lambda \boldsymbol{\hat{\tau}}$ does not affect the evolution of the path (as mentioned before, only the normal component is of relevance), it does not contribute to the normal velocity of the curve and in the actual algorithm the action of $\lambda \boldsymbol{\hat{\tau}}$ is not implemented directly but instead is effected by means of a simple interpolation as a reparameterisation step.

The convergence of the transition path to the MEP is achieved by evaluating the dynamics of the system, as its stationary states satisfy the condition in equation \ref{equation:stationary-states}, i.e. that the forces acting perpendicular to the transition path approach zero.

\section{Monte Carlo Methods}
\label{chapter:simulation_techniques:section:monte_carlo_methods}

\subsection{Rejection Sampling}

Molecular dynamics (as outlined in section \ref{chapter:simulation_techniques:section:molecular_dynamics}) has been an extremely successful method for obtaining thermodynamic and structural properties throughout the field of atomistic simulation. As long as the time step and trajectory length are carefully chosen it universally yields Boltzmann-weighted averages of these properties:

\begin{equation}
\langle A \rangle = \lim_{t \to \infty} \frac{1}{t} \int_{t} A(\tau) d\tau .
\end{equation}

\noindent However, in some cases it is either impractical, or even impossible, to carry out molecular dynamics simulations --- for example the problem of a variable volume simulation can become unstable in MD unless the simulation cell is sufficiently large.

Boltzmann-weighted averages of thermodynamic and structural properties of the system can, however, be obtained using a different method altogether --- namely through the application of Monte Carlo (MC) statistical mechanics. Unlike MD, where new configurations are generated through application of Newton's equations of motion over a small time step to determine the updated values of atomic positions and velocities, in MC a new configuration is instead generated through non-uniform, pseudo-random sampling of relevant phase-space dimensions, provided that the samples are distributed according to Boltzmann statistics:

\begin{equation}
P(E) \propto \exp( - \beta E),
\end{equation}

\noindent for the canonical ensemble, where the Boltzmann factor is given by the term:\footnote{The Boltzmann factor does not give a probability distribution by itself, since it is not normalised --- the normalisation is given by the inverse of the partition function, which is the sum of Boltzmann factors for all available states of the system. However, for our purposes this is not a practical issue, as we can sample probability distribution up to an unknown normalising constant.}

\begin{equation}
\exp( - \beta E) = \exp ( - \frac{E}{k_B T} ).
\end{equation}

\noindent In addition to dealing with situations where MD formulation of the problem is ill-defined, MC statistical mechanics can also carry other advantages, such as providing faster convergence of thermodynamic properties in certain situations (more detail in \cite{doi:10.1021/jp960880x}), or generating less correlated samples obtained at a similar computational cost.

Essentially all methods of sampling a non-uniform distribution are based on the availability of a pseudo-random number generator which provides uniformly distributed samples. The most common and simple algorithm used to manipulate a single, uniformly distributed random variable $X$, into variable $Y$ that obeys the required distribution is usually referred to as ``rejection sampling'', or ``acceptance-rejection sampling''.

Rejection sampling (more details in \cite{bishop2006pattern}) relies on the observation that one can sample a probability distribution $f(x)$ by sampling an instrumental distribution $g(x)$ that bounds $f(x)$ instead. In practical terms this means that to sample $f(x)$, which cannot be sampled directly, it suffices to uniformly sample $M g(x)$ (which for $M > 1$ it bounds $f(x)$), and probabilistically accept or reject the samples from $M g(x)$. The rejection sampling algorithm can be summarised up as follows:

\begin{enumerate*}
\item Generate sample $x_0$ from $g(x)$.
\item Generate sample $y_0$ from the uniform distribution $[0, M g(x_0)]$.
\item If $y_0 > f(x_0)$ the sample is rejected. Otherwise, sample $x_0$ is kept.
\item The set of kept (accepted) samples $\{x_i\}_i$ is distributed according to $f(x)$.
\end{enumerate*}

\noindent While rejection sampling has the advantage of being trivial to implement, the efficiency of this method largely depends on the ratio between the area underneath $g(x)$, to the area underneath $f(x)$, as demonstrated in figure \ref{figure:rejection_sampling} below.

\begin{figure}[H]
\begin{center}
\resizebox{12cm}{!}{\footnotesize{}\input{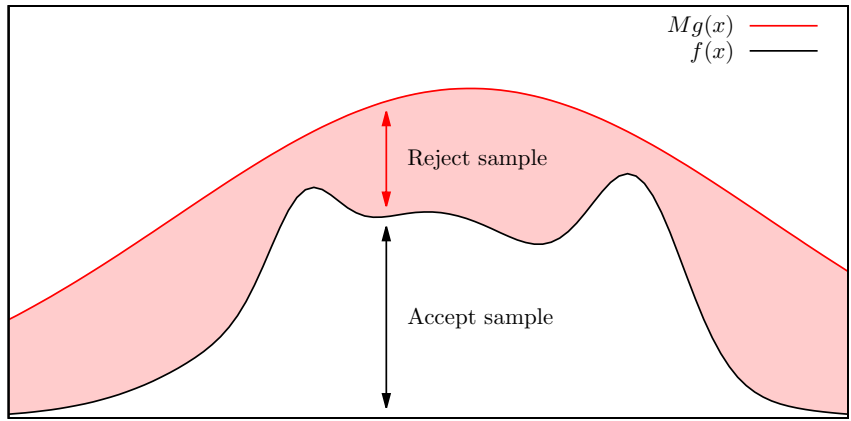}\normalsize{}}
\caption[Rejection sampling of $f(x)$ with distribution $M g(x)$.]{Example of distribution $M g(x)$, as used for sampling of distribution $f(x)$.}
\label{figure:rejection_sampling}
\end{center}
\end{figure}

Furthermore, the rejection sampling method becomes very inefficient when sampling multi-dimensional probability distributions. In multiple dimensions the acceptance rate decreases exponentially with the number of sampled dimensions (more details in \cite{bishop2006pattern}).

\subsection{Slice Sampling}

There is a wealth of sampling methods available in the literature that improves on the drawbacks of rejection sampling: adaptive rejection sampling (more details in \cite{1992}, \cite{1994}, \cite{1995}), the Metropolis-Hastings algorithm (more details in \cite{metropolis:1087}, \cite{HASTINGS01041970}) or Gibbs sampling (more details in \cite{4767596}) to name a few, but explaining all of them would be beyond the scope of this work. Instead, I will outline the background behind the slice sampling method, as proposed by Neal in the early 2000s (more details in \cite{NealSliceSampling}). Slice sampling shares many similarities with other Markov chain methods (such as Metropolis-Hastings and Gibbs sampling), but it also improves on them as it is capable of adjusting the step size automatically to match the local shape of the density function. Its implementation is also extremely straightforward. In its simplest form, slice sampling of a one-dimensional probability distribution $f(x)$ is achieved in the following way:

\begin{enumerate*}
\item Pick a starting point $x_0$ (any point underneath $f(x)$ is sufficient).
\item Fix $x_0$ and generate uniform sample $y_0$ from $[0, f(x_0)]$.
\item Fix $y_0$ and generate uniform sample $x_1$ from \emph{slice} $\{x: f(x) = y_0\}$.
\item Iteration consists of steps 2. and 3. --- all samples $\{x_i\}_i$ are accepted (and distributed according to $f(x)$).
\end{enumerate*}

Sampling $x$ from the \emph{slice} $\{x: f(x) = \text{const}\}$ can be achieved in a number of ways, the most common being stepping-out or doubling procedures. In the case of the stepping out procedure, given an estimate $w$ of the scale of the width of the slice, we proceed by finding bounds of the slice $(L,R)$ as follows:

\begin{enumerate*}
\item Pick an interval of size $w$ containing $x_0$.
\begin{enumerate*}
\item $L$ bound is given by $x_0 - w \times \text{Uniform}(0,1)$
\item $R$ bound is given by $L + w$
\end{enumerate*}
\item If $L$ or $R$ in slice, extend the bound by $w$ in that direction, until both $L$ and $R$ outside of slice.
\item Sample $x_1$ uniformly from $(L,R)$.
\item If $x_1$ in slice, accept the sample, otherwise use it to update $L$ or $R$ respectively.
\end{enumerate*}

\noindent The stepping-out procedure is appropriate for any distribution as long as an estimate $w$ of the scale of the width of the slice is known (the size of the interval can always be limited to $mw$ for any positive integer $m$). The doubling procedure can however expand the interval faster than the stepping-out procedure, and it might be more appropriate. This method works as follows:

\begin{enumerate*}
\item Pick an interval of size $w$ containing $x_0$.
\begin{enumerate*}
\item $L$ bound is given by $x_0 - w \times \text{Uniform}(0,1)$
\item $R$ bound is given by $L + w$
\end{enumerate*}
\item If $L$ or $R$ in slice, extend the bound by $(L - R)$ in the random direction\footnote{Please note that it is essential for the correctness of the method that the slice is extended in the random direction, as it produces a final interval that is the same as one that could be obtained from a different sample $x_0$ (more details in \cite{NealSliceSampling}).}, until both $L$ and $R$ outside of slice.
\item Sample $x_1$ uniformly from $(L,R)$.
\item If $x_1$ in slice, accept the sample, otherwise use it to update $L$ or $R$ respectively.
\end{enumerate*}

\noindent The general concept of the slice sampling algorithm is summarised in figure \ref{figure:slice_sampling} below.

\begin{figure}[H]
\begin{center}
\resizebox{12cm}{!}{\footnotesize{}\input{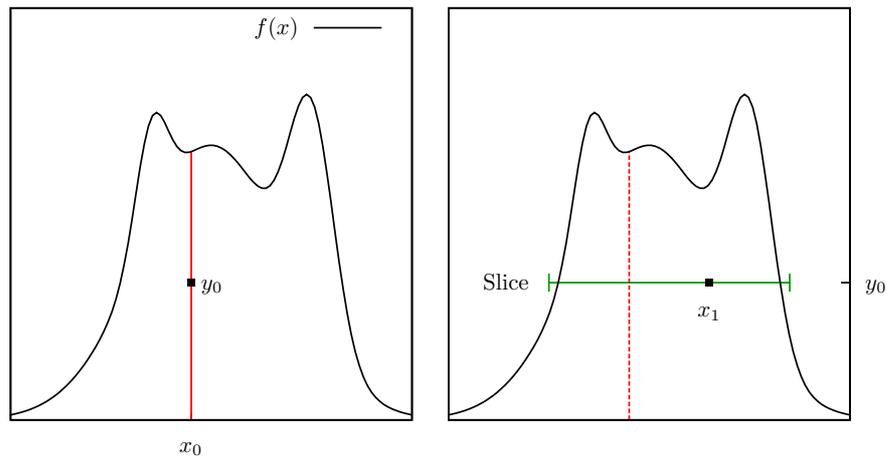}\normalsize{}}
\caption{Summary of the slice sampling algorithm.}
\label{figure:slice_sampling}
\end{center}
\end{figure}

Slice sampling algorithm can be extended trivially to multiple dimensions by sampling each dimension in turn repeatedly (as in Gibbs sampling). The major benefit of this method (apart from ease of implementation) is that unlike Metropolis-Hastings algorithm or Gibbs sampling, it is not sensitive to the step size (which if too small causes slow decorrelation of the random walk, and if too large leads to a high rejection rate). In effect the step size is automatically adjusted to match the shape of the sampled density function.

\cleardoublepage

\chapter{Gaussian Approximation Potential}
\label{chapter:gaussian_approximation_potential}

\section{Introduction}

In this chapter I introduce the theoretical background that underpins the Gaussian Approximation Potential (GAP) --- a new class of interatomic potentials that can be derived from energy, force and stress data and which is computed using explicitly quantum-mechanical \emph{ab initio} methods, although it is not limited to them as data obtained from classical calculations can be used equally well. It is often impossible or impractical to develop a physical model of the studied phenomena in a closed functional form of an interatomic potential that relies on a fixed number of fitted parameters. Consequently, the Gaussian Approximation Potential relies on a non-parametric approach to multidimensional regression, usually referred to as Gaussian process regression.

I begin by describing the most common (weight-space view) formulation of Gaussian process regression in section \ref{chapter:gaussian_approximation_potential:section:gaussian_process_regression}, demonstrating how it can be used for inference of continuous and differentiable functions in multiple dimensions. We follow this in section \ref{chapter:gaussian_approximation_potential:section:interatomic_potential} by outlining how this methodology can be used for fitting of potential energy surfaces. In section \ref{chapter:gaussian_approximation_potential:section:sparsification} I discuss the computational issues of GAP, that are important in the context of atomistic simulations. Finally, I finish this chapter with an in-depth discussion of atomic environments, and how they are relevant in the context of GAP potential in section \ref{chapter:gaussian_approximation_potential:section:description_of_atomic_environments}.

\section{Gaussian Process Regression}
\label{chapter:gaussian_approximation_potential:section:gaussian_process_regression}

The problem of finding a classical atomic energy function which reproduces the quantum-mechanical potential energy, is equivalent to the problem of supervised learning --- a machine learning technique for inferring a function from a training data set. In our case, the atomic energy function is a continuous value function and therefore we can classify it as a regression problem.

Following the analysis in \cite{mackay2003information}, in the Bayesian interpretation of the regression problem a non-linear function \(y(\mathbf{x})\) is assumed to underlie the data \(\{\mathbf{x}^{(n)}, t_n\}_{n = 1}^N\), where the set of input vectors is given by \(\mathbf{X}_N = \{\mathbf{x}^{(n)}\}_{n = 1}^N\) and the set of corresponding target values is given by \(\mathbf{t}_N = \{t_n\}_{n = 1}^N\). The inference of the function \(y(\mathbf{x})\) is described by the posterior probability distribution using Bayes theorem:

\begin{equation}
P(y(\mathbf{x})|\mathbf{t}_N, \mathbf{X}_N) = \frac{P(\mathbf{t}_N|y(\mathbf{x}), \mathbf{X}_N) P(y(\mathbf{x}))}{P(\mathbf{t}_N|\mathbf{X}_N)}.
\end{equation}

\noindent A common approach when dealing with the regression problems is to parameterise the function \(y(\mathbf{x})\) by restricting it to some well defined class of functions that we consider, where the prior distribution on functions \(P(y(\mathbf{x}))\) is implicit in the choice of the parametric model. However, in practice, the parameterisation of the function \(y(\mathbf{x})\) is irrelevant to the prediction of future values of \(t_{N + 1}\), given the input vector \(\mathbf{x}^{(N + 1)}\) and the data \(\{\mathbf{x}^{(n)}, t_n\}_{n = 1}^N\). All that is relevant is the assumed prior distribution \(P(y(\mathbf{x}))\) and the assumed noise \(P(\mathbf{t}_N|y(\mathbf{x}), \mathbf{X}_N)\).

A more general approach is to give a prior probability to every possible function by placing the prior probability distribution \(P(y(\mathbf{x}))\) directly over the space of functions. The simplest type of prior over functions is the Gaussian process which is a generalisation of the Gaussian probability distribution over a vector space of finite size to a distribution over an infinite function space. Although it may appear that the computational complexity associated with the inference of a function from a space of infinite size is intractable, it is possible to make predictions of future target values \(t_{N + 1}\) with finite computational resources as both conditional and marginal distributions of a multivariate Gaussian distribution (or a Gaussian process) are Gaussian as well.

We begin by expanding the function \(y(\mathbf{x})\) in an infinite set of basis functions \(\boldsymbol{\phi}(\mathbf{x}) = \{\phi_\mathbf{h}(\mathbf{x})\}_\mathbf{h}\):

\begin{equation}
y(\mathbf{x}) = \sum_\mathbf{h} w_\mathbf{h} \phi_\mathbf{h}(\mathbf{x}).
\label{equation:expansion}
\end{equation}

\noindent Assuming that the distribution of \(\mathbf{w} = \{w_\mathbf{h}\}_\mathbf{h}\) is a Gaussian, we notice that $y$, being a linear function of $w$, must also be Gaussian distributed, and following the derivation in \cite{mackay2003information} or \cite{rasmussen2006gaussian}, it can be shown that:

\begin{equation}
P(\mathbf{y}_N) \propto \exp\left(-\frac{1}{2} \mathbf{y}_N^T (\sigma_w^2 \Phi \Phi^T)^{-1} \mathbf{y}_N\right),
\end{equation}

\noindent which is a joint, multivariate Gaussian distribution (elements of the covariance matrix \(\sigma_w^2 \Phi \Phi^T\) can be calculated by integrating over all values of \(h\)) and where the set of function values corresponding to input \(\mathbf{X}_N\) is given by \(\mathbf{y}_N = \{y_n\}_{n = 1}^N\). The elements of the covariance matrix are given by:

\[
\Phi =
\begin{pmatrix}
\phi_1(\mathbf{x}^{(1)}) & \phi_2(\mathbf{x}^{(1)}) & \cdots \\
\phi_1(\mathbf{x}^{(2)}) & \phi_2(\mathbf{x}^{(2)}) & \cdots \\
\vdots & \vdots & \ddots \\
\phi_1(\mathbf{x}^{(N)}) & \phi_2(\mathbf{x}^{(N)}) & \cdots
\end{pmatrix} =
\begin{pmatrix}
\boldsymbol{\phi}(\mathbf{x}^{(1)}) \\
\boldsymbol{\phi}(\mathbf{x}^{(2)}) \\
\vdots \\
\boldsymbol{\phi}(\mathbf{x}^{(N)})
\end{pmatrix}.
\]

\noindent This result is the defining property of a Gaussian process --- the probability distribution of a function \(y(\mathbf{x})\) is a Gaussian process, if for any finite set of points \(\mathbf{X}_N\) the density \(P(\mathbf{y}_N)\) is a Gaussian.

Assuming that each target value \(t_n\) differs from the corresponding function value by additive Gaussian noise of variance \(\sigma_\nu^2\), \(t_n\) is also Gaussian, and we can further show that:

\begin{equation}
P(\mathbf{t}_N) \propto \exp\left(-\frac{1}{2} \mathbf{t}_N^T (\sigma_w^2 \Phi \Phi^T + \sigma_\nu^2 \mathbf{I})^{-1} \mathbf{t}_N\right).
\end{equation}

Having defined the probability $P(\mathbf{t}_N)$, we can now define the probability of inferring observation $t_{N+1}$, given the observed vector $\mathbf{t}_N$ --- the last necessary step for performing regression. Since the joint probability $P(t_{N+1}, \mathbf{t}_N)$ must be Gaussian, so is the conditional probability:

\begin{equation}
P(t_{N + 1} | \mathbf{t}_N) = \frac{P(t_{N+1}, \mathbf{t}_N)}{P(\mathbf{t}_N)}.
\end{equation}

\noindent By rewriting the new covariance matrix in terms of the covariance matrix \(\sigma_w^2 \Phi \Phi^T\) (according to the analysis outlined in \cite{mackay2003information} or \cite{rasmussen2006gaussian}), and substituting it into the above equation, one can obtain the expression for conditional probability:

\begin{equation}
P(t_{N + 1}|\mathbf{t}_N, \mathbf{X}_N) \propto \exp\left(-\frac{(t_{N + 1} - \bar{t}_{N + 1})^2}{2 \sigma_{\bar{t}_{N + 1}}^2}\right),
\end{equation}

\noindent where:

\begin{equation}
\bar{t}_{N + 1} = \sigma_w^2 \boldsymbol{\phi}(\mathbf{x}^{(N + 1)}) \Phi^T \cdot (\sigma_w^2 \Phi \Phi^T + \sigma_\nu^2 \mathbf{I})^{-1} \cdot \mathbf{t}_N
\label{equation:mean}
\end{equation}

\begin{align}
\sigma_{\bar{t}_{N + 1}}^2 &= \sigma_w^2 \boldsymbol{\phi}(\mathbf{x}^{(N + 1)}) \boldsymbol{\phi}^T(\mathbf{x}^{(N + 1)}) \nonumber \\
&- \sigma_w^2 \boldsymbol{\phi}(\mathbf{x}^{(N + 1)}) \Phi^T \cdot (\sigma_w^2 \Phi \Phi^T + \sigma_\nu^2 \mathbf{I})^{-1} \cdot \sigma_w^2 \Phi \boldsymbol{\phi}^T(\mathbf{x}^{(N + 1)}),
\label{equation:variance}
\end{align}

\noindent and consequently one can immediately identify \(\bar{t}_{N + 1}\) as the predictive mean at point \(\mathbf{x}^{(N + 1)}\), with \(\sigma_{\bar{t}_{N + 1}}\) as its corresponding error.

It is worth noting that in order to predict multiple future target values, the covariance matrix needs to be computed and inverted only once. Consequently, the computation of the covariance matrix elements, which involves integration over all basis functions, is only performed during the teaching process. By combining the tools of Bayesian inference and the Gaussian process, we obtain a non-linear, non-parametric method of solving multidimensional regression problems. Our choice of the set of basis functions (or the corresponding covariance function), imposes the prior directly over the infinite space of functions. This allows us to predict the future values in a very general and rigorous way, which correspond to a model with an infinite number of parameters. At the same time, the Gaussian process regression is easy to implement and extend, and it remains computationally tractable, allowing us to compute predictions at \(M\) new points with a computational cost that scales as \(\mathcal{O}(N M)\) for the predictive mean, \(\mathcal{O}(N^2 M)\) for the corresponding error and \(\mathcal{O}(N^3)\) for training (where \(N\) is the number of the teaching points).

\subsection{Covariance Function}

In the above treatment of Gaussian process regression we have deliberately left out the issue of calculating the covariance matrix \(\sigma_w^2 \Phi \Phi^T\) explicitly, as it requires some further discussion. If the underlying function $y(\mathbf{x})$ that we are trying to infer is expanded in an infinite set of basis functions $\phi(\mathbf{x})$ (as in equation \ref{equation:expansion}), one must wonder how this computation is numerically tractable. In practice it turns out that if the summation is replaced by integration, and the limits are taken to be $\pm \infty$, one can evaluate that integral analytically.

In the most straightforward case of a one-dimensional regression, we can demonstrate this using Gaussian radial basis functions as an example:

\begin{equation}
y(x) = \sum_h w_h \phi_h(x) = \sum_h w_h \exp \left[ - \frac{(x - h)^2}{2 r^2} \right].
\end{equation}

\noindent Substituting this expression into the covariance matrix \(\sigma_w^2 \Phi \Phi^T\) and taking the summation limit to approach infinity, one obtains the following expression for the element $ij$ of the covariance matrix:\footnote{An in-depth discussion of this derivation, as well as discussion of the weight-space and function-space formulations of the Gaussian process regression can be found in MacKay's (more details in \cite{mackay2003information}, Part V, Chapter 45) and Rasmussen's (more details in \cite{rasmussen2006gaussian}, Chapter 2) books.}

\begin{align}
\left( \sigma_w^2 \Phi \Phi^T \right)_{ij} &\propto \int_{h_{min}}^{h_{max}} \phi_h(x_i) \phi_h(x_j) dh \nonumber \\
&\propto \int_{h_{min}}^{h_{max}} \exp \left[ - \frac{(x_i - h)^2}{2 r^2} \right] \exp \left[ - \frac{(x_j - h)^2}{2 r^2} \right] dh .
\end{align}

\noindent Finally, taking the limits of the integration $h_{min} \to - \infty$ and $h_{max} \to + \infty$, the above integral can be solved analytically:

\begin{equation}
\left( \sigma_w^2 \Phi \Phi^T \right)_{ij} \propto \exp \left[ - \frac{(x_j - x_i)^2}{4 r^2} \right],
\end{equation}

\noindent and we can incorporate the normalising constant inside the $\sigma_w^2$, thus obtaining the following square-exponential covariance function:

\begin{equation}
\left( \Phi \Phi^T \right)_{ij} = \exp \left[ - \frac{(x_j - x_i)^2}{4 r^2} \right] = C(x_i, x_j).
\end{equation}

The above treatment can be directly extended to the multiple-dimensional case with ease and it demonstrates that Gaussian process regression can be considered from a different perspective altogether. Instead of specifying the prior distribution in terms of basis functions, it can be redefined in terms of the covariance function instead. This is the function-space view. Within the Gaussian Process regression formalism the only constraint on the choice of the covariance function is that it must correspond to a non-negative-definite covariance matrix (more details in \cite{mackay2003information}).

Table \ref{table:covariance_functions} below presents the most common covariance functions used for Gaussian Process regression (more details in \cite{rasmussen2006gaussian}):

\begin{table}[H]
\vspace{0.5cm}
\begin{center}
\begin{tabular}{ l c }
\multicolumn{2}{c}{Stationary covariance functions:\tablefootnote{Stationary covariance functions are invariant to translations of the input coordinates.} $C(x_i, x_j) = C(x_i - x_j)$} \\
\midrule
\emph{Constant} & $\theta$ \\
\midrule
\emph{Square-Exponential} & $\exp \left[ - \frac{(x_i - x_j)^2}{2 \theta^2} \right]$ \\
\midrule
\emph{Exponential} & $\exp \left[ - \frac{x_i - x_j}{\theta} \right]$ \\
\midrule
\emph{Gamma-Exponential} & $\exp \left[ - \left( \frac{x_i - x_j}{\theta} \right)^{\gamma} \right]$ \\
\midrule
\emph{Mat\'{e}rn Class} & $\frac{1}{2^{\nu - 1} \Gamma(\nu)} \left( \frac{\sqrt{2 \nu}}{l} (x_i - x_j) \right)^{\nu} K_{\nu} \left( \frac{\sqrt{2 \nu}}{l} (x_i - x_j) \right)$ \\
\midrule
\emph{Rational Quadratic} & $\left( 1 + \frac{(x_i - x_j)^2}{2 \alpha \theta^2} \right)^{- \alpha}$ \\
\midrule
& \\
\multicolumn{2}{c}{Non-stationary covariance functions:} \\
\midrule
\emph{Dot Product}\tablefootnote{Dot product covariance function is invariant to rotation of the input coordinates, but not to translations.} & $\theta + \mathbf{x}_i \cdot \mathbf{x}_j$ \\
\midrule
\emph{Polynomial} & $(\theta + \mathbf{x}_i \cdot \mathbf{x}_j)^p$ \\
\midrule
\end{tabular}
\caption{The most commonly used covariance functions.}
\label{table:covariance_functions}
\end{center}
\end{table}

\noindent It is also worth mentioning that new covariance functions can be created from the existing ones, as the sum of two kernels is also a kernel and the product of two kernels is also a kernel, etc. (more details in \cite{rasmussen2006gaussian}).

\subsection{Hyperparameters}

The regression parameters \(\sigma_w\), \(\sigma_\nu\), and any other adjustable parameters appearing in the covariance expression (for example \(\theta\) in square-exponential covariance in table \ref{table:covariance_functions}) are usually referred to as hyperparameters and the choice of their value depends on the prior knowledge of the dataset. We can think of these parameters as having the following physical meaning:

\begin{itemize*}
\item $\sigma_w \to$ prior knowledge of the variance of parameters $\{w_{\mathbf{h}}\}_{\mathbf{h}}$.
\item $\theta \to$ prior knowledge of the length-scale (width) of the basis functions.
\item $\sigma_{\nu} \to$ noise in the data measurement.
\end{itemize*}

In the above scenario, making a prediction of $t_{N + 1}$ would be ideally performed by integrating over all the available values of the hyperparameters:

\begin{align}
P(t_{N + 1}|\mathbf{t}_N, \mathbf{X}_N) = \int & P(t_{N + 1} | \mathbf{t}_N, \mathbf{X}_N, \sigma_w, \theta, \sigma_{\nu}) P(\sigma_w | \mathbf{t}_N, \mathbf{X}_N) \nonumber \\
& P(\theta | \mathbf{t}_N, \mathbf{X}_N) P(\sigma_{\nu} | \mathbf{t}_N, \mathbf{X}_N) \, d\sigma_w \, d\theta \, d\sigma_{\nu} .
\end{align}

\noindent However, it is usually impossible to evaluate such an integral analytically. Even if it is possible to carry out such integration by means of numerical methods in principle, for example using Markov chain Monte Carlo, it is usually not a practical solution. Instead, one usually relies on the maximum likelihood principle, effectively assuming that $P(\sigma_w | \mathbf{t}_N, \mathbf{X}_N)$, $P(\theta | \mathbf{t}_N, \mathbf{X}_N)$ and $P(\sigma_{\nu} | \mathbf{t}_N, \mathbf{X}_N)$ are delta functions, which simplifies the above expression to:

\begin{equation}
P(t_{N + 1}|\mathbf{t}_N, \mathbf{X}_N) = P(t_{N + 1} | \mathbf{t}_N, \mathbf{X}_N, \sigma_w^{max}, \theta^{max}, \sigma_{\nu}^{max}) ,
\end{equation}

\noindent where the hyperparameters are usually selected by hand based on the prior knowledge of the known features of the data. Alternatively, in the absence of any prior knowledge of their values, one can attempt to infer them from the available data by optimising:

\begin{equation}
P(\sigma_w, \theta, \sigma_{\nu} | \mathbf{t}_N, \mathbf{X}_N) \propto P(\mathbf{t}_N, \mathbf{X}_N | \sigma_w, \theta, \sigma_{\nu}) P(\sigma_w, \theta, \sigma_{\nu}),
\end{equation}

\noindent which assuming uniform prior on hyperparameters $P(\sigma_w, \theta, \sigma_{\nu})$ can be achieved by maximising the likelihood $P(\mathbf{t}_N, \mathbf{X}_N | \sigma_w, \theta, \sigma_{\nu})$. This is usually carried out in the logarithm space, as this simplifies the problem analytically, using any of the available optimisation methods (such as those outlined in section \ref{chapter:simulation_techniques:section:geometry_optimisation}).

\section{Interatomic Potential}
\label{chapter:gaussian_approximation_potential:section:interatomic_potential}

Throughout this work we are interested in either improving an existing interatomic potentials or creating a completely new one by applying Gaussian process regression to include information computed directly using quantum-mechanical methods. Each of these two approaches have their merits. Using an existing, simple potential (such as Lennard-Jones, Finnis-Sinclair or EAM) which already contains an accurate description of the physical system in its equilibrium configuration means that we only need to train the energy correction in the regions of phase space where the original, underlying potential differs from the quantum-mechanical description. However, as we will explore in chapters \ref{chapter:bispectrum-gap_potential_for_tungsten} and \ref{chapter:soap-gap_potential_for_tungsten}, this approach has its disadvantages as the potential energy landscape of energy correction might be more complicated than the original, underlying energy landscape. On the other hand, using Gaussian process regression to find a new potential altogether can make the training process significantly more complex as physical behaviour that we take for granted (such as atomic repulsion for example) needs to be included explicitly.

In this section we present a formalism that can be applied to either of the above two cases --- assuming that $E_{new} = E_{core} + (E_{QM} - E_{core})$, creating an energy correction to an existing potential corresponds to a non-zero $E_{core}$ term, whereas for the purpose of creating a new-potential altogether, we simply take $E_{core} = 0$. Consequently, our new interatomic potential becomes:

\begin{align}
E_{new} &= E_{core} + (E_{QM} - E_{core}) \nonumber \\
&= E_{core} + E_{GAP} \nonumber \\
&= \sum_{i = 1}^N \epsilon_i^{(core)} + \sum_{i = 1}^N \epsilon_i^{(GAP)},
\end{align}

\noindent and we need to find the atomic energy function \(\epsilon^{(GAP)}\), such that:

\begin{equation}
\sum_{i = 1}^N \epsilon_i^{(GAP)} = E_{GAP} = E_{QM} - E_{core} ,
\end{equation}

\noindent where the individual, reference function values $\epsilon$ are not available when training a potential derived from data computed using explicitly quantum-mechanical method, and the training is from the total energies $E$ instead.

In any computer simulation which involves molecular dynamics, we are also interested in the forces acting on the atoms. Hence, we also need to be able to accurately predict forces:

\begin{align}
\mathbf{f}^{(i)(new)} &= \mathbf{f}^{(i)(core)} + (\mathbf{f}^{(i)(QM)} - \mathbf{f}^{(i)(core)}) \nonumber \\
&= \mathbf{f}^{(i)(core)} + \mathbf{f}^{(i)(GAP)} \nonumber \\
&= - \boldsymbol{\nabla}^{(i)} E_{core} - \boldsymbol{\nabla}^{(i)} E_{GAP} \nonumber \\
&= - \boldsymbol{\nabla}^{(i)} \sum_{j = 1}^N \epsilon_{j}^{(core)} - \boldsymbol{\nabla}^{(i)} \sum_{j = 1}^N \epsilon_{j}^{(GAP)}.
\end{align}

\noindent At the same time, in order to be able to extract the maximum amount of information from our training data we want to be able to infer the atomic energy function \(\epsilon\) from the forces acting on the atoms in our training configurations:

\begin{equation}
- \boldsymbol{\nabla}^{(i)} \sum_{j = 1}^N \epsilon_{j}^{(GAP)} = \mathbf{f}^{(i)(GAP)} = \mathbf{f}^{(i)(QM)} - \mathbf{f}^{(i)(core)}.
\end{equation}

\noindent as well as from the virial stress tensor which is the stress acting on the simulation cell (more details in \cite{JCC:JCC10070}). It can be expressed as a linear combination of atomic positions and derivatives of the local energy $\epsilon$:

\begin{equation}
- \sum_{i = 1}^N x^{(i)}_\alpha \frac{\partial}{\partial x^{(i)}_\beta} \sum_{j = 1}^N \epsilon^{(GAP)}_j = \sum_{i = 1}^N x^{(i)}_\alpha f^{(i)(GAP)}_\beta = \tau^{(GAP)}_{\alpha \beta} = \tau^{(QM)}_{\alpha \beta} - \tau^{(core)}_{\alpha \beta}.
\end{equation}

Consequently, our training data is given by \(E_{GAP}\), \(\{\mathbf{f}^{(i)(GAP)}\}_{i = 1}^N\) and \(\tau^{(GAP)}_{\alpha \beta}\) and it follows that we need to be able to train the atomic energy function \(\epsilon^{(GAP)}\) from the sum of its values when we are training from total energies, from the sum of its derivatives when we are training from forces and from the linear combination of atomic positions and its derivatives when training from virial stress tensor. Additionally, if we are to use the atomic energy function in molecular dynamics to predict the forces on atoms, we need to be able to calculate its exact derivative. All of these requirements can be satisfied by adapting the Gaussian process regression formalism accordingly.

\subsection{Total Energies}

Following the work in \cite{2005PhDT}, Gaussian process regression can be easily extended to train from the sum of function values. We expand the sum of functions in an infinite set of basis functions:

\begin{equation}
E_{GAP} = \sum_{i = 1}^N \epsilon_i^{(GAP)} = \sum_{i = 1}^N \epsilon^{(GAP)}(\mathbf{q}^{(i)}) = \sum_\mathbf{h} w_\mathbf{h} \sum_{i = 1}^N \phi_\mathbf{h}(\mathbf{q}^{(i)}),
\end{equation}

\noindent and assume that the distribution of \(\mathbf{w} = \{w_\mathbf{h}\}_\mathbf{h}\) is a Gaussian. Following the same derivation as in section \ref{chapter:gaussian_approximation_potential:section:gaussian_process_regression}, the elements of the covariance matrix \(\sigma_w^2 \Phi \Phi^T\) now become:

\[
\Phi =
\begin{pmatrix}
\sum_{i = 1}^N \boldsymbol{\phi}(\mathbf{q}^{(i)}) \\
\sum_{i = 1}^N \boldsymbol{\phi}(\mathbf{q}^{(i)}) \\
\vdots
\end{pmatrix}
\begin{array}{l l}
\longleftarrow & \text{\footnotesize{teaching configuration 1}} \\
\longleftarrow & \text{\footnotesize{teaching configuration 2}} \\
 & \text{\footnotesize{etc.}}
\end{array}
\]

\noindent The remaining part of the derivation is the same, except for the computation of the covariance matrix, which now involves the additional cross terms. Consequently, we can find the value of the atomic energy function for atom \(i\) (and its error) by simply adapting equations \ref{equation:mean} and \ref{equation:variance}:

\begin{equation}
\bar{\epsilon}_i = \sigma_w^2 \boldsymbol{\phi}(\mathbf{q}^{(i)}) \Phi^T \cdot (\sigma_w^2 \Phi \Phi^T + \sigma_\nu^2 \mathbf{I})^{-1} \cdot \mathbf{E}^{(\text{train})}
\end{equation}

\begin{align}
\sigma_{\bar{\epsilon}_i}^2 &= \sigma_w^2 \boldsymbol{\phi}(\mathbf{q}^{(i)}) \boldsymbol{\phi}^T(\mathbf{q}^{(i)}) \nonumber \\
&- \sigma_w^2 \boldsymbol{\phi}(\mathbf{q}^{(i)}) \Phi^T \cdot (\sigma_w^2 \Phi \Phi^T + \sigma_\nu^2 \mathbf{I})^{-1} \cdot \sigma_w^2 \Phi \boldsymbol{\phi}^T(\mathbf{q}^{(i)}),
\end{align}

\noindent where \(\mathbf{E}^{(\text{train})}\) is the vector of the total energy corrections.

\subsection{Forces and Stresses}

In order to train the atomic energy function from the sum of its derivatives we again follow the work in \cite{2005PhDT}. We expand the sum of derivatives in an infinite set of basis functions:

\begin{align}
- f_\alpha^{(i)(GAP)} &= \sum_{j = 1}^N \frac{\partial \epsilon_j^{(GAP)}}{\partial x_\alpha^{(i)}} = \sum_{j = 1}^N \frac{\partial \epsilon^{(GAP)}(\mathbf{q}^{(j)})}{\partial x_\alpha^{(i)}} \nonumber \\
&= \sum_\mathbf{h} w_\mathbf{h} \sum_{j = 1}^N \frac{\partial \phi_\mathbf{h}(\mathbf{q}^{(j)})}{\partial x_\alpha^{(i)}} = \sum_\mathbf{h} w_\mathbf{h} \sum_{i = j}^N \psi_{\mathbf{h}, \alpha}^{(i)}(\mathbf{q}^{(j)}).
\end{align}

\noindent and assume that the distribution of \(\mathbf{w} = \{w_\mathbf{h}\}_\mathbf{h}\) is a Gaussian. Adapting the derivation in section \ref{chapter:gaussian_approximation_potential:section:gaussian_process_regression} according to \cite{2005PhDT}, the covariance matrix now becomes \(\sigma_w^2 \Psi \Psi^T\), with elements:

\[
\Psi =
\begin{pmatrix}
\sum_{j = 1}^N \boldsymbol{\psi}_x^{(i)}(\mathbf{q}^{(j)}) \\
\sum_{j = 1}^N \boldsymbol{\psi}_y^{(i)}(\mathbf{q}^{(j)}) \\
\sum_{j = 1}^N \boldsymbol{\psi}_z^{(i)}(\mathbf{q}^{(j)}) \\
\vdots
\end{pmatrix}
\begin{array}{l l}
 & \text{\footnotesize{teaching configuration 1,}} \\
\longleftarrow & \text{\footnotesize{atom i,}} \\
 & \text{\footnotesize{x, y, z components}} \\
 & \text{\footnotesize{etc.}}
\end{array}
\]

\noindent The computation of the covariance matrix again involves additional cross terms, and we can find the value of the atomic energy function for atom \(i\) (and its error) by adapting equations \ref{equation:mean} and \ref{equation:variance}:

\begin{equation}
\bar{\epsilon}_i = \sigma_w^2 \boldsymbol{\phi}(\mathbf{q}^{(i)}) \Psi^T \cdot (\sigma_w^2 \Psi \Psi^T + \sigma_\nu^2 \mathbf{I})^{-1} \cdot - \mathbf{f}^{(\text{train})}
\end{equation}

\begin{align}
\sigma_{\bar{\epsilon}_i}^2 &= \sigma_w^2 \boldsymbol{\phi}(\mathbf{q}^{(i)}) \boldsymbol{\phi}^T(\mathbf{q}^{(i)}) \nonumber \\
&- \sigma_w^2 \boldsymbol{\phi}(\mathbf{q}^{(i)}) \Psi^T \cdot (\sigma_w^2 \Psi \Psi^T + \sigma_\nu^2 \mathbf{I})^{-1} \cdot \sigma_w^2 \Psi \boldsymbol{\phi}^T(\mathbf{q}^{(i)}),
\end{align}

\noindent where \(\mathbf{f}^{(\text{train})}\) is the vector of the force corrections.

Since the virial stress tensor is simply a linear combination of atomic positions and derivatives of the local energy, the above methodology extends straightforwardly and the atomic energy function can be inferred from the virial stress tensor by combining the above two results.

Finally, as described in \cite{2005PhDT}, calculating an exact derivative of the atomic energy function is equivalent to another Gaussian process and we can find it (and its error) by computing:

\begin{align}
\frac{\partial \bar{\epsilon}_j}{\partial x_\alpha^{(i)}} &= \sigma_w^2 \boldsymbol{\psi}_\alpha^{(i)}(\mathbf{q}^{(j)}) \Phi^T \cdot (\sigma_w^2 \Phi \Phi^T + \sigma_\nu^2 \mathbf{I})^{-1} \cdot \mathbf{E}^{(\text{train})} \nonumber \\
&= \sigma_w^2 \boldsymbol{\psi}_\alpha^{(i)}(\mathbf{q}^{(j)}) \Psi^T \cdot (\sigma_w^2 \Psi \Psi^T + \sigma_\nu^2 \mathbf{I})^{-1} \cdot - \mathbf{f}^{(\text{train})}
\end{align}

\begin{align}
\sigma_{\frac{\partial \bar{\epsilon}_j}{\partial x_\alpha^{(i)}}}^2 &= \sigma_w^2 \boldsymbol{\psi}_\alpha^{(i)}(\mathbf{q}^{(j)}) \boldsymbol{\psi}_\alpha^{(i)T}(\mathbf{q}^{(j)}) \nonumber \\
&- \sigma_w^2 \boldsymbol{\psi}_\alpha^{(i)}(\mathbf{q}^{(j)}) \Phi^T \cdot (\sigma_w^2 \Phi \Phi^T + \sigma_\nu^2 \mathbf{I})^{-1} \cdot \sigma_w^2 \Phi \boldsymbol{\psi}_\alpha^{(i)T}(\mathbf{q}^{(j)}) \nonumber \\
&= \sigma_w^2 \boldsymbol{\psi}_\alpha^{(i)}(\mathbf{q}^{(j)}) \boldsymbol{\psi}_\alpha^{(i)T}(\mathbf{q}^{(j)}) \nonumber \\
&- \sigma_w^2 \boldsymbol{\psi}_\alpha^{(i)}(\mathbf{q}^{(j)}) \Psi^T \cdot (\sigma_w^2 \Psi \Psi^T + \sigma_\nu^2 \mathbf{I})^{-1} \cdot \sigma_w^2 \Psi \boldsymbol{\psi}_\alpha^{(i)T}(\mathbf{q}^{(j)})
\end{align}

\noindent where \(\mathbf{E}^{(\text{train})}\) is the vector of the total energy corrections and \(\mathbf{f}^{(\text{train})}\) is the vector of the force corrections.

\section{Sparsification}
\label{chapter:gaussian_approximation_potential:section:sparsification}

As outlined in section \ref{chapter:gaussian_approximation_potential:section:gaussian_process_regression}, Gaussian process regression allows us to compute predictions at \(M\) new points with a computational cost that scales as \(\mathcal{O}(N M)\) for the predictive mean, \(\mathcal{O}(N^2 M)\) for the corresponding error and \(\mathcal{O}(N^3)\) for training (where \(N\) is the number of teaching points). While this method remains computationally tractable for training sets consisting of several thousands of input points, this limit can be easily exceeded if one needs to train the atomic energy function from the sums of function values or from the sums of function derivatives (because of the resultant cross terms). Furthermore, our training data is very often correlated especially when training from forces, so if we are to train an atomic energy function that remains accurate in a wide variety of atomic configurations, the training data must include a wide variety of configurations. We need to use a sparse model that preserves the desirable properties of the full Gaussian process regression to maximise accuracy, but at the same time involves a minimal number of input points to minimise the computational cost.

In recent years there have been many attempts to make sparse approximations to the original Gaussian process regression (more details in \cite{QuinoneroUnifyingView}). One which is especially useful for our application was proposed by Snelson and Ghahramani in \cite{NIPS2005_543}. It uses the covariance, which is parameterised by \(S\) pseudo-input points (such that \(S \ll N\)), consisting of \(S\) pseudo-input vectors and \(S\) corresponding pseudo-input targets.

Following the original derivation and using the notation from section \ref{chapter:gaussian_approximation_potential:section:gaussian_process_regression} we can define the covariance elements corresponding to the sparse pseudo-inputs as:

\[
\Phi' =
\begin{pmatrix}
\boldsymbol{\phi}'(\mathbf{x}'^{(1)}) \\
\boldsymbol{\phi}'(\mathbf{x}'^{(2)}) \\
\vdots \\
\boldsymbol{\phi}'(\mathbf{x}'^{(S)})
\end{pmatrix} ,
\]

\noindent and our original equations \ref{equation:mean} and \ref{equation:variance} for predictive mean and its corresponding error become:

\begin{align}
\bar{t}_{N + 1} &= \sigma_w^2 \boldsymbol{\phi}(\mathbf{x}^{(N + 1)}) \Phi^T \cdot (\sigma_w^2 \Phi \Phi^T + \sigma_\nu^2 \mathbf{I})^{-1} \cdot \mathbf{t}_N \nonumber \\
&\approx \sigma_w^2 \boldsymbol{\phi}(\mathbf{x}^{(N + 1)}) \Phi'^T \cdot \mathbf{Q}_{S}^{-1} \cdot \mathbf{t}'_S
\label{equation:sparse_mean}
\end{align}

\begin{align}
\sigma_{\bar{t}_{N + 1}}^2 &\approx \sigma_w^2 \boldsymbol{\phi}(\mathbf{x}^{(N + 1)}) \boldsymbol{\phi}^T(\mathbf{x}^{(N + 1)}) \nonumber \\
&- \sigma_w^2 \boldsymbol{\phi}(\mathbf{x}^{(N + 1)}) \Phi'^T \cdot \left( (\sigma_w^2 \Phi' \Phi'^T)^{-1} - \mathbf{Q}_{S}^{-1} \right) \cdot \sigma_w^2 \Phi' \boldsymbol{\phi}^T(\mathbf{x}^{(N + 1)}) \nonumber \\
&+ \sigma_\nu^2 ,
\label{equation:sparse_variance}
\end{align}

\noindent where the original covariance matrix $(\sigma_w^2 \Phi \Phi^T + \sigma_\nu^2 \mathbf{I})$ is replaced by $\mathbf{Q}_{S}$ and the vector of target values $\mathbf{t}_N$ is replaced by the pseudo-input targets $\mathbf{t}'_S$. They can be computed by evaluating:

\begin{equation}
\mathbf{t}'_S = \sigma_w^2 \Phi' \Phi^T (\boldsymbol{\Lambda} + \sigma_\nu^2 \mathbf{I})^{-1} \mathbf{t}_N ,
\end{equation}

\noindent where $\boldsymbol{\Lambda} = \text{diag} (\lambda_n)$ is a diagonal matrix constructed from the elements:

\begin{align}
\lambda_n &= (\sigma_w^2 \Phi \Phi^T)_{nn} \nonumber \\
&+ (\sigma_w^2 \boldsymbol{\phi}(\mathbf{x}^{(n)}) \Phi'^T)^T \cdot (\sigma_w^2 \Phi' \Phi'^T)^{-1} \cdot (\sigma_w^2 \boldsymbol{\phi}(\mathbf{x}^{(n)}) \Phi'^T) ,
\end{align}

\noindent and the new covariance matrix $\mathbf{Q}_{S}$ is given by:

\begin{equation}
\mathbf{Q}_{S} = (\sigma_w^2 \Phi' \Phi'^T) + (\sigma_w^2 \Phi' \Phi^T) \cdot (\boldsymbol{\Lambda} + \sigma_\nu^2 \mathbf{I})^{-1} \cdot (\sigma_w^2 \Phi \Phi'^T) .
\end{equation}

\noindent By looking at equations \ref{equation:sparse_mean} and \ref{equation:sparse_variance} we can immediately notice that evaluating the matrix product no longer corresponds to the sum of $N$ elements, but instead it is replaced by $S$ terms (where \(S \ll N\)). Furthermore, the pseudo-input targets always correspond to individual atomic energies as opposed to a linear combination of them (as in the case of total energies or forces or stresses).

The covariance can then be optimised using a gradient-based method, which optimises the hyperparameters of the covariance function and the locations of the pseudo-input points (in terms of the input coordinates) in the same joint optimisation. Without going into too much mathematical detail (the derivation can be found in \cite{NIPS2005_543}), it suffices to say that our optimisation is achieved by maximising the marginal likelihood, which corresponds to finding both optimal values of the hyperparameters and the locations of pseudo-input points that best reproduce the original input data. We can assess how well our sparse pseudo-inputs reproduce the original data by comparing our value of the marginal likelihood with that of the original data with minimised hyperparameters. If they are close enough for our purposes, we can use our optimised pseudo-inputs to approximate the Gaussian process regression using the full, non-sparsified original data.

Consequently, by applying the sparse Gaussian process using pseudo-inputs, we can eliminate problems arising from both the use of large numbers of input points due to training from sums and from highly correlated input data. This allows us to compute predictions at \(M\) new points with a computational cost that scales as \(\mathcal{O}(S M)\) for the predictive mean, \(\mathcal{O}(S^2 M)\) for the corresponding error and \(\mathcal{O}(S^2 N)\) for training, where \(N\) is the number of the original input points and \(S\) is the number of sparse pseudo-input points, such that \(S \ll N\).

\section{Description of Atomic Environments}
\label{chapter:gaussian_approximation_potential:section:description_of_atomic_environments}

Although one could try to compute the Gaussian process regression for the atomic energy function \(\epsilon\) using atomic coordinates \(\{\mathbf{x}^{(i)}\}_{i = 1}^N\) as input, such an approach would be both impractical and computationally expensive. To do this one would need to ensure that all structures related by simple symmetry transformations are explicitly included in the training dataset (structures related by a rotation, translation or reflection often correspond to the same energy, as dictated by the symmetry of the system). At the same time, the computational cost associated with performing Gaussian process regression can be significantly decreased by reducing the dimensionality of the input. Hence, in order to reduce the dimensionality and simplify the process of training the atomic energy function from quantum-mechanical information, we should use a set of invariants as input instead. These need to represents the atomic neighbourhood of atoms accurately and also occupy a space of fewer dimensions. Consequently, we can approximate the atomic energy function \(\epsilon\) as:

\begin{equation}
\epsilon_i = \epsilon(\{\mathbf{x}^{(j)} - \mathbf{x}^{(i)}\}_{j = 1}^N) \to \epsilon(\mathbf{q}^{(i)}),
\end{equation}

\noindent where we define the atomic neighbourhood as a set of atoms with coordinates \(\{\mathbf{x}^{(j)}\}_{j = 1}^{N}\), such that the energy $\epsilon_i$ obeys:

\begin{equation}
\frac{\partial \epsilon_i}{\partial \mathbf{x}^{(j)}} \neq 0,
\end{equation}

\noindent and where \(\mathbf{q}^{(i)}\) is a vector of parameters associated with the environment of atom \(i\). If \(\mathbf{q}^{(i)}\) provides a complete description of the atomic environment, the mapping is one-to one and the equality between the two sides of our approximation holds.

The problem of finding a mapping $\{\mathbf{x}^{(j)} - \mathbf{x}^{(i)}\}_{j = 1}^N \to \mathbf{q}^{(i)}$ (which we will refer to as the ``descriptor'') that is both complete and invariant to the relevant symmetry transformations is not a trivial one. Consequently, we present a more in-depth discussion of this problem, as well as potential solutions, below.

\subsection{Rotational and Permutational Invariance}

For the purpose of Gaussian process regression and its application in the Gaussian Approximation Potential, a good descriptor should not only provide a faithful representation of the atomic environment, ideally, retaining the completeness of the Cartesian representation. It should also contain all the appropriate symmetries such as rotation, translation and reflection, and furthermore it should also provide permutational invariance with respect to ordering of the atoms. In the most straightforward example of using Cartesian coordinates as a descriptor, even if one ensures a consistent method determining the order of the atoms, any change in the neighbour positions that affects this ordering would lead to discontinuities in the atomic energy function. This by itself would be unphysical, and make computation of forces which are derivatives of the atomic energy function ill-defined.

The most common method of providing a rotationally invariant descriptor is based on calculating a set of geometric parameters describing the system, such as bond lengths, bond angles, tetrahedral angles, etc. It has rotational symmetry built in but the size of a complete set of such parameters is highly impractical as it grows as $\mathcal{O}{(\exp (N))}$ with the number of bonds $N$ surrounding the central atom (more details in \cite{2010PhDT}). It is also easy to see that, due to its size, the information contained in such a set is highly redundant. There is, however, no systematic way of reducing its size without an associated loss of accuracy.

A more practical, rotationally invariant representation of atomic environment, where the positions of $N$ neighbours relative to the central atom are given by $\{\mathbf{r}_i\}_{i=1}^N$, can be constructed by computing the symmetric matrix:

\begin{equation}
\boldsymbol{\Sigma} =
\begin{pmatrix}
\mathbf{r}_1 \cdot \mathbf{r}_1 & \mathbf{r}_1 \cdot \mathbf{r}_2 & \dots & \mathbf{r}_1 \cdot \mathbf{r}_N \\
\mathbf{r}_2 \cdot \mathbf{r}_1 & \mathbf{r}_2 \cdot \mathbf{r}_2 & \dots & \mathbf{r}_2 \cdot \mathbf{r}_N \\
\vdots & \vdots & \ddots & \vdots \\
\mathbf{r}_N \cdot \mathbf{r}_1 & \mathbf{r}_N \cdot \mathbf{r}_2 & \dots & \mathbf{r}_N \cdot \mathbf{r}_N
\end{pmatrix}
\end{equation}

\noindent where the diagonal elements correspond to the bond lengths, and the off-diagonal elements are related to bond angles (although scaled by the bond lengths). This representation can be shown to be complete (more details in \cite{weyl1950theory}, \cite{2010PhDT} and \cite{PhysRevB.87.184115}) up to an arbitrary rotation and reflection. For a representation that is complete up to an arbitrary rotation alone, matrix $\boldsymbol{\Sigma}$ needs to be complemented with the appropriate quadrant information:

\begin{gather}
\boldsymbol{\Sigma}^{*} = \nonumber \\
\begin{pmatrix}
r_1 r_1 (\cos \theta_{11}, \sin \theta_{11}) & r_1 r_2 (\cos \theta_{12}, \sin \theta_{12}) & \dots & r_1 r_N (\cos \theta_{1N}, \sin \theta_{1N}) \\
r_2 r_1 (\cos \theta_{21}, \sin \theta_{21}) & r_2 r_2 (\cos \theta_{22}, \sin \theta_{22}) & \dots & r_2 r_N (\cos \theta_{2N}, \sin \theta_{2N}) \\
\vdots & \vdots & \ddots & \vdots \\
r_N r_1 (\cos \theta_{N1}, \sin \theta_{N1}) & r_N r_2 (\cos \theta_{N2}, \sin \theta_{N2}) & \dots & r_N r_N (\cos \theta_{NN}, \sin \theta_{NN})
\end{pmatrix}
\end{gather}

\noindent which provides a more compact representation than the set of bond lengths, bond angles, tetrahedral angles, etc., although it is still vastly redundant.

Unfortunately, all of the above solutions suffer from the fact that permutational invariance cannot be readily included. Permuting the neighbouring atoms shuffles the columns and rows of matrix $\boldsymbol{\Sigma}$ and although one could attempt to compare two structures using a distance metric:

\begin{equation}
d = \min_{\mathbf{P}} | \boldsymbol{\Sigma} - \mathbf{P} \boldsymbol{\Sigma} \mathbf{P}^T | ,
\end{equation}

\noindent where $\mathbf{P}$ is a general permutation operator, and we minimise over all possible permutations, this metric is not differentiable at locations where the permutation operator changes (more details in \cite{PhysRevB.87.184115}).

One way of achieving permutational invariance is through the use of symmetric polynomials:

\begin{equation}
\Pi_k (x_1, x_2, \dots, x_N) = \Pi_k (x_{\mathbf{P} 1}, x_{\mathbf{P} 2}, \dots, x_{\mathbf{P} N}),
\end{equation}

\noindent where $k$ corresponds to the degree of the polynomial. In the most straightforward case for $k = 1$:

\begin{equation}
\Pi_1 (x_1, x_2, \dots, x_N) = \sum_{i = 1}^N x_i.
\end{equation}

\noindent This representation, however, is not rotationally invariant.

\subsection{Bond-Order Parameters}

The most commonly used set of parameters that is both rotationally and permutationally invariant are the bond-order parameters originally introduced in \cite{PhysRevB.28.784}, which are widely used to analyse the atomic structure of solids in the field of computational chemistry (more details in section \ref{chapter:classical_and_quantum_simulation_of_solids:section:interatomic_potentials:subsection:atomic_environments}). However, they do not provide a complete representation of the system (i.e. the mapping is not one-to-one). They are nevertheless a good starting point in our analysis that should lead to a continuous, differentiable set of parameters to accurately describe atomic configurations that we can use with Gaussian process regression.

To begin, the atomic environment can be approximated by a three-dimensional density function $\rho$ and in the simplest possible case the neighbouring atoms can be approximated by point-masses:

\begin{equation}
\rho_i = \rho(\{\mathbf{x}^{(j)} - \mathbf{x}^{(i)}\}_{j = 1}^N) = \sum_{j = 1}^N \alpha_j \delta(\mathbf{x} - (\mathbf{x}^{(j)} - \mathbf{x}^{(i)})),
\label{equation:density}
\end{equation}

\noindent where $\alpha_{j}$ is a weight associated with atomic species $j$ (although for a single-species case it can be assumed that $\alpha = 1$), and $\delta$ is the three-dimensional Dirac-delta function.

If the atomic density function is projected onto the surface of a sphere, it can be expanded using a spherical harmonics basis:

\begin{equation}
\rho(\theta, \phi) = \sum_{l = 0}^{\infty} \sum_{m = -l}^{l} Q_{lm} Y_{lm}(\theta, \phi),
\end{equation}

\noindent where:

\begin{equation}
Q_{lm} = \int \rho(\theta, \phi) Y_{lm}(\theta, \phi) \sin \theta d\theta d\phi,
\end{equation}

\noindent and $Y_{lm}(\theta, \phi)$ are orthonormalised spherical harmonic functions and \(\theta\) and \(\phi\) are the polar and azimuthal angles measured with respect to an arbitrary reference frame.

For the atomic density function $\rho_i$ of atom $i$, composed of a sum of (weighted) Dirac-delta functions, this expansion becomes:

\begin{equation}
Q_{lm}^{(i)} = \sum_{j =1 }^N \alpha_j Y_{lm}(\theta(\mathbf{x}^{(j)} - \mathbf{x}^{(i)}), \phi(\mathbf{x}^{(j)} - \mathbf{x}^{(i)})),
\end{equation}

\noindent and consequently for any neighbouring atom \(j\), we can define a set of numbers:

\begin{equation}
Q_{lm}^{(i)(j)} = Y_{lm}(\theta(\mathbf{x}^{(j)} - \mathbf{x}^{(i)}), \phi(\mathbf{x}^{(j)} - \mathbf{x}^{(i)})),
\end{equation}

\noindent In the original formulation (more details in \cite{PhysRevB.28.784}), Steinhardt defines a quantity:

\begin{equation}
\bar{Q}_{lm}^{(i)} = \frac{1}{N_i} \sum_{j = 1}^{N_i} Q_{lm}^{(i)(j)},
\end{equation}

\noindent where the sum is over all atoms in the neighbourhood of atom \(i\), and it is normalised by the number of neighbours $N_i$.

Although the parameters \(\bar{Q}_{lm}^{(i)}\) depend on the choice of the reference frame, the following rotationally invariant combinations can be constructed from them:

\begin{equation}
Q_l^{(i)} = \left(\frac{4 \pi}{2 l + 1} \sum_{m = -l}^l |\bar{Q}_{lm}^{(i)}|^2\right)^{^1 / _2}
\end{equation}

\begin{equation}
W_l^{(i)} = \sum_{\substack{m_1, m_2, m_3 \\ m_1 + m_2 + m_3 = 0}} \begin{pmatrix} l & l & l \\ m_1 & m_2 & m_3 \end{pmatrix} \bar{Q}_{l m_1}^{(i)} \bar{Q}_{l m_2}^{(i)} \bar{Q}_{l m_3}^{(i)},
\end{equation}

\noindent which are the second-order and third-order invariants respectively \cite{duijneveldt:4655} and the term in brackets is a Wigner \(3j\) symbol. Finally, it is possible to define a reduced order parameter \(\hat{W}_l^{(i)}\), which is almost insensitive to the precise definition of a neighbour:

\begin{equation}
\hat{W}_l^{(i)} = W_l^{(i)} / \left(\sum_{m = -l}^l |\bar{Q}_{lm}^{(i)}|^2\right)^{^3 / _2}.
\end{equation}

\noindent Any combination of \(Q_l^{(i)}\), \(W_l^{(i)}\) or \(\hat{W}_l^{(i)}\) can be used as a set of rotationally and permutationally invariant parameters that can be expanded or contracted, depending on how precisely we want to describe the atomic neighbourhood of the atom \(i\). Elements with odd values of $l$ can also be skipped, or one can take its absolute value, in order to impose reflection symmetry. However, the set of bond-order parameters is also a highly incomplete descriptor --- not only is the angular representation incomplete (more on that in the following section), but the representation of any radial information is missing altogether. Furthermore, in this original formulation, assuming a finite neighbourhood cutoff distance, this descriptor has a discontinuity at the neighbourhood cutoff. Both the radial sensitivity and continuity properties can be, however, easily fixed by inclusion of radial information using radial basis functions.

In order to extrapolate the original formulation of the bond order parameters into a continuous and differentiable descriptor we introduce a continuous and differentiable radial weight function \(w_{ij}\) and modify the quantities \(Q_{lm}^{(i)(j)}\) and \(\bar{Q}_{lm}^{(i)}\) accordingly:

\begin{equation}
Q_{lm}^{(i)(j)} = w_{ij}(|\mathbf{x}^{(j)} - \mathbf{x}^{(i)}|) Y_{lm}(\theta(\mathbf{x}^{(j)} - \mathbf{x}^{(i)}), \phi(\mathbf{x}^{(j)} - \mathbf{x}^{(i)}))
\end{equation}

\begin{equation}
\bar{Q}_{lm}^{(i)} = \frac{1}{\sum_{j = 1}^{N_i} w_{ij}} \sum_{j = 1}^{N_i} Q_{lm}^{(i)(j)}.
\label{equation:q_bar}
\end{equation}

\noindent Since, in practice, we use a radial function with a finite range cutoff, we therefore limit the range of the bond order parameters and the atomic energy function \(\epsilon\) which uses the bond order parameters as its coordinates to this cutoff.

Although there is a lot of freedom in the choice of the functional form of the radial basis (specified by the set of weight functions \(w_{ij}\)) we use to calculate the bond order parameters, there are some aspects that require careful consideration. As already mentioned, our radial basis functions need to be continuous and differentiable and decay to zero at some finite cutoff if our bond order parameters are to remain continuous. Furthermore, within the Gaussian process regression formalism, the force on an atom is a function of bond order parameter derivative so, if we are to avoid unwanted force impulses we require the bond order derivatives to be continuous and to also decay to zero at the cutoff. Note that the requirement of bond order derivatives being continuous is then satisfied as our bond order parameters need to be differentiable in the first place.

A more detailed analysis of the bond order parameters also reveals that they correspond to expanding our atomic density function (the function consisting of Dirac delta functions centred at atomic positions) in a basis of spherical harmonics and our radial basis and than calculating the corresponding invariants. Sturm-Liouville theory and, in particular, the theory of the associated eigenfunctions and their completeness, ensures that we can expand the atomic density function without any loss of information as long as we choose a suitable radial basis such as a set of spherical Bessel functions. Then any loss of information about the atomic environment is attributed entirely to the computation of the associated invariants.

However, since our expansion becomes complete only as the number of elements in the basis approaches infinity, we should also consider radial bases that are not complete but which nevertheless contain sufficient information about our system when the basis is sparse (i.e. the number of basis elements is small). If the radial basis functions are, in addition, easy to evaluate, these factors might indeed be of greater importance --- there is then a direct trade-off between the accuracy of our atomic environment representation and its dimensionality.

An example of the first few elements of a simple radial basis, with adjustable parameters \(r_{cut}\) and \(r_0\), that is continuous and differentiable and is guaranteed to decay to zero as $r$ approaches $r_{cut}$ is:

\begin{equation}
R_1(r) = \left\{ \begin{array}{l l} 1 & \quad 0 \leq r < r_{cut} - r_0 \\
\cos^2(\frac{\pi}{2} \frac{r - r_{cut} + r_0}{r_0}) & \quad r_{cut} - r_0 \leq r \leq r_{cut} \\ \end{array} \right. ,
\label{equation:radial_basis_1}
\end{equation}

\begin{equation}
R_2(r) = \left\{ \begin{array}{l l} 0 & \quad 0 \leq r < r_{cut} - 2 r_0 \\
\cos^2(\frac{\pi}{2} \frac{r - r_{cut} + r_0}{r_0}) & \quad r_{cut} - 2 r_0 \leq r \leq r_{cut} \\ \end{array} \right. ,
\label{equation:radial_basis_2}
\end{equation}

\begin{equation}
R_3(r) = \left\{ \begin{array}{l l} \cos^2(\frac{\pi}{2} \frac{r - r_{cut} + r_0}{r_{cut} - r_0}) & \quad 0 \leq r < r_{cut} - r_0 \\
\cos^2(\frac{\pi}{2} \frac{r - r_{cut} + r_0}{r_0}) & \quad r_{cut} - r_0 \leq r \leq r_{cut} \\ \end{array} \right. .
\label{equation:radial_basis_3}
\end{equation}

\begin{figure}[H]
\begin{center}
\resizebox{12cm}{!}{\footnotesize{}\input{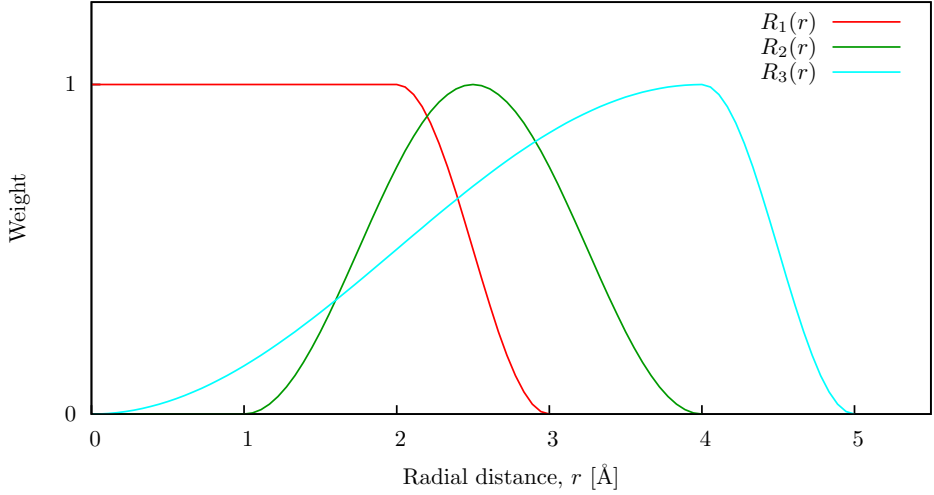}\normalsize{}}
\vspace{0.75cm}
\caption[Bond-order parameters radial basis functions.]{Bond-order parameters radial basis functions specified in equations \ref{equation:radial_basis_1}, \ref{equation:radial_basis_2} and \ref{equation:radial_basis_3}.}
\label{fig:weight_functions}
\end{center}
\end{figure}

\noindent A more in depth discussion of radial basis functions including an example of an orthonormalised basis set can be found in \cite{PhysRevB.87.184115}.

Finally, we demonstrate how one can calculate the derivatives of the bond order parameters with respect to the Cartesian coordinates. One needs to be able to calculate the derivatives in order to learn an energy function $\epsilon$ from its derivatives (forces). This can be achieved using an expression for regular solid harmonics \(R_{lm}\) in the Cartesian coordinates from \cite{varsalovic1989quantum}:

\begin{align}
R_{lm} &= \sqrt{\frac{4 \pi}{2 l + 1}} r^l Y_{lm} \nonumber \\
&= \sqrt{(l + m)! (l - m)!} \sum_{\substack{p, q, s \\ p + q + s = l \\ p - q = m}} \frac{1}{p! q! s!} \left(- \frac{x + i y}{2}\right)^p \left(\frac{x - i y}{2}\right)^q z^s,
\end{align}

\noindent where \(p\), \(q\) and \(s\) are all positive integers. Consequently, the spherical harmonics can be rewritten in terms of Cartesian coordinates as:

\begin{align}
Y_{lm} = & \sqrt{\frac{2 l + 1}{4 \pi}} \left(x^2 + y^2 + z^2\right)^{^{- l} / 2} \nonumber \\
& \sqrt{(l + m)! (l - m)!} \sum_{\substack{p, q, s \\ p + q + s = l \\ p - q = m}} \frac{1}{p! q! s!} \left(- \frac{x + i y}{2}\right)^p \left(\frac{x - i y}{2}\right)^q z^s,
\label{equation:spherical_harmonics}
\end{align}

\noindent and finding their derivatives with respect to the Cartesian coordinates is now equivalent to differentiating a polynomial. In order to find the derivatives of the bond order parameters with respect to the Cartesian coordinates, one needs to apply an appropriate coordinate transformation:

\begin{align}
\frac{\partial Q_{lm}^{(j)(k)}}{\partial x_\alpha^{(i)}} &= \frac{\partial Q_{lm}^{(j)(k)}(\mathbf{x}^{(k)} - \mathbf{x}^{(j)})}{\partial x_\alpha^{(i)}} = \frac{\partial Q_{lm}^{(j)(k)}(\mathbf{x}^{(k)} - \mathbf{x}^{(j)})}{\partial (x_\alpha^{(k)} - x_\alpha^{(j)})} \frac{\partial x_\alpha^{(k)} - x_\alpha^{(j)}}{\partial x_\alpha^{(i)}} \nonumber \\
&= \frac{\partial Q_{lm}^{(j)(k)}(\mathbf{x})}{\partial x_\alpha} \frac{\partial x_\alpha^{(k)} - x_\alpha^{(j)}}{\partial x_\alpha^{(i)}} = (\delta_{ik} - \delta_{ij}) \frac{\partial Q_{lm}^{(j)(k)}(\mathbf{x})}{\partial x_\alpha},
\label{equation:transformation}
\end{align}

\noindent and the problem simply becomes a matter of applying the chain rule a sufficient number of times.

\subsection{Power Spectrum and Bispectrum}

In the above analysis of bond-order parameters it is critical to remember that even as \(l\) approaches infinity the bond order parameters do not give a complete description of the atomic neighbourhood. Although in some situations they do provide sufficient information about the environment of the atom, their accuracy can be vastly improved by the application of representation theory concepts, as carried out by Bart\'{o}k-P\'{a}rtay, Kondor and Cs\'{a}nyi (more details in \cite{2010PhDT}, \cite{PhysRevLett.104.136403} and \cite{PhysRevB.87.184115}).

Using the bra-ket notation and Einstein summation convention for simplicity, we can expand the atomic density function $\rho$ in terms of a spherical harmonics basis:

\begin{equation}
| \rho \rangle = c_{lm} | Y_{lm} \rangle ,
\end{equation}

\noindent where the basis functions form an orthonormal set:

\begin{equation}
\langle Y_{l'm'} | Y_{lm} \rangle = \delta_{l'l} \delta_{m'm} ,
\end{equation}

\noindent and where the inner product is defined as:

\begin{equation}
\langle f | g \rangle = \int f^* g \sin \theta d \theta d \phi .
\end{equation}

An arbitrary rotation $\hat{R}$ transforms the spherical harmonics basis functions by expanding them into a linear combination of spherical harmonics with the same $l$ index:

\begin{equation}
\hat{R} | Y_{lm} \rangle = D_{m m'}^l (\hat{R}) | Y_{lm'} \rangle ,
\end{equation}

\noindent where $\mathbf{D}^l (\hat{R})$ is the Wigner matrix (more details in \cite{PhysRevB.87.184115}) and its elements can be computed by evaluating:

\begin{equation}
D_{m m'}^l (\hat{R}) = \langle Y_{lm'} | \hat{R} | Y_{lm} \rangle .
\end{equation}

Consequently, we can again expand the atomic density function $\rho$ under an arbitrary rotation $\hat{R}$:

\begin{equation}
\hat{R} | \rho \rangle = \hat{R} \left( c_{lm} | Y_{lm} \rangle \right) = c_{lm} \hat{R} | Y_{lm} \rangle = c_{lm} D_{m m'}^l (\hat{R}) | Y_{lm'} \rangle ,
\end{equation}

\noindent and so we can observe that under an arbitrary rotation $\hat{R}$, the vector $\mathbf{c}_{l}$ transforms according to:

\begin{equation}
\mathbf{c}_{l} \xrightarrow{\hat{R}} \mathbf{D}^l (\hat{R}) \mathbf{c}_{l} .
\end{equation}

Now, exploiting the property that Wigner matrices are unitary (more details in \cite{PhysRevB.87.184115}):

\begin{equation}
\left( \mathbf{D}^l \right)^* \mathbf{D}^l = \mathbf{I} ,
\end{equation}

\noindent the simplest rotationally invariant parameter is given by:

\begin{equation}
p_l = \mathbf{c}_{l}^* \mathbf{c}_{l} ,
\end{equation}

\noindent where we can immediately observe that its transformation under an arbitrary rotation $\hat{R}$ is given by:

\begin{equation}
\mathbf{c}_{l}^* \mathbf{c}_{l} \xrightarrow{\hat{R}} \mathbf{c}_{l}^* \left( \mathbf{D}^l \right)^* \mathbf{D}^l \mathbf{c}_{l} = \mathbf{c}_{l}^* \mathbf{c}_{l} .
\end{equation}

\noindent We refer to $p_l$ as the rotational power spectrum.

A finite set of rotational power spectrum parameters is unlikely to provide a complete description of the atomic density function $\rho$ (in fact it is far from being complete) but the same formalism can be applied to couple multiple angular momentum channels and therefore obtain a larger, more complete set of rotationally invariant parameters. Again, following the work of Bart\'{o}k-P\'{a}rtay, Kondor and Cs\'{a}nyi (more details in \cite{2010PhDT}, \cite{PhysRevLett.104.136403} and \cite{PhysRevB.87.184115}), one can define a tensor $\mathbf{g}_{l_1 l_2 l}$:

\begin{equation}
\bigoplus_{l = | l_1 - l_2 |}^{l_1 + l_2} \mathbf{g}_{l_1 l_2 l} = \mathbf{C}^{l_1 l_2} \left( \mathbf{c}_{l_1} \otimes \mathbf{c}_{l_2} \right) ,
\label{equation:bispectrum}
\end{equation}

\noindent where $\mathbf{c}_{l_1} \otimes \mathbf{c}_{l_2}$ is the direct product of $\mathbf{c}_{l_1}$ and $\mathbf{c}_{l_2}$, and $\mathbf{C}^{l_1 l_2}$ is a tensor of Clebsch-Gordan coefficients (which can be thought of as coupling constants). By construction, tensor $\mathbf{g}_{l_1 l_2 l}$ transforms under an arbitrary rotation $\hat{R}$ according to:

\begin{equation}
\mathbf{g}_{l_1 l_2 l} \xrightarrow{\hat{R}} \mathbf{D}^l (\hat{R}) \mathbf{g}_{l_1 l_2 l} ,
\end{equation}

\noindent and consequently we can construct a parameter of the next order, that couples multiple angular momentum channels:

\begin{equation}
b_{l \, l_1 l_2} = \mathbf{c}_{l}^* \mathbf{g}_{l_1 l_2 l} .
\end{equation}

\noindent We refer to $b_{l \, l_1 l_2}$ as the rotational bispectrum. It is trivial to show that it is invariant under an arbitrary rotation $\hat{R}$:

\begin{equation}
\mathbf{c}_{l}^* \mathbf{g}_{l_1 l_2 l} \xrightarrow{\hat{R}} \mathbf{c}_{l}^* \left( \mathbf{D}^l \right)^* \mathbf{D}^l \mathbf{g}_{l_1 l_2 l} = \mathbf{c}_{l}^* \mathbf{g}_{l_1 l_2 l} .
\end{equation}

Finally, we can rewrite the bispectrum formula in terms of Clebsch-Gordan coefficients:

\begin{equation}
b_{l \, l_1 l_2} = c_{lm}^* C_{m m_1 m_2}^{l \, l_1 l_2} c_{l_1 m_1} c_{l_2 m_2} ,
\end{equation}

\noindent from which it becomes apparent that Steinhardt bond-order parameters, or the rotational power spectrum, are in fact a subset of rotational bispectrum parameters:

\begin{equation}
p_l \propto \left( Q_l \right)^2 ,
\end{equation}

\begin{equation}
b_{l \, 0 \, l} = N c_{lm}^* \delta_{m m_2} c_{lm_2} = N c_{lm}^* c_{lm} \propto p_l \propto \left( Q_l \right)^2 ,
\end{equation}

\begin{equation}
b_{l \, l \, l} = c_{lm}^* C_{m \, m_1 m_2}^{l \, l \, l} c_{l m_1} c_{l m_2} = (-1)^m C_{m \, m_1 m_2}^{l \, l \, l} c_{lm} c_{l m_1} c_{l m_2} \propto W_l ,
\end{equation}

\noindent since the Clebsch-Gordan coefficients are related to the Wigner \(3j\) symbol by:

\begin{equation}
\begin{pmatrix} l_1 & l_2 & l_3 \\ m_1 & m_2 & m_3 \end{pmatrix} = \frac{(-1)^{l_1 - l_2 - m_3}}{\sqrt{2 l_3 + 1}} C_{m_1 m_2 -m_3}^{l_1 l_2 l_3} ,
\end{equation}

\noindent and:

\begin{equation}
c_{lm} = (-1)^m c_{lm}^* .
\end{equation}

The rotational bispectrum parameters can be expanded to include radial information using a treatment analogous to that presented with bond-order parameters in the previous section. Consequently, the expansion of the atomic density function $\rho$ becomes:

\begin{equation}
| \rho \rangle = c_{nlm} | w_n , Y_{lm} \rangle .
\end{equation}

\noindent If the radial basis is orthonormal, we exploit the property:

\begin{equation}
\langle w_{n'} , Y_{l'm'} | w_{n} , Y_{lm} \rangle = \delta_{n' n} \delta_{l'l} \delta_{m'm} ,
\end{equation}

\noindent and the rotational power spectrum and bispectrum parameters become:

\begin{equation}
p_{n \, l} = c_{nlm}^* c_{nlm} ,
\end{equation}

\begin{equation}
b_{n \, l \, l_1 l_2} = c_{nlm}^* C_{m m_1 m_2}^{l \, l_1 l_2} c_{n \, l_1 m_1} c_{n \, l_2 m_2} .
\end{equation}

\noindent A more in depth discussion of radial basis functions, including discussion of the treatment when the radial basis functions are not orthogonal, is given in \cite{PhysRevB.87.184115}.

As in the case of the bond-order parameters, if our rotational bispectrum parameters are to remain continuous we require the radial basis set $\{w_n\}$ to be continuous and differentiable and to decay to zero at some finite cutoff. It is also important to ensure that individual radial basis functions are sufficiently coupled --- having weakly coupled functions without sufficient overlap can lead to unphysical rotational invariance of rotating subsets of atoms occupying shells at similar distance from the origin. 

One way of ensuring radial basis coupling is through the selection of basis functions that cover a wide range of distances --- this approach, however, has the disadvantage of reducing sensitivity to radial information at a specific distance (usually selected to correspond to the distance of a nearest neighbour shell). Different radial channels can be also coupled explicitly, although at the cost of increasing the number of invariant parameters:

\begin{equation}
p_{n_1 n_2 \, l} = c_{n_1lm}^* c_{n_2lm} ,
\end{equation}

\begin{equation}
b_{n \, n_1 n_2 l \, l_1 l_2} = c_{n l m}^* C_{m m_1 m_2}^{l \, l_1 l_2} c_{n_1 l_1 m_1} c_{n_2 l_2 m_2} .
\end{equation}

Finally, we note that, as in the case of bond-order parameters, elements $c_{lm}$ transform under the reflection about the origin as:

\begin{equation}
c_{lm} \xrightarrow{\text{reflection}} (-1)^l c_{lm} .
\end{equation}

\noindent Consequently in order to impose reflection symmetry we compute absolute value of the elements with odd values of $l$ or we skip them altogether.

\subsection{4-dimensional Bispectrum}

An alternative method of including radial information in rotational bispectrum parameters has been suggested by Bart\'{o}k-P\'{a}rtay, Kondor and Cs\'{a}nyi (more details in \cite{2010PhDT}, \cite{PhysRevLett.104.136403} and \cite{PhysRevB.87.184115}). This does not require explicit introduction of a radial basis set but it still provides representation of a three-dimensional atomic density function. One can define a four-dimensional sphere $S^3$ with radius $r_0$, where the surface is defined as a set of points $\mathbf{s} \in \mathbb{R}$, such that:

\begin{equation}
s_1^2 + s_2^2 + s_3^2 + s_4^2 = r_0^2 ,
\end{equation}

\noindent and the polar angles $\phi$, $\theta$ and $\theta_0$ are defined as:

\begin{align}
s_1 &= r_0 \cos \theta_0 \nonumber \\
s_2 &= r_0 \sin \theta_0 \cos \theta \nonumber \\
s_3 &= r_0 \sin \theta_0 \sin \theta \cos \phi \nonumber \\
s_4 &= r_0 \sin \theta_0 \sin \theta \sin \phi .
\end{align}

\noindent We use the projection from three-dimensional space onto the surface of a four-dimensional sphere defined by:

\begin{equation}
\begin{pmatrix} x \\ y \\ z \end{pmatrix} \to \begin{pmatrix} \phi &=& \arctan \left(\frac{y}{x}\right) \\ \theta &=& \arccos \left(\frac{z}{r}\right) \\ \theta_0 &=& \pi \frac{r}{r_0} \end{pmatrix} .
\end{equation}

An arbitrary density function $\rho$ can now be expanded on the surface of a four-dimensional sphere in terms of a (four-dimensional) hyper-spherical harmonics basis:

\begin{equation}
| \rho \rangle = c_{m' m}^j | U_{m' m}^j \rangle ,
\end{equation}

\noindent where the basis functions form an orthonormal set:

\begin{equation}
\langle U_{m_1' m_1}^{j_1} | U_{m_2' m_2}^{j_2} \rangle = \delta_{j_1 j_2} \delta_{m_1' m_2'} \delta_{m_1 m_2} ,
\end{equation}

\noindent and where the inner product is defined as:

\begin{equation}
\langle f | g \rangle = \int f^* g \sin^2 \theta_0 d \theta_0 \sin \theta d \theta d \phi .
\end{equation}

The reminder of the analysis is analogous to that for the three-dimensional bispectrum, with Wigner matrices $\mathbf{D}^l (\hat{R})$ having four-dimensional equivalents (that are also unitary):

\begin{equation}
D_{m m'}^l (\hat{R}) \to R_{m_1' m_1 m_2' m_2}^j (\hat{R}) = \langle U_{m_1' m_1}^{j_1} | \hat{R} | U_{m_2' m_2}^{j_2} \rangle .
\end{equation}

\noindent The four-dimensional equivalents of Clebsch-Gordan coefficients can be expressed in terms of the three-dimensional one:

\begin{equation}
C_{m \, m_1 m_2}^{l \, l_1 l_2} \to H_{j_1 m_1 m_1' j_2 m_2 m_2'}^{j \, m \, m'} = C_{m \, m_1 m_2}^{j \, j_1 j_2} C_{m' \, m_1' m_2'}^{j \, j_1 j_2} ,
\end{equation}

\noindent and the four-dimensional analogue of the equation \ref{equation:bispectrum} is given by:

\begin{equation}
\bigoplus_{j = | j_1 - j_2 |}^{j_1 + j_2} \mathbf{g}_{j_1 j_2 j} = \mathbf{H}^{j_1 j_2} \left( \mathbf{c}_{j_1} \otimes \mathbf{c}_{j_2} \right) ,
\end{equation}

Consequently, the four-dimensional equivalents of the rotational power spectrum and bispectrum are given by:

\begin{equation}
p_l = \mathbf{c}_{j}^* \mathbf{c}_{j} ,
\end{equation}

\begin{equation}
b_{j \, j_1 j_2} = \mathbf{c}_{j}^* \mathbf{g}_{j_1 j_2 j} = \left(c_{m' m}^j\right)^* C_{m \, m_1 m_2}^{j \, j_1 j_2} C_{m' \, m_1' m_2'}^{j \, j_1 j_2} c_{m_1' m_1}^{j_1} c_{m_2' m_2}^{j_2} .
\end{equation}

Finally, in order to eliminate the invariance with respect to the third polar angle (which corresponds to translational invariance with respect to the origin), we can modify the atomic density function by the addition of a Dirac-delta function corresponding to the central atom as a fixed reference point at $(0,0,0)$:

\begin{equation}
\rho \to \rho' = \delta(\mathbf{0}) + \rho .
\end{equation}

\noindent The four-dimensional rotational bispectrum components corresponding to half-valued $j_1 + j_2 + j$ again correspond to terms that change their sign under reflection, and consequently we either skip them or take absolute values of them in order to enforce reflection symmetry.

It is also worth noticing that the four-dimensional rotational bispectrum parameters have only three indices while containing both angular and radial information (unlike the three-dimensional case where the radial basis introduces a fourth index). There is also no ambiguity in selecting an appropriate radial basis and the only adjustable parameter is that of $r_0$. Consequently, the four-dimensional version of the rotational bispectrum provides much more elegant solution to the descriptor problem in the context of GAP.

\subsection{Descriptors and Invariance of Covariance Function}

So far in our analysis we have considered the problem of potential energy surface fitting (outlined in sections \ref{chapter:gaussian_approximation_potential:section:gaussian_process_regression} and \ref{chapter:gaussian_approximation_potential:section:interatomic_potential}), and the problem of finding a faithful representation of an atomic environment (outlined in this section so far) completely independently. However, by simplifying equation \ref{equation:mean} into:

\begin{equation}
\epsilon_i = \epsilon(\mathbf{q}^{(i)}) = \sum_{j = 1}^M \alpha_j K(\mathbf{q}^{(i)}, \mathbf{q}^{(j)}) ,
\end{equation}

\noindent where $K$ is the covariance function, $\{\alpha_j\}_{j = 1}^M$ are the coefficients determined by the Gaussian Process regression fitting procedure, $\{\mathbf{q}_j\}_{j = 1}^M$ are the descriptor coordinates of the training data set and $\mathbf{q}_i$ are the descriptor coordinates of the environment of atom $i$, obtained by a mapping:

\begin{equation}
\mathbf{q}^{(i)} = \mathbf{q}(\{\mathbf{x}^{(j)} - \mathbf{x}^{(i)}\}_{j = 1}^N) ,
\end{equation}

\noindent we should realise that it is not the choice of a descriptor mapping that is fundamental for the purpose of potential energy surface fitting but, the choice of the covariance function that is constructed from the descriptors that is of critical importance. In fact, one can incorporate the descriptor mapping inside the similarity measure $K$ directly and bypass the idea of a descriptor altogether:

\begin{equation}
\begin{Bmatrix} K(\mathbf{q}, \mathbf{q}') \\ \mathbf{q}(\{\mathbf{x}^{(j)} - \mathbf{x}\}_{j = 1}^N) \end{Bmatrix} \to K'(\{\mathbf{x}^{(j)} - \mathbf{x}\}_{j = 1}^N, \{\mathbf{x}^{(j)} - \mathbf{x}'\}_{j = 1}^{N'}) .
\end{equation}

\noindent This approach not only gives a better control of the symmetries built inside the covariance function but also provides a controlled and systematic way of ensuring that the covariance function changes smoothly with the Cartesian coordinates.

\subsection{Smooth Overlap of Atomic Positions (SOAP)}

The similarity of two atomic environments can be defined as the overlap between their corresponding atomic density functions $\rho$ and $\rho'$ computed according to:

\begin{equation}
S(\rho, \rho') = \int \rho(\mathbf{x}) \rho'(\mathbf{x}) d \mathbf{x}.
\end{equation}

\noindent Consequently, one can propose a similarity kernel (more details in \cite{PhysRevB.87.184115}):

\begin{equation}
k(\rho, \rho') = \int \left| S(\rho, \hat{R} \rho') \right|^n d \hat{R},
\label{equation:kernel}
\end{equation}

\noindent where we integrate a simple function of the overlap of two atomic environments over all possible rotations defined by operator $\hat{R}$. While it is easy to see that integrating over all arbitrary rotations ensures rotational invariance, the definition of the atomic density function from equation \ref{equation:density} clearly satisfies the permutational invariance as the ordering of the elements in the sum does not matter.

However, in the context of computing an atomic density overlap, retaining the definition of the atomic density expressed as a sum of Dirac-delta functions is extremely impractical. It is not an efficient method of capturing the similarity of two atomic environments with atomic positions that are very close to each other but not identical. Furthermore, it would result in a kernel $k(\rho, \rho')$ that is both discontinuous and non-differentiable. Consequently, we modify the equation \ref{equation:density} by expanding the atomic densities in terms of three-dimensional Gaussian functions instead:

\begin{equation}
\rho_i = \rho(\{\mathbf{x}^{(j)} - \mathbf{x}^{(i)}\}_{j = 1}^N) = \sum_{j = 1}^N \alpha_j \exp \left(- \frac{ | \mathbf{x} - (\mathbf{x}^{(j)} - \mathbf{x}^{(i)}) |^2 }{2 \theta_j^2}\right) ,
\end{equation}

\noindent where the Dirac-delta function result can be recovered in the limit as $\theta_j \to 0$, and $\theta_j$ (the width of Gaussians corresponding to atomic species $j$) can be used to control the smoothness of the kernel $k(\rho, \rho')$ corresponding to the change in Cartesian coordinates of the atomic positions.

The obvious difficulty in evaluating the kernel $k(\rho, \rho')$ is performing the integration over all possible rotations $\hat{R}$ analytically. However, this can be achieved by expanding the Gaussian functions using a spherical harmonics basis:

\begin{equation}
\exp \left( - \frac{ | \mathbf{x}_1 - \mathbf{x}_2 |^2 }{2 \theta^2} \right) = 4 \pi \left( - \frac{r_1^2 + r_2^2}{2 \theta^2} \right) \sum_{lm} i_l \left( \frac{r_1 r_1}{\theta^2} \right) Y_{lm} (\mathbf{\hat{r}}_1) Y_{lm}^* (\mathbf{\hat{r}}_2) ,
\end{equation}

\noindent where $i_l$ are the modified spherical Bessel functions of the first kind. Consequently, the atomic density function can be expanded as:

\begin{equation}
\rho_i = \sum_j^N \sum_{lm} c_{lm}^j (r) Y_{lm} (\mathbf{\hat{r}}) ,
\label{equation:expansion2}
\end{equation}

\noindent where:

\begin{equation}
c_{lm}^j (r) = 4 \pi \exp \left( - \frac{r^2 + r_{ij}^2}{2 \theta_j^2} \right) i_l \left( \frac{r r_{ij}}{\theta_j^2} \right) Y_{lm}^* (\mathbf{\hat{r}}_{ij}) .
\end{equation}

Exploiting the property that an arbitrary rotation $\hat{R}$ transforms the spherical harmonics basis functions in terms of a linear combination of spherical harmonics with the same index $l$ and expansion coefficients given by the Wigner matrix $\mathbf{D}^l (\hat{R})$ (as outlined in the previous section), the overlap of two atomic environments subject to an arbitrary rotation $\hat{R}$ is given by:

\begin{align}
S(\rho, \hat{R} \rho') &= \sum_{i \, i'} \sum_{\substack{l \, m \\ l' m' m''}} D_{m' m''}^{l'} (\hat{R}) \int \left( c_{lm}^i (r) \right)^* c_{l' m'}^{i'} (r) dr \nonumber \\
&\times \int \left( Y_{lm} (\theta, \phi) \right)^* Y_{l' m''} (\theta, \phi) \sin \theta \, d \theta \, d \phi \nonumber \\
&= \sum_{i \, i'} \sum_{l \, m \, m'} \tilde{I}_{l \, m \, m'}^{i \, i'} D_{m' m''}^{l'} (\hat{R}) \nonumber \\
&= \sum_{l \, m \, m'} I_{l \, m \, m'} D_{m' m''}^{l'} (\hat{R}) ,
\end{align}

\noindent where:

\begin{equation}
\tilde{I}_{l \, m \, m'}^{i \, i'} = 4 \pi \exp \left( - \frac{r_{i}^2 + r_{i'}^2}{4 \theta^2} \right) i_l \left( \frac{r_{i} r_{i'}}{\theta^2} \right) Y_{lm} (\mathbf{\hat{r}}_{i}) Y_{lm'}^* (\mathbf{\hat{r}}_{i'}) ,
\end{equation}

\noindent and:

\begin{equation}
I_{l \, m \, m'} = \sum_{i \, i'} \tilde{I}_{l \, m \, m'}^{i \, i'} .
\end{equation}

In order to evaluate the rotationally invariant kernel $k(\rho, \rho')$ we rely on the property that the direct product of two Wigner matrices can be decomposed in terms of a direct sum of Wigner matrices and Clebsch-Gordan coefficients (more details in \cite{PhysRevB.87.184115}). Consequently, by combining the above result with equation \ref{equation:kernel} for $n = 2$ we obtain (we ignore the case of $n = 1$ as we recognise that for $n = 1$ the order of integration can be exchanged and therefore no angular information is included):

\begin{align}
k(\rho, \rho') |_{n = 2} &= \int S^* (\rho, \hat{R} \rho') S (\rho, \hat{R} \rho') d \hat{R} \nonumber \\
&= \sum_{\substack{l \, m \, m' \\ \lambda \, \mu \, \mu'}} I_{l \, m \, m'}^* I_{\lambda \, \mu \, \mu'} \int \left( D_{m m'}^{l} (\hat{R}) \right)^* D_{\mu \mu'}^{\lambda} (\hat{R}) d \hat{R} \nonumber \\
&= \sum_{l \, m \, m'} I_{l \, m \, m'}^* I_{l \, m \, m'} .
\end{align}

\noindent An analogous result for $n = 3$ is given by:

\begin{equation}
k(\rho, \rho') |_{n = 3} = \sum_{\substack{l \, m \, m' \\ l_1 m_1 m_1' \\ l_2 m_2 m_2'}} I_{l \, m \, m'}^* C_{l_1 m_1 l_2 m_2}^{l \, m} C_{l_1 m_1' l_2 m_2'}^{l \, m'} I_{l_1 m_1 m_1'} I_{l_2 m_2 m_2'} ,
\end{equation}

\noindent where conceptual similarities to the rotational power spectrum and the bispectrum should already become obvious.

In practical terms, computation of elements $I_{l \, m \, m'}$ involves summation of $\tilde{I}_{l \, m \, m'}^{i \, i'}$ terms over all possible pairs of atoms $i$ and $i'$. This becomes an increasingly computationally intensive task in situations where the central atom is surrounded by a large number of neighbours. To overcome this problem the atomic density function can be expanded using radial basis functions instead:

\begin{align}
\rho_i &= \sum_{j = 1}^N \alpha_j \exp \left(- \frac{ | \mathbf{x} - (\mathbf{x}^{(j)} - \mathbf{x}^{(i)}) |^2 }{2 \theta_j^2}\right) \nonumber \\
&= \sum_{nlm} c_{nlm} w_n (r) Y_{lm} (\mathbf{\hat{r}}) ,
\end{align}

\noindent which eliminates the summation over neighbouring atoms from equation \ref{equation:expansion2}. If the radial basis is orthonormal we observe that:

\begin{equation}
\int w_{n_1}^* (r) w_{n_2} (r) dr = \delta_{n_1 n_2} ,
\end{equation}

\noindent and the terms $I_{l \, m \, m'}$ become:

\begin{equation}
I_{l \, m \, m'} = \sum_n c_{nlm} \left( c_{nlm'}' \right)^* .
\end{equation}

\noindent Substituting the above result into $k(\rho, \rho') |_{n = 2}$ and $k(\rho, \rho') |_{n = 3}$, we obtain:

\begin{equation}
k(\rho, \rho') |_{n = 2} = \sum_{n_1 n_2 l \, m \, m'} c_{n_1 l m} \left( c_{n_1 l m'}' \right)^* \left( c_{n_2 l m} \right)^* c_{n_2 l m} = \sum_{n_1 n_2 l} p_{n_1 n_2 l} p_{n_1 n_2 l}' ,
\end{equation}

\noindent where $p_{n_1 n_2 l}$ is the rotational power spectrum defined in the previous section and:

\begin{equation}
k(\rho, \rho') |_{n = 3} = \sum_{\substack{n \, n_1 n_2 \\ l \, l_1 l_2}} b_{n \, n_1 n_2 l \, l_1 l_2} b_{n \, n_1 n_2 l \, l_1 l_2}' ,
\end{equation}

\noindent where $b_{n \, n_1 n_2 l \, l_1 l_2}$ is the rotational bispectrum defined in the previous section. Consequently, we can recognise that a Smooth Overlap of Atomic Positions kernel is equivalent to a three-dimensional rotational power spectrum and bispectrum generated by Gaussian atomic density functions and a dot-product covariance function.

Finally, the final form of the SOAP covariance function is obtained by scaling it by a normalising factor (as suggested in \cite{rasmussen2006gaussian}), and raising it to a positive power $\zeta \geq 2$: 

\begin{equation}
K(\rho, \rho') = \left( \frac{k(\rho, \rho')}{\sqrt{k(\rho, \rho)} \sqrt{k(\rho', \rho')}} \right)^{\zeta} .
\end{equation}

\noindent This increases the sensitivity of the covariance function to pairs of atomic environments with significant overlap.

\section{Implementation}
\label{chapter:gaussian_approximation_potential:section:implementation}

For the purpose of this work we use an implementation of the Gaussian process regression developed within the \texttt{libAtoms} \cite{libatoms} software library for the purpose of carrying out molecular dynamics simulations, for which author of this work is one of the contributors. The implementation includes all the necessary modifications, as outlined in section \ref{chapter:gaussian_approximation_potential:section:gaussian_process_regression} in order to fit the atomic energy function \(\epsilon^{(GAP)}\) from the data consisting of total energies, atomic forces and stress virials as input.

Since the amount of noise present in the energy, force and/or stress viral observations usually differs significantly, the computation of the covariance matrix has been modified such that each observation can correspond to an independent value of the noise parameter:

\begin{equation}
\sigma_\nu \to \left\{ \begin{array}{l} \sigma_\nu^{(\text{energy})} N \\ \sigma_\nu^{(\text{force})} \\ \sigma_\nu^{(\text{virial})} N \end{array} \right.
\end{equation}

\noindent where $N$ corresponds to the number of atoms in the simulation cell in the total energy and stress virial calculations.

Throughout this work we investigate the problem of the atomic energy function \(\epsilon^{(GAP)}\) fitting using a number of descriptors (as outlined in section \ref{chapter:gaussian_approximation_potential:section:description_of_atomic_environments}). However, since the bond-order parameters and the rotational power spectrum constitute a subset of the bispectrum parameters, we only need to outline the cases of bispectrum and SOAP below.

For the bispectrum descriptor we use a square-exponential covariance function, defined as:

\begin{equation}
k(\mathbf{q}^{(i)}, \mathbf{q}^{(j)}) = \exp \left( - \frac{(\mathbf{q}^{(i)} - \mathbf{q}^{(j)})^2}{2 \theta^2} \right) .
\end{equation}

\noindent However, this form assumes that the characteristic length-scale of the bispectrum phase space is isotropic, which is often not the case. Consequently, an anisotropic version of square-exponential covariance is used instead:

\begin{equation}
k(\mathbf{q}^{(i)}, \mathbf{q}^{(j)}) = \exp \left( - \frac{1}{2} (\mathbf{q}^{(i)} - \mathbf{q}^{(j)})^T \mathbf{\Sigma} (\mathbf{q}^{(i)} - \mathbf{q}^{(j)}) \right) ,
\end{equation}

\noindent where $\mathbf{\Sigma}$ is a diagonal matrix of hyperparameters with each element of the diagonal corresponding to a different characteristic length-scale of the appropriate dimension of the bispectrum:

\begin{equation}
\theta \to \mathbf{\Sigma} = \begin{pmatrix} \frac{1}{\theta_1^2} & & \\ & \frac{1}{\theta_2^2} & \\ & & \ddots \end{pmatrix} .
\end{equation}

\noindent The corresponding basis functions are given by:

\begin{equation}
\phi_\mathbf{h}(\mathbf{q}^{(i)}) = \exp\left(- (\mathbf{q}^{(i)} - \mathbf{h})^T \mathbf{\Sigma} (\mathbf{q}^{(i)} - \mathbf{h}) \right) .
\end{equation}

In the case of Smooth Overlap of Atomic Positions we use polynomial covariance:

\begin{equation}
k(\mathbf{q}^{(i)}, \mathbf{q}^{(j)}) = \left( \mathbf{q}^{(i)} \cdot \mathbf{q}^{(j)} \right)^{\zeta} ,
\end{equation}

\noindent which has no adjustable parameters that correspond to the characteristic length-scale of the descriptor, but $\theta$ in this case is related to the width of Gaussian functions representing atoms in the atomic density function instead and it can be adjusted for systematic control of covariance smoothness. Additionally, one can tune the sensitivity of the covariance function using a new hyperparameter $\zeta$.

As demonstrated in \cite{rasmussen2006gaussian} polynomial covariance of degree $\zeta$ corresponds to polynomial basis functions of the form:

\begin{equation}
\phi_\mathbf{h}(\mathbf{q}^{(i)}) = \boldsymbol{\varphi}(\mathbf{h}) \cdot \boldsymbol{\varphi}(\mathbf{q}^{(i)}) ,
\end{equation}

\noindent where the vector $\boldsymbol{\varphi}(\mathbf{q})$ is constructed from the vector $\mathbf{q}$ according to:

\begin{equation}
\boldsymbol{\varphi}(\mathbf{q}) = \begin{Bmatrix} \sqrt{\frac{\zeta !}{m_1 ! \dots m_D !}} q_1^{m_1} \dots q_D^{m_D} \\ \sqrt{\frac{\zeta !}{m_1 ! \dots m_D !}} q_1^{m_1} \dots q_D^{m_D} \\ \vdots \end{Bmatrix} ,
\end{equation}

\noindent for all possible combinations of $\{m_i\}_{i = 1}^D$ such that $\sum_{i = 1}^D m_i = \zeta$ where $D$ corresponds to the dimensionality of vector $\mathbf{q}$ and vector elements $m_i$ specify the degree of the polynomial.

Finally, our implementation uses the sparsification scheme based on pseudo-inputs as outlined in section \ref{chapter:gaussian_approximation_potential:section:sparsification}. This effectively allows for deconvolution of teaching information during the fitting process as the weight assigned to the sparse pseudo-input point with atomic environment $\mathbf{q}$ corresponds to the atomic energy $\epsilon^{(GAP)}(\mathbf{q})$ and not to the linear combination of atomic environments (as in the case of fitting from total energies) or a linear combination of atomic environments and their derivatives (as in the case of fitting from forces and stress virials). This property has an extremely favourable effect on the computational cost of evaluating atomic energies and their derivatives with the GAP interatomic potential.

\cleardoublepage

\newgeometry{top=2.75cm,bottom=2.75cm,left=3.5cm,right=3cm} 

\chapter{Bulk Properties and Lattice Defects in Tungsten}
\label{chapter:bulk_properties_and_lattice_defects_in_tungsten}

\section{Introduction}
\label{chapter:of_bulk_properties_and_lattice_defects_in_tungsten:section:introduction}

All the methodology outlined so far in chapters \ref{chapter:classical_and_quantum_simulation_of_solids}, \ref{chapter:simulation_techniques} and \ref{chapter:gaussian_approximation_potential} has been very general, and can be universally applied to the simulation of any class of solids. It also serves as a review of the existing research available in the literature.

In this chapter I focus on the details of simulating properties of tungsten --- a transition metal that was selected as a ``testing ground'' for our GAP potential for metals. It marks the beginning of the second part of this thesis where I give an account of my own work which starts with preliminary calculations of tungsten properties with the existing, well established models and associated testing for convergence of these results.

I begin in section \ref{chapter:bulk_properties_and_lattice_defects_in_tungsten:section:bcc_lattice} with a brief outline of the basic properties of tungsten, focusing on the features that are of particular interest from the perspective of developing an interatomic potential. We follow, in section \ref{chapter:bulk_properties_and_lattice_defects_in_tungsten:section:convergence_of_dft_calculations}, by outlining how quantum-mechanical methods such as density functional theory can be employed to predict these properties and what precautions need to be taken to ensure convergence of the results. Finally, I demonstrate the relevant methods and techniques employed to compute bulk properties, such as elastic constants or phonon spectrum, and various lattice defects. Whenever appropriate, I present the results of these calculations for tungsten using both classical and quantum-mechanical models.

\section{BCC Lattice}
\label{chapter:bulk_properties_and_lattice_defects_in_tungsten:section:bcc_lattice}

The body-centred cubic (bcc) structure is a very common crystal structure in nature. Examples of metals that naturally form bcc crystals include iron, chromium, molybdenum, tungsten, vanadium, niobium and tantalum and their technological prominence is well established. They have been extensively used by humankind since the Iron Age. While there are some considerable differences among these metals --- description of iron, for example, is extremely complicated with at least four allotropic forms and complex magnetic behaviour --- it is, nevertheless, established that some of these properties, such as plasticity in particular, are generic among most bcc metals and can be attributed to the common lattice crystallography.

Unlike face-centred cubic (fcc) or hexagonal close-packed (hcp) structures, in bcc crystals there are no truly close packed planes. Slip can occur in the direction of the shortest Burgers vector $\langle 111 \rangle$ which contains the nearest neighbour. In principle any plane containing a $\langle 111 \rangle$ direction can be a potential slip plane. In practice, however, heat is required to overcome the activation energy for slip to occur and the activation barrier usually correlates closely with how densely constituent atoms are packed within the slip plane.

The most densely packed planes of the $\langle 111 \rangle$ zone are the $\{110\}$ planes. There are six $\{110\}$ slip planes each with two possible $\langle 111 \rangle$ directions giving 12 possible slip systems in total. The second most densely packed slip planes are the $\{112\}$ planes forming another 12 possible slip systems and their activation energies are usually close to those of the $\{110\}$ planes. There are also a further 24 $\langle 111 \rangle \{123\}$ slip systems and going even further yet another 24 $\langle 111 \rangle \{134\}$ slip systems. These are all, however, significantly less densely packed and consequently they do not play an important role in the description of plasticity in bcc systems. The separation of the consecutive planes (in terms of a conventional cell lattice constant $a$) of the slip systems mentioned above are given in table \ref{table:plane_distance} below:

\begin{table}[H]
\vspace{0.5cm}
\begin{center}
\begin{tabular}{ c c c }
\midrule
$\langle 111 \rangle \{110\}$ & $\to$ & $\frac{1}{\sqrt{2}} \, a$ \\
\midrule
$\langle 111 \rangle \{112\}$ & $\to$ & $\frac{1}{\sqrt{6}} \, a$ \\
\midrule
$\langle 111 \rangle \{123\}$ & $\to$ & $\frac{1}{\sqrt{14}} \, a$ \\
\midrule
$\langle 111 \rangle \{134\}$ & $\to$ & $\frac{1}{\sqrt{26}} \, a$ \\
\midrule
\end{tabular}
\caption[Separation distance of slip planes in bcc $\langle 111 \rangle$ zone.]{Separation distance of high symmetry slip planes in the bcc $\langle 111 \rangle$ zone.}
\label{table:plane_distance}
\end{center}
\end{table}

\noindent It can be easily computed for any other arbitrary plane $\{hkl\}$ in a cubic system with lattice constant $a$ using the equation:

\begin{equation}
\frac{1}{d^2} = \frac{h^2 + k^2 + l^2}{a^2} .
\end{equation}

This ordering of slip systems in terms of physical significance should not come as a surprise. Physical intuition dictates that as the separation of the planes decreases the activation energy of a slip system is anticipated to increase in order to overcome the electrostatic repulsion of atoms in the neighbouring planes as they get close to each other. A diagram representing the $\langle 111 \rangle$ zone with the most common slip systems of physical significance in bcc systems is given in figure \ref{figure:slip_systems} below.

\begin{figure}[H]
\begin{center}
\vspace{0.25cm}
\resizebox{12cm}{!}{\footnotesize{}\input{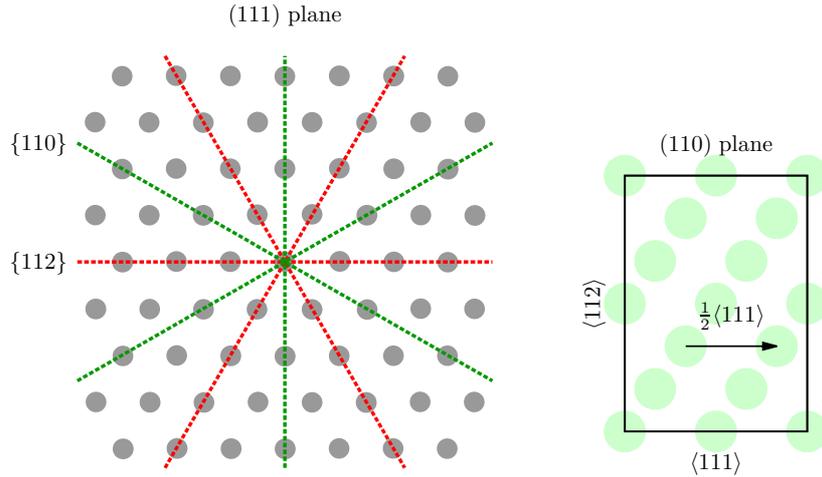}\normalsize{}}
\caption{Slip systems of bcc $\langle 111 \rangle$ zone.}
\label{figure:slip_systems}
\end{center}
\end{figure}

\noindent The above analysis suggests that in our investigation of the properties of tungsten we should pay particular attention to the high symmetry crystallographic directions of $\langle 110 \rangle$, $\langle 111 \rangle$ and $\langle 112 \rangle$ and the corresponding planes.

\section{Classical and \emph{Ab Initio} Calculations}
\label{chapter:bulk_properties_and_lattice_defects_in_tungsten:section:convergence_of_dft_calculations}

In the past, interatomic potentials have been almost exclusively employed to model bcc metals. Even today, in simulations involving more than a few hundreds of atoms one is still limited to a Finnis-Sinclair potential (more detail in \cite{doi:10.1080/01418618408244210}), the embedded-atom method (more details in \cite{PhysRevB.29.6443}) or more recently developed bond-order potentials (more details in \cite{PhysRevB.69.094115}, \cite{PhysRevB.75.104119} and \cite{PhysRevB.74.174117}). Unfortunately, all of these potentials have their shortcomings. While Finnis-Sinclair potentials and the embedded-atom method are computationally simple to evaluate and they often provide good qualitative description of the system, neither of them is capable of reproducing the properties of crystal defects to the quantum-mechanical degree of accuracy in quantitative terms (more details in \cite{PhysRevB.75.104119}).

Some of the limitations of the second-moment EAM- or FS-type interatomic potentials have been overcome by deriving the analytic form of the potential directly using perturbation theory with respect to the underlying electronic structure, as in the generalised perturbation theory (GPT) potentials developed by Moriarty (more details in \cite{PhysRevB.38.3199} and \cite{PhysRevB.42.1609}). These potentials have been very successful in modelling the behaviour of period four and five transition metals (more details in \cite{PhysRevB.49.12431} and \cite{0953-8984-14-11-305}). At the same time, although the bond-order potential improves significantly on the accuracy of EAM and FS, it is computationally much more complex and in its non-analytic form it is only capable of computing forces that approximate the derivatives of the total energy of the system (as outlined in section \ref{chapter:classical_and_quantum_simulation_of_solids:section:interatomic_potentials:bond_order_potential}, unless analytic BOP is used the Hellmann-Feynman forces only become exact as the bond orders converge to their exact values).

More recently, various quantum-mechanical schemes have been used to compute the properties of bcc metals. However, these investigations were limited by the system size and therefore they could not reproduce the large-scale phenomena that directly influence plasticity behaviour such as dislocation glide or brittle fracture (for details see \cite{Woodward200559}, \cite{Hafner200071} and \cite{Finnis20041}). They have, however, been very successful in predicting elastic, vibrational and even lattice defect properties that can be simulated in cells consisting of up to a few hundreds of atoms.

In the next few sections we will investigate the most important elastic, vibrational and lattice defect properties of tungsten using both classical and \emph{ab initio} schemes. This allows us to systematically assess the limitations of both classical and quantum-mechanical methods in quantitative terms. At the same time, it will allow us to investigate how training data for the development of GAP potential for tungsten can be obtained.

Since our investigation into the GAP methodology involves exploring the possibility of developing a quantum-mechanical correction to the existing interatomic potential, we decided to use Finnis-Sinclair interatomic potential as an example classical method. It has the advantages of computational simplicity and it has been widely used in existing studies of bcc systems. Although there exists an EAM potential for tungsten, we find that conceptually it is not radically different from the Finnis-Sinclair potential. At the same time, although the bond-order potential offers improved accuracy over both the EAM method and the Finnis-Sinclair potential, its computational complexity and issues related to the computation of forces make it unsuitable as the core potential in our GAP methodology.

For the purpose of performing \emph{ab initio} calculations we use the CASTEP package for the first principles electronic structure calculations (more details in \cite{clark05-zkryst}). CASTEP uses density functional theory to determine the ground state electronic structure of the system and in all of our DFT calculations we use the Perdew-Burke-Ernzerhof exchange-correlation functional (more details in \cite{PhysRevLett.77.3865}).

\begin{sloppypar} 
We use Finnis-Sinclair potential implementation developed within the \texttt{libAtoms} \cite{libatoms} software library, for which author of this work is one of the contributors. We have also developed a driver that allows us to use the CASTEP package from within the same library. All of the data can then be analysed using \texttt{quippy} \cite{quippy} Python interface to the \texttt{libAtoms/QUIP} molecular dynamics framework.
\end{sloppypar} 

\subsection{Convergence of DFT Calculations}

When performing DFT calculations it is critical to ensure that all quantities of interest (in our case total energy, forces and stresses) are converged with respect to any adjustable parameters. In principle, this is as simple as running a series of calculations while changing a single parameter at a time, and inspecting the quantities of interest. The most straightforward test case for a bcc system that provides for a careful inspection of total energy, forces and stresses involves:

\begin{enumerate*}
\item A simulation cell containing at least two atoms as a primitive unit cell will have zero forces.
\item Randomised atomic positions and lattice vectors in order to avoid zero forces and stresses.
\end{enumerate*}

\noindent In density functional theory calculations involving periodic supercells, the electronic wavefunctions are expanded in terms of a discrete set of plane waves where a carefully chosen set of $k$-points can be used to accurately represent the wavefunction at all $k$-points. Furthermore, the basis set is truncated by omitting plane waves with kinetic energies higher than a predefined maximum cutoff energy. This necessitates a careful analysis of how all quantities of interest converge with the plane-wave cutoff and the $k$-point sampling.

Furthermore, when simulating metals using density functional theory partial band occupancies need to be introduced in order to eliminate the discontinuous changes in total energy that occur when an energy band crosses Fermi level. An electronic temperature is introduced through a Gaussian-like smearing of each energy level. While the energy calculation can be corrected for the effects of finite electronic temperature using appropriate correction (which permits use of large smearing widths up to $1 \, \text{eV}$; more details in the CASTEP software user documentation), there is no corresponding expression for a similar correction of forces or stresses. Consequently, one should analyse the effect of finite electronic temperature on the calculated values of forces and stresses as there is a trade-off between accuracy and instability due to reordering of the bands.

\restoregeometry 

\clearpage

\thispagestyle{empty}

\begin{sidewaysfigure}
\begin{center}
\resizebox{24cm}{!}{\footnotesize{}\input{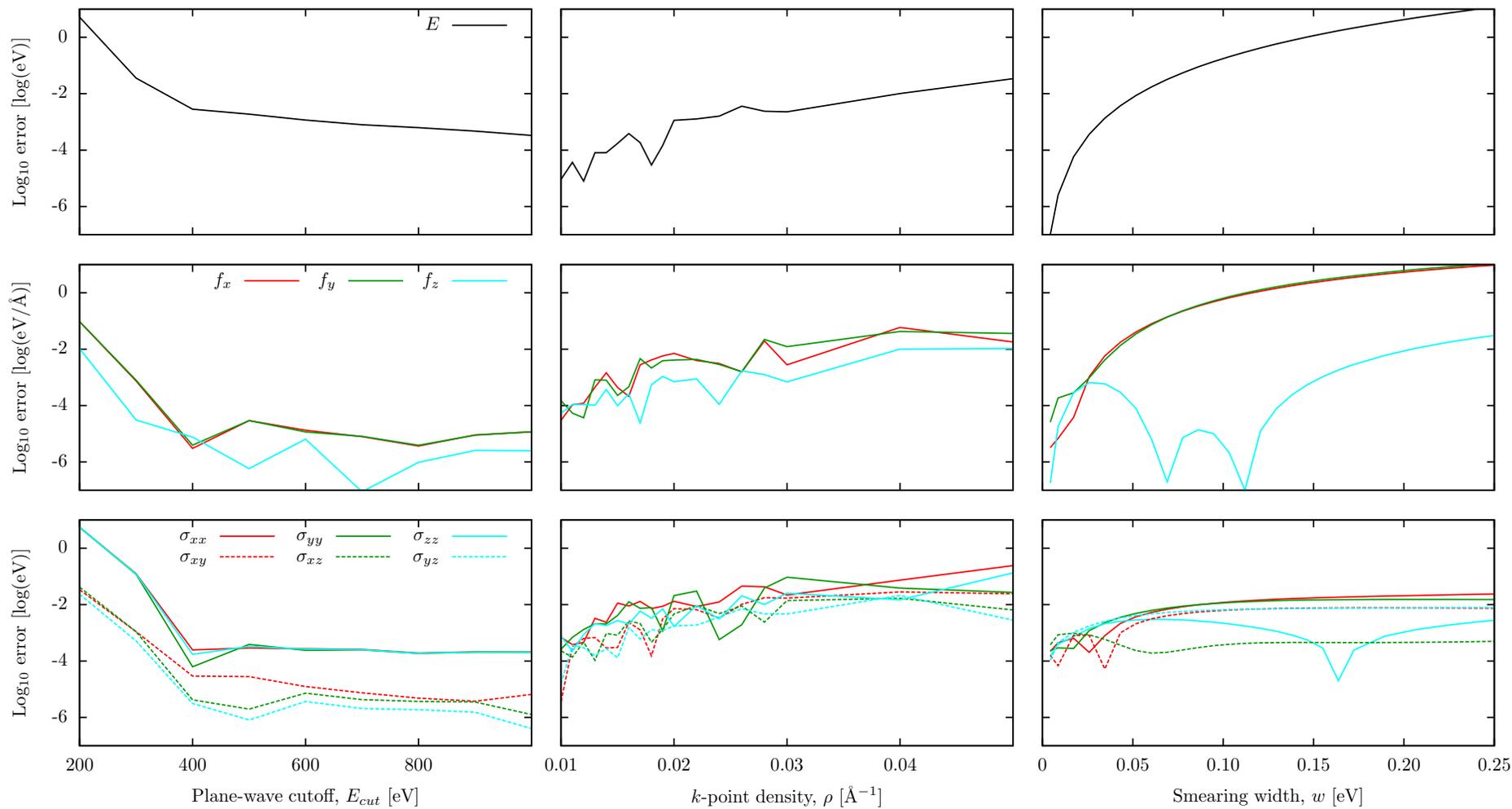}\normalsize{}}
\vspace{0.75cm}
\captionsetup{margin=4cm}
\caption[Convergence of DFT parameters.]{Convergence of energy, force and stress virial calculations as a function of plane-wave energy cutoff, $k$-point sampling density and smearing width parameters.}
\captionsetup{margin=1cm}
\label{figure:dft_convergence}
\end{center}
\end{sidewaysfigure}

\clearpage

We present the outcome of convergence calculations for energies, forces and stresses as a function of plane-wave cutoff, $k$-point sampling density and finite electronic temperature smearing in figure \ref{figure:dft_convergence}. Since it is vital for the purpose of GAP methodology that the training data is both accurate and consistent, our (conservative) choice of parameters for our subsequent work is summarised in table \ref{table:convergence} below:

\begin{table}[H]
\vspace{0.5cm}
\begin{center}
\begin{tabular}{ c c c r l }
\midrule
plane-wave energy cutoff, & $E_{cut}$ & $\to$ & $600$ & $\text{eV}$ \\
\midrule
$k$-point sampling density, & $\rho$ & $\to$ & $0.015$ & $\text{\AA}^{-1}$ \\
\midrule
smearing width, & $w$ & $\to$ & $0.1$ & $\text{eV}$ \\
\midrule
\end{tabular}
\caption{Converged values of DFT parameters.}
\label{table:convergence}
\end{center}
\end{table}

\vspace{-0.5cm} 

\section{Lattice Constant and Elastic Properties}
\label{chapter:bulk_properties_and_lattice_defects_in_tungsten:section:elastic_properties}

We begin the quantitative assessment of the Finnis-Sinclair interatomic potential and density functional theory results with a calculation of the tungsten bcc lattice constant. This is easily achieved using a geometry optimisation approach (as outlined in section \ref{chapter:simulation_techniques:section:geometry_optimisation}) but one can also obtain it by computing phase energy-volume curves and reading the lattice parameter corresponding to the ground state energy. The energy-volume curve of bcc tungsten is given in figure \ref{figure:energy-volume} below.

\vspace{-0.5cm} 

\begin{figure}[H]
\begin{center}
\resizebox{12cm}{!}{\footnotesize{}\input{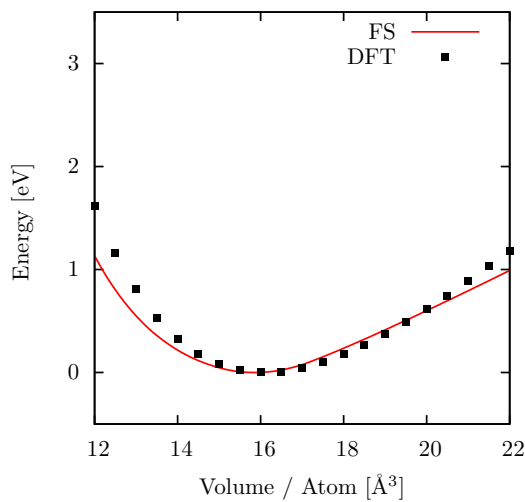}\normalsize{}}
\vspace{0.75cm}
\caption[Energy-volume curve of bcc tungsten.]{Energy-volume curve of bcc tungsten evaluated using FS and DFT models.}
\label{figure:energy-volume}
\end{center}
\end{figure}

\noindent The energy-volume phase diagram of tungsten for other common crystal phases is included in appendix \ref{appendix:tungsten_energy-volume_phase_diagram}.

For the purpose of this work it is beneficial to rescale the Finnis-Sinclair interatomic potential so that it has a matching lattice constant to that of DFT calculations. This way comparing the results obtained using classical and quantum-mechanical approaches is easier and the original result can be always reclaimed by scaling the lattice back to its original value. Furthermore, since we are planning to use the Finnis-Sinclair potential as core potential, it should simplify the fitting process in the development of our GAP method. In fact, for the purpose of fitting a GAP correction with the Finnis-Sinclair potential core, one can match the harmonic regime of the energy-volume curves obtained using Finnis-Sinclair and DFT methods by applying the transformation:

\begin{align}
r_{FS} &\to r_{FS}' = \alpha r_{FS} , \nonumber \\
E_{FS} (r_{FS}) &\to E_{FS}' (r_{FS}) = \beta E_{FS} (\alpha r_{FS}) ,
\end{align}

\noindent where the coefficients $\alpha$ and $\beta$ are defined as:

\begin{gather}
\alpha = \left( \frac{V_0^{(DFT)}}{V_0^{(FS)}} \right)^\frac{1}{3} = \frac{a_0^{(DFT)}}{a_0^{(FS)}} , \nonumber \\
\beta = \frac{B^{(DFT)}}{B^{(FS)} \alpha^3} .
\end{gather}

\noindent The parameter $B$ is the bulk modulus. It measures the resistance of the substance to uniform compression and it effectively corresponds to the quadratic coefficient of the energy-volume curve. If one defines bulk modulus as $B = B_0 + B_0' P$, finding the value of $B$ corresponds to fitting the energy as a function of volume in the Birch-Murnaghan equation of state (more details in \cite{PhysRev.71.809}):

\begin{align}
E (V) &= E_0 \nonumber \\
&+ \frac{9 V_0 B_0}{16} \left( \left[ \left( \frac{V_0}{V} \right)^{^2 / _3} - 1 \right]^3 B_0' + \left[ \left( \frac{V_0}{V} \right)^{^2 / _3} - 1 \right]^2 \left[ 6 - 4 \left( \frac{V_0}{V} \right)^{^2 / _3} \right] \right) ,
\end{align}

\noindent and the problem is equivalent to that of fitting a quadratic polynomial.

Finally, the rescaled forces can be easily obtained as:

\begin{equation}
\mathbf{f}_{FS} \to \mathbf{f}_{FS}' (r_{FS}) = - \boldsymbol{\nabla}_{FS} E_{FS}' (r_{FS}) = \alpha \beta \mathbf{f}_{FS} (r_{FS}) ,
\end{equation}

\noindent and stress virials being a linear combination of forces and atomic positions are obtained using the same method.

The resulting lattice constant and bulk modulus from the DFT calculation and corresponding scaling factors for the Finnis-Sinclair potential are given in table \ref{table:lattice} below:

\begin{table}[H]
\vspace{0.5cm}
\begin{center}
\begin{tabular}{ c c r l c c c r }
\midrule
$a_0$ & $=$ & $3.1805$ & $\text{\AA}$ & , & $\alpha$ & $=$ & $0.99519$ \\
\midrule
$B$ & $=$ & $304.59$ & $\text{GPa}$ & , & $\beta$ & $=$ & $0.99302$ \\
\midrule
\end{tabular}
\caption[Tungsten lattice parameter and bulk modulus.]{Tungsten DFT lattice parameter and bulk modulus, and corresponding Finnis-Sinclair potential scaling factors.}
\label{table:lattice}
\end{center}
\end{table}

\subsection{Linear Elasticity Theory}

In the linear limit of continuous elasticity theory, the relationship between stress and strain is given by Hooke's law:

\begin{equation}
\sigma_{ij} = - \sum_{k = 1}^3 \sum_{l = 1}^3 c_{ijkl} \epsilon_{kl} ,
\end{equation}

\noindent where $c_{ijkl}$ is the stiffness tensor. For anisotropic cubic structures the tensor $c_{ijkl}$ has only three independent elements and the above equation reduces to (more details in \cite{atanackovic2000theory}):

\begin{equation}
\begin{pmatrix}\sigma_{xx} \\ \sigma_{yy} \\ \sigma_{zz} \\ \sigma_{yz} \\ \sigma_{zx} \\ \sigma_{xy} \end{pmatrix} = \begin{pmatrix}C_{11} & C_{12} & C_{12} \\ C_{12} & C_{11} & C_{12} \\ C_{12} & C_{12} & C_{11} \\ & & & C_{44} \\ & & & & C_{44} \\ & & & & & C_{44} \end{pmatrix} \begin{pmatrix}\epsilon_{xx} \\ \epsilon_{yy} \\ \epsilon_{zz} \\ \epsilon_{yz} \\ \epsilon_{zx} \\ \epsilon_{xy} \end{pmatrix} ,
\end{equation}

\noindent and for isotropic materials this further reduces to only two independent elements (more details in \cite{atanackovic2000theory}), since:

\begin{equation}
C_{44} = \frac{C_{11} - C_{12}}{2} .
\end{equation}

\newgeometry{top=2.75cm,bottom=2.75cm,left=3.5cm,right=3cm} 

By evaluating stresses for small strains $\epsilon_{xx}$ and $\epsilon_{yz}$ we can compute all three elastic constants $C_{11}$, $C_{12}$ and $C_{44}$ that determine the elastic properties of bcc tungsten. In order to remain within the linear regime we only use strains of up to $1\%$ in these calculations. The results for DFT and FS interatomic potential are given in table \ref{table:elastic_constants} below:

\begin{table}[H]
\vspace{0.5cm}
\begin{center}
\begin{tabular}{ c c c }
& DFT & FS \\
\midrule
$C_{11} \, [\text{GPa}]$ & $516.86$ & $514.23$ \\
\midrule
$C_{12} \, [\text{GPa}]$ & $198.18$ & $200.12$ \\
\midrule
$C_{44} \, [\text{GPa}]$ & $142.30$ & $157.21$ \\
\midrule
\end{tabular}
\caption{Tungsten elastic constants.}
\label{table:elastic_constants}
\end{center}
\end{table}

In the table above it should come as no surprise that the $C_{11}$ and $C_{12}$ parameters obtained using DFT and the FS interatomic potential match closely. We have rescaled our FS potential bulk modulus to match that of DFT and these are related by $B = \frac{1}{3} (C_{11} + 2 C_{12})$. However, there is approximately a $10\%$ error in the $C_{44}$ parameter which is related to shear modulus. This is a manifestation of a common behaviour often observed while fitting conventional interatomic potentials with fixed number of parameters. All the other elastic parameters such as Young's modulus or Poisson's ratio can also be expressed in terms of $C_{11}$, $C_{12}$ and $C_{44}$ (more details in \cite{atanackovic2000theory}).

\subsection{Anharmonic Regime}

We now explore the behaviour of DFT and the FS interatomic potential in the anharmonic regime. We again evaluate the stresses but this time for a large spectrum of strains ranging from $-10\%$ to $+10\%$. We are interested in three stress-strain curves in particular:

\begin{itemize*}
\item $\sigma_{xx}$ vs. $\epsilon_{xx} \to$ which corresponds to longitudinal compression (the slope is equal to $C_{11}$ in the linear regime).
\item $\sigma_{yy}$ vs. $\epsilon_{xx} \to$ which corresponds to transverse expansion (the slope is equal to $C_{12}$ in the linear regime).
\item $\sigma_{yz}$ vs. $\epsilon_{yz} \to$ which corresponds to shearing (the slope is equal to $C_{44}$ in the linear regime).
\end{itemize*}

\noindent Plots of the above three curves are given in figure \ref{figure:fs_stress-strain} below.

\restoregeometry 

\begin{figure}[H]
\begin{center}
\resizebox{12cm}{!}{\footnotesize{}\input{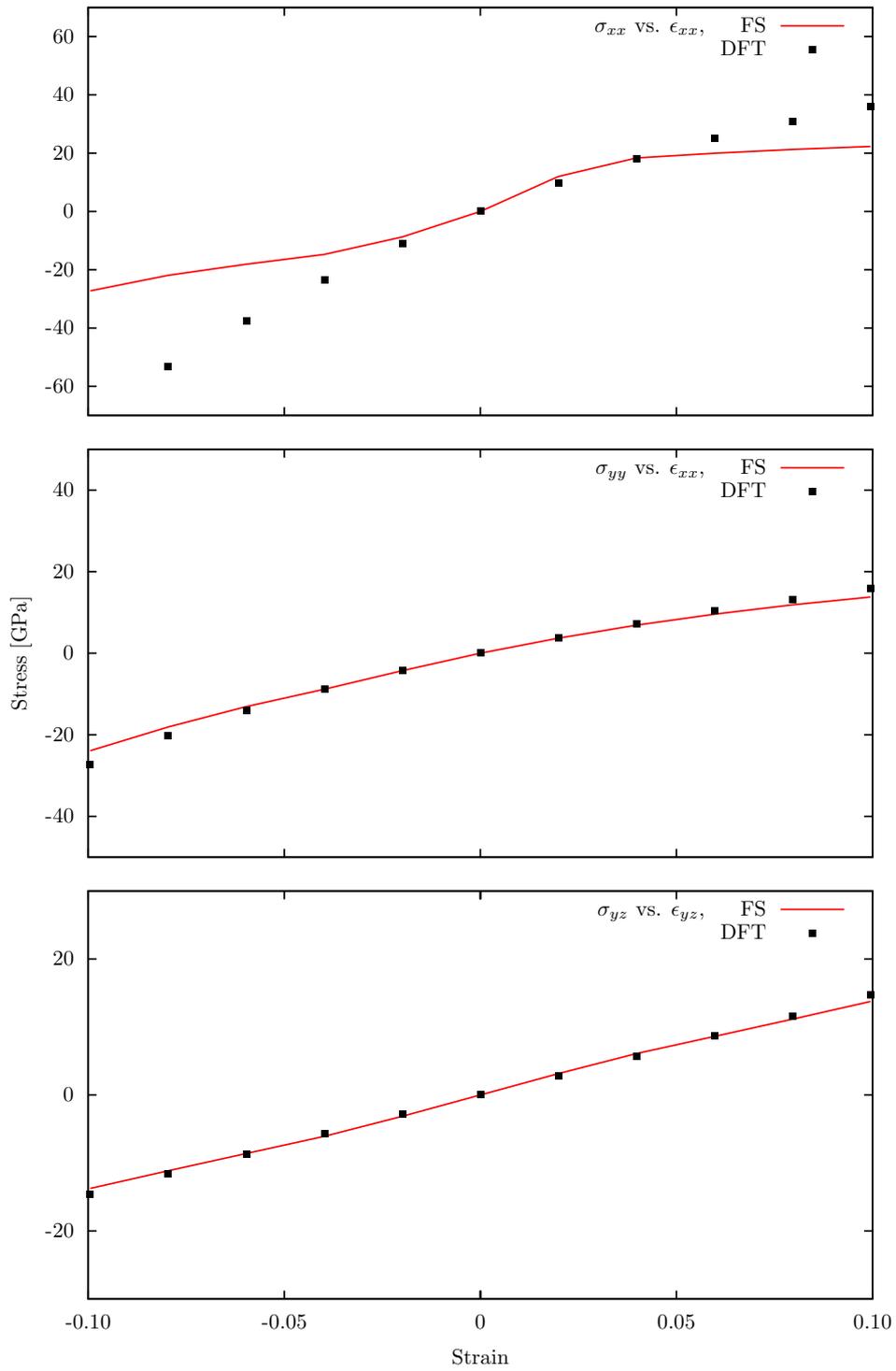}\normalsize{}}
\vspace{0.75cm}
\caption[Stress-strain curves of bcc tungsten.]{Stress-strain curves of bcc tungsten for a range of strains from $-10\%$ to $+10\%$.}
\label{figure:fs_stress-strain}
\end{center}
\end{figure}

\clearpage 

\newgeometry{top=2.75cm,bottom=2.75cm,left=3.5cm,right=3cm} 

As expected, in the harmonic regime the FS stress-strain curve matches that of DFT closely which is not surprising since the slope at zero strain corresponds to $C_{11}$, $C_{12}$ and $C_{44}$ elastic constants respectively. However, for longitudinal compression, there is a significant deviation between the two for strains above $2.5\%$ and for transverse expansion above $6\%$, which can be explained by the onset of non-linearity. This behaviour is something we will attempt to describe more accurately with our GAP potential.

\section{Phonon Spectrum}
\label{chapter:bulk_properties_and_lattice_defects_in_tungsten:section:phonon_spectrum}

At non-zero temperatures the atoms that compose the crystal lattice fluctuate randomly around their lattice sites (this random motion corresponds to heat). Consequently the position of atom $i$ can be written as:

\begin{equation}
\mathbf{x}_i = \mathbf{R}_{\mathbf{l}} + \mathbf{x}_i^0 + \mathbf{u}_i = \mathbf{x}_{\mathbf{l}, i}^0 + \mathbf{u}_i ,
\end{equation}

\noindent where $\mathbf{R}$ represents the lattice vector and $\mathbf{u}$ is the displacement away from equilibrium. One can Taylor expand the potential energy of the system around these equilibrium lattice sites:

\begin{equation}
E = E_0 + \frac{1}{2} \sum_{\substack{\mathbf{l}, \mathbf{l}' \\ i, i' \\ \alpha, \alpha'}} \frac{\partial^2 E}{\partial u_{\mathbf{l}, i, \alpha} \partial u_{\mathbf{l}', i', \alpha'}} u_{\mathbf{l}, i, \alpha} u_{\mathbf{l}', i', \alpha'} .
\end{equation}

\noindent We have ignored the first-order term as the expansion is around the equilibrium and the expansion terminates after the second-order term as we are only approximating the harmonic regime.

Consequently, the dynamics of the system is described by a set of coupled equations of motion:

\begin{equation}
m_i \frac{\partial^2 u_{\mathbf{l}, i, \alpha}}{\partial t^2} = \sum_{\substack{\mathbf{l}' \\ i' \\ \alpha'}} \frac{\partial^2 E}{\partial u_{\mathbf{l}, i, \alpha} \partial u_{\mathbf{l}', i', \alpha'}} u_{\mathbf{l}', i', \alpha'} ,
\label{equation:phonons}
\end{equation}

\noindent which, in a solid, have wave-like solutions:

\begin{equation}
\mathbf{u}_{\mathbf{l}, i} = \frac{1}{\sqrt{N m_i}} \sum_{\mathbf{k}, \beta} A(\mathbf{k}, \beta) \exp \left( i (\mathbf{k} \cdot \mathbf{x}_{\mathbf{l}, i}^0 - \omega(\mathbf{k}, \beta) t ) \right) \mathbf{e} (\mathbf{k}, \beta, i) .
\end{equation}

\noindent Substituting the above solution into the equation of motion \ref{equation:phonons} one obtains a system of linear equations that can be solved using the usual means employed for the treatment of coupled harmonic oscillators (more details in \cite{dove1993introduction}). Hence, one can obtain $\omega$ as a function of wavevector $\mathbf{k}$ and polarisation $\beta)$ which gives the dispersion relation.

In our analysis of the vibrational properties of bcc tungsten we compute the dispersion relation for all polarisation modes (two transverse and one longitudinal), using the most general wavevector path which exploits the symmetries of the bcc Brillouin zone $\{\Gamma-\text{H}-\text{N}-\Gamma-\text{P}-\text{H} | \text{P}-\text{N}\}$ (more details in \cite{Setyawan2010299}). For both the Finnis-Sinclair interatomic potential and DFT we use a finite displacement method to calculate phonon spectrum. This involves the calculation of forces on the atoms in the perturbed supercell where the force constant matrix is approximated through numerical differentiation. This procedure can be performed for any classical or quantum-mechanical method that delivers accurate forces, although, a careful analysis of the required supercell size is necessary. The requirement for a large supercell usually results in a large computational cost for the DFT phonon spectrum calculations.\footnote{It can be avoided for insulators by using Density Functional Perturbation Theory, however, at the time when this work was carried out this method was not available for metallic systems.}

The resulting phonon spectrum of bcc tungsten computed using DFT and the Finnis-Sinclair potential is shown in figure \ref{figure:fs_phonons} below.

\begin{figure}[H]
\begin{center}
\resizebox{12cm}{!}{\footnotesize{}\input{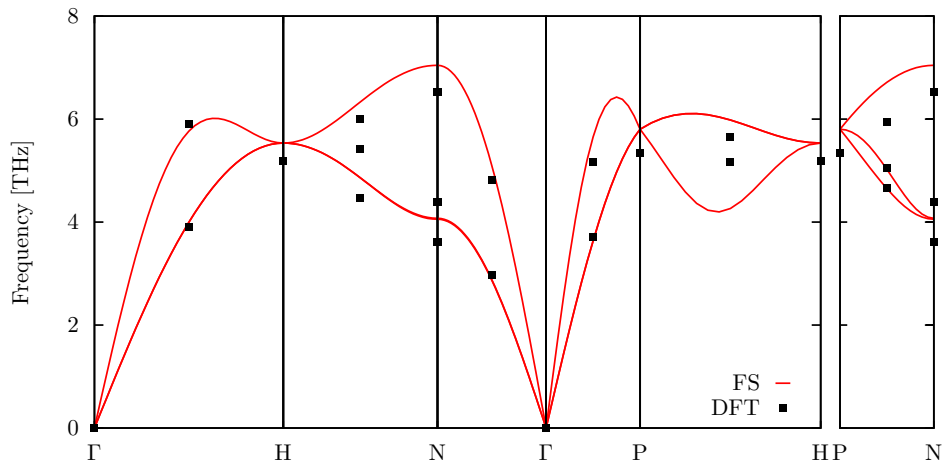}\normalsize{}}
\vspace{0.5cm}
\caption{Phonon spectrum of bcc tungsten.}
\label{figure:fs_phonons}
\end{center}
\end{figure}

While the FS potential provides a qualitatively good description of the vibrational properties of tungsten, in some of the high-symmetry directions in the Brillouin zone it fails to fully reproduce the DFT result. In particular, description of the transverse modes of vibration along the path $\{\text{H}-\text{N}-\Gamma\}$ does not fully match the predictions of the DFT model with FS modes appearing degenerate.

\restoregeometry 

\section{Vacancy}
\label{chapter:bulk_properties_and_lattice_defects_in_tungsten:section:vacancy}

The simplest lattice defect one can simulate is that of an isolated vacancy. It corresponds to removing one atom from its lattice site and optimising the positions of all the surrounding atoms while looking at the resulting energy change of the system which gives the vacancy formation energy $E_f^{(\text{vac.})}$. This quantity is of physical significance, since at a finite temperature all materials contain vacancies. The vacancy density is proportional\footnote{The entropy of vacancy formation also enters this expression but at low temperatures the formation energy term is of greater importance.} to $\exp( -E_f^{(\text{vac.})} / k_B T)$ where $T$ is the system temperature and $k_B$ is the Boltzmann constant. The vacancy formation energy is therefore expected to be of fundamental importance in many processes involving dislocation nucleation and migration.

Given the bcc lattice ground state energy per atom $E_0$ we can compute the vacancy formation energy at a constant pressure (fixed volume) as:

\begin{equation}
E_f^{(\text{vac.})} = \min_{\mathbf{x}_i \dots \mathbf{x}_N} ( E^{(\text{vac.})} ) - N E_0 ,
\end{equation}

\noindent using the supercell method. In practice one should vary the system size in order to ensure the convergence of $E_f^{(\text{vac.})}$ with supercell size to ensure that the interaction between the neighbouring vacancy images is negligible. As the supercell size approaches infinity and calculation is performed at the ground state lattice constant one can intuitively predict that the boundary conditions corresponds to a zero pressure situation. However, since the calculation requires a large cubic simulation cell, as the stress field of a point defect has spherical symmetry, for DFT calculations computational complexity imposes a limit on the accessible supercell size which leaves us far from the zero pressure limit. Therefore, the convergence rate to the zero pressure result can be improved by optimising our simulation energy in terms of the lattice vectors as well as the atomic positions. Hence we compute: 

\begin{equation}
E_f^{(\text{vac.})} |_{P = 0} = \min_{\substack{\mathbf{x}_i \dots \mathbf{x}_N \\ V}} ( E^{(\text{vac.})} ) - N E_0 .
\end{equation}

The formation energy as a function of system size at both, fixed and variable cell volume, calculated using the Finnis-Sinclair interatomic potential is shown in figure \ref{figure:vacancy_convergence} below. As anticipated, the convergence rate of the vacancy formation energy is better in the zero pressure calculations and consequently we find that we can obtain accurate estimates of vacancy formation energy (to within $0.01 \, \text{eV}$) in a simulation cell of 53 atoms. A brief discussion and calculation of the formation energies of tungsten di- and tri-vacancies is given in appendix \ref{appendix:tungsten_di-_and_tri-vacancies}.

\vspace{-0.25cm} 

\begin{figure}[H]
\begin{center}
\resizebox{12cm}{!}{\footnotesize{}\input{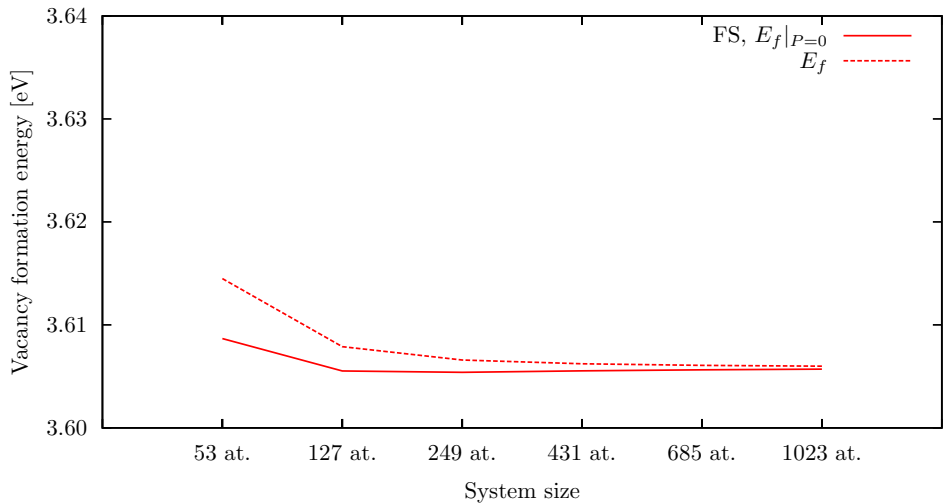}\normalsize{}}
\vspace{0.5cm} 
\caption[Vacancy formation energy as a function of system size.]{Convergence of vacancy formation energy with the system size for simulation cells with and without lattice relaxation.}
\label{figure:vacancy_convergence}
\end{center}
\end{figure}

\vspace{-1.0cm} 

\section{Surfaces}
\label{chapter:bulk_properties_and_lattice_defects_in_tungsten:section:surfaces}

We use the same methodology to calculate the formation energy of surfaces as we did for vacancies in the previous section. However, when dealing with plane defects the preparation of the simulation cell requires further consideration. Within periodic boundary approach convention introducing a free surface with part of the simulation cell occupied by vacuum corresponds to simulation of a series of slabs with finite thickness.

\begin{figure}[H]
\begin{center}
\resizebox{12cm}{!}{\footnotesize{}\input{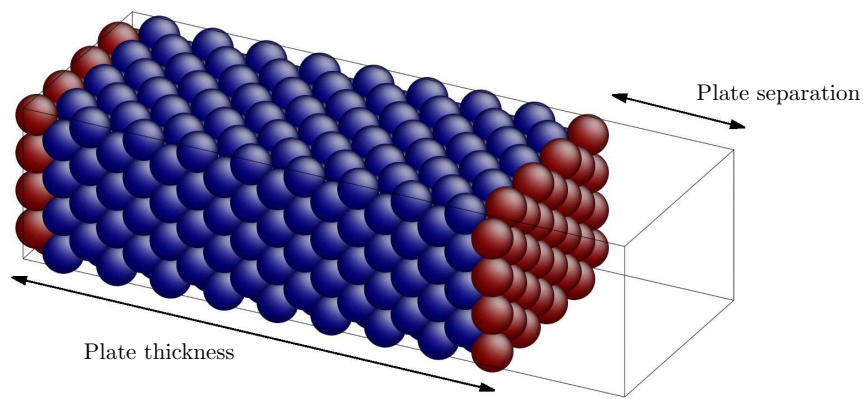}\normalsize{}}
\vspace{0.25cm}
\caption{Surface simulation cell.}
\label{figure:surface_cell}
\end{center}
\end{figure}

\noindent An example of a free surface simulation cell where the surface atoms have been highlighted is shown in figure \ref{figure:surface_cell}.\footnote{We use \texttt{AtomEye} atomistic configuration viewer which is also available within \texttt{quippy} Python interface and \texttt{libAtoms/QUIP} molecular dynamics framework for visualisation of simulation cells (more details in \cite{0965-0393-11-2-305}).}

One should observe that there are always two surfaces per simulation cell and consequently the formation energy becomes:

\begin{equation}
E_f^{(\text{surf.})} = \frac{1}{2} \left( \min_{\mathbf{x}_i \dots \mathbf{x}_N} ( E^{(\text{surf.})} ) - N E_0 \right) ,
\end{equation}

\noindent and the surface energy is usually given per unit area. The simulation cell size in the plane of the surface should not affect the value of the formation energy, although it does need to be large enough to allow for surface reconstruction. Care should also be taken in order to ensure that the slab is of sufficient thickness and distance between the two surfaces should be increased until convergence is achieved.

In our investigation of surface energies we find it interesting to investigate how classical and quantum-mechanical methods describe the process of pulling of the two surfaces apart which creates free surfaces separated by an increasing amount of vacuum. This provides an interesting insight into the description of surface behaviour and also gives an indication of the length-scale of the range of the interactions between the surfaces.

We compute the surface formation energy as a function of slab separation for the four high symmetry surfaces $(100)$, $(110)$, $(111)$ and $(112)$ (the choice of surfaces being dictated by our discussion of the crystallographic directions of physical importance in section \ref{chapter:bulk_properties_and_lattice_defects_in_tungsten:section:bcc_lattice}). The results are shown in figure \ref{figure:fs_surfaces} below.

We find that while the FS description of surfaces is qualitatively correct, the calculated surface energies differ from $10\%$ for the $(111)$ surface, to almost $30\%$ for the $(100)$ surface as compared to the DFT result. The ordering of the surface energies is also different --- while DFT predicts that the $(110)$ surface is energetically the most favourable and $(100)$ the least favourable, the FS potential predicts the $(100)$ surface as the one with the lowest surface energy. Since an accurate description of surface energies is critical for the modelling of phenomena such as crack propagation, for instance Griffith's criterion for the growth of crack involves balance between elastic and surface energies, we will attempt to include an accurate description of free surfaces in our GAP potential for tungsten.

\begin{figure}[H]
\begin{center}
\resizebox{12cm}{!}{\footnotesize{}\input{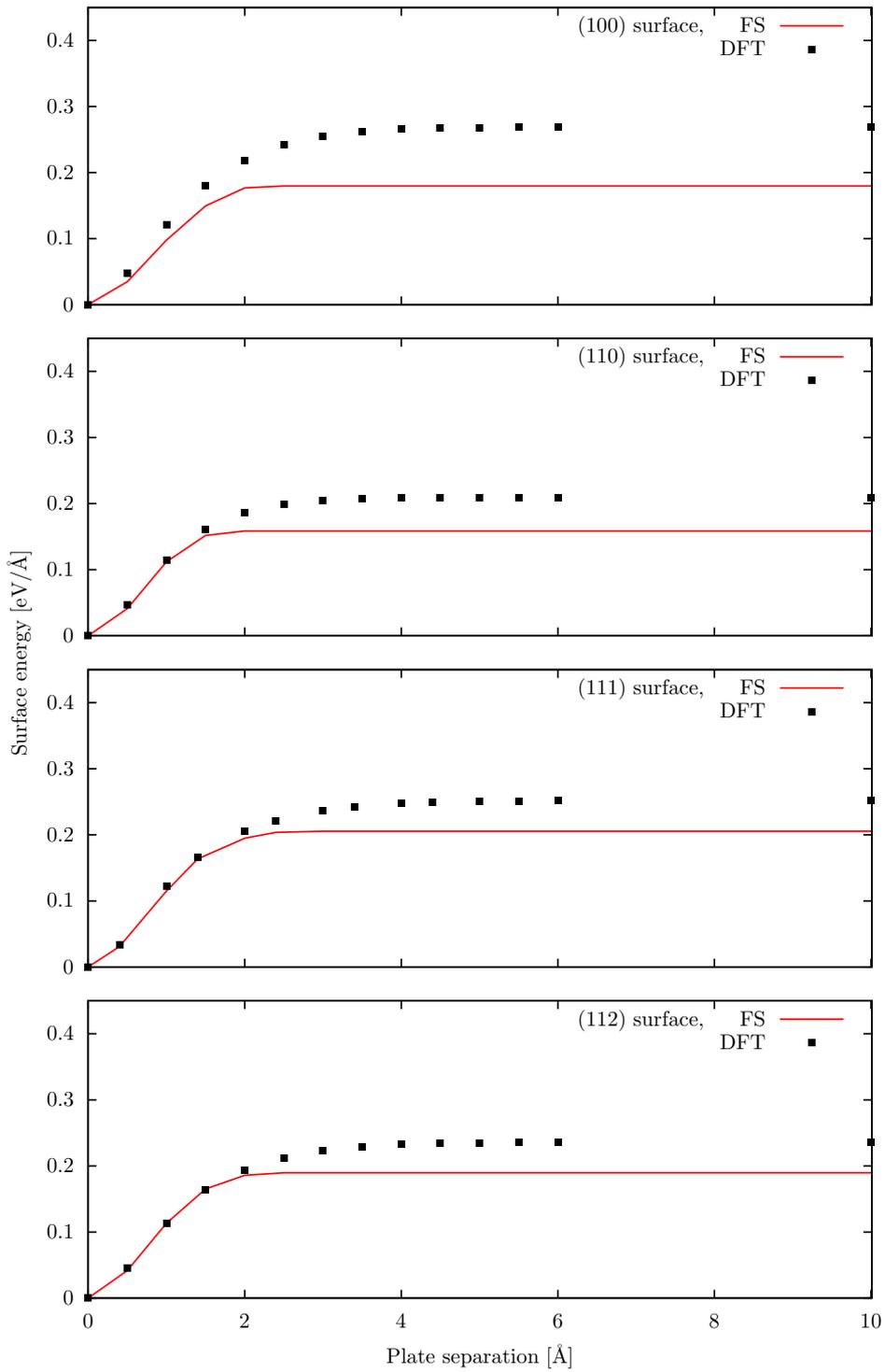}\normalsize{}}
\vspace{0.75cm}
\caption[Surface energy as a function of plate separation.]{Surface energy for the high symmetry surfaces $(100)$, $(110)$, $(111)$ and $(112)$ computed with the Finnis-Sinclair potential and DFT method.}
\label{figure:fs_surfaces}
\end{center}
\end{figure}

\section[Gamma Surfaces (Generalised Stacking Faults)]{\texorpdfstring{Gamma Surfaces \\ (Generalised Stacking Faults)}{Gamma Surfaces (Generalised Stacking Faults)}}
\label{chapter:bulk_properties_and_lattice_defects_in_tungsten:section:gamma_surfaces}

The gamma surfaces --- a theoretical construct introduced by Vitek in late 1960s (more details in \cite{doi:10.1080/14786436808227500}) --- are two-dimensional energy surfaces that give the variation of energy on displacing the two parts of the crystal relative to each other along a crystal plane. Since the displacement vector is periodic with the lattice, one obtains a two-dimensional energy surface bound by the lattice vectors and formed by all the unique combinations of the relative displacement vector. The concept was originally introduced as a means of finding potential stacking faults in metals. This is because a local minimum in the gamma surface corresponds to a metastable stacking fault and, hence, the concept of a generalised stacking fault. Furthermore, we also find that together with the data obtained in the previous section by pulling two surfaces apart (an ``orthogonal'' concept to that of a gamma surface) it gives a further insight into the assessment of the accuracy of interatomic potentials.

We compute the gamma surface of tungsten by adding the relative displacement vector to the lattice vector perpendicular to the gamma surface. This effectively shears the simulation cell but since the shear is not applied to the atomic positions, one could visualise this as shearing the simulation cell and moving the atoms so that there is just one gamma surface per cell. This is the most efficient method of computing an arbitrary point on the gamma surface as the simulation cell size can be kept to a minimum. Also, as was the case for free surface, care should be taken in order to ensure that the simulation cell height and, hence, the distance separating two adjacent gamma surfaces is sufficiently large. An example of the gamma surface simulation cell is shown in figure \ref{figure:gamma_surface_cell} below.

While the gamma surface could be investigated in its unrelaxed form, performing relaxation of the atomic positions in the direction perpendicular to the gamma surface provides a greater physical insight. Relaxation in the directions parallel to the gamma surface does not make sense as such computation would bring all points of the gamma surface to one of the metastable (or stable) configurations. Consequently, we can compute the gamma surface energy according to:

\begin{equation}
E_f^{(\gamma \, \text{surf.})} = \min_{\mathbf{x}_i^{\perp} \dots \mathbf{x}_N^{\perp}} ( E^{(\gamma \, \text{surf.})} ) - N E_0 .
\end{equation}

\begin{figure}[H]
\begin{center}
\resizebox{12cm}{!}{\footnotesize{}\input{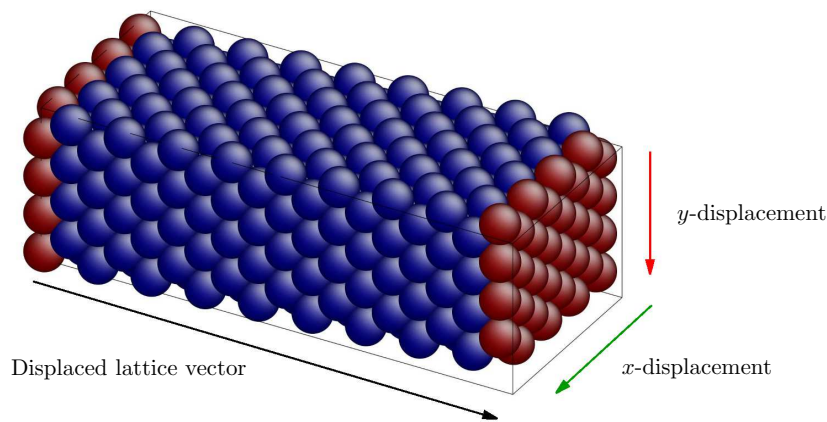}\normalsize{}}
\vspace{0.25cm}
\caption{Gamma surface simulation cell.}
\label{figure:gamma_surface_cell}
\end{center}
\end{figure}

\noindent We will discuss the physical meaning of the relaxed gamma surface and its relevance to screw dislocations in more detail in the following section.

Even though the only slip planes of physical significance in bcc systems are those of $(110)$ and $(112)$ (as outlined in section \ref{chapter:bulk_properties_and_lattice_defects_in_tungsten:section:bcc_lattice}) for reasons of completeness and to verify that our theoretical analysis is indeed correct we compute the unrelaxed gamma surfaces for the four high symmetry surfaces $(100)$, $(110)$, $(111)$ and $(112)$ (as in the case of free surfaces). The results are shown in figure \ref{figure:fs_gamma_surfaces} below.

As in the case of the free surfaces we find that the FS description of the gamma surfaces is qualitatively correct and the shape of the surface is in agreement with that predicted using DFT method. However, we anticipate that this shape is largely determined by the arrangement of atoms in the bcc lattice and the resulting atomic repulsion. Consequently, a detailed comparison of the FS and DFT results reveals that in quantitative terms the FS interatomic potential underestimates the energy of some of the regions of the gamma surface as compared to the DFT method.

An accurate description of gamma surfaces is critical for the accurate description of dislocation structure (any dissociation into partial dislocations involves stacking faults and the partials separation distance is determined by the balance between elastic and stacking fault energies). Hence, we anticipate that our GAP potential for tungsten will need to provide a quantitatively more accurate description of the gamma surface energies than that of the FS interatomic potential.

\clearpage

\thispagestyle{empty}

\begin{sidewaysfigure}
\begin{center}
\resizebox{24cm}{!}{\footnotesize{}\input{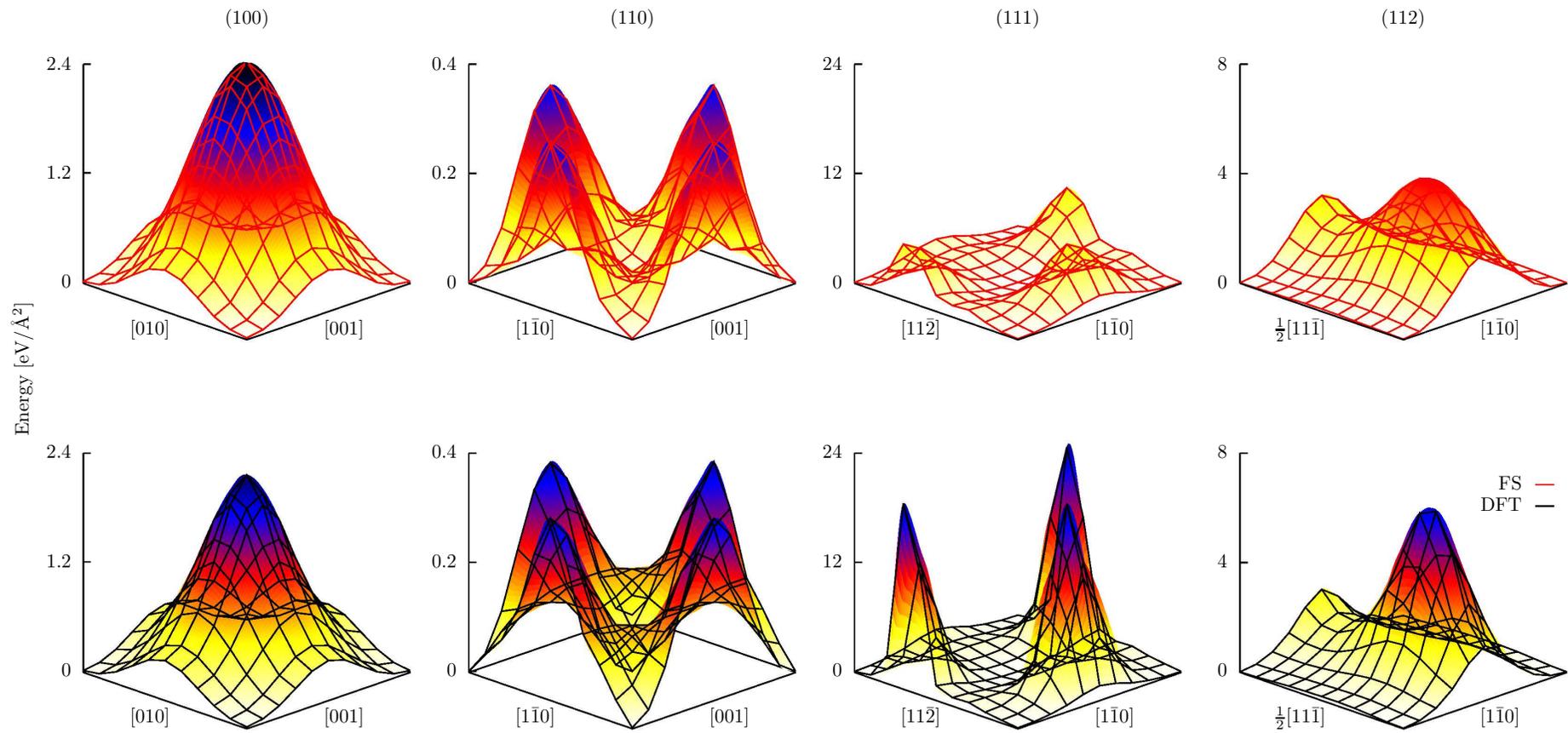}\normalsize{}}
\vspace{0.25cm}
\captionsetup{margin=4cm}
\caption[Gamma surface energies.]{Gamma surface energy for the high symmetry surfaces $(100)$, $(110)$, $(111)$ and $(112)$ computed with the Finnis-Sinclair potential and DFT method.}
\captionsetup{margin=1cm}
\label{figure:fs_gamma_surfaces}
\end{center}
\end{sidewaysfigure}

\clearpage

The computational cost of calculating relaxed gamma surfaces is significantly higher due to the need for an independent, constrained geometry optimisation for each point of the gamma surface. Hence, we only compute the relaxed gamma surfaces for the two slip planes of physical significance, i.e. $(110)$ and $(112)$ gamma surfaces. The results are shown in figure \ref{figure:fs_gamma_surfaces_relaxed} below.

\begin{figure}[H]
\begin{center}
\resizebox{12cm}{!}{\footnotesize{}\input{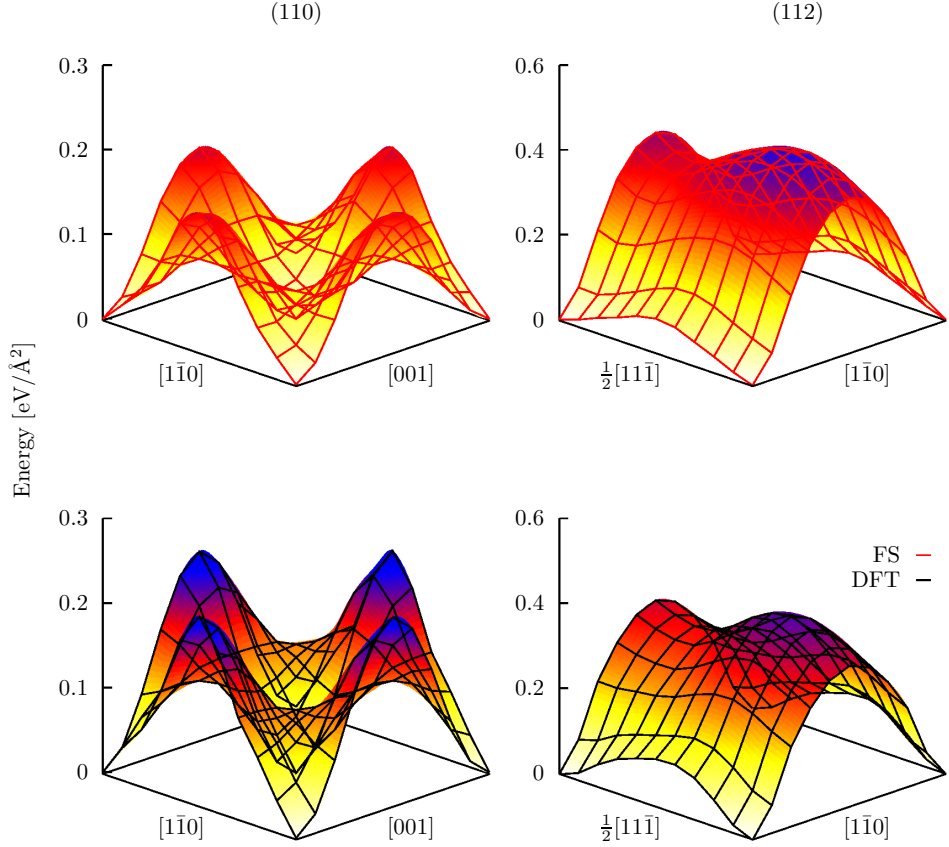}\normalsize{}}
\vspace{0.25cm}
\caption[Relaxed gamma surface energies.]{Relaxed gamma surface energy for the high symmetry surfaces $(110)$ and $(112)$ computed with the Finnis-Sinclair potential and DFT method.}
\label{figure:fs_gamma_surfaces_relaxed}
\end{center}
\end{figure}

\section{Dislocations}
\label{chapter:bulk_properties_and_lattice_defects_in_tungsten:section:dislocations}

The material properties that are most important for the simulation of plasticity in metals are directly related to the production, mobility and evolution of dislocations. In most metals plastic deformation is controlled by the interaction of dislocations with the underlying lattice and its defects (be it other dislocations, solutes or grain boundaries) and by the applied stress.

While there are some variations in the plasticity behaviour of different bcc metals, as outlined in section \ref{chapter:bulk_properties_and_lattice_defects_in_tungsten:section:bcc_lattice}, their common behaviour can be attributed to their lattice crystallography. In particular, it is widely believed that the existence of a ductile-to-brittle transition at low temperatures, in some bcc metals, is a manifestation of the inability of the dislocations to move at the required rate at low temperatures in order to relieve stress concentrations. Consequently, an investigation of the quantum-mechanical and classical description of dislocations is critical in our development of GAP potential for tungsten.

The dominant type of dislocation observed in bcc metals has a Burgers vector $\frac{1}{2} \langle 111 \rangle$ which is also the distance of the first nearest neighbour and the shortest lattice vector of the bcc lattice. $\langle 100 \rangle$ dislocations with Burgers vector corresponding to the second nearest neighbour distance have been observed in some bcc metals but they are believed to be the products of reactions between $\frac{1}{2} \langle 111 \rangle$ dislocations (more details in \cite{bulatov2006computer}). As mentioned in section \ref{chapter:bulk_properties_and_lattice_defects_in_tungsten:section:bcc_lattice} there are two distinct slip planes of physical significance in the $\langle 111 \rangle$ zone (i.e. $\{110\}$ and $\{112\}$ planes) but they differ in activation energy and the $\{110\}$ plane is energetically the most favourable one. The zonal characteristics of dislocation slip systems and corresponding TEM observations explain the prominent role of $\frac{1}{2} \langle 111 \rangle$ screw dislocations in bcc metals plasticity. Kinematically, at low and moderate temperatures the non-screw dislocations behave as ``slaves'' to the dominant screw dislocations (more details in \cite{Cai20051}).

\subsection{Long Range Behaviour (Linear Elasticity Theory)}

The theory of elasticity allows us to treat dislocations in the regime far away from the dislocation line as a continuous medium. At all regions in the crystal, apart from very close to the dislocation core, the stress is small enough to be treated by linear elasticity theory. Assuming no grain boundaries this allows us to derive simple analytical solutions for the elastic energy stored in the displacement field of a dislocation.

We begin by representing the displacement field in terms of a cylinder of elastic material modelled by the Volterra deformation (more details in \cite{hull2001introduction}). While this treatment assumes that the underlying elastic properties of the continuous medium are isotropic, unusually, it is a good first approximation for tungsten properties. Although we will only attempt to simulate screw dislocations in tungsten, the formalism for edge dislocations is nearly identical and we will present the corresponding formulas for the elastic energy stored in an edge dislocation in this section for reasons of completeness.

\begin{figure}[H]
\begin{center}
\resizebox{12cm}{!}{\footnotesize{}\input{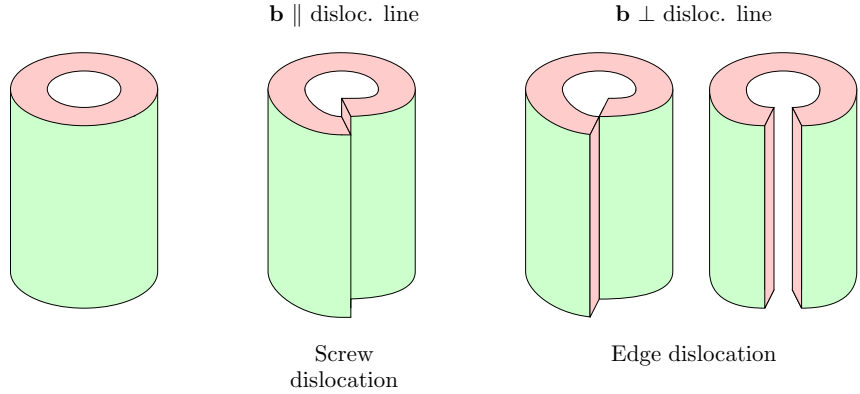}\normalsize{}}
\caption{Screw and edge dislocations described in terms of Volterra's tube.}
\label{figure:volterras_cylinder}
\end{center}
\end{figure}

\noindent As shown in figure \ref{figure:volterras_cylinder} a radial slit has been cut in the cylinder parallel to the $z$-axis and the free surfaces have been displaced with respect to each other by the distance $b$, the Burgers vector. The deformation in the cylinder is small and determined by the periodicity of the lattice everywhere apart from the dislocation core where the strain is very large and where linear elasticity theory is no longer appropriate.

It is easy to see that for a screw dislocation the displacement field is given by:

\begin{equation}
u_{z} = \frac{b \times \theta}{2 \pi} = \frac{b}{2 \pi} \arctan\left(\frac{y}{x}\right),
\end{equation}

\noindent where the dislocation line is parallel to the $z$-axis. The displacement field around the edge dislocation is more complex since the lattice is not deformed in the $z$-direction but the strain is found in the $x$-$y$ plane instead. It is given by (the derivation can be found in \cite{hirth1982theory}):

\begin{align}
u_{x} &= \frac{b}{2 \pi} \left( \arctan\left(\frac{y}{x}\right) + \frac{x y}{2 (1 - \nu) (x^{2} + y^{2})} \right), \\
u_{y} &= \frac{b}{2 \pi} \left( \frac{1 - 2 \nu}{4 (1 - \nu)} + \frac{x y}{2 (1 - \nu) (x^{2} + y^{2})} \right),
\end{align}

\noindent where $\nu$ is the Poisson ratio of the material.

Since the local strain is defined by $\varepsilon_{ij} = \frac{1}{2} ( \frac{du_{x_{i}}}{dx_{j}} + \frac{du_{x_{j}}}{dx_{i}} )$, we can obtain the strain field by direct substitution. Furthermore, by applying Hooke's law we obtain the stress field of a screw dislocation:

\begin{align}
\sigma_{xz} = \sigma_{zx} &= - \frac{\mu \times b}{2 \pi} \frac{\sin \theta}{r}, \\
\sigma_{yz} = \sigma_{zy} &= \frac{\mu \times b}{2 \pi} \frac{\cos \theta}{r},
\end{align}

\noindent and similarly for an edge dislocation:

\begin{align}
\sigma_{xx} &= - \frac{\mu \times b}{2 \pi (1 - \nu)} y \frac{3x^{2} + y^{2}}{(x^{2} + y^{2})^{2}}, \\
\sigma_{yy} &= \frac{\mu \times b}{2 \pi (1 - \nu)} y \frac{x^{2} - y^{2}}{(x^{2} + y^{2})^{2}}, \\
\sigma_{xy} = \sigma_{yx} &= \frac{\mu \times b}{2 \pi (1 - \nu)} x \frac{x^{2} - y^{2}}{(x^{2} + y^{2})^{2}},
\end{align}

\noindent where $\mu$ is the shear modulus of the material.

Finally, since the elastic energy stored in a material under strain $\varepsilon$ is given by $dE = \frac{1}{2} \sum \sigma_{ij} \times \varepsilon_{ij} dV$ the elastic energy per unit length due to the dislocation is given by:

\begin{align}
E_{screw} = \frac{\mu \times b^{2}}{4 \pi} \ln \frac{R}{r},
\end{align}

\noindent for screw dislocation, and similarly:

\begin{align}
E_{edge} = \frac{\mu \times b^{2}}{4 \pi (1 - \nu)} \ln \frac{R}{r},
\end{align}

\noindent for an edge dislocation, where $R$ is the outer, external radius which is in practice determined by the grain size and $r$ is the core radius.

\subsection{Dislocation Core}

Linear elasticity theory alone is not sufficient to describe the structure of the dislocation core. It is found that core radii are of the order of lattice spacings and, hence, one needs to take the underlying lattice into account. The structure of the core is no longer purely dictated by minimising the elastic energy but instead there is a trade-off between elastic energy and chemical energy of the bonds involved and the spreading of the dislocation line onto neighbouring atoms is observed. We refer to the resulting structure as a dislocation core.

Early hypotheses suggested fcc-like planar splitting of bcc screw dislocation core to explain the observed slip systems of bcc metals (more details in \cite{Cohen1962894}). However, it was not until it was suggested that dissociation into equivalent $\{110\}$ planes of the $\langle 111 \rangle$ zone is possible, even though no planes in the bcc $\langle 111 \rangle$ zone contain stable stacking faults, that the observed high Peierls barrier and strong temperature dependence of the yield stress could be explained (more details in \cite{Cai20051}, \cite{Duesbery19981481}; representation of $\langle 111 \rangle$ zone is shown in figure \ref{figure:slip_systems}).

At this point we should clarify that although the term ``dissociation'' might imply splitting of the dislocation line into multiple partial dislocations this is not the case for the bcc screw dislocation core. It is perhaps more appropriate to refer to the particular core reconstruction as ``polarisation'' (more details in \cite{Cai20051}).

In practice, when carrying out atomistic simulations the core structure of a screw dislocation is determined by two factors: the properties of the interatomic potential used and the boundary conditions applied to the simulation cell. The most commonly observed core structures of screw dislocation in bcc systems are the three-fold structure (non-symmetric, or polarisable) and the six-fold structure (symmetric, or non-polarisable), as shown in figure \ref{figure:dislocation_structure} below. We will refer to them as non-symmetric and symmetric cores\footnote{The description of symmetric core as six-fold symmetric, and non-symmetric core as three-fold symmetric is misleading, since the symmetry around the $\{111\}$ axis remains three-fold --- it is the 180$^{\circ}$ rotation symmetry around $\{110\}$ axes of the $\langle 111 \rangle$ zone that is broken.} as explained in \cite{Cai20051}.

\vspace{0.5cm} 

\begin{figure}[H]
\begin{center}
\resizebox{12cm}{!}{\footnotesize{}\input{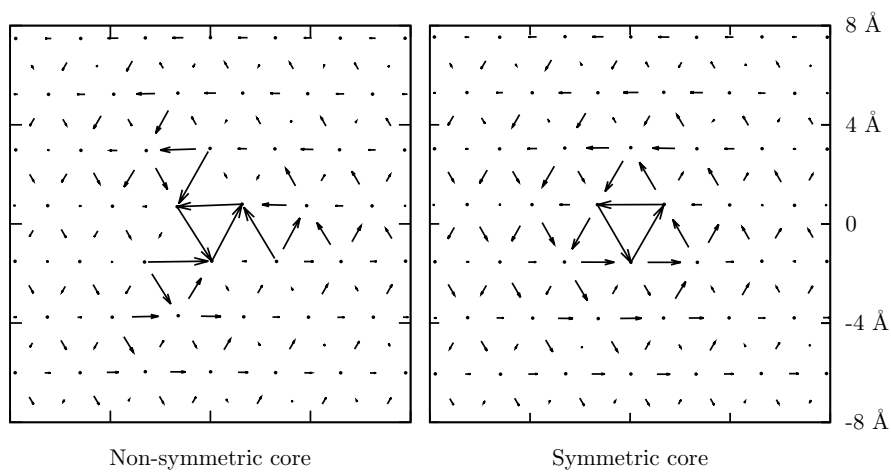}\normalsize{}}
\vspace{0.5cm}
\caption[$\frac{1}{2} \langle 111 \rangle$ screw dislocation core structure.]{Two most commonly observed core structures of $\frac{1}{2} \langle 111 \rangle$ screw dislocation in bcc metals.}
\label{figure:dislocation_structure}
\end{center}
\end{figure}

\begin{sloppypar} 
The effects of core structure relaxation for $\frac{1}{2} \langle 111 \rangle$ screw dislocations in bcc transition metals (usually molybdenum or tantalum) have been throughly examined using atomistic simulations employing simple interatomic potentials. While most classical methods predict a polarised core structure, more recent studies employing quantum-mechanical methods have called these results into question (more details in \cite{PhysRevB.54.6941}, \cite{PhysRevLett.84.1499}, \cite{Woodward200559}, \cite{Duesbery19981481}). The mobility of the screw dislocation is believed to be a direct consequence of the amount of spreading of the dislocation line into the slip planes of the $\langle 111 \rangle$ zone. In this model, the dislocation effectively anchors itself to the particular lattice site. In order to transition onto the neighbouring site a significant amount of energy is required to retract its movement. Consequently, it is widely believed that the atomic rearrangement in the dislocation core affects the lattice resistance to the dislocation motion, i.e. the Peierels stress of the screw dislocation in bcc metals is related to the non-planar character of the core structure (more details in \cite{Cai20051}). A precise description of the dislocation core is therefore necessary for an accurate prediction of dislocation properties in bcc tungsten.
\end{sloppypar} 

\subsection{Gamma Surfaces and Screw Dislocation}

As proposed by Vitek (more details in \cite{doi:10.1080/14786436908217784}, \cite{Duesbery19981481}), the calculated core structure of $\frac{1}{2} \langle 111 \rangle$ screw dislocations in bcc transition metals can be rationalised in terms of the strictly planar gamma surface concept. Assuming that the symmetric core corresponds to the screw dislocation spreading onto all six $(110)$ planes of the $\langle 111 \rangle$ zone, while the non-symmetric core only spreads onto three of them, one can compute the $(110)$ gamma surface energies associated with the displacement vector along the $\langle 111 \rangle$ direction. The symmetric or non-symmetric structures are then expected to be energetically favourable according to the following criterion:

\begin{itemize}
\item $6 E_f^{(\gamma \, \text{surf.})} (\frac{1}{6} \mathbf{b}) < 3 E_f^{(\gamma \, \text{surf.})} (\frac{1}{3} \mathbf{b}) \to$ symmetric core.
\item $6 E_f^{(\gamma \, \text{surf.})} (\frac{1}{6} \mathbf{b}) > 3 E_f^{(\gamma \, \text{surf.})} (\frac{1}{3} \mathbf{b}) \to$ non-symmetric core.
\end{itemize}

Based on our earlier calculations in section \ref{chapter:bulk_properties_and_lattice_defects_in_tungsten:section:gamma_surfaces}, we can plot the gamma surface energies along the $\langle 111 \rangle$ lattice vector using DFT and the Finnis-Sinclair interatomic potential for tungsten. The results are shown in figure \ref{figure:fs_gamma_surface_111} below.

\begin{figure}[H]
\begin{center}
\resizebox{12cm}{!}{\footnotesize{}\input{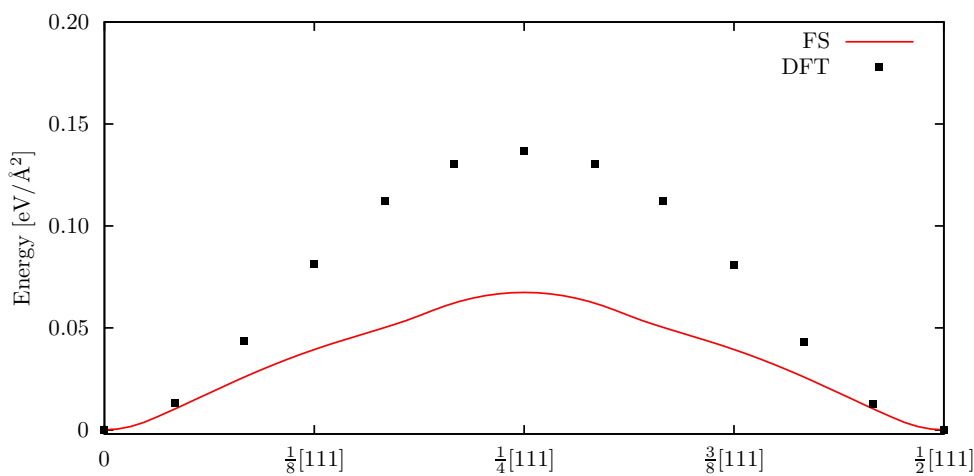}\normalsize{}}
\vspace{0.5cm}
\caption[$\langle 111 \rangle$ cross-section of $(110)$ gamma surface energy.]{$\langle 111 \rangle$ cross-section of $(110)$ gamma surface energy for Finnis-Sinclair and DFT models.}
\label{figure:fs_gamma_surface_111}
\end{center}
\end{figure}

\subsection{Visualisation of Dislocation Core Structure}

Since the displacement field of a screw dislocation is in the direction parallel to the dislocation line (and therefore impossible to visualise by looking at the atomic positions alone), one way of quantifying the displacement field of a dislocation is by using the dislocation displacement maps proposed by Vitek (more details in \cite{doi:10.1080/14786437008238490}). An example of the screw component of the dislocation displacement map is given in figure \ref{figure:dislocation_structure}. When looking at the plane perpendicular to the dislocation line each dot represents a column of atoms. The dislocation displacement map is constructed by computing the displacement of an atom from the reference lattice and the arrows indicate the difference between the displacements of the neighbouring atoms, i.e. the length of the arrow is proportional to the difference in the magnitude of the displacement.

In the case of a screw dislocation, the characteristics of the calculated core structure are most easily visualised by plotting the screw component of the dislocation displacement map, i.e. relative displacement of the neighbouring atoms due to the dislocation in the direction parallel to the dislocation line. By convention the differential displacement is always mapped into the domain of $(-\frac{1}{2}b, \frac{1}{2}b)$ by adding or subtracting the required multiple of $b$. The arrow lengths are then normalised by $\frac{1}{3}b$, which is the magnitude of the separation of the neighbouring atoms in the $\langle 111 \rangle$ direction of the bcc lattice. Finally, since the arrows representing the screw component of the dislocation displacement map correspond to the displacement that is strictly out of the plane, the direction of the arrow is always such that it connects two neighbouring atoms.

Some of the qualitative aspects of the calculated core structure of a screw dislocation are also captured by the edge component of the dislocation displacement map which shows the relative displacement of the neighbouring atoms due to the dislocation in the direction perpendicular to the dislocation line. For the screw dislocation these displacements are usually found to be of the order of $10-100$ times smaller than that of the screw components (they are zero for a perfect screw dislocation as described by the elasticity theory). Consequently, the scaling of the arrows is usually adjusted and the direction of the arrows corresponds to the direction of displacement that is projected onto the plane perpendicular to the dislocation line.

Dislocation displacement maps, while extremely useful for the description of the qualitative aspects of simple dislocation core structures, can sometimes be cumbersome when analysing multiple dissociation schemes. Furthermore, in the case of mixed dislocations with both screw and edge components and, especially, when there is no prior knowledge of the Burgers vector, construction of a dislocation displacement map can be a difficult task. It is therefore desirable to have a robust and automated procedure for visualising the screw and edge aspects of the dislocation structure in a quantitative manner.

A more general concept that extends the ideas of dislocation displacement maps relies on the fact that atomic misfit associated with a dislocation can be quantified using the Nye tensor (more details in \cite{Hartley20051313}). The Nye tensor is calculated from the atomic positions of the dislocated crystal which are compared to the reference lattice as in the case of dislocation displacement maps. It then describes the distribution of the resultant Burgers vector in terms of contour plots.

In order to compute the Nye tensor for each atom of the dislocated lattice we identify its nearest neighbour atoms as those lying within a sphere of radius $R = \frac{1}{2}(R_1 + R_2)$ where $R_1$ and $R_2$ are the first and second coordination radii of the reference lattice. We then define $\mathbf{Q}^{(\gamma)}$ as the radius vectors of the nearest neighbours (where $\gamma$ indexes over the neighbours) and compare them to $\mathbf{P}^{(\beta)}$ which is the equivalent set of radius vectors for the reference lattice. The reference vector $\mathbf{P}^{(\beta)}$ with the smallest deviation angle $|\theta_{\gamma \beta}|$ is recognised as the one corresponding to $\mathbf{Q}^{(\gamma)}$. This procedure is followed to establish the correspondence between all dislocated vectors $\mathbf{Q}^{(\gamma)}$ and reference vectors $\mathbf{P}^{(\beta)}$. If two reference lattice vectors can be associated with a given bond vector the latter is rejected. It is also suggested that bonds with a deviation angle (angle between dislocated bond vector and that of a reference lattice) exceeding a critical value $\theta_{max}$ should be rejected in order to tune the sensitivity of the resulting Nye tensor for certain lattice misfit features (it is recommended that $\theta_{max}$ is equal to $27^{\circ}$ for fcc and $15^{\circ}$ for bcc lattices; more details in \cite{Hartley20051313}, \cite{doi:10.1080/14786430600660849}). For example the value of $\theta_{max}$ equal to $27^{\circ}$ reveals Shockley partials but not stacking faults in the fcc lattice.

Once the association between dislocated vectors $\mathbf{Q}^{(\gamma)}$ and reference vectors $\mathbf{P}^{(\gamma)}$ is established, the correspondence tensor $\mathbf{G}$ is constructed for each atom of the dislocated lattice. The system of $3\gamma$ equations is written in the matrix form:

\begin{equation}
\mathbf{P} = \mathbf{Q} \mathbf{G} .
\end{equation}

\noindent The mean-squares solution of the correspondence tensor $\mathbf{G}$ is given by:

\begin{equation}
\mathbf{G} = \mathbf{Q}^{+} \mathbf{P} ,
\end{equation}

\noindent where $\mathbf{Q}^{+}$ is the generalised inverse (Moore-Penrose matrix). It is defined as:

\begin{equation}
\mathbf{Q}^{+} = (\mathbf{Q}^{T} \mathbf{Q})^{-1} \mathbf{Q}^{T} .
\label{equation:moore-penrose}
\end{equation}

The Nye tensor $\boldsymbol{\alpha}$ is computed from the spatial derivatives of $\mathbf{G}$ by the means of finite differences for each atom of the dislocated lattice. We define vector $\mathbf{A}$ as:

\begin{equation}
A(ij)_k = \frac{\partial G_{ij}}{\partial x_k} ,
\end{equation}

\noindent and the finite differences equations can be expressed in a matrix form:

\begin{equation}
\Delta_k G_{ij} = Q_{kl} A(ij)_l .
\end{equation}

\noindent We can compute vector $\mathbf{A}$ using equation \ref{equation:moore-penrose}:

\begin{equation}
A(ij)_k = Q^{+}_{kl} \Delta_l G_{ij} ,
\end{equation}

\noindent and by repeating this procedure for all nine components the value of the Nye tensor is obtained as:

\begin{equation}
\alpha_{ij} = - \epsilon_{ikl} \frac{\partial G_{lj}}{\partial x_k} = - \epsilon_{ikl} A(lj)_k .
\end{equation}

In order to obtain a contour plot of Nye tensor values in the plane perpendicular to the dislocation line we use bicubic interpolation to resample the values obtained for the lattice points. For a screw dislocation with a dislocation line along the $z$-direction we are particularly interested in the $\alpha_{13}$, $\alpha_{23}$ and $\alpha_{33}$ components of the Nye tensor. The screw component is captured by $\alpha_{33}$, while the $x$-$y$ edge components are captured by $\alpha_{13}$ and $\alpha_{23}$ respectively. The overall edge component can be visualised by computing $\sqrt{\alpha_{13}^2 + \alpha_{23}^2}$.

We demonstrate the Nye tensor and dislocation displacement map visualisations of the polarised screw dislocation core computed using the Finnis-Sinclair interatomic potential in figure \ref{figure:dislocation_structure_visualisation} below.

\begin{figure}[H]
\begin{center}
\resizebox{12cm}{!}{\footnotesize{}\input{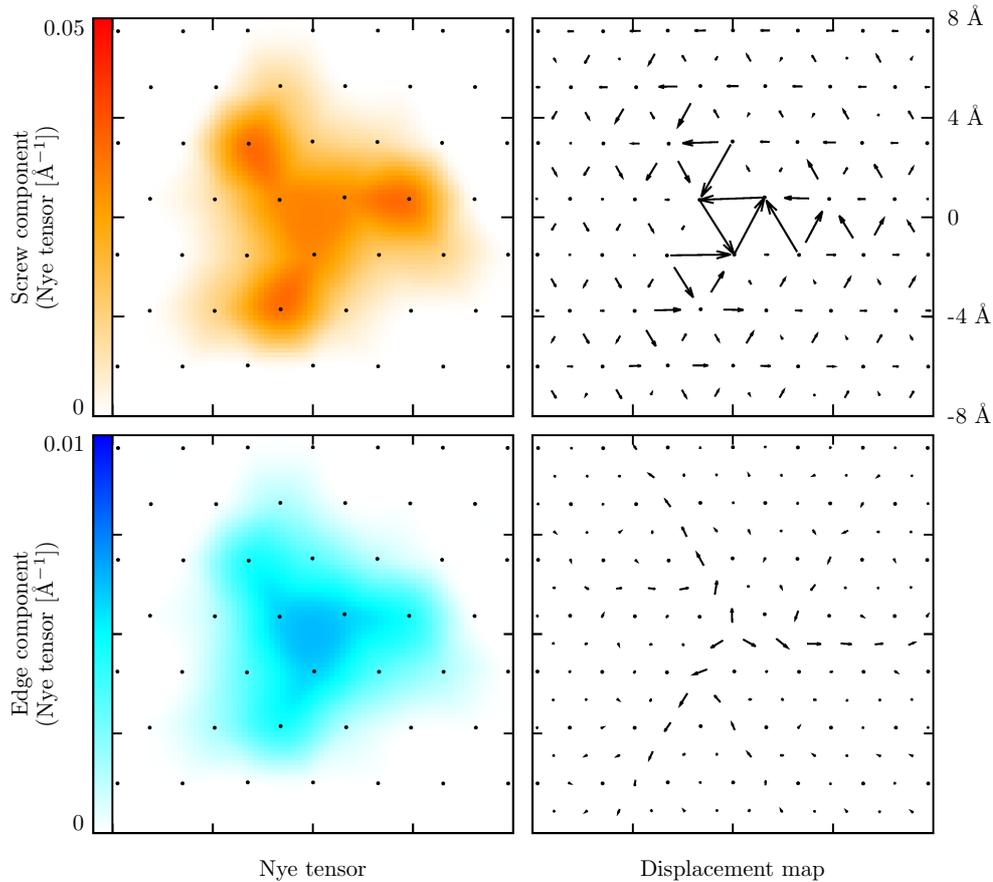}\normalsize{}}
\vspace{0.5cm}
\caption[Visualisation of screw dislocation core.]{Nye tensor and dislocation displacement map visualisation of non-symmetric screw dislocation core structure computed using the Finnis-Sinclair interatomic potential.}
\label{figure:dislocation_structure_visualisation}
\end{center}
\end{figure}

\subsection{Simulation Approaches}

The symmetry of a single dislocation is not compatible with periodic boundary conditions of the usual simulation cell. The displacement field introduced by a dislocation line makes it impossible to match the opposite boundaries of a simulation cell without introducing artificial stresses reminiscent of grain boundaries (more details in \cite{Woodward200559}). Consequently, the simulation of dislocations requires careful preparation of the simulation environment and two independent approaches are usually employed:

\begin{enumerate*}
\item Simulation of dislocation dipoles --- dipoles are arranged in a way that superposition of the strain fields cancels out at the cell boundaries.
\item Simulation of isolated dislocations --- suitable boundary conditions and a vacuum region is used to terminate the dislocation.
\end{enumerate*}

The first method was originally employed for DFT treatment of dislocations in silicon (more details in \cite{PhysRevLett.69.2224}) but the same ideas have been more recently used to investigate $\frac{1}{2} \langle 111 \rangle$ screw dislocations in molybdenum, tantalum and iron (more details in \cite{PhysRevB.54.6941}, \cite{PhysRevLett.84.1499}, \cite{doi:10.1080/0141861021000034568}, \cite{PhysRevB.68.014104}, \cite{doi:10.1080/14786430310001611644}, \cite{PhysRevB.67.140101}, \cite{PhysRevB.70.104113}). In our study we use a quadrupolar periodic arrangement of screw dislocations which, with an appropriate choice of lattice vectors, can be reduced to a cell of half the original size which contains only two dislocations with opposite Burgers vectors. A schematic representation of the simulation cell is shown in figure \ref{figure:dislocation_dipole} below.

\begin{figure}[H]
\begin{center}
\resizebox{12cm}{!}{\footnotesize{}\input{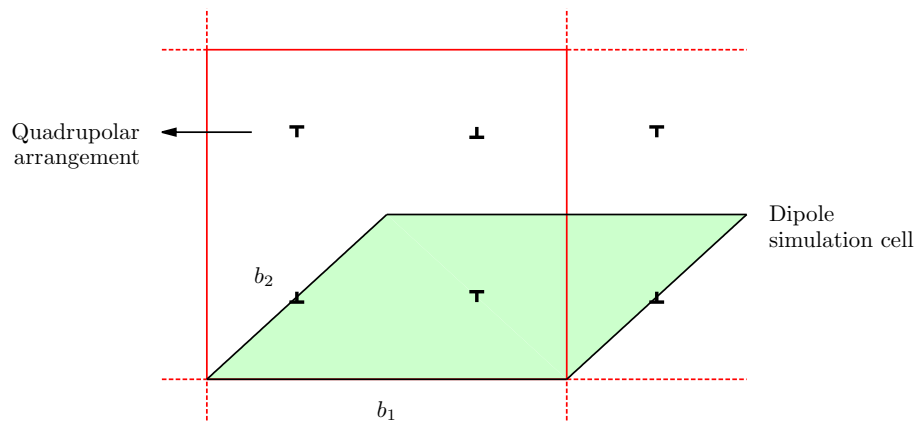}\normalsize{}}
\caption{Schematic representation of dislocation dipole simulation cell.}
\label{figure:dislocation_dipole}
\end{center}
\end{figure}

\noindent The choice of lattice vectors, dislocation separation and the resulting simulation cell size for our choice of simulation cells is given in table \ref{table:simulation_cells} below. Note that lattice vector $b_1$ contains a $z$-component which is necessary for the superposition of dislocation strain fields to cancel out at the cell boundaries:

\begin{table}[H]
\vspace{0.5cm}
\begin{center}
\begin{tabular}{ r l c c c }
\midrule
135 at. & (19.85 \r{A}): & $b_1 = 5 u_1$ & $b_2 = \frac{5}{2} u_1 + \frac{9}{2} u_2 + \frac{1}{3} u_3$ & $b_3 = u_3$ \\
\midrule
459 at. & (36.61 \r{A}): & $b_1 = 9 u_1$ & $b_2 = \frac{9}{2} u_1 + \frac{17}{2} u_2 + \frac{1}{3} u_3$ & $b_3 = u_3$ \\
\midrule
1215 at. & (59.56 \r{A}): & $b_1 = 15 u_1$ & $b_2 = \frac{15}{2} u_1 + \frac{27}{2} u_2 + \frac{1}{3} u_3$ & $b_3 = u_3$ \\
\midrule
1995 at. & (76.33 \r{A}): & $b_1 = 19 u_1$ & $b_2 = \frac{19}{2} u_1 + \frac{35}{2} u_2 + \frac{1}{3} u_3$ & $b_3 = u_3$ \\
\midrule
3375 at. & (99.27 \r{A}): & $b_1 = 25 u_1$ & $b_2 = \frac{25}{2} u_1 + \frac{45}{2} u_2 + \frac{1}{3} u_3$ & $b_3 = u_3$ \\
\midrule
7839 at. & (151.30 \r{A}): & $b_1 = 39 u_1$ & $b_2 = \frac{39}{2} u_1 + \frac{67}{2} u_2 + \frac{1}{3} u_3$ & $b_3 = u_3$ \\
\midrule
\end{tabular}
\caption[Dislocation dipole simulation cell configurations.]{Dislocation dipole simulation cell configurations and corresponding dislocation separation distances. Lattice vectors are expressed in terms of $u_1 = [1 1 \bar{2}]$, $u_2 = [\bar{1} 1 0]$ and $u_3 = \frac{1}{2}[1 1 1]$.}
\label{table:simulation_cells}
\end{center}
\end{table}

The second approach to the simulation of dislocations involves a single, isolated dislocation. In our study we prepare our simulation cell by creating a cylinder that is periodic in the direction parallel to the dislocation line and terminated by a vacuum in the perpendicular directions. We introduce the strain field of an ideal dislocation inside the cylinder according to linear elasticity theory and subsequently divide the cylinder into an active region (inner cylinder) and inactive region (outer annulus surrounding the active region). The atomic positions inside the inactive region are fixed which imposes the boundary conditions equivalent to that of an idealised dislocation in an infinitely large crystal. When the size of inactive region is sufficiently large there is no need to worry about dislocation images due to free surfaces. However, care needs to be taken so that the size of both active and inactive regions is sufficiently large. The thickness of the inactive region should be greater than the effective range of the interatomic interactions while the size of the active region needs to be selected so that the stress field of our dislocation approximates the long range stress field predicted by the linear elasticity theory in the inactive region. A schematic representation of the simulation cell used in this approach is shown in figure \ref{figure:isolated_dislocation} below.

\begin{figure}[H]
\begin{center}
\resizebox{12cm}{!}{\footnotesize{}\input{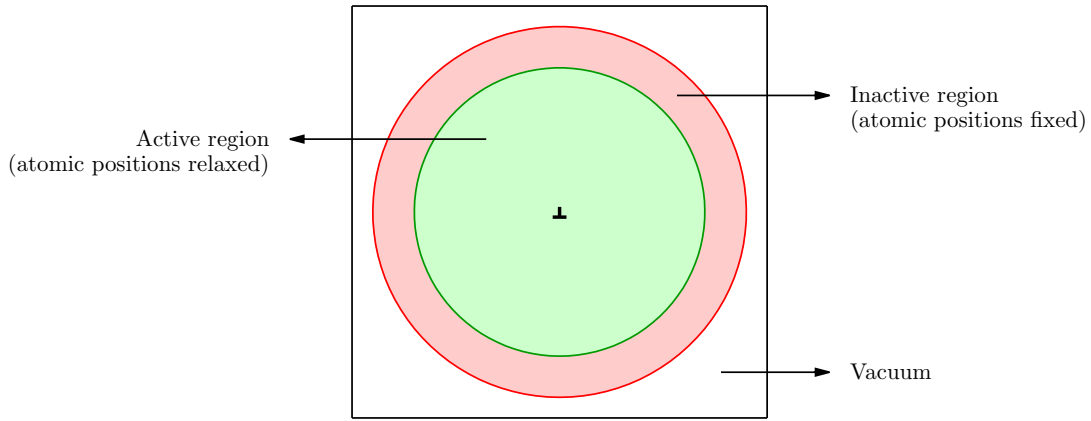}\normalsize{}}
\caption{Schematic representation of isolated dislocation simulation cell.}
\label{figure:isolated_dislocation}
\end{center}
\end{figure}

In order to investigate the convergence of the system size for our simulations of dislocations using either of the above mentioned methods we use the Finnis-Sinclair interatomic potential to perform geometry optimisation with respect to atomic positions. We also optimise the lattice vectors for the dislocation dipole simulation cells. This allows us to compute the local energy of the dislocation core which we defined as the local energy of atoms inside a cylinder with a radius equivalent to two lattice constants. The error in the dislocation core local energy as a function of system size is shown in figure \ref{figure:fs_dislocation_size_convergence} below.

\begin{figure}[H]
\begin{center}
\vspace{0.25cm}
\resizebox{12cm}{!}{\footnotesize{}\input{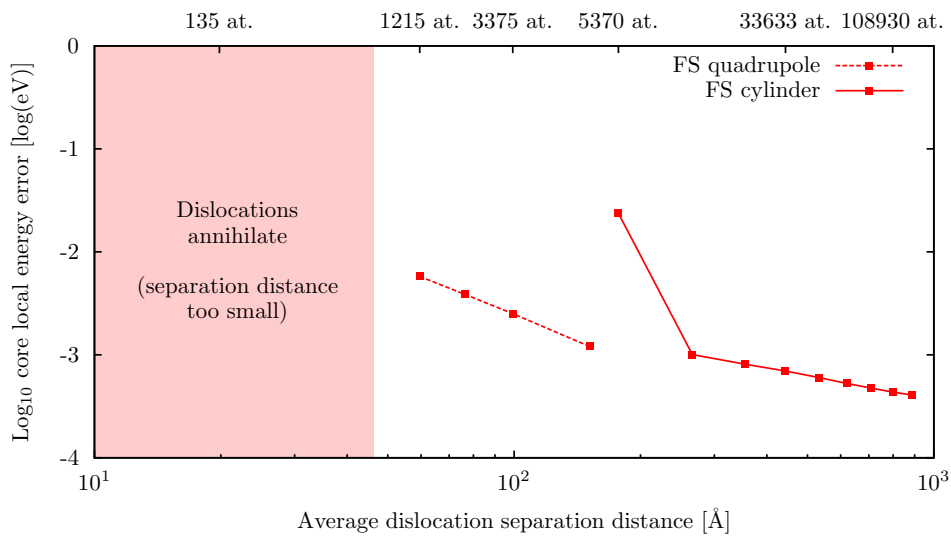}\normalsize{}}
\vspace{0.75cm}
\caption[Dislocation core local energy error as a function of system size.]{Convergence of dislocation core local energy error with the system size for dislocation quadrupole and isolated dislocation.}
\label{figure:fs_dislocation_size_convergence}
\end{center}
\end{figure}

It is clear that the system size required to simulate a truly isolated dislocation can be only achieved by the means of interatomic potentials. Even the dislocation dipole method requires substantial computational effort for the larger simulation cells which turns out to be prohibitive for the DFT method. We find that the largest system size for which we can evaluate single-point DFT energies and forces (at a substantial computational cost) corresponds to the smallest, 135 atom dislocation dipole simulation cell (this figure will, of course, increase over time).

As demonstrated in figure \ref{figure:fs_dislocation_size_convergence} the non-symmetric core when modelled by the FS interatomic potential cannot be simulated in the 135 atom dislocation dipole simulation cell unless the core atoms are constrained. However, based on the results of the DFT calculations available in the literature for other bcc transition metals (more details in \cite{PhysRevB.54.6941}, \cite{PhysRevLett.84.1499}, \cite{doi:10.1080/0141861021000034568}, \cite{PhysRevB.68.014104}, \cite{doi:10.1080/14786430310001611644}, \cite{PhysRevB.67.140101}, \cite{PhysRevB.70.104113}) we anticipate that for the symmetric core structure the amount of spreading of the dislocation into the slip planes of the $\langle 111 \rangle$ zone is smaller and therefore a system size of 135 atoms might prove sufficient for verification of our results by means of a DFT calculation.

Consequently, our modus operandi is as follows: even though a single evaluation of energy and forces using DFT method in 135 atom dislocation dipole cell is computationally tractable we find that carrying out a series of calculations that would be necessary to perform a geometry optimisation would be highly impractical. The calculations would take weeks even while running on hundreds of computing cores in parallel. Hence, to verify our predictions of dislocation behaviour obtained using our GAP potential for tungsten, we will carry out geometry optimisation using the GAP potential, and verify the resulting configurations with single-point DFT calculations. Hence, we can use the resulting DFT energies and forces to benchmark our GAP model.

\cleardoublepage

\chapter[Bispectrum-GAP Potential for Tungsten]{\texorpdfstring{Bispectrum-GAP Potential \\ for Tungsten}{Bispectrum-GAP Potential for Tungsten}}
\label{chapter:bispectrum-gap_potential_for_tungsten}

\section{Introduction}
\label{chapter:bispectrum-gap_potential_for_tungsten:section:introduction}

In this chapter I outline our first attempt at training a GAP interatomic potential for tungsten based on the 4-dimensional bispectrum descriptor of the atomic environment and the square-exponential covariance function. In order to simplify the training process and to reduce the size of the required training database, we begin by training a quantum-mechanical based GAP correction to the existing Finnis-Sinclair interatomic potential (as outlined in section \ref{chapter:gaussian_approximation_potential:section:interatomic_potential}).

In order to generate the DFT training data we combine the sampling techniques and simulation principles outlined in the previous chapters \ref{chapter:simulation_techniques} and \ref{chapter:bulk_properties_and_lattice_defects_in_tungsten}. I will discuss these procedures in more detail in section \ref{chapter:bispectrum-gap_potential_for_tungsten:section:training_protocol_and_dataset}. In section \ref{chapter:bispectrum-gap_potential_for_tungsten:section:results} I present the results obtained with the resulting FS/bispectrum-GAP potential for tungsten, where we quantitatively assess its performance and present its prediction of the screw dislocation core structure. I follow this with a discussion of our investigation into the convergence of the hyperparameters.

Finally, in section \ref{chapter:bispectrum-gap_potential_for_tungsten:section:discussion}, I finish this chapter with a brief discussion of the results obtained using the bispectrum-GAP potential compared against the Finnis-Sinclair interatomic potential and the DFT method and analyse its shortcomings.

\section{Training Protocol and Dataset}
\label{chapter:bispectrum-gap_potential_for_tungsten:section:training_protocol_and_dataset}

\subsection{Elastic Constants}

In our first attempt at training a quantum-mechanical based GAP correction to the FS potential, we begin by investigating whether one can generate a more accurate description of the elastic properties of the material than that provided by the underlying core potential. Although the lattice constant and bulk modulus in the FS potential can be set to any required value by a simple rescaling (as demonstrated in section \ref{chapter:bulk_properties_and_lattice_defects_in_tungsten:section:elastic_properties}), this is not enough to reproduce all the elastic constants predicted using the DFT method. Furthermore, as we showed in section \ref{chapter:bulk_properties_and_lattice_defects_in_tungsten:section:elastic_properties} in order to predict stress-strain curves in the anharmonic regime, significant corrections to the potential energy surface are required.

Since all the elastic properties of tungsten can be calculated by computing simulation stresses as the primitive unit cell is strained (strictly speaking they can be computed from energies as well, but with lower accuracy\footnote{Calculation from energies involves fitting a quadratic curve, as opposed to fitting a line in the case of stresses.}), we find that the most efficient way of training elastic constants is by training from the DFT values of these quantities calculated for randomly strained primitive unit cells. Since we are training the GAP correction from stresses, it is vital that the DFT $k$-point sampling density is sufficiently converged so that all the resulting stress data is reliable and consistent. Since the bcc lattice primitive unit cell contains just a single atom, our problem of generating training data is equivalent to the problem of sampling the energy landscape in the phase-space of the lattice parameters. In principle, this could be achieved by means of a fixed pressure, variable volume molecular dynamics simulation. However, the concept of MD simulation of a single atom would be ill-defined as the forces on an atom in a primitive unit cell are always zero. Consequently, we employ slice sampling (as outlined in section \ref{chapter:simulation_techniques:section:monte_carlo_methods}) to sample the energy landscape in the space of the lattice parameters directly. Since the energy is invariant to rotations of the simulation cell, the sampled space is six-dimensional.

\begin{figure}[H]
\begin{center}
\resizebox{12cm}{!}{\footnotesize{}\input{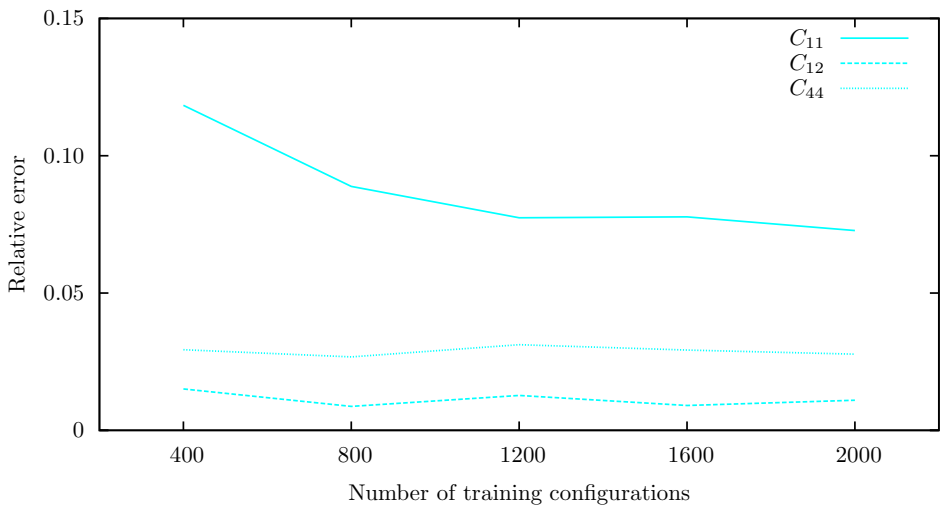}\normalsize{}}
\vspace{0.75cm}
\caption[Convergence of elastic constants with training data.]{Convergence of the three independent elastic constants $C_{11}$, $C_{12}$ and $C_{44}$ with the volume of training data.}
\label{figure:convergence_elastic_constants}
\end{center}
\end{figure}

Figure \ref{figure:convergence_elastic_constants} shows the convergence of the three independent elastic constants $C_{11}$, $C_{12}$ and $C_{44}$, as the volume of training data increases.

\subsection{Phonon Spectrum}

As demonstrated in figure \ref{figure:fs_phonons}, the phonon spectrum predicted by the Finnis-Sinclair interatomic potential does not account for some of the acoustic modes of vibration predicted using DFT method (in particular in the $\{\text{H}-\text{N}\}$ and $\{\text{N}-\Gamma\}$ parts of the spectrum). Although the FS potential, overall, provides a description of the phonon spectrum that in a qualitative agreement with DFT, we also find that some of the phonon frequencies do not match the DFT prediction.

The phonon spectrum describes a collective motion of atoms, which apart from some highly symmetrical points in the Brillouin zone cannot be captured within a small simulation cell under periodic boundary conditions. Hence, we find that in order to represent normal modes of vibration one requires a simulation cell of a sufficient size. The training data needs to be obtained using simulation cells whose size reflects that of the simulation cells required to compute the phonon spectrum using finite displacement method.

We find that when performing DFT calculations with the convergence criteria described in section \ref{chapter:bulk_properties_and_lattice_defects_in_tungsten:section:convergence_of_dft_calculations}, we are limited to only very small simulation cells (of the order of 50 atoms) due to the computational cost associated with extremely high $k$-point sampling density required to converge stresses. However, stresses are the most efficient method of providing training data only for small simulation cells. So as long as we can provide accurate forces which are key for reconstruction of the dynamical matrix necessary for an accurate description of the phonon spectrum, we can decrease the density of $k$-point sampling and simply reject the resulting stresses from our phonon spectrum training dataset.

When we decrease linear $k$-point sampling density to $0.03 \, \text{\r{A}}^{-1}$, we find that we can generate training data using 128 atom cubic cells (4 x 4 x 4 supercell of conventional bcc unit cell) at an acceptable computational cost. However, even with a unit cell of constant volume, this corresponds to sampling of 381-dimensional phase space. In practice we anticipate that the only configurations that are physically relevant for the purpose of describing phonon spectrum are the ones that are easily accessible when the system is evolved dynamically using Newton's equations of motion. Consequently, we sample the accessible states of the system by the means of a molecular dynamics simulation (as outlined in section \ref{chapter:simulation_techniques:section:molecular_dynamics}).

Even at our reduced $k$-point sampling density, while it is computationally feasible to compute energies and forces for a few tens of 128 atom configurations, generating long MD trajectories necessary to obtain uncorrelated training data is highly impractical. Consequently, we explore two ways of generating MD trajectories for the purpose of generating our training dataset:

\begin{enumerate*}
\item \begin{sloppypar}The MD trajectory is computed using the Finnis-Sinclair interatomic potential, with snapshots of the trajectory selected as training samples and the atomic forces recomputed using DFT method.\end{sloppypar} 
\item The MD trajectory is computed using DFT method at an even further reduced $k$-point sampling and plane-wave energy cutoff. Snapshots are taken from this trajectory and forces recomputed using converged values for these parameters.
\end{enumerate*}

While the first method is computationally less expensive, it relies on the fact that the potential energy surface of the Finnis-Sinclair potential is similar to that obtained by the means of DFT calculations. Our sampling will correspond to physically relevant configurations only if there is sufficient overlap between the two potential energy surfaces. In practice, we find that although near the ground state the Finnis-Sinclair PES approximates that of the DFT method reasonably well, they diverge away from the harmonic regime (as demonstrated in section \ref{chapter:bulk_properties_and_lattice_defects_in_tungsten:section:elastic_properties}). 

Consequently, in order to generate a meaningful phonon spectrum training dataset we compute our MD trajectories using DFT at ``under-converged'' values of $k$-point sampling and plane-wave energy cutoff. Due to the computational cost associated with equilibration of the thermostats, we carry out our simulations at constant volume and energy over a range of volumes around $\pm 1\%$ of the ground state volume and temperatures of $300 \, \text{K}$ and $1000 \, \text{K}$.

\subsection{Lattice Defects}

We follow the same procedure as outlined above to obtain a training dataset for the lattice defects --- i.e. we compute an MD trajectory using DFT at ``under-converged'' values of $k$-point sampling and plane-wave energy cutoff. We then recompute snapshots from this trajectory as training samples at converged values of these parameters. We minimise the potential energy stored in the lattice by the means of a geometry optimisation before the start of the MD simulation. The initial state of the trajectory is generated by randomising the kinetic energies of the atoms, such that they are Boltzmann distributed with the velocities being a function of simulation temperature. To reduce the computational cost, we again use a linear $k$-point sampling density of $0.03 \, \text{\r{A}}^{-1}$ to obtain converged values of energy and forces and we discard the stresses.

In order to reproduce the DFT value of vacancy formation energy, we begin by training with data using 53 atom cubic cells (3 x 3 x 3 supercell). However, we find that this data alone (which corresponds to a vacancy separation distance of $9.52 \, \text{\r{A}}$) is not sufficient. In order to account for lattice relaxation, we also carry out simulations over a range of volumes around $\pm 1\%$ of the volume corresponding to a single vacancy in a simulation cell with relaxed lattice vectors. Our MD runs are carried out at temperatures of $300 \, \text{K}$ and $1000 \, \text{K}$, and we also include training data for different vacancy density and so a limited number of 127 atom cubic cells (4 x 4 x 4 supercell giving a vacancy separation distance of $12.68 \, \text{\r{A}}$) are also computed.

We find that generating a GAP potential that reproduces the formation energies of free surfaces is significantly simpler than doing the same for vacancies. This can be accounted for by the fact that the atomic environment of the surface atoms is radically different to that of other lattice defects inside the bulk. Hence, by considering atomic coordination alone, an atom next to a vacancy might have its coordination number reduced by a small fraction, but a surface atom has its coordination effectively halved. Since the bispectrum descriptor can distinguish such configurations with ease, a reduced volume of training data is required to fit the GAP potential for description of surfaces. It is then in agreement with our expectations that the training process is greatly simplified.

Finally, in order to reproduce gamma surface energies we sample the gamma surface using a 10 x 10 regular grid of points along the lattice vectors. We use configurations in which the atomic positions are relaxed in the direction perpendicular to the surface as the initial configurations. We then carry out short trajectory MD simulations over a range of volumes ($\pm 1\%$) from the ground state volume. Due to the significant computational cost associated with these calculations, we only sample the $(110)$ and $(112)$ gamma surfaces in our MD simulations (as we believe these are the two slip planes with the most physical significance). We also restrict ourselves to trajectories at a single temperature of $300 \, \text{K}$. If we need to describe high temperature processes in the future, our training dataset can always be expanded with data obtained at other temperatures.

\section{Results}
\label{chapter:bispectrum-gap_potential_for_tungsten:section:results}

In order to demonstrate how the bispectrum-GAP interatomic potential can be systematically improved, we repeatedly carry out our training procedure while increasing the size of the training database. At the same time we monitor the performance of the potential by calculating the RMS energy and force errors for the training datasets, and we verify the predicted values of lattice constants, elastic constants, formation energies of isolated vacancy and surfaces and the RMS error in the phonon spectrum. A summary of the training databases and performance of the associated bispectrum-GAP potential is given in table \ref{table:bispectrum-gap_results} below.

We find that as the number of training configurations in the training database increases the overall performance of the resulting bispectrum-GAP correction to the Finnis-Sinclair interatomic potential improves. However, we also observe that when the bispectrum-GAP correction is fitted using an incomplete training database and benchmarked against the configurations that were not included in the training process, the performance of the resulting potential can be worse than that of the FS interatomic potential alone. For example, column FS/b-GAP$_2$ in table \ref{table:bispectrum-gap_results} corresponds to a bispectrum-GAP potential where no lattice defects were included in the training database. While this potential improves the description of bulk properties compared to the FS interatomic potential, we also observe that the RMS energy and force errors for configurations with lattice defects are greater than the corresponding errors for the FS interatomic potential alone. This behaviour, namely that the bispectrum-GAP correction can decrease the performance of the resulting potential in the extrapolation regime, is against our initial expectations. However, we offer an in-depth explanation of this phenomenon in section \ref{chapter:bispectrum-gap_potential_for_tungsten:section:discussion} of this chapter.

\begin{sloppypar} 
To investigate the elastic properties of our bispectrum-GAP interatomic potential in the anharmonic regime (from now on we are using the most complete bispectrum-GAP potential, designated as FS/b-GAP in table \ref{table:bispectrum-gap_results}), we compute the stress-strain curves corresponding to longitudinal compression, transverse expansion and shearing for a range of strains from $-10\%$ to $+10\%$. The results are shown in figure \ref{figure:bispectrum-gap_stress-strain} below.
\end{sloppypar} 

\clearpage

\newgeometry{top=1.5cm,bottom=1.5cm,left=3.5cm,right=1.5cm}

\thispagestyle{empty}

\begin{sidewaystable}
\begin{center}
\scriptsize{}
\begin{tabular}{ l | c | c | c | c | c || c | c | }
& FS/b-GAP$_{1}$ & FS/b-GAP$_{2}$ & FS/b-GAP$_{3}$ & FS/b-GAP$_{4}$ & FS/b-GAP & FS & DFT \\
\hline
\multicolumn{8}{ l }{} \\
\multicolumn{8}{ l }{\emph{Training database errors:}} \\
\cline{1-7}
RMS energy error per atom [eV] & 0.2093 & 0.0999 & 0.0631 & 0.0251 & 0.0016 & 0.0112 & \multicolumn{1}{ c }{} \\
RMS force error [eV/\r{A}] & 5.2245 & 4.8777 & 2.0360 & 1.4322 & 0.1906 & 0.7411 & \multicolumn{1}{ c }{} \\
\cline{1-7}
\multicolumn{8}{ l }{} \\
\multicolumn{8}{ l }{\emph{Number of atomic environments in training database:}} \\
\cline{1-6}
bcc primitive cells (MCMC, 2000 $\times$ 1 at.) & 2000 & 2000 & 2000 & 2000 & 2000 & \multicolumn{2}{ c }{} \\
bcc bulk (MD, 60 $\times$ 128 at.) & --- & 7680 & 7680 & 7680 & 7680 & \multicolumn{2}{ c }{} \\
vacancy (MD, 400 $\times$ 53 at., 20 $\times$ 127 at.) & --- & --- & 23740 & 23740 & 23740 & \multicolumn{2}{ c }{} \\
100, 110, 111, 112 surfaces (MD, 180 $\times$ 12 at.) & --- & --- & --- & 2160 & 2160 & \multicolumn{2}{ c }{} \\
110, 112 gamma surfaces (MD, 6183 $\times$ 12 at.) & --- & --- & --- & --- & 74196 & \multicolumn{2}{ c }{} \\
\cline{1-6}
\multicolumn{8}{ l }{} \\
\multicolumn{8}{ l }{\emph{RMS energy error per atom:} [eV]} \\
\cline{1-7}
bcc primitive cells & 0.0006 & 0.0007 & 0.0009 & 0.0011 & 0.0016 & 0.0158 & \multicolumn{1}{ c }{} \\
bcc bulk & \emph{0.0038} & 0.0001 & 0.0001 & 0.0001 & 0.0001 & 0.0002 & \multicolumn{1}{ c }{} \\
vacancy & \emph{0.0441} & \emph{0.0278} & 0.0001 & 0.0001 & 0.0002 & 0.0013 & \multicolumn{1}{ c }{} \\
100, 110, 111, 112 surfaces & \emph{0.3120} & \emph{0.3246} & \emph{0.3062} & 0.0003 & 0.0004 & 0.0233 & \multicolumn{1}{ c }{} \\
110, 112 gamma surfaces & \emph{0.2477} & \emph{0.1071} & \emph{0.0562} & \emph{0.0306} & 0.0019 & 0.0127 & \multicolumn{1}{ c }{} \\
\cline{1-7}
\multicolumn{8}{ l }{} \\
\multicolumn{8}{ l }{\emph{RMS force error:} [eV/\r{A}]} \\
\cline{1-7}
bcc primitive cells & --- & --- & --- & --- & --- & --- & \multicolumn{1}{ c }{} \\
bcc bulk & \emph{1.8516} & 0.1097 & 0.0867 & 0.0955 & 0.0902 & 0.1460 & \multicolumn{1}{ c }{} \\
vacancy & \emph{4.0107} & \emph{2.0216} & 0.1025 & 0.1207 & 0.1267 & 0.2415 & \multicolumn{1}{ c }{} \\
100, 110, 111, 112 surfaces & \emph{1.6530} & \emph{1.9752} & \emph{2.6188} & 0.1101 & 0.1424 & 0.5706 & \multicolumn{1}{ c }{} \\
110, 112 gamma surfaces & \emph{5.8994} & \emph{5.8119} & \emph{2.4350} & \emph{1.7404} & 0.2172 & 0.8845 & \multicolumn{1}{ c }{} \\
\cline{1-7}
\multicolumn{8}{ l }{} \\
\hline
lattice const. [\r{A}] & 3.1799 & 3.1804 & 3.1805 & 3.1810 & 3.1812 & 3.1805 & 3.1805 \\
C11 elastic constant [GPa] & 479.26 & 471.52 & 475.30 & 475.02 & 481.34 & 514.23 & 516.86 \\
C12 elastic constant [GPa] & 200.35 & 198.02 & 200.69 & 199.45 & 199.89 & 200.12 & 198.18 \\
bulk modulus [GPa] & 293.32 & 289.18 & 292.22 & 291.31 & 293.71 & 304.83 & 304.41 \\
shear modulus / C44 elastic constant [GPa] & 146.25 & 147.36 & 148.56 & 148.84 & 150.05 & 157.21 & 142.30 \\
\hline
RMS phonon spectrum error [THz] & --- & 0.232 & 0.215 & 0.273 & 0.342 & 0.392 & \multicolumn{1}{ c }{} \\
\hline
vacancy energy [eV] & --- & --- & 3.25 & 3.24 & 3.19 & 3.61 & 3.27 \\
\hline
100 surface energy [eV / \r{A}$^2$] & 0.302 & 0.367 & 0.321 & 0.251 & 0.250 & 0.179 & 0.251 \\
110 surface energy [eV / \r{A}$^2$] & 0.354 & 0.437 & 0.360 & 0.204 & 0.205 & 0.158 & 0.204 \\
111 surface energy [eV / \r{A}$^2$] & 0.421 & 0.205 & 0.218 & 0.222 & 0.222 & 0.202 & 0.222 \\
112 surface energy [eV / \r{A}$^2$] & 0.263 & 0.509 & 0.284 & 0.215 & 0.216 & 0.187 & 0.216 \\
\hline
RMS $\{110\} \langle 111 \rangle$ gamma surface energy error [eV] & 1.045 & 3.359 & 0.104 & 0.081 & 0.162 & 0.695 & \multicolumn{1}{ c }{} \\
\cline{1-7}
RMS dislocation energy error [eV] & 2.000 & 5.599 & 0.548 & 0.576 & 0.179 & 1.265 & \multicolumn{1}{ c }{} \\
\cline{1-7}
\end{tabular}
\normalsize{}
\end{center}
\captionsetup{margin=4cm}
\caption[Bispectrum-GAP training database and potential summary.]{Summary of the training databases and performance of the associated bispectrum-GAP potential.}
\captionsetup{margin=1cm}
\label{table:bispectrum-gap_results}
\end{sidewaystable}

\restoregeometry

\clearpage

\begin{figure}[H]
\begin{center}
\resizebox{12cm}{!}{\footnotesize{}\input{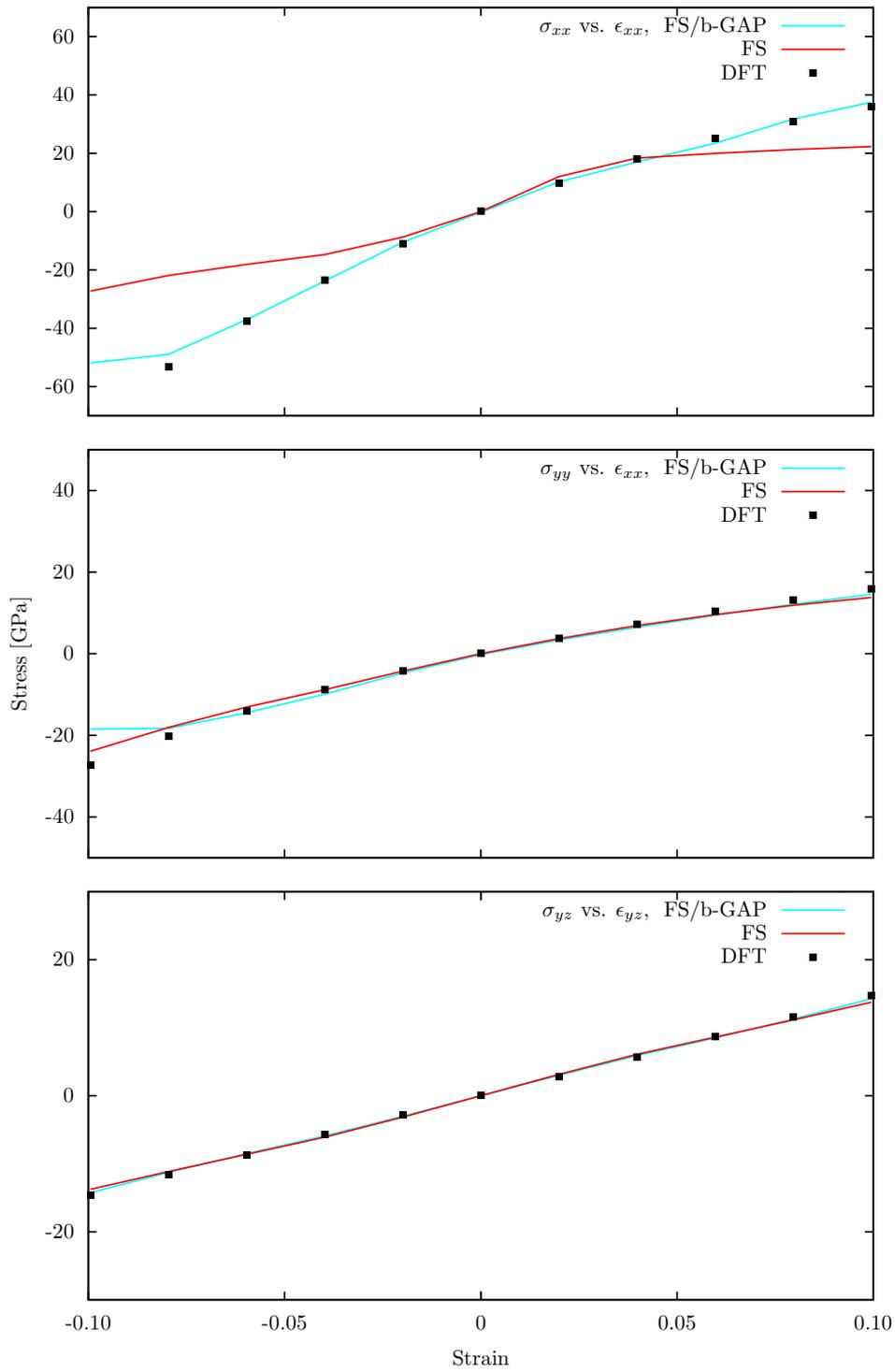}\normalsize{}}
\vspace{0.75cm}
\caption[FS/Bispectrum-GAP stress-strain curves.]{FS/Bispectrum-GAP stress-strain curves of bcc tungsten for a range of strains from $-10\%$ to $+10\%$.}
\label{figure:bispectrum-gap_stress-strain}
\end{center}
\end{figure}

\clearpage 

The stress-strain curves computed using the FS/bispectrum-GAP interatomic potential account for the non-linearity of the longitudinal compression and transverse expansion and they offer a much improved description of the elastic properties in the anharmonic regime as compared to the FS potential. However, we can also observe that for compressive strains larger than $8\%$, the description provided by the FS/bispectrum-GAP starts to break down. Investigation of the underlying data in detail reveals that this corresponds to an increase in energy over the ground state of $\sim 0.2 \, \text{eV}$ per atom. We find that the error in this energy is related to the fact that our training data was generated only at temperatures of up to $1000 \, \text{K}$. It is another demonstration of a general property that we observe with the FS/bispectrum-GAP potential, namely, that it provides good accuracy in the interpolation regime but its extrapolative powers are limited.

The phonon spectrum of bcc tungsten computed using the bispectrum-GAP interatomic potential is shown in figure \ref{figure:bispectrum-gap_phonons} below.

\begin{figure}[H]
\begin{center}
\resizebox{12cm}{!}{\footnotesize{}\input{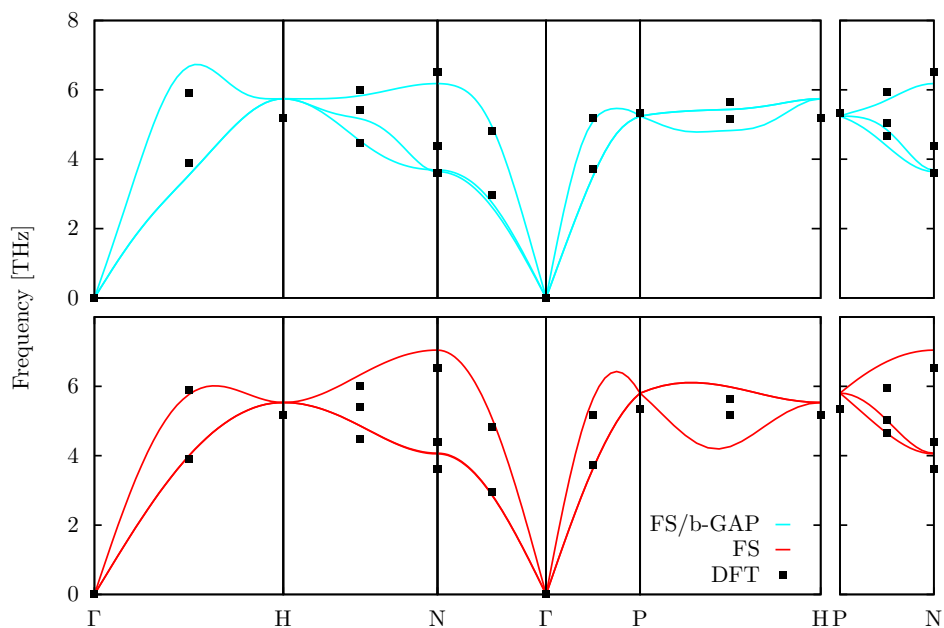}\normalsize{}}
\vspace{0.5cm}
\caption[FS/Bispectrum-GAP phonon spectrum.]{FS/Bispectrum-GAP phonon spectrum of bcc tungsten.}
\label{figure:bispectrum-gap_phonons}
\end{center}
\end{figure}

\noindent While the FS/bispectrum-GAP interatomic potential improves on the description of the phonon spectrum compared to the FS model (for instance it reproduces the transverse modes along the $\{\text{H}-\text{N}\}$ path), it fails to reproduce all of the non-degenerate modes along the $\{\text{N}-\Gamma\}$ path. The RMS error in phonon frequencies is nevertheless reduced as indicated in table \ref{table:bispectrum-gap_results}.

A cross-section of $(110)$ gamma surface energies along the $\langle 111 \rangle$ lattice vector computed using the FS/bispectrum-GAP interatomic potential is shown in figure \ref{figure:bispectrum-gap_gamma_surface_111} below. We observe a significant increase in accuracy as compared to the Finnis-Sinclair model and, since we anticipate that the dislocation structure is dictated by the energetics of $(110)$ and $(112)$ gamma surfaces, we expect to be able to predict the $\frac{1}{2} \langle 111 \rangle$ screw dislocation core structure with much improved accuracy using the FS/bispectrum-GAP potential.

\begin{figure}[H]
\begin{center}
\resizebox{12cm}{!}{\footnotesize{}\input{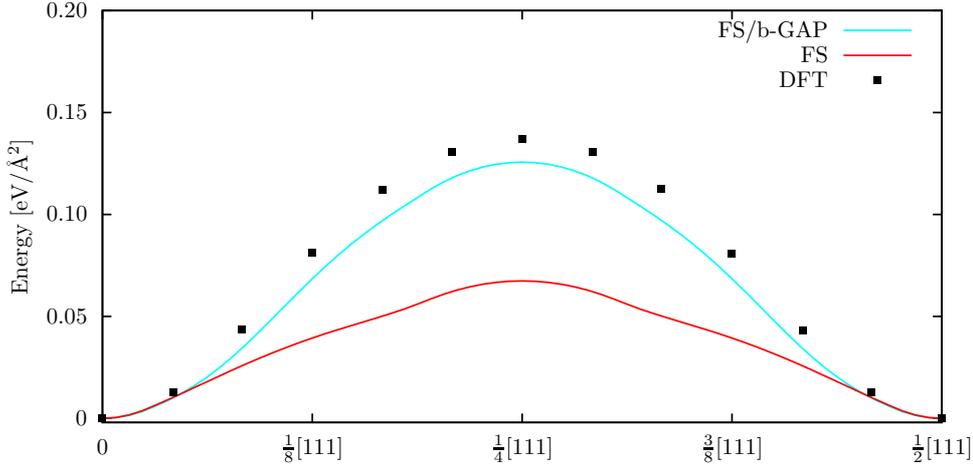}\normalsize{}}
\vspace{0.5cm}
\caption[$\langle 111 \rangle$ cross-section of $(110)$ gamma surface energy.]{$\langle 111 \rangle$ cross-section of $(110)$ gamma surface energy for GAP, FS and DFT models.}
\label{figure:bispectrum-gap_gamma_surface_111}
\end{center}
\end{figure}

As an independent test of our training methodology we also investigate how our FS/bispectrum-GAP interatomic potential copes with predicting the energy of an idealised, unrelaxed structure of a screw dislocation. With the smallest dislocation dipole simulation cell consisting of 135 atoms (which corresponds to the upper limit in terms of the system size that we can evaluate using the DFT method due to computational complexity), we investigate the relationship between the unrelaxed energy of such dislocation dipole system as a function of the Burgers vector. Since these are unrelaxed configurations, we only needed to evaluate ten 135 atom cells using DFT in order to verify the FS/bispectrum-GAP predictions against the target DFT values.\footnote{For Burgers vector values different from $\pm \frac{n}{2} \langle 111 \rangle$ and an integer $n$, the dislocation displacement field does not match the lattice periodicity, which results in a stacking fault connecting the two dislocation lines pointing in the opposite directions. While somewhat unusual, we find this test to be a good predictor of the overall capabilities of the potential since it also benchmarks the accuracy of gamma surface energies.} The values obtained are plotted in figure \ref{figure:bispectrum-gap_dislocation_dipoles} below.

\begin{figure}[H]
\begin{center}
\resizebox{12cm}{!}{\footnotesize{}\input{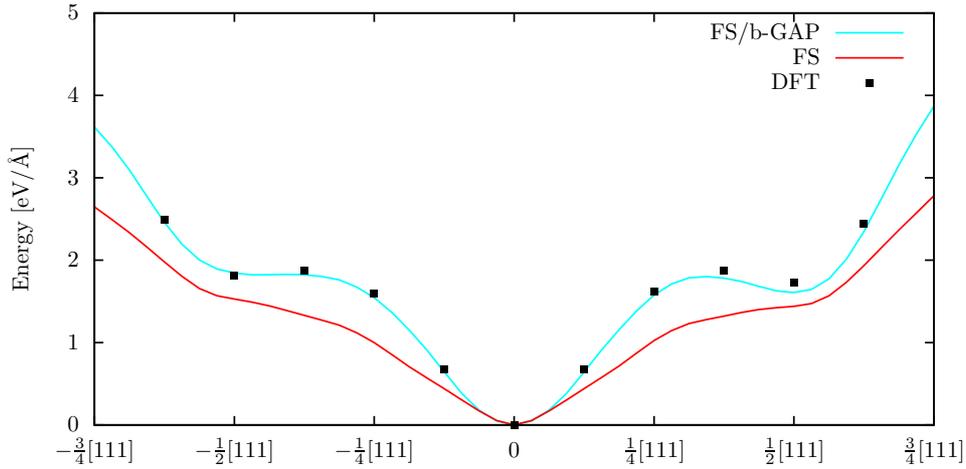}\normalsize{}}
\vspace{0.5cm}
\caption[Unrelaxed dislocation dipole energy.]{Unrelaxed dislocation dipole energy as a function of the Burgers vector for GAP, FS and DFT models.}
\label{figure:bispectrum-gap_dislocation_dipoles}
\end{center}
\end{figure}

\subsection{Screw Dislocation Core Structure}

As demonstrated in table \ref{table:bispectrum-gap_results} and in figure \ref{figure:bispectrum-gap_stress-strain} above, the Gaussian process regression based potential cannot be used to extrapolate energies associated with atomic environments that are radically different to those included in the training database. Nevertheless, even though our training database consists of atomic configurations in small unit cells exclusively, it enables us to carry out simulations in large systems, provided that the individual atomic environments are familiar.

As discussed in section \ref{chapter:bulk_properties_and_lattice_defects_in_tungsten:section:dislocations}, it is believed that the core structure of $\frac{1}{2} \langle 111 \rangle$ screw dislocation in tungsten can be rationalised in terms of the properties of the strictly planar gamma surfaces. We will now demonstrate that by including gamma surfaces in our training dataset, our bispectrum-GAP correction to the Finnis-Sinclair interatomic potential is capable of predicting the core structure of $\frac{1}{2} \langle 111 \rangle$ screw dislocations, even though no dislocation configurations are included in our training database explicitly.

We begin by investigating the convergence of dislocation core local energy with the system size for our dislocation simulations using both dislocation dipole and isolated dislocation methods. The error in the dislocation core local energy, as a function of system size, is given in figure \ref{figure:bispectrum-gap_dislocation_size_convergence} below.

\begin{figure}[H]
\begin{center}
\vspace{0.25cm}
\resizebox{12cm}{!}{\footnotesize{}\input{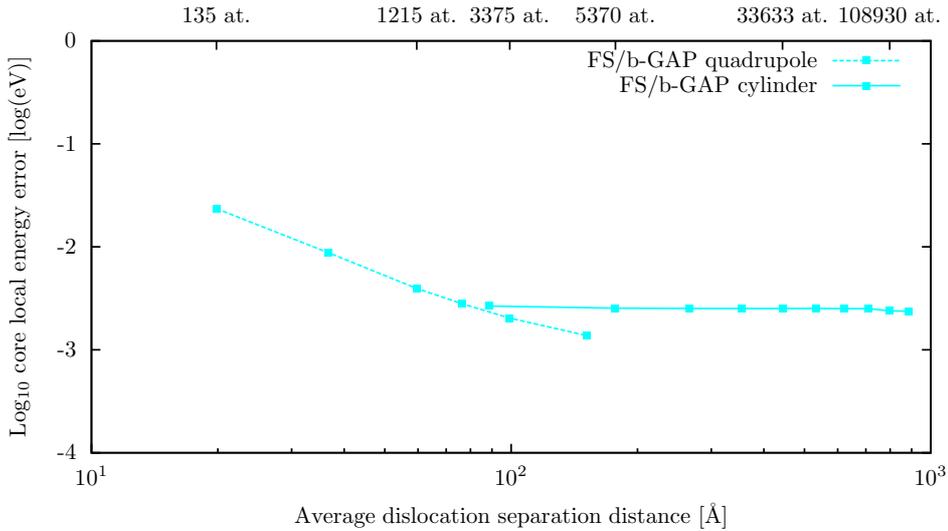}\normalsize{}}
\vspace{0.75cm}
\caption[Dislocation core local energy error as a function of system size.]{Convergence of dislocation core local energy error with the system size for dislocation quadrupole and isolated dislocation.}
\label{figure:bispectrum-gap_dislocation_size_convergence}
\end{center}
\end{figure}

\begin{sloppypar} 
Finally, we investigate the screw dislocation core structure obtained by means of geometry optimisation using the FS/bispectrum-GAP interatomic potential. In order to validate our results, we use the final atomic coordinates of the geometry optimisation performed with the FS/bispectrum-GAP interatomic potential as a starting point of a geometry optimisation using the DFT method in a 135 atom dislocation dipole simulation cell. The $\frac{1}{2} \langle 111 \rangle$ screw dislocation core structures computed using the bispectrum-GAP, Finnis-Sinclair and DFT methods are presented in figure \ref{figure:bispectrum-gap_dislocation_structure} below, where we characterise the dislocation structures using the Nye tensor (as outlined in section \ref{chapter:bulk_properties_and_lattice_defects_in_tungsten:section:dislocations}). For the Finnis-Sinclair and FS/bispectrum-GAP we also compute local energies of individual atoms.
\end{sloppypar} 

We find that the dislocation core structure predicted by the FS/bispectrum-GAP significantly improves on the description of the Finnis-Sinclair potential alone. Both FS/bispectrum-GAP and DFT predict a symmetric core (while Finnis-Sinclair predicts non-symmetric core) and the screw component (corresponding to out-of-plane displacement of atoms) of FS/bispectrum-GAP and DFT matches perfectly. However, there is a small difference in the edge component of the Nye tensor between the FS/bispectrum-GAP and DFT. This suggests that the two structures are not in full agreement and the displacement of atoms within the $(111)$ plane is not exactly the same though the differences are small.

When we verify the final atomic coordinates of the geometry optimisation performed with the FS/bispectrum-GAP interatomic potential by performing a single-point energy and force evaluation using DFT, we find that the maximum force error between FS/bispectrum-GAP and DFT methods is $0.62 \, \text{eV/\r{A}}$. While this is of the same order of magnitude as the RMS force error of the overall training database ($0.19 \, \text{eV/\r{A}}$), we feel that in order to achieve a more accurate description of the dislocation core structure, the force errors of the FS/bispectrum-GAP potential need to be reduced further.

\begin{figure}[H]
\begin{center}
\vspace{0.25cm}
\resizebox{12cm}{!}{\footnotesize{}\input{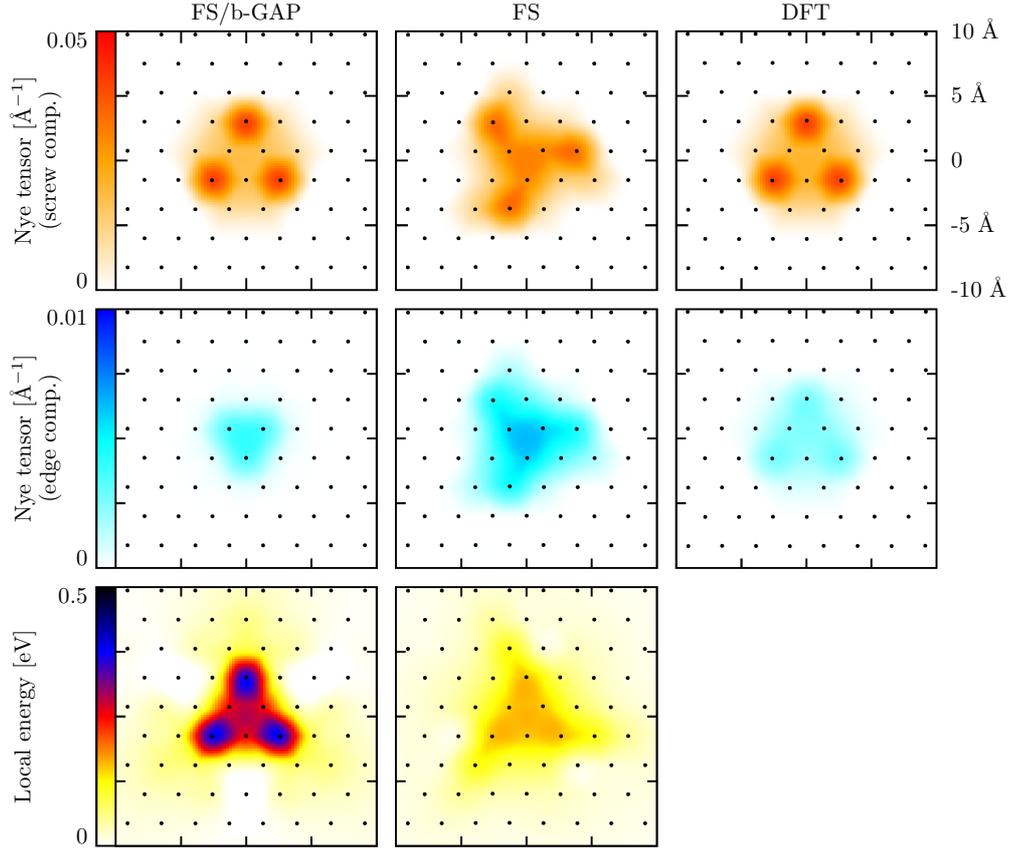}\normalsize{}}
\caption[$\frac{1}{2} \langle 111 \rangle$ screw dislocation core structure.]{$\frac{1}{2} \langle 111 \rangle$ screw dislocation core structures evaluated using bispectrum-GAP, FS and DFT models.}
\label{figure:bispectrum-gap_dislocation_structure}
\end{center}
\end{figure}

\subsection{Hyperparameters}

In GAP methodology hyperparameters are the adjustable parameters of the covariance function that reflect the prior knowledge of the dataset (as outlined in chapter \ref{chapter:gaussian_approximation_potential}). Consequently, for the bispectrum-GAP interatomic potential our fitting procedure can be tuned with the following hyperparameters:

\clearpage 

\begin{itemize*}
\item noise in the training data $\to \{\sigma_\nu^{(\text{energy})} , \sigma_\nu^{(\text{force})} , \sigma_\nu^{(\text{virial})} \}$
\item characteristic length-scale of the atomic descriptor $\to \mathbf{\Sigma} = \{ \frac{1}{\theta_1^2} , \frac{1}{\theta_2^2} , \dots \}$
\item scale of energy variation in potential energy surface $\to \sigma_w$
\item 4-dimensional bispectrum expansion cutoff $\to j_{max}$
\item potential cutoff distance $\to r_{cut}$
\end{itemize*}

\noindent We have a prior knowledge of the noise in the training data from our investigation of the convergence of energies, forces and stress virials as a function of plane-wave energy cutoff, $k$-point sampling and smearing width. Consequently, we set $\sigma_\nu^{(\text{energy})}$ to $0.001 \, \text{eV/atom}$ and $\sigma_\nu^{(\text{force})}$ to $0.1 \, \text{eV/\r{A}}$ when the $k$-point sampling density is equal to $0.03 \, \text{\r{A}}^{-1}$ and $\sigma_\nu^{(\text{energy})}$ to $0.0001 \, \text{eV/atom}$, $\sigma_\nu^{(\text{force})}$ to $0.01 \, \text{eV/\r{A}}$ and $\sigma_\nu^{(\text{virial})}$ to $0.01 \, \text{eV/atom}$ when the $k$-point sampling density is equal to $0.015 \, \text{\r{A}}^{-1}$.

We find that the training outcome is not very sensitive to changes in the value of the scale of energy variation in the potential energy surface hyperparameter $\sigma_w$, as long as it approximates the scale of energy variation of the underlying potential. Consequently, we find that setting it to $0.5-1.0 \, \text{eV}$ usually gives good results and the training outcome does not change significantly when its value is kept close to this range.

Establishing the characteristic length-scale of the atomic descriptor hyperparameter is more complicated. In principle, the length-scale parameter for each of the bispectrum dimensions can be set independently. In practice, we find, however, that the best way of investigating its effect on the training outcome is by deriving its value from the training data explicitly --- if our training data corresponds to $N$ observations, and observation $i$ corresponds to bispectrum descriptor vector $\mathbf{h}^{(i)}$, then we can express $\mathbf{\Sigma}$ in terms of a new parameter $\theta_{factor}$ such that:

\begin{equation}
\theta_i = \frac{\max( \{ h_i^{(j)} \}_{j = 1}^N ) - \min( \{ h_i^{(j)} \}_{j = 1}^N )}{\theta_{factor}} ,
\end{equation}

\noindent and the best value of $\theta_{factor}$ is found empirically.

Finally, the choice of the $j_{max}$ parameter is dictated by the accuracy required of the 4-d bispectrum descriptor. In practice this is a trade-off between computational cost and descriptor sensitivity, and the choice of $r_{cut}$ is solely dictated by the physics of the system investigated. We summarise the results of our investigation into finding suitable values of hyperparameters $\theta_{factor}$, $j_{max}$ and $r_{cut}$ in table \ref{table:bispectrum-gap_hyperparameters} below:

\clearpage

\newgeometry{top=1.5cm,bottom=1.5cm,left=3.5cm,right=1.5cm}

\thispagestyle{empty}

\begin{sidewaystable}
\begin{center}
\scriptsize{}
\begin{tabular}{ l | c | c | c | c | c | c | c || c | c | c || c | c | c |}
& \multicolumn{7}{ c ||}{$j_{max}$ convergence} & \multicolumn{3}{ c ||}{cutoff convergence} & \multicolumn{3}{ c |}{$\theta_{factor}$ convergence} \\
\cline{2-14}
\emph{FS/b-GAP$_{3}$ training database} & 2 & 4 & 6* & 8 & 10 & 12 & 14 & 4.0 \AA & 5.0 \AA* & 6.0 \AA & 1.0 & 2.0* & 3.0 \\
\hline
\multicolumn{14}{ l }{} \\
\multicolumn{14}{ l }{\emph{Training database errors:}} \\
\hline
RMS energy error per atom [eV] & 0.1010 & 0.0760 & 0.0631 & 0.0532 & 0.0729 & 0.0831 & 0.0989 & 0.0477 & 0.0631 & 0.0601 & 0.0887 & 0.0631 & 0.0392 \\
RMS force error [eV/\r{A}] & 5.3524 & 3.7042 & 2.0360 & 2.0295 & 2.4895 & 2.8476 & 3.0338 & 2.8660 & 2.0360 & 1.9760 & 2.9537 & 2.0360 & 2.2185 \\
\hline
\multicolumn{14}{ l }{} \\
\multicolumn{14}{ l }{\emph{RMS energy error per atom:} [eV]} \\
\hline
bcc primitive cells & 0.0089 & 0.0027 & 0.0009 & 0.0006 & 0.0004 & 0.0003 & 0.0002 & 0.0008 & 0.0009 & 0.0008 & 0.0006 & 0.0009 & 0.0011 \\
bcc bulk & 0.0010 & 0.0003 & 0.0001 & 0.0001 & 0.0001 & 0.0001 & 0.0001 & 0.0000 & 0.0001 & 0.0001 & 0.0001 & 0.0001 & 0.0001 \\
vacancy & 0.0005 & 0.0003 & 0.0001 & 0.0000 & 0.0000 & 0.0000 & 0.0000 & 0.0001 & 0.0001 & 0.0001 & 0.0001 & 0.0001 & 0.0001 \\
100, 110, 111, 112 surfaces & \emph{0.4708} & \emph{0.0987} & \emph{0.3062} & \emph{0.1976} & \emph{0.2163} & \emph{0.2018} & \emph{0.2114} & \emph{0.1997} & \emph{0.3062} & \emph{0.2521} & \emph{0.2232} & \emph{0.3062} & \emph{0.1108} \\
110, 112 gamma surfaces & \emph{0.0930} & \emph{0.0909} & \emph{0.0562} & \emph{0.0552} & \emph{0.0806} & \emph{0.0950} & \emph{0.1148} & \emph{0.0470} & \emph{0.0562} & \emph{0.0592} & \emph{0.1009} & \emph{0.0562} & \emph{0.0438} \\
\hline
\multicolumn{14}{ l }{} \\
\multicolumn{14}{ l }{\emph{RMS force error:} [eV/\r{A}]} \\
\hline
bcc primitive cells & --- & --- & --- & --- & --- & --- & --- & --- & --- & --- & --- & --- & --- \\
bcc bulk & 0.3584 & 0.2347 & 0.0867 & 0.0592 & 0.0470 & 0.0436 & 0.0416 & 0.0755 & 0.0867 & 0.1090 & 0.0777 & 0.0867 & 0.0962 \\
vacancy & 0.4147 & 0.2738 & 0.1025 & 0.0637 & 0.0504 & 0.0444 & 0.0451 & 0.0999 & 0.1025 & 0.1136 & 0.0823 & 0.1025 & 0.1215 \\
100, 110, 111, 112 surfaces & \emph{21.7883} & \emph{6.4410} & \emph{2.6188} & \emph{1.5256} & \emph{1.4970} & \emph{1.5692} & \emph{1.3366} & \emph{4.9638} & \emph{2.6188} & \emph{3.3376} & \emph{1.8579} & \emph{2.6188} & \emph{3.8316} \\
110, 112 gamma surfaces & \emph{5.3383} & \emph{4.3662} & \emph{2.4350} & \emph{2.4545} & \emph{3.0171} & \emph{3.4533} & \emph{3.6830} & \emph{3.3811} & \emph{2.4350} & \emph{2.3339} & \emph{3.5784} & \emph{2.4350} & \emph{2.6170} \\
\hline
\multicolumn{14}{ l }{} \\
\hline
lattice const. [\r{A}] & 3.1842 & 3.1817 & 3.1805 & 3.1800 & 3.1799 & 3.1800 & 3.1801 & 3.1806 & 3.1805 & 3.1802 & 3.1800 & 3.1805 & 3.1810 \\
C11 elastic constant [GPa] & 554.76 & 517.46 & 475.30 & 480.44 & 491.34 & 497.49 & 511.86 & 486.37 & 475.30 & 477.39 & 480.54 & 475.30 & 475.94 \\
C12 elastic constant [GPa] & 150.84 & 193.96 & 200.69 & 199.65 & 198.92 & 200.47 & 201.49 & 200.66 & 200.69 & 201.51 & 198.96 & 200.69 & 199.30 \\
bulk modulus [GPa] & 285.48 & 301.80 & 292.22 & 293.25 & 296.39 & 299.48 & 304.94 & 295.89 & 292.22 & 293.47 & 292.82 & 292.22 & 291.51 \\
shear modulus / C44 elastic constant [GPa] & 145.86 & 140.97 & 148.56 & 146.84 & 143.53 & 143.95 & 140.68 & 146.10 & 148.56 & 146.77 & 146.11 & 148.56 & 148.78 \\
\hline
RMS phonon spectrum error [THz] & --- & 0.576 & 0.215 & 0.202 & 0.143 & 0.173 & 0.203 & 0.299 & 0.215 & 0.338 & 0.228 & 0.215 & 0.261 \\
\hline
vacancy energy [eV] & 3.75 & 3.05 & 3.25 & 3.25 & 3.26 & 3.26 & 3.27 & 3.28 & 3.25 & 3.29 & 3.26 & 3.25 & 3.26 \\
\hline
100 surface energy [eV / \r{A}$^2$] & --- & --- & 0.321 & 0.338 & 0.355 & 0.350 & 0.346 & --- & 0.321 & --- & 0.360 & 0.321 & 0.162 \\
110 surface energy [eV / \r{A}$^2$] & --- & --- & 0.360 & 0.378 & 0.375 & 0.378 & 0.413 & --- & 0.360 & --- & 0.435 & 0.360 & 0.224 \\
111 surface energy [eV / \r{A}$^2$] & --- & --- & 0.218 & 0.314 & 0.320 & 0.341 & 0.374 & --- & 0.218 & --- & 0.398 & 0.218 & 0.114 \\
112 surface energy [eV / \r{A}$^2$] & --- & --- & 0.284 & 0.301 & 0.318 & 0.324 & 0.352 & --- & 0.284 & --- & 0.337 & 0.284 & -0.447 \\
\hline
RMS $\{110\} \langle 111 \rangle$ gamma surface energy error [eV] & 0.557 & 0.111 & 0.104 & 1.120 & 0.160 & 0.235 & 0.180 & 1.423 & 0.104 & 0.480 & 0.494 & 0.104 & 0.127 \\
\hline
RMS dislocation energy error [eV] & 0.686 & 0.382 & 0.548 & 1.723 & 0.305 & 0.928 & 0.494 & 2.484 & 0.548 & 0.804 & 1.143 & 0.548 & 0.882 \\
\hline
\end{tabular}
\normalsize{}
\end{center}
\captionsetup{margin=4cm}
\caption[Convergence of hyperparameters.]{Convergence of hyperparameters for the bispectrum-GAP potential, where we investigate bispectrum 4-dimensional bispectrum expansion cutoff $j_{max}$, potential cutoff distance $r_{cut}$ and characteristic length-scale $\theta_{factor}$.}
\captionsetup{margin=1cm}
\label{table:bispectrum-gap_hyperparameters}
\end{sidewaystable}

\restoregeometry

\clearpage

\section{Discussion}
\label{chapter:bispectrum-gap_potential_for_tungsten:section:discussion}

While the bispectrum-GAP correction to the Finnis-Sinclair interatomic potential improves the energetics of all lattice defects that were included in the training data set, we also find that the accuracy of the resulting potential is decreased for lattice features that were not explicitly trained --- i.e. the FS/bispectrum-GAP potential works very well within the interpolation regime, however the accuracy decreases very abruptly as soon as one enters the extrapolative regime. This behaviour is not only clearly demonstrated in table \ref{table:bispectrum-gap_results}, where we analyse RMS force and energy errors as we add data to our training data set, but it also manifests itself in the stress-strain relationships in the anharmonic regime (see figure \ref{figure:bispectrum-gap_stress-strain}).

Our description of the $\frac{1}{2} \langle 111 \rangle$ screw dislocation core structure relies on the assumption that it can be rationalised in terms of the properties of the strictly planar gamma surfaces. While this proves correct to some extent --- our GAP potential predicts the symmetric core structure, which matches the DFT model --- we also find that the edge component of the dislocation structure is not in the full agreement with the DFT predictions. This discrepancy can be accounted for by the fact that there is indeed some degree of extrapolation (in the sense of atomic configuration space) involved when predicting the screw dislocation core structure from gamma surface energetics alone.

An investigation of the dependence of the bispectrum-GAP potential on the hyperparameters reveals what we believe to be a major cause of the failure of our potential in the extrapolative regime. From the theoretical formulation of the bispectrum descriptor, one expects that the accuracy of the potential can be systematically improved as the value of $j_{max}$ increases (as the precision of the atomic representation improves). However, what we find instead is that we obtain the best accuracy for $j_{max}$ values in the range of 6-8, and increasing it beyond this range results in an increase of RMS force and energy errors (as demonstrated in table \ref{table:bispectrum-gap_hyperparameters}).

Our explanation of this behaviour is as follows: computing the bispectrum parameters involves expansion of the atomic density in terms of spherical harmonics, which can be thought of as calculating a Fourier series for a system that is rotationally, instead of translationally periodic. However, our atomic density is composed of Dirac delta functions and, consequently, the Fourier representation does not converge as the frequency increases because the Fourier transform of Dirac delta function spans the entire frequency domain. Since we need to truncate the components of the bispectrum expansion at a finite value of $j_{max}$, this effectively corresponds to truncating the representation of the Dirac delta function in the frequency domain. The result of truncating a representation of the Dirac delta function is demonstrated in figure \ref{figure:fourier_representation_dirac} below.

\begin{figure}[H]
\begin{center}
\resizebox{12cm}{!}{\footnotesize{}\input{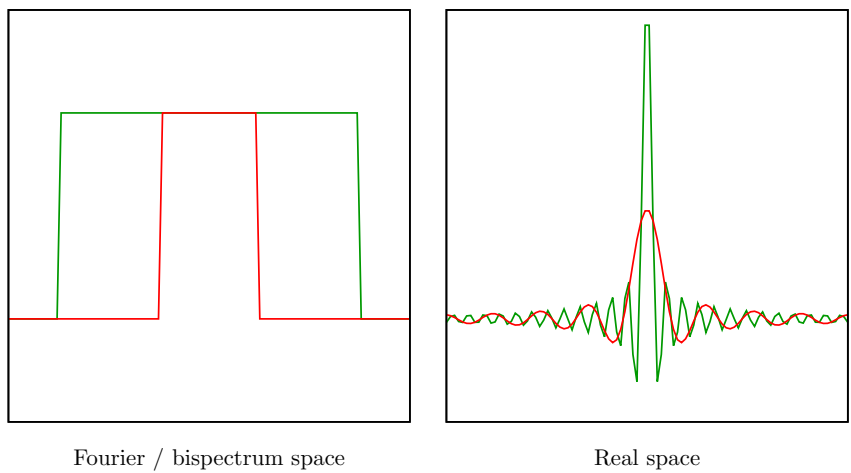}\normalsize{}}
\vspace{0.5cm}
\caption[Truncating Fourier expansion of Dirac delta function.]{Representation of a Fourier expansion of Dirac delta function which was truncated in frequency space.}
\label{figure:fourier_representation_dirac}
\end{center}
\end{figure}

\noindent Consequently, by making an analogy between the bispectrum and a Fourier series, we can identify the major shortcoming of the bispectrum descriptor of the atomic environment when used for the purpose of Gaussian Approximation Potential --- including high frequency components of bispectrum representation results in high frequency oscillations in the representation of the atomic density function when the potential energy surface is trained. This accounts for both: decreased accuracy of the bispectrum-GAP potential as we increase the value of $j_{max}$; and limited extrapolative power. This ``noisyness'' of the coordinates of the potential energy surface, while present in any GAP potential that is based on the bispectrum descriptor, is especially problematic in the systems with high coordination numbers. While one requires spherical harmonics of higher degree in order to account for angular dependence in systems with many nearest neighbours (such as metals), there is a clear trade-off between the noise in the resulting PES and the accuracy of the descriptor.

Finally, we finish this chapter with a summary of the protocol that we developed for the purpose of training GAP based interatomic potential for bcc transition metals, based on our results for tungsten. We will use this protocol again in the next chapter in order to develop a GAP potential based on the Smooth Overlap of Atomic Positions descriptor, which we believe provides a solution to the problems of the bispectrum based potentials mentioned above.

\begin{table}[H]
\vspace{1cm}
\begin{center}
\begin{tabular}{ r l c l }
\midrule
1. & Elastic constants & $\to$ & MC sampling in the lattice space \\
& 2000 environments & & temperature: $300 \, \text{K}$ \\
& & & $\bullet \, $ slice sampling algorithm \\
& & & $\bullet \, $ primitive unit cell \\
& & & $\bullet \, $ training from energies and stresses \\
\midrule
2. & Phonon spectrum & $\to$ & MD, no defects \\
& 7680 environments & & temperature: $300, 1000 \, \text{K}$ \\
& & & volumes: ground state, $\pm 1\%$ \\
& & & $\bullet \, $ 128 at. simulation cell \\
& & & $\bullet \, $ training from energies and forces \\
\midrule
3. & Vacancy & $\to$ & MD, isolated monovacancy \\
& 23740 environments & & temperature: $300, 1000 \, \text{K}$ \\
& & & volumes: ground state, $\pm 1\%$ \\
& & & $\bullet \, $ 53 and 127 at. simulation cell \\
& & & $\bullet \, $ training from energies and forces \\
\midrule
4. & Surfaces & $\to$ & MD, $(100)$, $(110)$, $(111)$, $(112)$ \\
& 2160 environments & & temperature: $300 \, \text{K}$ \\
& & & volumes: ground state \\
& & & $\bullet \, $ 12 at. simulation cell \\
& & & $\bullet \, $ training from energies and forces \\
\midrule
5. & Gamma surfaces & $\to$ & MD, $(110)$, $(112)$ \\
& 74196 environments & & temperature: $300 \, \text{K}$ \\
& & & volumes: ground state, $\pm 1\%$ \\
& & & $\bullet \, $ 12 at. simulation cell \\
& & & $\bullet \, $ training from energies and forces \\
\midrule
\end{tabular}
\caption[Summary of GAP training protocol.]{Summary of the protocol for generating training database for GAP based interatomic potential for tungsten.}
\label{table:training_protocol}
\end{center}
\end{table}

\cleardoublepage

\chapter{SOAP-GAP Potential for Tungsten}
\label{chapter:soap-gap_potential_for_tungsten}

\section{Introduction}
\label{chapter:soap-gap_potential_for_tungsten:section:introduction}

In this chapter I expand on the results obtained while training the FS/bispectrum-GAP interatomic potential for tungsten. We use the data and protocols developed in the previous chapter but we improve our methodology. Instead of using the 4-dimensional bispectrum and the square-exponential covariance function we use the Smooth Overlap of Atomic Positions (SOAP) kernel in order to overcome the limitations of the bispectrum based potentials. We also explore the idea of fitting a SOAP-GAP interatomic potential from scratch, without using the Finnis-Sinclair interatomic potential as a core potential.

In addition to using the training dataset generated for the bispectrum-GAP potentials, in this chapter we also investigate the interactions between different types of lattice defects. Consequently, in section \ref{chapter:soap-gap_potential_for_tungsten:section:training_protocol_and_dataset} I discuss how appropriate training data can be generated. We follow this with a brief discussion of how iterative GAP potentials can be trained.

In section \ref{chapter:soap-gap_potential_for_tungsten:section:results} I present the results obtained with the SOAP-GAP potential for tungsten, trained with and without the Finnis-Sinclair interatomic potential core. We also demonstrate how the iterative-SOAP-GAP interatomic potential offers an improved description of the screw dislocation core structure compared to the bispectrum based GAP potentials. We then use the iterative-SOAP-GAP potential to calculate the Peierls barrier of an isolated screw dislocation and dislocation-vacancy interaction map. I finish this chapter with a discussion of the convergence of the hyperparameters and, in section \ref{chapter:soap-gap_potential_for_tungsten:section:discussion}, I discuss the results obtained with the SOAP-GAP interatomic potential for tungsten.

\section{Training Protocol and Dataset}
\label{chapter:soap-gap_potential_for_tungsten:section:training_protocol_and_dataset}

\subsection{Lattice Defects Interaction}

In our first attempt at training a SOAP based GAP interatomic potential for tungsten we reuse all of the training data generated for the purpose of fitting the bispectrum-GAP potentials. We anticipate that the existing training dataset is capable of describing an isolated vacancy far away from a screw dislocation where the effects of the dislocation strain field are negligible. However, in order to accurately predict the behaviour of a vacancy in the neighbourhood of the screw dislocation core we need to expand our training dataset. It should include configurations that are representative of the atomic environments that might be encountered when the vacancy and dislocation interact.

We again rely on the idea that the $(110)$ gamma surface can be used to rationalise the $\frac{1}{2} \langle 111 \rangle$ screw dislocation core structure (as the core spreads along the $(110)$ planes). Hence, we generate our training data by introducing a vacancy inside the gamma surface simulation cell either at the first, second or third layer of atoms from the gamma surface. We also increase the size of the simulation cell from 12 to 48 atoms as we need to account for the interaction of the vacancy with its periodic images.

We use the protocol introduced in section \ref{chapter:bispectrum-gap_potential_for_tungsten:section:training_protocol_and_dataset} for generating MD trajectories. We begin by minimising the potential energy stored in the lattice by means of geometry optimisation before the start of the MD simulation. The initial state of the trajectory is generated by randomising the kinetic energies of the atoms, such that they are Boltzmann distributed with velocities being a function of simulation temperature. To reduce the computational cost, we again use a linear $k$-point sampling density of $0.03 \, \text{\r{A}}^{-1}$ to obtain converged values of energy and forces and we discard the stresses.

Due to the considerable computational cost associated with these simulations (the simulation cell is four times the size of the simulation cell of a gamma surface), we sample the gamma surface with vacancy at the first, second and third layer of atoms using a 5 x 5 regular grid at the ground state volume and a single temperature of $300 \, \text{K}$ only.

\newgeometry{top=2.5cm,bottom=2.5cm,left=3.5cm,right=3cm} 

\subsection{Iterative GAP}

As demonstrated in chapter \ref{chapter:bispectrum-gap_potential_for_tungsten}, a 135 atom dislocation dipole simulation cell is the smallest system which can reproduce a screw dislocation, and at the same time it also corresponds to the largest system for which we can evaluate single-point DFT energies and forces (although at a substantial computational cost). Computing an MD trajectory or carrying out a large number of geometry optimisation iterations using the DFT method would be highly impractical for this system as the calculations would take weeks even while running on hundreds of compute cores in parallel. However, one can use an existing GAP potential to generate MD and geometry optimisation trajectories. If the GAP potential can reproduce the DFT potential energy surface with sufficient accuracy, snapshots of the trajectory can be taken and energies and forces can be recomputed using the DFT method. One can then train an improved GAP potential based on the recomputed data and the potential can be iteratively improved with training data that could not be obtained otherwise.

\begin{sloppypar} 
By computing the $\frac{1}{2} \langle 111 \rangle$ screw dislocation core structure with both bispectrum-GAP and SOAP-GAP potentials we find that inclusion of gamma surface training data in the training dataset is sufficient to predict a symmetric core structure which is in qualitative agreement with the DFT result. However, this by itself does not provide a sufficiently accurate description for a quantitative study of the dislocation mobility processes (as outlined in the previous chapter). Hence, in order to overcome this limitation we apply the methodology of ``iterative'' training outlined above --- from now onwards, we will refer to the resulting potential as the iterative-SOAP-GAP potential.
\end{sloppypar} 

\vspace{-0.25cm} 

\begin{figure}[H]
\begin{center}
\resizebox{12cm}{!}{\footnotesize{}\input{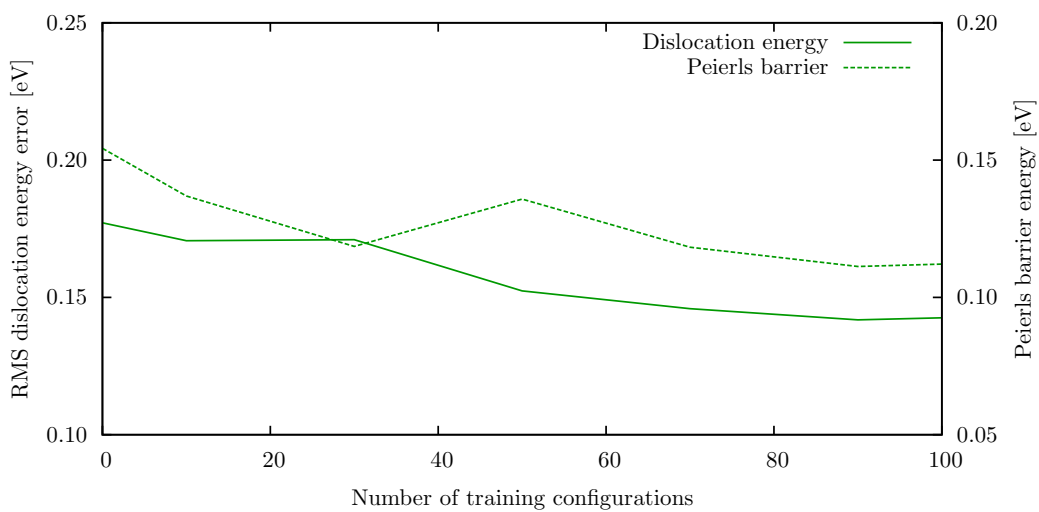}\normalsize{}}
\vspace{0.75cm}
\caption[Convergence of dislocation energy with training data.]{Convergence of the RMS energy error of an unrelaxed dislocation and Peierls barrier with the volume of training data.}
\label{figure:convergence_dislocation}
\end{center}
\end{figure}

\restoregeometry 

To obtain our iterative potential we generate MD trajectories of a screw dislocation in a 135 atom dislocation dipole simulation cell using the SOAP-GAP interatomic potential. According to the protocol developed in the previous chapter for generating training data, we carry out the MD simulations at temperatures of $300 \, \text{K}$ and $1000 \, \text{K}$. Snapshots of the trajectory are then selected as training samples and to reduce the computational cost of DFT calculations they are recomputed at a linear $k$-point sampling density of $0.03 \, \text{\r{A}}^{-1}$. We again discard the stresses.

In order to assess the accuracy of our description of the screw dislocation in tungsten, and verify whether our iterative scheme works, we compute the RMS energy error of an unrelaxed dislocation in the 135 atom dislocation dipole simulation cell (we use ten DFT values computed in figure \ref{figure:bispectrum-gap_dislocation_dipoles} as reference). We also compute the Peierls barrier (energy required to migrate the dislocation to a neighbour lattice site) using the string method (outlined in section \ref{chapter:simulation_techniques:section:transition_state_search}) in the same 135 atom dislocation dipole simulation cell. The convergence of the RMS energy error of an unrelaxed dislocation and Peierls barrier, computed using the iterative-SOAP-GAP potential, as the volume of training data increases is shown in figure \ref{figure:convergence_dislocation}.

\section{Results}
\label{chapter:soap-gap_potential_for_tungsten:section:results}

Similarly to the bispectrum based GAP potentials, in order to demonstrate how the SOAP-GAP interatomic potential can be systematically improved, we repeatedly carry out our training procedure while increasing the size of the training database. At the same time we monitor the performance of the potential.

We verify the predicted values of lattice constants, elastic constants, formation energies of the isolated vacancy and surfaces and the RMS error in the phonon spectrum for the SOAP based potentials, trained with and without the Finnis-Sinclair interatomic potential core. We also include the results for the iterative-SOAP-GAP potential. A summary of the training databases and performance of the associated SOAP-GAP potential is given in table \ref{table:soap-gap_results} below.

\clearpage

\newgeometry{top=1.5cm,bottom=1.5cm,left=3.5cm,right=1.5cm}

\thispagestyle{empty}

\begin{sidewaystable}
\begin{center}
\scriptsize{}
\begin{tabular}{ l || c | c | c || c | c | c | c | c | c || c | c | }
& FS/S-GAP$_{1}$ & FS/S-GAP$_{2}$ & FS/S-GAP$_{3}$ & S-GAP$_{1}$ & S-GAP$_{2}$ & S-GAP$_{3}$ & S-GAP$_{4}$ & S-GAP & I-S-GAP & FS & DFT \\
\hline
\multicolumn{12}{ l }{} \\
\multicolumn{12}{ l }{\emph{Training database errors:}} \\
\cline{1-11}
RMS energy error per atom [eV] & 0.0225 & 0.0554 & 0.0411 & 0.0391 & 0.0142 & 0.0110 & 0.0004 & 0.0003 & 0.0003 & 0.0095 & \multicolumn{1}{ c }{} \\
RMS force error [eV/\r{A}] & 0.7783 & 1.3053 & 1.0422 & 0.8941 & 0.4575 & 0.3480 & 0.0768 & 0.0629 & 0.0629 & 0.6492 & \multicolumn{1}{ c }{} \\
\cline{1-11}
\multicolumn{12}{ l }{} \\
\multicolumn{12}{ l }{\emph{Number of atomic environments in training database:}} \\
\cline{1-10}
bcc primitive cells (MCMC, 2000 $\times$ 1 at.) & 2000 & 2000 & 2000 & 2000 & 2000 & 2000 & 2000 & 2000 & 2000 & \multicolumn{2}{ c }{} \\
bcc bulk (MD, 60 $\times$ 128 at.) & --- & 7680 & 7680 & --- & 7680 & 7680 & 7680 & 7680 & 7680 & \multicolumn{2}{ c }{} \\
vacancy (MD, 400 $\times$ 53 at., 20 $\times$ 127 at.) & --- & --- & 23740 & --- & --- & 23740 & 23740 & 23740 & 23740 & \multicolumn{2}{ c }{} \\
100, 110, 111, 112 surfaces (MD, 180 $\times$ 12 at.) & --- & --- & --- & --- & --- & --- & 2160 & 2160 & 2160 & \multicolumn{2}{ c }{} \\
110, 112 gamma surfaces (MD, 6183 $\times$ 12 at.) & --- & --- & --- & --- & --- & --- & 74196 & 74196 & 74196 & \multicolumn{2}{ c }{} \\
110 gamma surface + vacancy (MD, 750 $\times$ 47 at.) & --- & --- & --- & --- & --- & --- & --- & 35250 & 35250 & \multicolumn{2}{ c }{} \\
screw dislocation quadrupole (MD, 100 $\times$ 135 at.) & --- & --- & --- & --- & --- & --- & --- & --- & 13500 & \multicolumn{2}{ c }{} \\
\cline{1-10}
\multicolumn{12}{ l }{} \\
\multicolumn{12}{ l }{\emph{RMS energy error per atom:} [eV]} \\
\cline{1-11}
bcc primitive cells & 0.0010 & 0.0009 & 0.0009 & 0.0001 & 0.0001 & 0.0001 & 0.0001 & 0.0001 & 0.0001 & 0.0158 & \multicolumn{1}{ c }{} \\
bcc bulk & \emph{0.0001} & 0.0001 & 0.0001 & \emph{0.0004} & 0.0001 & 0.0001 & 0.0001 & 0.0001 & 0.0001 & 0.0002 & \multicolumn{1}{ c }{} \\
vacancy & \emph{0.0009} & \emph{0.0100} & 0.0001 & \emph{0.0065} & \emph{0.0031} & 0.0000 & 0.0001 & 0.0001 & 0.0001 & 0.0013 & \multicolumn{1}{ c }{} \\
100, 110, 111, 112 surfaces & \emph{0.1195} & \emph{0.0658} & \emph{0.0456} & \emph{0.0773} & \emph{0.1005} & \emph{0.0845} & 0.0001 & 0.0001 & 0.0001 & 0.0233 & \multicolumn{1}{ c }{} \\
110, 112 gamma surfaces & \emph{0.0252} & \emph{0.0783} & \emph{0.0569} & \emph{0.0536} & \emph{0.0113} & \emph{0.0070} & 0.0004 & 0.0005 & 0.0005 & 0.0127 & \multicolumn{1}{ c }{} \\
110 gamma surface + vacancy & \emph{0.0082} & \emph{0.0241} & \emph{0.0258} & \emph{0.0209} & \emph{0.0032} & \emph{0.0016} & \emph{0.0005} & 0.0002 & 0.0002 & 0.0045 & \multicolumn{1}{ c }{} \\
screw dislocation quadrupole & \emph{0.0003} & \emph{0.0026} & \emph{0.0023} & \emph{0.0033} & \emph{0.0002} & \emph{0.0001} & \emph{0.0002} & \emph{0.0002} & 0.0002 & 0.0015 & \multicolumn{1}{ c }{} \\
\cline{1-11}
\multicolumn{12}{ l }{} \\
\multicolumn{12}{ l }{\emph{RMS force error:} [eV/\r{A}]} \\
\cline{1-11}
bcc primitive cells & --- & --- & --- & --- & --- & --- & --- & --- & --- & --- & \multicolumn{1}{ c }{} \\
bcc bulk & \emph{0.1993} & 0.0615 & 0.0476 & \emph{0.1635} & 0.0203 & 0.0228 & 0.0284 & 0.0279 & 0.0278 & 0.1460 & \multicolumn{1}{ c }{} \\
vacancy & \emph{0.3852} & \emph{0.5824} & 0.0528 & \emph{0.2364} & \emph{0.1442} & 0.0228 & 0.0302 & 0.0294 & 0.0294 & 0.2415 & \multicolumn{1}{ c }{} \\
100, 110, 111, 112 surfaces & \emph{1.0101} & \emph{1.5858} & \emph{0.9157} & \emph{1.1757} & \emph{0.9613} & \emph{0.3310} & 0.0482 & 0.0508 & 0.0509 & 0.5706 & \multicolumn{1}{ c }{} \\
110, 112 gamma surfaces & \emph{0.9770} & \emph{1.5896} & \emph{1.3566} & \emph{1.0823} & \emph{0.5995} & \emph{0.4868} & 0.0661 & 0.0684 & 0.0690 & 0.8845 & \multicolumn{1}{ c }{} \\
110 gamma surface + vacancy & \emph{0.7096} & \emph{1.3774} & \emph{0.9531} & \emph{0.9753} & \emph{0.3337} & \emph{0.1915} & \emph{0.1248} & 0.0785 & 0.0793 & 0.3963 & \multicolumn{1}{ c }{} \\
screw dislocation quadrupole & \emph{0.3262} & \emph{0.4082} & \emph{0.3560} & \emph{0.3626} & \emph{0.0863} & \emph{0.0733} & \emph{0.0469} & \emph{0.0469} & 0.0383 & 0.2702 & \multicolumn{1}{ c }{} \\
\cline{1-11}
\multicolumn{12}{ l }{} \\
\hline
lattice const. [\r{A}] & 3.1809 & 3.1808 & 3.1808 & 3.1803 & 3.1803 & 3.1803 & 3.1803 & 3.1803 & 3.1803 & 3.1805 & 3.1805 \\
C11 elastic constant [GPa] & 478.66 & 476.74 & 475.37 & 517.74 & 517.69 & 517.75 & 517.68 & 518.30 & 518.03 & 514.23 & 516.86 \\
C12 elastic constant [GPa] & 203.94 & 204.00 & 202.46 & 198.67 & 198.68 & 198.88 & 198.41 & 198.61 & 198.46 & 200.12 & 198.18 \\
bulk modulus [GPa] & 295.51 & 294.91 & 293.43 & 305.02 & 305.02 & 305.17 & 304.83 & 305.17 & 304.98 & 304.83 & 304.41 \\
shear modulus / C44 elastic constant [GPa] & 142.98 & 142.69 & 142.83 & 142.69 & 142.70 & 142.73 & 142.97 & 142.68 & 142.98 & 157.21 & 142.30 \\
\hline
RMS phonon spectrum error [THz] & 0.962 & 0.167 & 0.197 & 0.583 & 0.146 & 0.142 & 0.138 & 0.126 & 0.129 & 0.392 & \multicolumn{1}{ c }{} \\
\hline
vacancy energy [eV] & 2.86 & --- & 3.23 & 0.42 & 1.86 & 3.26 & 3.27 & 3.28 & 3.29 & 3.61 & 3.27 \\
\hline
100 surface energy [eV / \r{A}$^2$] & 0.231 & 0.064 & 0.057 & 0.076 & 0.068 & 0.145 & 0.252 & 0.252 & 0.252 & 0.179 & 0.251 \\
110 surface energy [eV / \r{A}$^2$] & 0.214 & 0.073 & 0.126 & 0.064 & 0.055 & 0.117 & 0.204 & 0.204 & 0.204 & 0.158 & 0.204 \\
111 surface energy [eV / \r{A}$^2$] & 0.300 & 0.085 & 0.120 & 0.095 & 0.088 & 0.122 & 0.222 & 0.222 & 0.222 & 0.202 & 0.222 \\
112 surface energy [eV / \r{A}$^2$] & 0.265 & 0.078 & 0.153 & 0.082 & 0.079 & 0.135 & 0.216 & 0.216 & 0.216 & 0.187 & 0.216 \\
\hline
RMS $\{110\} \langle 111 \rangle$ gamma surface energy error [eV] & 0.157 & 1.767 & 1.130 & 1.295 & 0.097 & 0.116 & 0.047 & 0.042 & 0.045 & 0.695 & \multicolumn{1}{ c }{} \\
\cline{1-11}
RMS dislocation energy error [eV] & 0.327 & 2.975 & 2.217 & 2.683 & 0.036 & 0.065 & 0.166 & 0.177 & 0.143 & 1.265 & \multicolumn{1}{ c }{} \\
\cline{1-11}
\end{tabular}
\normalsize{}
\end{center}
\captionsetup{margin=4cm}
\caption[SOAP-GAP training database and potential summary.]{Summary of the training databases and performance of the associated SOAP-GAP potential.}
\captionsetup{margin=1cm}
\label{table:soap-gap_results}
\end{sidewaystable}

\restoregeometry

\clearpage

\begin{figure}[H]
\begin{center}
\resizebox{12cm}{!}{\footnotesize{}\input{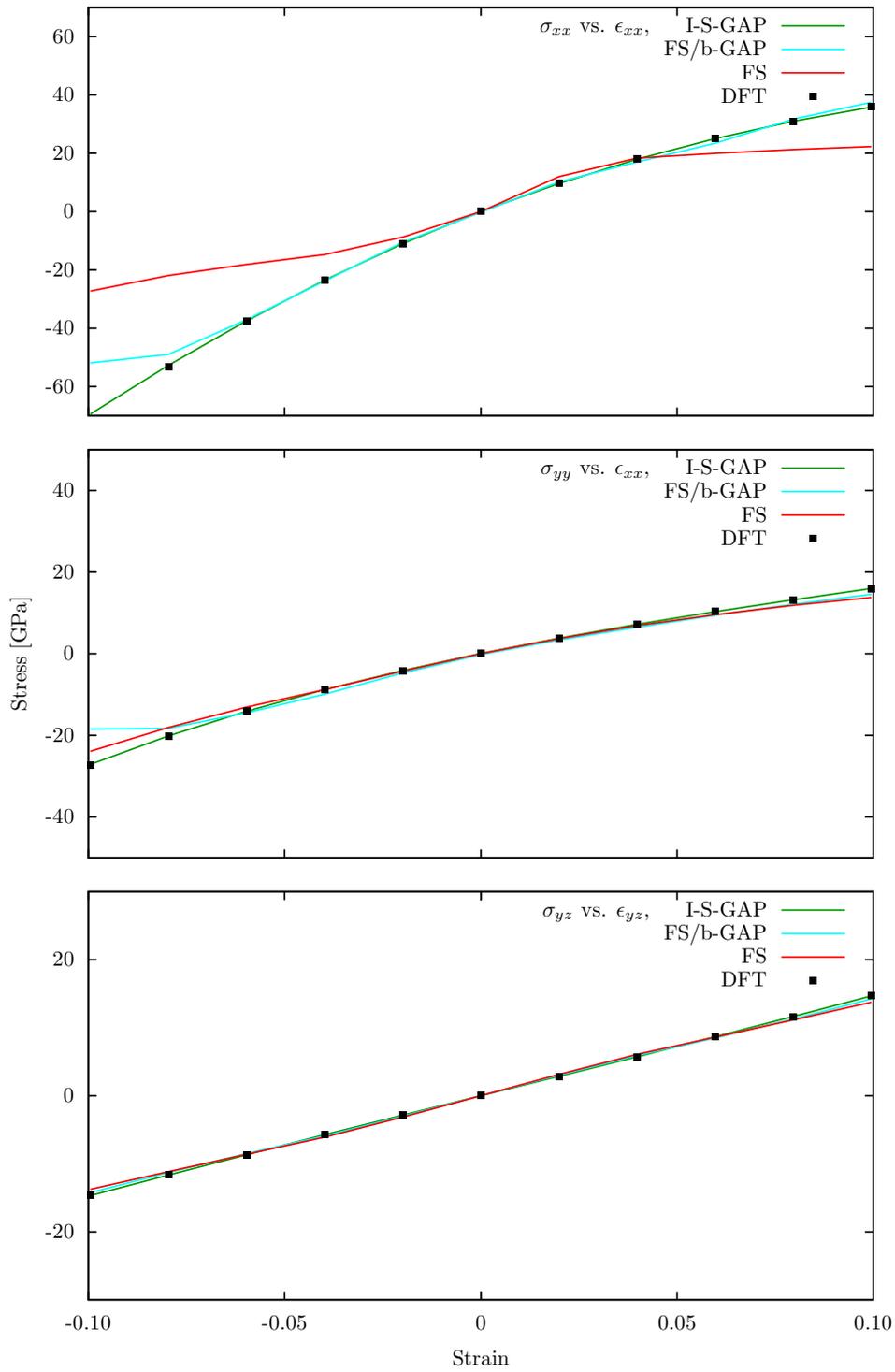}\normalsize{}}
\vspace{0.75cm}
\caption[Iterative-SOAP-GAP stress-strain curves.]{Iterative-SOAP-GAP stress-strain curves of bcc tungsten for a range of strains from $-10\%$ to $+10\%$.}
\label{figure:soap-gap_stress-strain}
\end{center}
\end{figure}

\newgeometry{top=2.5cm,bottom=2.5cm,left=3.5cm,right=3cm} 

\begin{sloppypar}
We verify the elastic properties of the iterative-SOAP-GAP interatomic potential in the anharmonic regime (from now on we are using the most complete SOAP-GAP potential, namely the iterative-SOAP-GAP potential, designated as I-S-GAP in table \ref{table:soap-gap_results}) by computing the stress-strain curves corresponding to longitudinal compression, transverse expansion and shearing for a range of strains from $-10\%$ to $+10\%$. The results are shown in figure \ref{figure:soap-gap_stress-strain}.
\end{sloppypar}

We find that the elastic behaviour predicted by the SOAP based GAP interatomic potentials is in a perfect agreement with the DFT model. SOAP-GAP potentials reproduce the stress-strain behaviour correctly even for the range of strains where the bispectrum-GAP potentials fail. Since both GAP potentials use the same elasticity training data, this indicates that the SOAP-GAP potentials are capable of providing an improved description of the system, even in the extrapolative regime.

The phonon spectrum of bcc tungsten computed using the iterative-SOAP-GAP interatomic potential is shown in figure \ref{figure:soap-gap_phonons} below.

\begin{figure}[H]
\begin{center}
\resizebox{12cm}{!}{\footnotesize{}\input{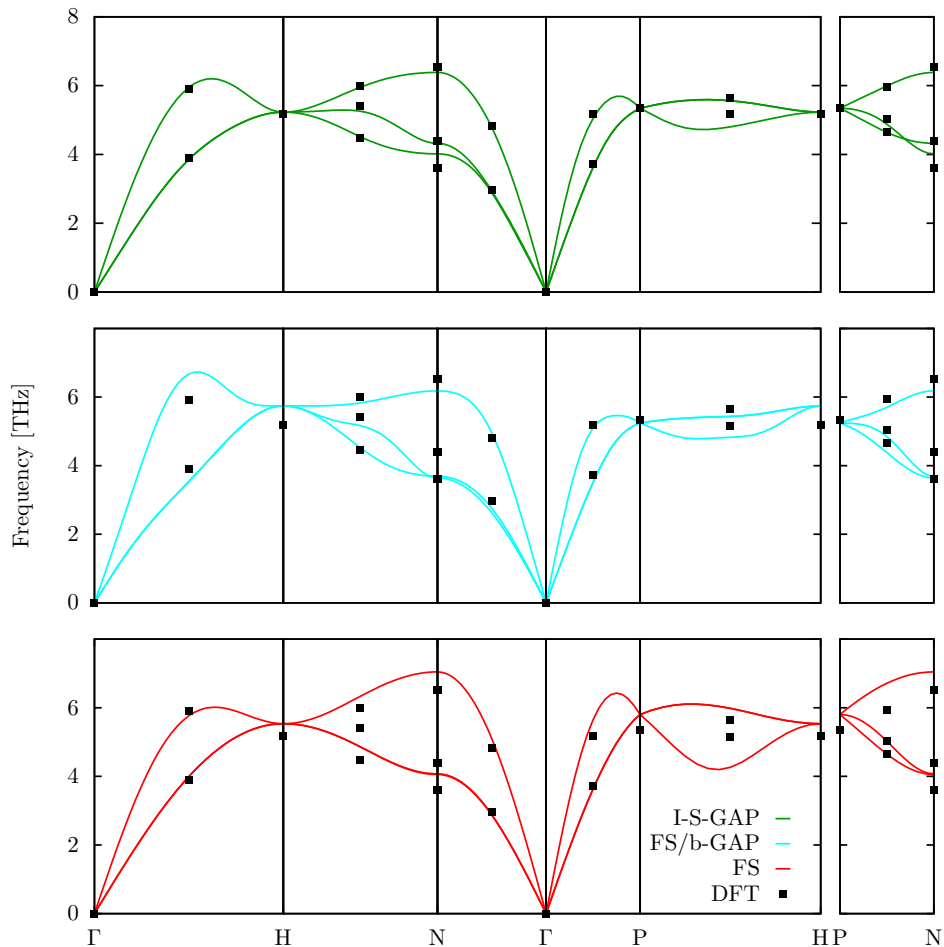}\normalsize{}}
\vspace{0.5cm}
\caption[Iterative-SOAP-GAP phonon spectrum.]{Iterative-SOAP-GAP phonon spectrum of bcc tungsten.}
\label{figure:soap-gap_phonons}
\end{center}
\end{figure}

\restoregeometry 

\noindent While both FS and bispectrum-GAP models fail to reproduce some of the transverse modes of vibration along the $\{\text{N}-\Gamma\}$ path of the phonon spectrum, we find that the SOAP-GAP potential offers an improved description of tungsten vibrational properties, both at a qualitative and quantitative level. The SOAP-GAP potential not only accounts for the non-degenerate transverse modes, but the RMS error in phonon frequencies is also significantly reduced (as demonstrated in table \ref{table:soap-gap_results}).

A cross-section of $(110)$ gamma surface energies along the $\langle 111 \rangle$ lattice vector computed using the iterative-SOAP-GAP interatomic potential is shown in figure \ref{figure:soap-gap_gamma_surface_111} below. Again, we observe a significant improvement compared to both FS and bispectrum-GAP models. While bispectrum-GAP offers a good description of gamma surfaces on the qualitative level, SOAP-GAP provides accuracy closely resembling that of DFT model.

\begin{figure}[H]
\begin{center}
\resizebox{12cm}{!}{\footnotesize{}\input{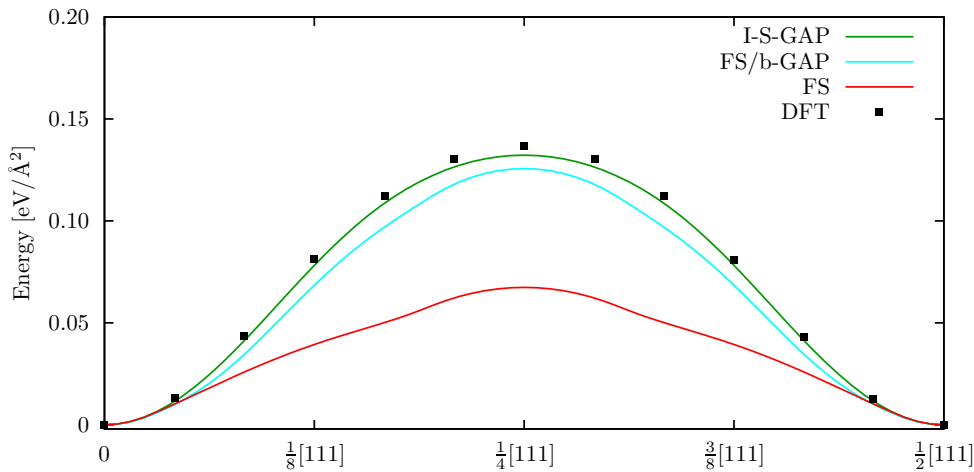}\normalsize{}}
\vspace{0.5cm}
\caption[$\langle 111 \rangle$ cross-section of $(110)$ gamma surface energy.]{$\langle 111 \rangle$ cross-section of $(110)$ gamma surface energy for GAP, FS and DFT models.}
\label{figure:soap-gap_gamma_surface_111}
\end{center}
\end{figure}

Finally, the energies of an unrelaxed dislocation dipole system as a function of the Burgers vector computed using iterative-SOAP-GAP interatomic potential are shown in figure \ref{figure:soap-gap_dislocation_dipoles} below. Again, as was the case with gamma surface energies, we find that SOAP-GAP provides a further improvement in accuracy over both FS and bispectrum-GAP models and closely resembles the DFT result.

\begin{figure}[H]
\begin{center}
\resizebox{12cm}{!}{\footnotesize{}\input{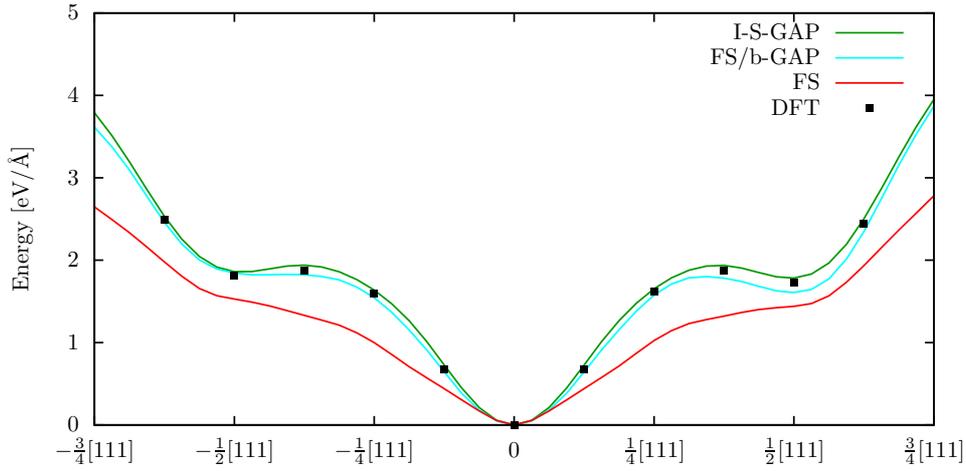}\normalsize{}}
\vspace{0.5cm}
\caption[Unrelaxed dislocation dipole energy.]{Unrelaxed dislocation dipole energy as a function of the Burgers vector for GAP, FS and DFT models.}
\label{figure:soap-gap_dislocation_dipoles}
\end{center}
\end{figure}

\subsection{Screw Dislocation Core Structure}

We begin our investigation of the properties of the $\frac{1}{2} \langle 111 \rangle$ screw dislocation by verifying the convergence of dislocation core local energy with the system size for our dislocation simulations using both dislocation dipole and isolated dislocation methods. The error in the dislocation core local energy, as a function of system size, is given in figure \ref{figure:soap-gap_dislocation_size_convergence} below.

\begin{figure}[H]
\begin{center}
\vspace{0.25cm}
\resizebox{12cm}{!}{\footnotesize{}\input{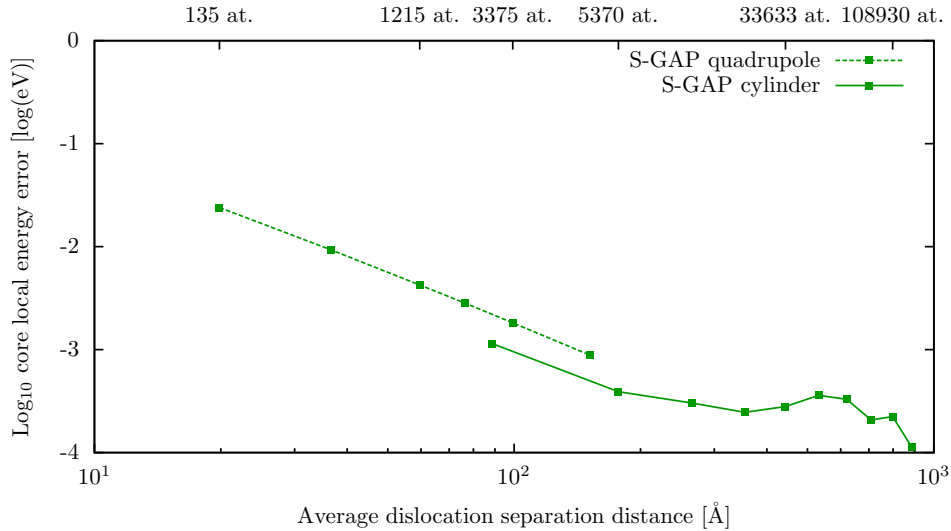}\normalsize{}}
\vspace{0.75cm}
\caption[Dislocation core local energy error as a function of system size.]{Convergence of dislocation core local energy error with the system size for dislocation quadrupole and isolated dislocation.}
\label{figure:soap-gap_dislocation_size_convergence}
\end{center}
\end{figure}

$\frac{1}{2} \langle 111 \rangle$ screw dislocation core structures computed by the means of geometry optimisation using SOAP-GAP, bispectrum-GAP, Finnis-Sinclair and DFT methods are presented in figure \ref{figure:soap-gap_dislocation_structure} below. We again characterise the dislocation structures using the Nye tensor (as outlined in section \ref{chapter:bulk_properties_and_lattice_defects_in_tungsten:section:dislocations}). For the GAP and Finnis-Sinclair models we also compute local energies of individual atoms.

We find that the dislocation core structure predicted by the SOAP-GAP is in perfect agreement with the DFT model. This is a clear improvement over the bispectrum-GAP. The bispectrum-GAP, although it does predict a symmetric core structure, fails to reproduce the edge component of the Nye tensor of the DFT result. We also discover that the local energy of core atoms is significantly different in SOAP-GAP and bispectrum-GAP predictions of the dislocation core structure. The SOAP-GAP result which reproduces the DFT structure in perfect detail suggests that the spreading of the dislocation in terms of local atomic energies is very limited. This is in contrast to both bispectrum-GAP and Finnis-Sinclair models --- the amount of the dislocation spreading is more confined than we previously anticipated.

As outlined in section \ref{chapter:bulk_properties_and_lattice_defects_in_tungsten:section:dislocations}, it is widely believed that the mobility of screw dislocations is influenced by spreading of the dislocation line into the slip planes of the $\langle 111 \rangle$ zone. This is because the dislocation effectively anchors itself to the particular lattice site. In order to transition onto the neighbouring site a significant amount of energy is required to retract its movement (more details in \cite{Cai20051}). Consequently, we anticipate that the precise structure of the dislocation core has an influence on the value of the Peierls barrier which we compute in the next section.

When we verify the final atomic coordinates of the geometry optimisation performed with the SOAP-GAP interatomic potential by performing a single-point energy and force evaluation using DFT, we find that the maximum force error between iterative-SOAP-GAP and DFT methods is $0.057 \, \text{eV/\r{A}}$ (as compared to $0.62 \, \text{eV/\r{A}}$ for the FS/bispectrum-GAP model). This is an order of magnitude reduction in force error and it is very similar in value to the RMS force error of the overall training database for the iterative-SOAP-GAP potential ($0.063 \, \text{eV/\r{A}}$). This gives us confidence that our model provides a quantitatively accurate description of the screw dislocation that can be used to investigate dislocation mobility and interactions with other lattice defects.

\clearpage

\thispagestyle{empty}

\begin{sidewaysfigure}
\begin{center}
\resizebox{18cm}{!}{\footnotesize{}\input{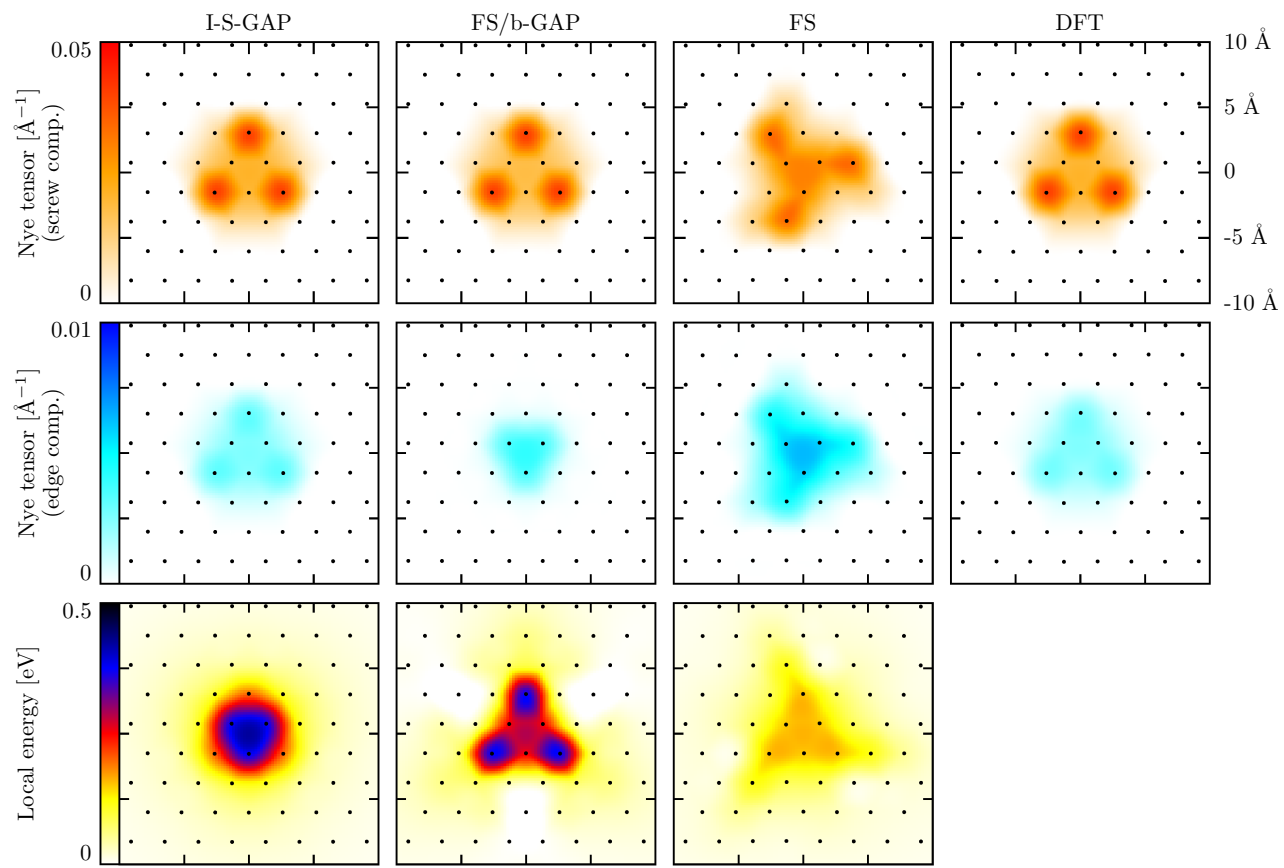}\normalsize{}}
\captionsetup{margin=4cm}
\caption[$\frac{1}{2} \langle 111 \rangle$ screw dislocation core structure.]{$\frac{1}{2} \langle 111 \rangle$ screw dislocation core structures evaluated using SOAP-GAP, bispectrum-GAP, FS and DFT models.}
\captionsetup{margin=1cm}
\label{figure:soap-gap_dislocation_structure}
\end{center}
\end{sidewaysfigure}

\clearpage

\subsection{Screw Dislocation Peierls Barrier}

We carry out our calculation of the Peierls barrier using a transition state searching implementation of the string method (more details in section \ref{chapter:simulation_techniques:section:transition_state_search}). The string method optimises the transition path in order to find the saddle points of the potential energy surface and we explore three different starting points for transition path optimisation, as demonstrated in figure \ref{figure:transition_state_cell} below.

\begin{figure}[H]
\begin{center}
\resizebox{12cm}{!}{\footnotesize{}\input{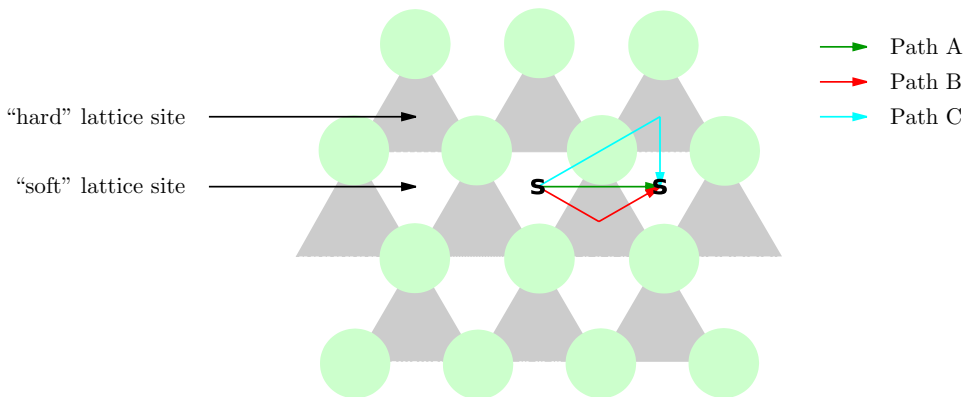}\normalsize{}}
\caption[Screw dislocation Peierls barrier simulation approach.]{Screw dislocation Peierls barrier simulation approach, where we investigate three different paths for dislocation migration.}
\label{figure:transition_state_cell}
\end{center}
\end{figure}

\vspace{-0.5cm} 

\noindent There are two types of dislocation sites in the $(111)$ plane of the bcc lattice and they have opposite chiralities. A screw dislocation with Burgers vector pointing out of the plane will produce a ``soft'' core (stable, corresponding to ground state structure), whereas the same dislocation in the other site will produce a ``hard'' core (sometimes metastable in some transition metals, corresponding to higher energy than ``soft'' core). On the other hand, a dislocation with Burgers vector pointing into the plane will produce a ``hard'' core in the first site and ``soft'' core in the second site (more details in \cite{PhysRevB.54.6941}, \cite{PhysRevB.68.014104}).  

We find that the ``hard'' core is not metastable in tungsten --- carrying out geometry optimisation of a ``hard'' core results in the dislocation line migrating to a neighbour lattice site that corresponds to the ``soft'' core structure. The ``hard'' core might, nevertheless, correspond to a saddle point in the potential energy surface and, consequently, might be a suitable transition state along the transition path. Hence, we construct an initial path A, which connects the two identical ``soft'' lattice sites directly, and further two alternative paths B and C, which explore the possibility of the transition path going through a transition state corresponding to a ``hard'' core configuration.

We carry out the transition state search using a string of 65 images and we find that all three paths A, B and C converge to the same minimum energy pathway (MEP). The MEP does not have a transition state that corresponds to a ``hard'' core structure. For the iterative-SOAP-GAP simulation in the 135 atom dislocation dipole simulation cell we also verify the energies of the resulting MEP by performing single-point calculations using DFT method for five points along the reaction coordinate. The results are shown in figure \ref{figure:soap-gap_transition_state} below.

\begin{figure}[H]
\begin{center}
\resizebox{12cm}{!}{\footnotesize{}\input{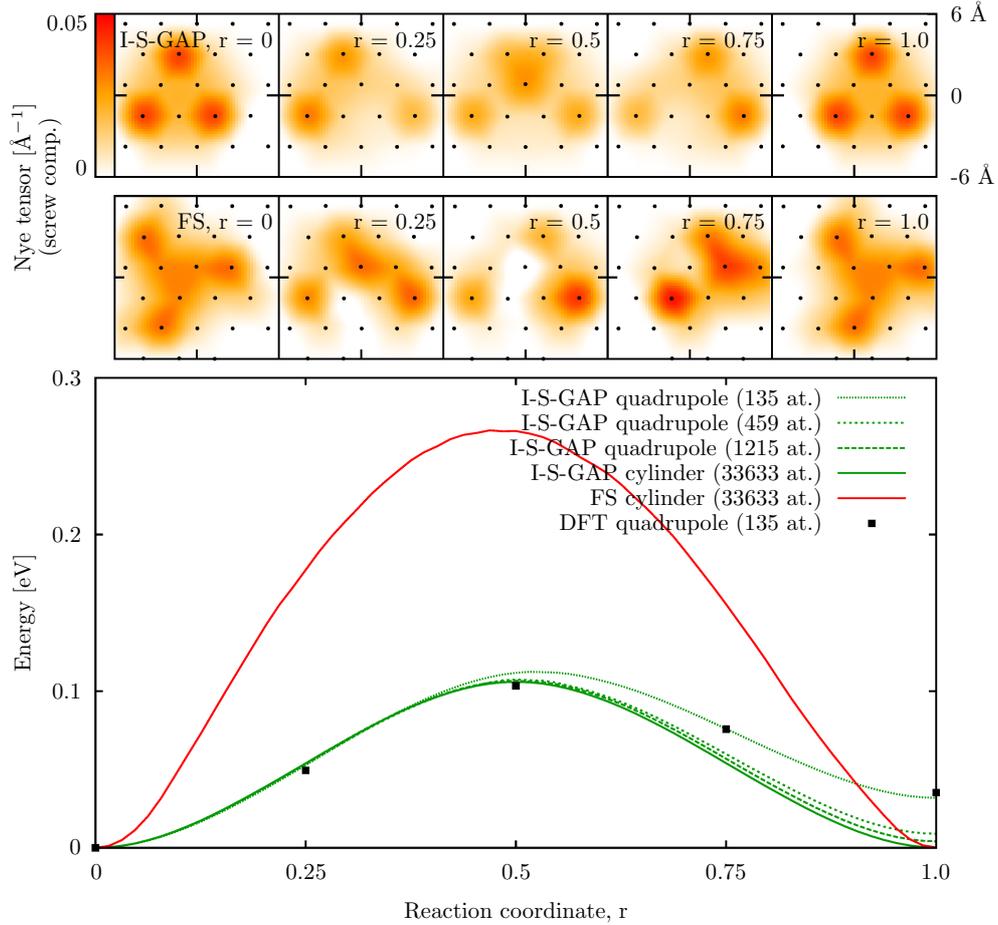}\normalsize{}}
\vspace{0.75cm}
\caption[$\frac{1}{2} \langle 111 \rangle$ screw dislocation Peierls barrier.]{$\frac{1}{2} \langle 111 \rangle$ screw dislocation Peierls barrier evaluated using both dislocation quadrupole and isolated dislocation approaches.}
\label{figure:soap-gap_transition_state}
\end{center}
\end{figure}

\vspace{-0.5cm} 

\noindent As anticipated, we find that the Peierls barrier of the SOAP-GAP is significantly lower than that of the Finnis-Sinclair potential. This is in agreement with the suggestions that the mobility of screw dislocations is affected by the amount of spreading of the dislocation core. By plotting the values of the Nye tensor along the reaction coordinate we also find that the transition path involves one of the neighbouring ``soft'' sites as opposed to the ``hard'' one --- at the mid-point the screw dislocation appears to be spread between three ``soft'' dislocation sites.

\newgeometry{top=2.4cm,bottom=2.4cm,left=3.5cm,right=3cm} 

\subsection{Dislocation-Vacancy Interactions}

One can calculate the binding energy between a vacancy and a screw dislocation by comparing the energies of two simulation cells $E^{(\text{disloc.})}$ and $E^{(\text{disloc.}+\text{vac.})}$ which have exactly the same geometry and differ by the presence of the vacancy alone. If we optimise the atomic positions in these configurations such that energy of the system is minimised, the dislocation-vacancy binding energy is given by:

\vspace{-0.25cm} 

\begin{equation}
E_b^{(\text{disloc.}+\text{vac.})} = \min_{\mathbf{x}_i \dots \mathbf{x}_N} ( E^{(\text{disloc.}+\text{vac.})} ) - \min_{\mathbf{x}_i \dots \mathbf{x}_M} ( E^{(\text{disloc.})} ) - E_f^{(\text{vac.})} + E_0 ,
\end{equation}

\noindent where $E_f^{(\text{vac.})}$ is the vacancy formation energy and $E_0$ is the ground state energy per atom of the perfect bcc lattice.

We begin by investigating the convergence of the dislocation-vacancy binding energy with the number of layers separating successive vacancies measured in multiples of the Burgers vector. We use an isolated dislocation approach in a system consisting of 33633 atoms for our simulations. The resulting value of the binding energy as a function of system depth is shown in figure \ref{figure:soap-gap_dislocation_depth_convergence} below.

\begin{figure}[H]
\begin{center}
\vspace{0.25cm}
\resizebox{12cm}{!}{\footnotesize{}\input{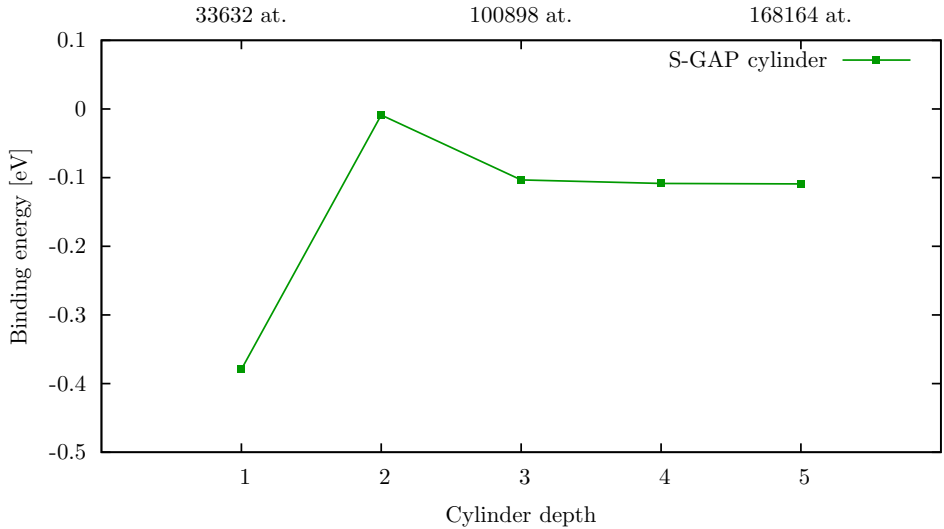}\normalsize{}}
\vspace{0.75cm}
\caption[Dislocation-vacancy binding energy as a function of system depth.]{Convergence of dislocation-vacancy binding energy with the number of layers separating successive vacancies measured in multiples of the Burgers vector.}
\label{figure:soap-gap_dislocation_depth_convergence}
\end{center}
\end{figure}

\vspace{-0.25cm} 

\noindent We find that the simulation cell consisting of three layers is sufficient to obtain converged values of the dislocation-vacancy binding energy, which is consistent with the results found in the literature (more details in \cite{0965-0393-19-7-074002}). Consequently, we proceed by computing the dislocation-vacancy interaction map in the region surrounding the dislocation core in a system consisting of 100898 atoms. This corresponds to carrying out a number of geometry optimisation simulations, each corresponding to vacancy position at a different lattice site. The results of our calculations are shown in figure \ref{figure:soap-gap_vacancy_map} below.

\begin{sloppypar} 
Due to the large simulation cell depth we find that verification of the dislocation-vacancy binding energies by the means of a single-point calculation using DFT method is not straightforward. Even in the smallest quadrupole configuration this corresponds to a simulation cell consisting of 404 atoms which is beyond our computational capabilities. We also anticipate that in such a small simulation cell dislocation-dislocation interactions would have a non-negligible effect on the dislocation-vacancy binding energy.
\end{sloppypar} 

\begin{figure}[H]
\begin{center}
\resizebox{12cm}{!}{\footnotesize{}\input{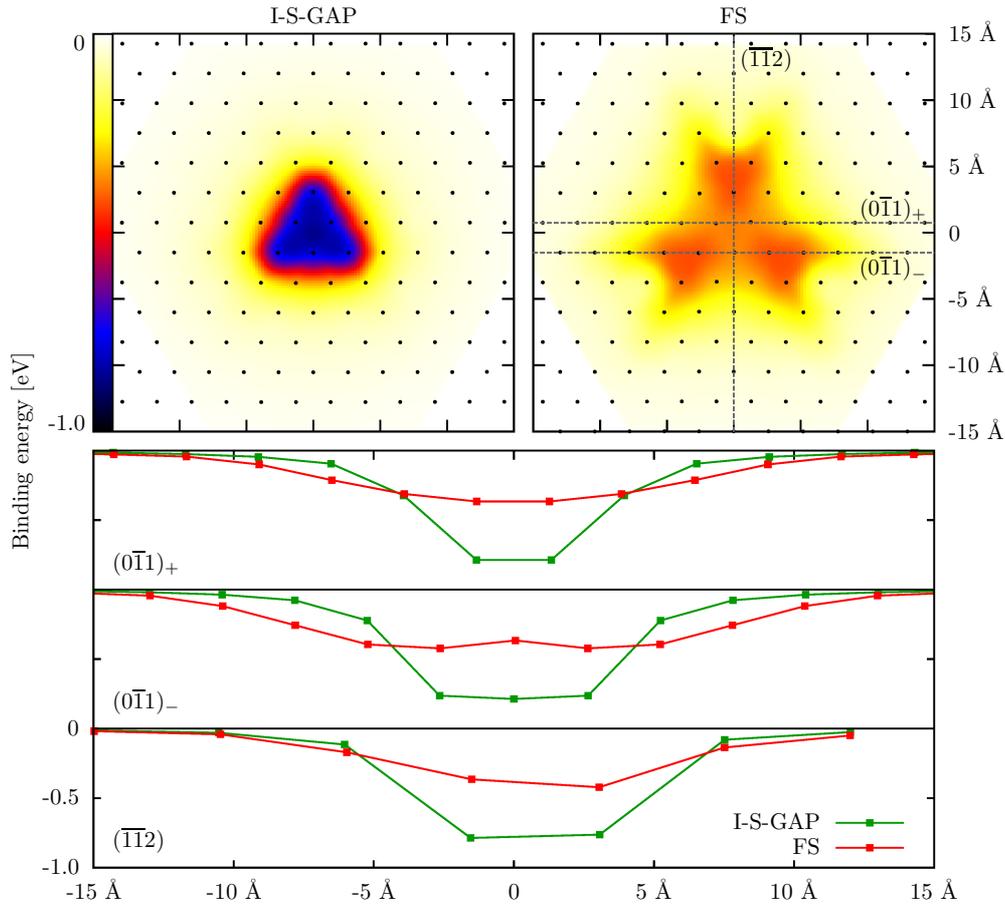}\normalsize{}}
\vspace{0.5cm}
\caption[Dislocation-vacancy interaction map.]{Dislocation-vacancy binding energy computed in an isolated dislocation system consisting of 100898 atoms.}
\label{figure:soap-gap_vacancy_map}
\end{center}
\end{figure}

\vspace{-0.25cm} 

While both Finnis-Sinclair and SOAP-GAP models predict attraction between vacancy and $\frac{1}{2} \langle 111 \rangle$ screw dislocation which can be regarded as a strong sink for vacancies, the above result suggest that the FS potential underestimates this attraction at the dislocation core. While it is unclear how the vacancy-dislocation interaction affects the mobility of the dislocation at this stage, this result demonstrates that such a study can be carried out with the SOAP-GAP interatomic potential in the future (in principle, the pinning or mediating effect of the vacancy on the screw dislocation could be investigated).

\subsection{Hyperparameters}

One of the benefits of the SOAP-GAP potential is that it reduces the number of adjustable hyperparameters as compared to the bispectrum-GAP potential. Instead of finding the vector of characteristic length-scales for each dimension of the bispectrum descriptor, the smoothness of the potential is instead adjusted with the width of a Gaussian used to represent the atomic density $\sigma_{atom}$ and the degree of the covariance function polynomial $\zeta$. Consequently, our procedure for training the SOAP-GAP interatomic potential depends only on the following hyperparameters:

\vspace{-0.2cm} 

\begin{itemize*}
\item noise in the training data $\to \{\sigma_\nu^{(\text{energy})} , \sigma_\nu^{(\text{force})} , \sigma_\nu^{(\text{virial})} \}$
\item width of Gaussian used to represent the atomic density $\to \sigma_{atom}$
\item degree of the covariance function polynomial $\to \zeta$
\item scale of energy variation in potential energy surface $\to \sigma_w$
\item SOAP radial and angular expansion cutoff $\to \{ n_{max}, l_{max} \}$
\item potential cutoff distance $\to r_{cut}$
\end{itemize*}

\vspace{-0.2cm} 

\noindent As in the case of the bispectrum-GAP potential, we have a prior knowledge of the noise in the training data from our investigation of the convergence of energies, forces and stress virials as a function of plane-wave energy cutoff, $k$-point sampling and smearing width. Consequently, we set $\sigma_\nu^{(\text{energy})}$ to $0.001 \, \text{eV/atom}$ and $\sigma_\nu^{(\text{force})}$ to $0.1 \, \text{eV/\r{A}}$ when the $k$-point sampling density is equal to $0.03 \, \text{\r{A}}^{-1}$, and $\sigma_\nu^{(\text{energy})}$ to $0.0001 \, \text{eV/atom}$, $\sigma_\nu^{(\text{force})}$ to $0.01 \, \text{eV/\r{A}}$ and $\sigma_\nu^{(\text{virial})}$ to $0.01 \, \text{eV/atom}$ when the $k$-point sampling density is equal to $0.015 \, \text{\r{A}}^{-1}$. We also set the hyperparameter corresponding to the scale of energy variation in the potential energy surface $\sigma_w$ to $1.0 \, \text{eV}$, which is the same value as the one we used for the bispectrum-GAP potential.

We find that the width of Gaussian functions that we use to represent the atomic density $\sigma_{atom}$ is dictated by the underlying physical properties of the system such as lattice constant and nearest neighbour distance. In the present case, we set it to the value of $0.5 \, \text{\r{A}}$.

The hyperparameter corresponding to the degree of the covariance function polynomial, $\zeta$, has the effect of increasing the sensitivity of the covariance function to change of the atomic positions. It was found empirically to work best with atomic systems when its value is equal to four or six (more details in \cite{PhysRevB.87.184115}). We summarise the results of our investigation into finding suitable values of $\{ n_{max}, l_{max} \}$ and $r_{cut}$ hyperparameters for the SOAP-GAP interatomic potential in table \ref{table:soap-gap_hyperparameters} below:

\restoregeometry 

\clearpage

\newgeometry{top=1.5cm,bottom=1.5cm,left=3.5cm,right=1.5cm}

\thispagestyle{empty}

\begin{sidewaystable}
\begin{center}
\scriptsize{}
\begin{tabular}{ l | c | c | c | c | c | c || c | c | c |}
& \multicolumn{6}{ c ||}{$n_{max}, l_{max}$ convergence} & \multicolumn{3}{ c |}{cutoff convergence} \\
\cline{2-10}
\emph{S-GAP$_{3}$ training database} & 6 & 8 & 10 & 12 & 14* & 16 & 4.0 \AA & 5.0 \AA* & 6.0 \AA \\
\hline
\multicolumn{10}{ l }{} \\
\multicolumn{10}{ l }{\emph{Training database errors:}} \\
\hline
RMS energy error per atom [eV] & 0.0125 & 0.0067 & 0.0078 & 0.0133 & 0.0110 & 0.0107 & 0.0073 & 0.0110 & 0.0063 \\
RMS force error [eV/\r{A}] & 0.5964 & 0.3966 & 0.3341 & 0.3386 & 0.3480 & 0.3480 & 0.3308 & 0.3480 & 0.3657 \\
\hline
\multicolumn{10}{ l }{} \\
\multicolumn{10}{ l }{\emph{RMS energy error per atom:} [eV]} \\
\hline
bcc primitive cells & 0.0002 & 0.0001 & 0.0001 & 0.0001 & 0.0001 & 0.0001 & 0.0001 & 0.0001 & 0.0001 \\
bcc bulk & 0.0001 & 0.0001 & 0.0001 & 0.0001 & 0.0001 & 0.0001 & 0.0001 & 0.0001 & 0.0001 \\
vacancy & 0.0001 & 0.0001 & 0.0000 & 0.0000 & 0.0000 & 0.0000 & 0.0001 & 0.0000 & 0.0000 \\
100, 110, 111, 112 surfaces & \emph{0.0408} & \emph{0.0257} & \emph{0.0556} & \emph{0.1069} & \emph{0.0845} & \emph{0.0819} & \emph{0.0419} & \emph{0.0845} & \emph{0.0140} \\
110, 112 gamma surfaces & \emph{0.0166} & \emph{0.0086} & \emph{0.0064} & \emph{0.0065} & \emph{0.0070} & \emph{0.0070} & \emph{0.0079} & \emph{0.0070} & \emph{0.0088} \\
110 gamma surface + vacancy & \emph{0.0039} & \emph{0.0018} & \emph{0.0018} & \emph{0.0015} & \emph{0.0016} & \emph{0.0015} & \emph{0.0018} & \emph{0.0016} & \emph{0.0014} \\
screw dislocation quadrupole & \emph{0.0002} & \emph{0.0002} & \emph{0.0003} & \emph{0.0001} & \emph{0.0001} & \emph{0.0001} & \emph{0.0002} & \emph{0.0001} & \emph{0.0002} \\
\hline
\multicolumn{10}{ l }{} \\
\multicolumn{10}{ l }{\emph{RMS force error:} [eV/\r{A}]} \\
\hline
bcc primitive cells & --- & --- & --- & --- & --- & --- & --- & --- & --- \\
bcc bulk & 0.0404 & 0.0251 & 0.0235 & 0.0231 & 0.0228 & 0.0228 & 0.0269 & 0.0228 & 0.0143 \\
vacancy & 0.0440 & 0.0280 & 0.0240 & 0.0231 & 0.0228 & 0.0227 & 0.0303 & 0.0228 & 0.0214 \\
100, 110, 111, 112 surfaces & \emph{0.5774} & \emph{0.6908} & \emph{0.3137} & \emph{0.3578} & \emph{0.3310} & \emph{0.3247} & \emph{0.3945} & \emph{0.3310} & \emph{0.5346} \\
110, 112 gamma surfaces & \emph{0.8277} & \emph{0.5400} & \emph{0.4624} & \emph{0.4711} & \emph{0.4868} & \emph{0.4865} & \emph{0.4521} & \emph{0.4868} & \emph{0.5045} \\
110 gamma surface + vacancy & \emph{0.3607} & \emph{0.2457} & \emph{0.2066} & \emph{0.1952} & \emph{0.1915} & \emph{0.1939} & \emph{0.2204} & \emph{0.1915} & \emph{0.2132} \\
screw dislocation quadrupole & \emph{0.1194} & \emph{0.0926} & \emph{0.0826} & \emph{0.0748} & \emph{0.0733} & \emph{0.0724} & \emph{0.0888} & \emph{0.0733} & \emph{0.0779} \\
\hline
\multicolumn{10}{ l }{} \\
\hline
lattice const. [\r{A}] & 3.1803 & 3.1803 & 3.1803 & 3.1803 & 3.1803 & 3.1803 & 3.1803 & 3.1803 & 3.1803 \\
C11 elastic constant [GPa] & 517.52 & 518.58 & 517.87 & 517.83 & 517.75 & 517.72 & 517.50 & 517.75 & 518.16 \\
C12 elastic constant [GPa] & 198.13 & 198.89 & 198.64 & 198.91 & 198.88 & 198.84 & 197.31 & 198.88 & 198.82 \\
bulk modulus [GPa] & 304.59 & 305.45 & 305.05 & 305.22 & 305.17 & 305.13 & 304.04 & 305.17 & 305.27 \\
shear modulus / C44 elastic constant [GPa] & 143.29 & 143.05 & 142.74 & 142.76 & 142.73 & 142.72 & 143.96 & 142.73 & 142.75 \\
\hline
RMS phonon spectrum error [THz] & 0.174 & 0.153 & 0.143 & 0.142 & 0.142 & 0.141 & 0.156 & 0.142 & 0.098 \\
\hline
vacancy energy [eV] & 3.25 & 3.26 & 3.25 & 3.26 & 3.26 & 3.26 & 3.25 & 3.26 & 3.25 \\
\hline
100 surface energy [eV / \r{A}$^2$] & 0.162 & 0.160 & 0.166 & 0.130 & 0.145 & 0.145 & 0.218 & 0.145 & 0.204 \\
110 surface energy [eV / \r{A}$^2$] & 0.229 & 0.132 & 0.138 & 0.096 & 0.117 & 0.120 & 0.189 & 0.117 & 0.195 \\
111 surface energy [eV / \r{A}$^2$] & --- & 0.132 & 0.155 & 0.100 & 0.122 & 0.157 & 0.157 & 0.122 & 0.080 \\
112 surface energy [eV / \r{A}$^2$] & 0.096 & 0.146 & 0.165 & 0.111 & 0.135 & 0.139 & 0.184 & 0.135 & 0.184 \\
\hline
RMS $\{110\} \langle 111 \rangle$ gamma surface energy error [eV] & 0.297 & 0.160 & 0.280 & 0.159 & 0.116 & 0.121 & 0.270 & 0.116 & 0.065 \\
\hline
RMS dislocation energy error [eV] & 0.316 & 0.167 & 0.314 & 0.062 & 0.065 & 0.058 & 0.229 & 0.065 & 0.216 \\
\hline
\end{tabular}
\normalsize{}
\end{center}
\captionsetup{margin=4cm}
\caption[Convergence of hyperparameters.]{Convergence of hyperparameters for the SOAP-GAP potential, where we investigate SOAP radial and angular expansion cutoff $\{ n_{max}, l_{max} \}$ and potential cutoff distance $r_{cut}$.}
\captionsetup{margin=1cm}
\label{table:soap-gap_hyperparameters}
\end{sidewaystable}

\restoregeometry

\clearpage

\section{Discussion}
\label{chapter:soap-gap_potential_for_tungsten:section:discussion}

We demonstrated in this chapter that the Smooth Overlap of Atomic Positions kernel overcomes some of the limitations that we encountered with the bispectrum-GAP interatomic potential. Our investigation of the dependance of the SOAP-GAP potential on the hyperparameters reveals that the accuracy of the SOAP based potentials can be systematically improved as the values of $\{ n_{max}, l_{max} \}$ increase (see table \ref{table:soap-gap_hyperparameters}) and we also find that the resulting potential is much better behaved in the extrapolative regime (as demonstrated for stress-strain curves in the anharmonic regime in figure \ref{figure:soap-gap_stress-strain}). We believe that this can be attributed to the fact that within the SOAP methodology the atomic density function is expanded using Gaussian functions instead of Dirac delta functions.

By extending the analysis carried out in section \ref{chapter:bispectrum-gap_potential_for_tungsten:section:discussion}, it is easy to see that the Fourier representation of a Gaussian function corresponds to another Gaussian function in real space (albeit with a different width). Consequently, while truncating some of the frequencies in the Fourier representation of a Gaussian still results in high frequency oscillations in the representation of the atomic density function, these are convoluted with a Gaussian envelope. Consequently, we anticipate these oscillations to be short-lived in real space and hence the ``noisyness'' of the coordinates of the potential energy surface associated with the bispectrum descriptor is significantly reduced.

\begin{sloppypar} 
Another aspect of GAP potential training that we investigated in this chapter is the inclusion of the Finnis-Sinclair interatomic potential core. Against our expectations we found that removing the core potential improves the accuracy of the resulting SOAP-GAP potential. Comparing the force errors of FS/SOAP-GAP$_3$ and SOAP-GAP$_3$ potentials from table \ref{table:soap-gap_results} demonstrates that in the absence of the core potential RMS force errors are reduced between two to three times, i.e. RMS force error for vacancy dataset corresponds to $0.023 \, \text{eV/\r{A}}$ for SOAP-GAP$_3$ and $0.053 \, \text{eV/\r{A}}$ for FS/SOAP-GAP$_3$ ($0.103 \, \text{eV/\r{A}}$ for FS/bispectrum-GAP$_3$). At the same time the total reduction of force errors between SOAP-GAP and FS/bispectrum-GAP potentials ranges from approximately five times in the interpolation regime to ten times or more in the extrapolation regime (as demonstrated for the relaxed structure of the screw dislocation where the force errors were reduced by over an order of magnitude).
\end{sloppypar} 

In spite of the increased complexity of the SOAP-GAP potential compared to the bispectrum-GAP, we believe that this improvement in accuracy is worth the cost as the computational time nevertheless scales linearly. While we only explored its applications in systems of up to $\sim 170,000$ atoms, simulation of bigger systems is a matter of parallelisation alone. We finish this chapter by presenting the comparison of the computational cost of the SOAP-GAP, bispectrum-GAP, Finnis-Sinclair and DFT models in figure \ref{figure:computational_cost} below.

\begin{figure}[H]
\begin{center}
\resizebox{12cm}{!}{\footnotesize{}\input{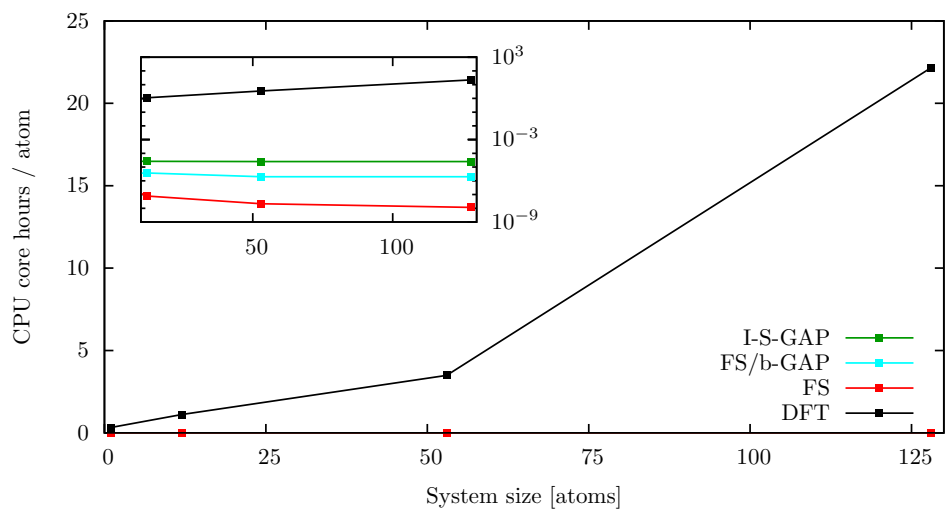}\normalsize{}}
\vspace{0.75cm}
\caption[Computational cost of GAP, FS and DFT models.]{Comparison of the computational cost of SOAP-GAP, bispectrum-GAP, FS and DFT models.}
\label{figure:computational_cost}
\end{center}
\end{figure}

\cleardoublepage

\chapter{Bond-based SOAP-GAP Potential}
\label{chapter:bond-based_soap-gap_potential}

\section{Introduction}
\label{chapter:bond-based_soap-gap_potential:section:introduction}

In this chapter I present the outcome my theoretical work on the bond-based covariance function for the GAP potential, which was carried out in parallel with the work on the Smooth Overlap of Atomic Positions (SOAP) kernel performed by members of my research group. In the previous chapter we decided to use the SOAP kernel in order to improve the bispectrum-GAP potential for tungsten as it builds directly on the bispectrum descriptor methodology. However, the analysis below offers a recipe for how a bond-based SOAP-GAP potential can be implemented and used efficiently for systems where the bond environment can be determined by the surrounding atoms that lie within a close neighbourhood of the bond.

The symmetry of the covariance function of a GAP interatomic potential is dictated by the spherical symmetry of an atom. The covariance is constructed by integrating over all possible rotations and in three dimensions this corresponds to three independent rotation directions. Carrying out this integration over all arbitrary rotations analytically is extremely difficult if at all possible. Consequently, as demonstrated in the previous chapters one needs to resort to expanding the atomic density function in terms of a spherical harmonics basis and this expansion needs to be truncated.

Unlike the interatomic potentials, the symmetry of the covariance function of a bond-based GAP potential is dictated by the cylindrical symmetry of a bond. There is only one rotation direction (along the axis of the bond), and we anticipate that integrating over all possible rotations can be achieved in some situations analytically. However, the idea of allocating local energies to bonds instead of atoms is also motivated by the success of tight binding and Bond Order Potential formalism where the most important contribution to the total energy comes from the interatomic matrix elements which directly correspond to bond energies.

I begin in section \ref{chapter:bond-based_soap-gap_potential:section:rotationally_invariant_bond_descriptor} by outlining the concept of a covariance function for the bond-based GAP potential. In section \ref{chapter:bond-based_soap-gap_potential:section:covariance_functions_for_smooth_atomic_density} I investigate the functional form of this covariance function. In section \ref{chapter:bond-based_soap-gap_potential:section:smooth_overlap_for_bond-based_gap_potential} I derive an analytic expression for a general, rotationally invariant covariance function between two bonds that is computed from smoothly changing atomic density functions. Finally, I finish this chapter with section \ref{chapter:bond-based_soap-gap_potential:section:implementation_considerations} by discussing some of the numerical and implementation issues that need addressing.

\section{Rotationally Invariant Bond Descriptor}
\label{chapter:bond-based_soap-gap_potential:section:rotationally_invariant_bond_descriptor}

In order to fit a bond-based GAP potential we need to be able to evaluate a covariance function $k(\rho, \rho')$ between two bonds where $\rho$ and $\rho'$ are real-valued, three-dimensional scalar fields (atomic density functions) describing the atomic environment of the bonds.

If we take the square-exponential covariance function as a starting point in our analysis:

\begin{equation}
k(r) = \exp \left[ - \frac{r^2}{2 \theta^2} \right] ,
\end{equation}

\noindent where $r$ is the ``distance'' between the two atomic environments and $\sigma$ is defined as the characteristic length-scale. We can define the ``distance'' between the two atomic environments $\rho$ and $\rho'$ by generalising the concept of the Euclidean distance:

\begin{equation}
r^2 = \int \left( \rho(\mathbf{x}) - \rho'(\mathbf{x}) \right)^2 d^3 \, \mathbf{x} ,
\end{equation}

\noindent where we align the bonds so that they are parallel to the $z$-axis and centred at the origin before we evaluate the above integral.

However, it is clear that our covariance $k(\rho, \rho')$ depends on the precise orientation of the bonds. Even when the bonds are aligned in the same direction, the covariance function is not invariant to individual rotations of any of the bonds about the $z$-axis (and our procedure used to align the bonds along the $z$-axis is completely arbitrary).

We can introduce rotational invariance by integrating the covariance function over all possible rotations about the $z$-axis. Using an arbitrary rotation operator $\hat{\mathbf{R}}$, we can redefine our covariance as follows:

\begin{equation}
k(r) = \int \exp \left[ - \frac{(\hat{\mathbf{R}} r)^2}{2 \theta^2} \right] d \, \hat{\mathbf{R}} ,
\end{equation}

\noindent where the ``distance'' between the atomic environments of the two bonds is given by:

\begin{equation}
(\hat{\mathbf{R}} r)^2 = \int \left( \rho(\mathbf{x}) - \rho'(\hat{\mathbf{R}} \mathbf{x}) \right)^2 d^3 \, \mathbf{x} .
\end{equation}

If the bonds in question are between atoms of the same species, the symmetry of the system also dictates that the covariance function should be invariant with respect to reflections about the $x$-$y$ plane. Reflection invariance can be easily included in the covariance function by summing over all possible mirror images. Consequently, we redefine our covariance function again by including the mirror image operator $\hat{\mathbf{M}}$ that changes the direction of the $z$-axis and we sum over the reflections about the $x$-$y$ plane:

\begin{equation}
k(r) = \sum_{\hat{\mathbf{M}}} \int \exp \left[ - \frac{(\hat{\mathbf{M}} \hat{\mathbf{R}} r)^2}{2 \theta^2} \right] d \, \hat{\mathbf{R}} .
\end{equation}

It is clear from the above analysis that in order to evaluate the covariance function $k(\rho, \rho')$ we need to carry out two integrations: first in real space over the real-valued, three-dimensional scalar fields describing the atomic environments; and second over all possible rotations about the $z$-axis. We find that in order to carry out this double integration analytically we need to approximate the functional form of the covariance function.

\section[Covariance Functions for Smooth Atomic Density]{\texorpdfstring{Covariance Functions \\ for Smooth Atomic Density}{Covariance Functions for Smooth Atomic Density}}
\label{chapter:bond-based_soap-gap_potential:section:covariance_functions_for_smooth_atomic_density}

Our definition of the atomic density function $\rho$ needs to fulfil three conditions:

\begin{itemize*}
\item It needs to be continuous and smooth in real space.
\item It needs to provide permutational invariance when the ordering of the atoms around the bond changes.
\item It needs to be integrable analytically.
\end{itemize*}

\noindent We find that the atomic density function that fulfils all of the above criteria is provided by the sum of three-dimensional Gaussian functions centred at the positions of the atoms that lie within the cutoff of the bond environment. This is also the same atomic density function as used for the purpose of the SOAP kernel (as outlined in section \ref{chapter:gaussian_approximation_potential:section:description_of_atomic_environments}). Consequently, the atomic density function $\rho$ describing the atomic environment of a bond is given by:

\begin{equation}
\rho = \sum_{i}^N R(\mathbf{r}_i) \exp \left[- \frac{(\mathbf{x} - \mathbf{r}_i)^2}{2 \sigma^2}\right] ,
\end{equation}

\noindent where $\sigma$ is the width of the Gaussian and $R(\mathbf{r}_i)$ is a scaling function that can be used to alter the height of the Gaussian (for example for a multi-species bond-based potential). At the same time it ensures that the descriptor is continuous at a finite cutoff by smoothly decaying to zero at the cutoff distance. Note that by definition $R(\mathbf{r}_i)$ has a radial symmetry about the axis parallel to the bond.

Since the density function consisting of a sum of Gaussian functions can be integrated analytically, the Euclidean distance $r_{\rho \hat{\mathbf{R}} \rho'}^2$ between atomic density field $\rho$, and field $\rho'$ under an arbitrary rotation $\hat{\mathbf{R}}$ is given by:

\scriptsize{}
\begin{align*}
& r_{\rho \hat{\mathbf{R}} \rho'}^2 = \int \Biggl( \rho(\mathbf{x}) - \rho'(\hat{\mathbf{R}} \mathbf{x}) \Biggr)^2 d^3 \, \mathbf{x} \nonumber \\ 
& = \int \Biggl( \sum_{i}^N R(\mathbf{r}_i) \exp \left[- \frac{(\mathbf{x} - \mathbf{r}_i)^2}{2 \sigma^2}\right] - \sum_{i'}^{N'} R(\hat{\mathbf{R}}\mathbf{r}_{i'}) \exp \left[- \frac{(\mathbf{x} - \hat{\mathbf{R}}\mathbf{r}_{i'})^2}{2 \sigma^2}\right] \Biggr)^2 d^3 \, \mathbf{x} \nonumber \\
& = \underbrace{\int \Biggl( \sum_{i}^N R^2(\mathbf{r}_i) \exp \left[- \frac{(\mathbf{x} - \mathbf{r}_i)^2}{\sigma^2}\right] + \sum_{i=2}^N \sum_{j < i}^N 2 R(\mathbf{r}_i) R(\mathbf{r}_j) \exp \left[- \frac{(\mathbf{x} - \mathbf{r}_i)^2 + (\mathbf{x} - \mathbf{r}_j)^2}{2 \sigma^2}\right] \Biggr) d^3 \, \mathbf{x}}_{C_{\rho \rho} \quad \text{(bond $\rho$ ``self'' overlap)}} \nonumber \\
& + \underbrace{\int \Biggl( \sum_{i'}^{N'} R^2(\hat{\mathbf{R}}\mathbf{r}_{i'}) \exp \left[- \frac{(\mathbf{x} - \hat{\mathbf{R}}\mathbf{r}_{i'})^2}{\sigma^2}\right] + \sum_{i'=2}^{N'} \sum_{j' < i'}^{N'} 2 R(\hat{\mathbf{R}}\mathbf{r}_{i'}) R(\hat{\mathbf{R}}\mathbf{r}_{j'}) \exp \left[- \frac{(\mathbf{x} - \hat{\mathbf{R}}\mathbf{r}_{i'})^2 + (\mathbf{x} - \hat{\mathbf{R}}\mathbf{r}_{j'})^2}{2 \sigma^2}\right] \Biggr) d^3 \, \mathbf{x}}_{C_{\hat{\mathbf{R}} \rho' \hat{\mathbf{R}} \rho'} \quad \text{(bond $\hat{\mathbf{R}}\rho'$ ``self'' overlap)}} \nonumber \\
& - 2 \underbrace{\int \Biggl( \sum_{i}^N \sum_{i'}^{N'} R(\mathbf{r}_i) R(\hat{\mathbf{R}}\mathbf{r}_{i'}) \exp \left[- \frac{(\mathbf{x} - \mathbf{r}_i)^2 + (\mathbf{x} - \hat{\mathbf{R}}\mathbf{r}_{i'})^2}{2 \sigma^2}\right] \Biggr) d^3 \, \mathbf{x}}_{C_{\rho \hat{\mathbf{R}} \rho'} \quad \text{(overlap between bonds $\rho$ and $\hat{\mathbf{R}}\rho'$)}}
\end{align*}
\normalsize{}

\noindent Exploiting the fact that $R(\mathbf{r}_{i'})$ is invariant to rotations about the axis parallel to the bond, we can immediately recognise that:

\clearpage 

\begin{align}
R(\hat{\mathbf{R}}\mathbf{r}_{i'}) &\to R(\mathbf{r}_{i'}) \nonumber \\[0.25cm]
C_{\hat{\mathbf{R}} \rho' \hat{\mathbf{R}} \rho'} &\to C_{\rho' \rho'} .
\end{align}

\noindent Consequently the distance $r_{\rho \hat{\mathbf{R}} \rho'}^2$ can be simplified as:

\begin{equation}
r_{\rho \hat{\mathbf{R}} \rho'}^2 = C_{\rho \rho} + C_{\rho' \rho'} - 2 C_{\rho \hat{\mathbf{R}} \rho'} ,
\end{equation}

\noindent where the overlap elements are given by:

\begin{align}
C_{\rho \rho} &= \sigma^3 \pi^{3 / 2} \left( \sum_i^N R^2(\mathbf{r}_i) + \sum_{i=2}^N \sum_{j < i}^N 2 R(\mathbf{r}_i) R(\mathbf{r}_j) \exp \left[- \frac{(\mathbf{r}_i - \mathbf{r}_j)^2}{4 \sigma^2}\right] \right) \nonumber \\[0.25cm]
C_{\rho' \rho'} &= \sigma^3 \pi^{3 / 2} \left( \sum_{i'}^{N'} R^2(\mathbf{r}_{i'}) + \sum_{i'=2}^{N'} \sum_{j' < i'}^{N'} 2 R(\mathbf{r}_{i'}) R(\mathbf{r}_{j'}) \exp \left[- \frac{(\mathbf{r}_{i'} - \mathbf{r}_{j'})^2}{4 \sigma^2}\right] \right) \nonumber \\[0.25cm]
C_{\rho \hat{\mathbf{R}} \rho'} &= \sigma^3 \pi^{3 / 2} \left( \sum_{i}^N \sum_{i'}^{N'} R(\mathbf{r}_i) R(\mathbf{r}_{i'}) \exp \left[- \frac{(\mathbf{r}_i - \hat{\mathbf{R}}\mathbf{r}_{i'})^2}{4 \sigma^2}\right] \right) .
\end{align}

\vspace{0.25cm} 

If the distance $r_{\rho \hat{\mathbf{R}} \rho'}^2$ between the two environments is defined to lie in the domain $\{ r \in \mathbb{R} | 0 \leq r \leq 1 \}$ where $0$ corresponds to no similarity and $1$ corresponds to identical environments, we can define the normalised distance $\hat{r}_{\rho \hat{\mathbf{R}} \rho'}^2$ as:

\begin{equation}
\hat{r}_{\rho \hat{\mathbf{R}} \rho'}^2 = \frac{r_{\rho \hat{\mathbf{R}} \rho'}^2}{C_{\rho \rho} + C_{\rho' \rho'}} = 1 - \frac{C_{\rho \hat{\mathbf{R}} \rho'}}{\frac{1}{2} ( C_{\rho \rho} + C_{\rho' \rho'} )} ,
\label{equation:normalised_distance}
\end{equation}

\vspace{0.25cm} 

\noindent and we can demonstrate that the characteristics of the square-exponential covariance function can be approximated by a polynomial covariance function:

\begin{equation}
k_{\rho \rho'} = (1 - \hat{r}_{\rho \rho'}^2)^n .
\end{equation}

\begin{sloppypar} 
\noindent A plot of the square-exponential covariance function and its polynomial approximation up to an order of $n = 6$ is shown in figure \ref{figure:covariance_function} below.
\end{sloppypar} 

\begin{figure}[H]
\begin{center}
\resizebox{12cm}{!}{\footnotesize{}\input{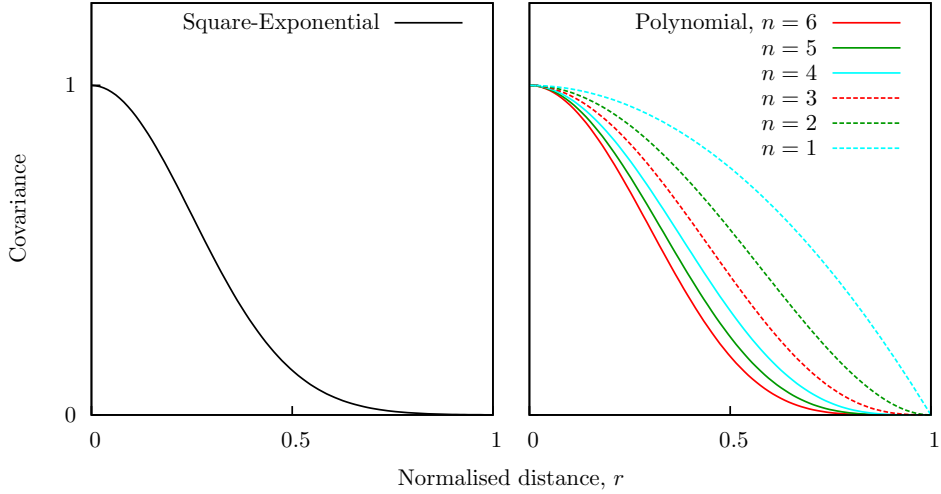}\normalsize{}}
\vspace{0.75cm}
\caption[Polynomial approximation to the square-exponential covariance.]{Polynomial approximation to the square-exponential covariance function up to an order of $n = 6$.}
\label{figure:covariance_function}
\end{center}
\end{figure}

\noindent Consequently, we can construct a rotationally invariant polynomial covariance function by integrating over all possible rotations $\hat{\mathbf{R}}$:

\begin{equation}
k_{\rho \rho'} = \int \left(1 - \hat{r}_{\rho \hat{\mathbf{R}} \rho'}^2\right)^n d \, \hat{\mathbf{R}} ,
\end{equation}

\noindent and by substituting equation \ref{equation:normalised_distance}, we obtain:

\begin{align}
k_{\rho \rho'} &= \int \left(\frac{C_{\rho \hat{\mathbf{R}} \rho'}}{\frac{1}{2} ( C_{\rho \rho} + C_{\rho' \rho'} )}\right)^n d \, \hat{\mathbf{R}} \nonumber \\
&= \left( \frac{2}{C_{\rho \rho} + C_{\rho' \rho'}} \right)^n \int \left( C_{\rho \hat{\mathbf{R}} \rho'} \right)^n d \, \hat{\mathbf{R}} \nonumber \\
&\propto \int \left( C_{\rho \hat{\mathbf{R}} \rho'} \right)^n d \, \hat{\mathbf{R}} ,
\label{equation:covariance}
\end{align}

\noindent which corresponds to the similarity kernel proposed by Bart\'{o}k-P\'{a}rtay, Kondor and Cs\'{a}nyi for the SOAP covariance function (more details in \cite{PhysRevB.87.184115}). The dot-product kernel of density overlap corresponds to a polynomial approximation of the square-exponential covariance function.

As is the case for the SOAP kernel, we do not need to worry about the normalisation constant $\left( \frac{2}{C_{i \, i} + C_{i' \, i'}} \right)^n$ as we can always renormalise our covariance function later by computing:

\begin{equation}
\hat{k}_{\rho \rho'} = \left( \frac{k_{\rho \rho'}}{\sqrt{k_{\rho \rho}} \sqrt{k_{\rho' \rho'}}} \right)^{\zeta} ,
\end{equation}

\noindent where $\zeta$ can be used to further tune the behaviour of $\hat{k}_{\rho \rho'}$.

Consequently, the problem of evaluating the covariance function $k_{\rho \rho'}$ is equivalent to that of evaluating $\int \left( C_{i \hat{\mathbf{R}} i'} \right)^n d \, \hat{\mathbf{R}}$. As we will demonstrate in the next section, for the case of a bond-based potential we can evaluate this integration for an arbitrary rotation $\hat{\mathbf{R}}$ analytically, for any order $n$ and with no need for expansion in a spherical harmonics basis (as is the case for the SOAP atomic descriptor; more details in section \ref{chapter:gaussian_approximation_potential:section:description_of_atomic_environments}).

\section[Smooth Overlap for Bond-based GAP Potential]{\texorpdfstring{Smooth Overlap for Bond-based \\ GAP Potential}{Smooth Overlap for Bond-based GAP Potential}}
\label{chapter:bond-based_soap-gap_potential:section:smooth_overlap_for_bond-based_gap_potential}

In order to evaluate $\int \left( C_{i \hat{\mathbf{R}} i'} \right)^n d \, \hat{\mathbf{R}}$ we expand $C_{\rho \hat{\mathbf{R}} \rho'}$ using the multinomial theorem:

\scriptsize{}
\begin{align*}
& \int \left( C_{i \hat{\mathbf{R}} i'} \right)^n d \, \hat{\mathbf{R}} = \sigma^{3n} \pi^{3n / 2} \int \left( \sum_{i}^N \sum_{i'}^{N'} R(\mathbf{r}_i) R(\mathbf{r}_{i'}) \exp \left[- \frac{(\mathbf{r}_i - \hat{\mathbf{R}}\mathbf{r}_{i'})^2}{4 \sigma^2}\right] \right)^n d \, \hat{\mathbf{R}} \nonumber \\
& \propto \sum_{i}^N \sum_{i'}^{N'} \left( R(\mathbf{r}_i) R(\mathbf{r}_{i'}) \right)^n \int \exp \left[- \frac{n (\mathbf{r}_i - \hat{\mathbf{R}}\mathbf{r}_{i'})^2}{4 \sigma^2}\right] d \, \hat{\mathbf{R}} \nonumber \\
& + \frac{n!}{1! (n - 1)!} \sum_{\substack{i \\ j \neq i}}^N \sum_{\substack{i' \\ j' \neq i'}}^{N'} \left( R(\mathbf{r}_i) R(\mathbf{r}_{i'}) \right)^{n - 1} \left( R(\mathbf{r}_j) R(\mathbf{r}_{j'}) \right) \int \exp \left[- \frac{(n - 1) (\mathbf{r}_i - \hat{\mathbf{R}}\mathbf{r}_{i'})^2 + (\mathbf{r}_j - \hat{\mathbf{R}}\mathbf{r}_{j'})^2}{4 \sigma^2}\right] d \, \hat{\mathbf{R}} \nonumber \\
& + \, \, \text{etc.}
\end{align*}
\normalsize{}

\noindent Since the rotation operator $\hat{\mathbf{R}}(\theta)$ can be defined in terms of the Cartesian coordinates as (up to an arbitrarily chosen angle $\theta_0$):

\begin{align}
x = r \cos(\theta_0) &\to \hat{\mathbf{R}}(\theta) x = r \cos(\theta_0 + \theta) \nonumber \\
y = r \sin(\theta_0) &\to \hat{\mathbf{R}}(\theta) y = r \sin(\theta_0 + \theta) \nonumber \\
z &\to \hat{\mathbf{R}}(\theta) z = z ,
\end{align}

\noindent we can rewrite the exponential terms that include the rotation operator $\hat{\mathbf{R}}(\theta)$ in the multinomial expansion of $C_{\rho \hat{\mathbf{R}} \rho'}$ as:

\scriptsize{}
\begin{align*}
& \exp \left[- \frac{n (\mathbf{r}_i - \hat{\mathbf{R}}\mathbf{r}_{i'})^2}{4 \sigma^2}\right] \nonumber \\
& \quad \quad = \exp \left[- \frac{n (r_i^2 + r_{i'}^2 - 2 z_i z_{i'})}{4 \sigma^2}\right] \exp \left[- \frac{n \left( x_i x_{i'} + y_i y_{i'} \right) \cos \theta + n \left( y_i x_{i'} - x_i y_{i'} \right) \sin \theta}{2 \sigma^2}\right] \\
\\
& \exp \left[- \frac{(n - 1) (\mathbf{r}_i - \hat{\mathbf{R}}\mathbf{r}_{i'})^2 + (\mathbf{r}_j - \hat{\mathbf{R}}\mathbf{r}_{j'})^2}{4 \sigma^2}\right] \nonumber \\
& \quad \quad = \exp \left[- \frac{(n - 1)(r_i^2 + r_{i'}^2 - 2 z_i z_{i'}) + (r_j^2 + r_{j'}^2 - 2 z_j z_{j'})}{4 \sigma^2}\right] \nonumber \\
& \quad \quad \times \exp \left[- \frac{ \left( (n - 1)(x_i x_{i'} + y_i y_{i'}) + (x_j x_{j'} + y_j y_{j'}) \right) \cos \theta + \left( (n - 1)(y_i x_{i'} - x_i y_{i'}) + (y_j x_{j'} - x_j y_{j'}) \right) \sin \theta}{2 \sigma^2}\right] \\
\\
& \, \, \text{etc.}
\end{align*}
\normalsize{}

\noindent Integration over all arbitrary rotations $\hat{\mathbf{R}}$ can be now evaluated analytically using a standard integral. Since the integration is over all possible angles:

\begin{equation}
\int_0^{2 \pi} \exp \left[ x \cos \theta + y \sin \theta \right] d \, \theta = 2 \pi I_0 \left( \sqrt{x^2 + y^2} \right) ,
\end{equation}

\noindent where $I_0$ is the modified Bessel function of the first kind.

Consequently we obtain:

\scriptsize{}
\begin{align*}
& \int \left( C_{i \hat{\mathbf{R}} i'} \right)^n d \, \hat{\mathbf{R}} = \sigma^{3n} \pi^{3n / 2} \int \left( \sum_{i}^N \sum_{i'}^{N'} R(\mathbf{r}_i) R(\mathbf{r}_{i'}) \exp \left[- \frac{(\mathbf{r}_i - \hat{\mathbf{R}}\mathbf{r}_{i'})^2}{4 \sigma^2}\right] \right)^n d \, \hat{\mathbf{R}} \nonumber \\
& \propto \sum_{i}^N \sum_{i'}^{N'} \left( R(\mathbf{r}_i) R(\mathbf{r}_{i'}) \right)^n \exp \left[- \frac{n (r_i^2 + r_{i'}^2 - 2 z_i z_{i'})}{4 \sigma^2}\right] I_0 \left( \frac{1}{2 \sigma^2} \sqrt{n^2 \left( x_i x_{i'} + y_i y_{i'} \right)^2 + n^2 \left( y_i x_{i'} - x_i y_{i'} \right)^2} \right) \nonumber \\
& + \sum_{\substack{i \\ j \neq i}}^N \sum_{\substack{i' \\ j' \neq i'}}^{N'} \frac{n!}{1! (n - 1)!} \left( R(\mathbf{r}_i) R(\mathbf{r}_{i'}) \right)^{n - 1} \left( R(\mathbf{r}_j) R(\mathbf{r}_{j'}) \right) \exp \left[- \frac{(n - 1)(r_i^2 + r_{i'}^2 - 2 z_i z_{i'}) + (r_j^2 + r_{j'}^2 - 2 z_j z_{j'})}{4 \sigma^2}\right] \nonumber \\
& \quad \quad \times I_0 \left( \frac{1}{2 \sigma^2} \sqrt{\left( (n - 1)(x_i x_{i'} + y_i y_{i'}) + (x_j x_{j'} + y_j y_{j'}) \right)^2 + \left( (n - 1)(y_i x_{i'} - x_i y_{i'}) + (y_j x_{j'} - x_j y_{j'}) \right)^2} \right) \nonumber \\
& + \, \, \text{etc.}
\end{align*}
\normalsize{}

\noindent where all the subsequent terms of the expansion are according to the multinomial theorem.

\clearpage 

While the above expression can be evaluated analytically, we need to be able to compute the derivatives of the covariance function with respect to the Cartesian coordinates of all atoms in order to train the bond-based GAP potential from energy derivatives (forces or stresses). In the above formulation we need to explicitly decide on the choice of $x$-$y$ coordinates when the bonds are aligned along the $z$-axis. This, however, turns out to be problematic --- while finding a suitable transformation matrix is a simple, well-defined procedure, this matrix turns out to be discontinuous with respect to the Cartesian coordinates of the atoms. Consequently, in its existing form the above expression cannot be differentiated.

One is however compelled to make the following observation --- since the similarity measure $\int \left( C_{i \hat{\mathbf{R}} i'} \right)^n d \, \hat{\mathbf{R}}$ is rotationally invariant, its arguments should not explicitly depend on the choice of the reference frame inside the atomic environments of bonds $\rho$ or $\rho'$, i.e. the choice of $x$-$y$ coordinates when the bonds are aligned along the $z$-axis is completely arbitrary. Consequently, it must be possible to rewrite the above expression in terms of bond radii and angles alone.

We begin by inspecting the above expression for the simple cases of $n = 1$, $n = 2$ and $n = 3$. We anticipate that only radial information is preserved for $n = 1$ as the order of integration in equation \ref{equation:covariance} can be exchanged. As a starting point, we rewrite the $x_i$ and $y_i$ coordinates in terms of:

\begin{align}
x_i^2 + y_i^2 &= r_i^2 \nonumber \\
x_i x_j + y_i y_j &= r_i r_j \cos \theta_{ij} \nonumber \\
y_i x_j - x_i y_j &= r_i r_j \sin \theta_{ij} ,
\end{align}

\noindent where $\theta_{ij}$ is an angle between atoms $i$ and $j$ projected onto the plane perpendicular to the bond axis. 

Carrying out the substitution and simplifying the resulting expression, for the simplest case of $n = 1$ we obtain:

\scriptsize{}
\begin{align*}
k_{\rho \rho'}|_{n=1} &= \sum_{i}^N \sum_{i'}^{N'} R(\mathbf{r}_i) R(\mathbf{r}_{i'}) \exp \left[- \frac{r_i^2 + r_{i'}^2 - 2 z_i z_{i'}}{4 \sigma^2}\right] \nonumber \\
& \quad \quad \times I_0 \left( \frac{1}{2 \sigma^2} \sqrt{\left( x_i x_{i'} + y_i y_{i'} \right)^2 + \left( y_i x_{i'} - x_i y_{i'} \right)^2} \right) \nonumber \\
\\
& = \sum_{i} \sum_{i'} R_i R_{i'} \exp \Bigg[\frac{z_i z_{i'}}{2 \sigma^2}\Bigg] \exp \Bigg[- \frac{r_i^2 + r_{i'}^2}{4 \sigma^2}\Bigg] I_0 \Bigg( \frac{r_i r_{i'}}{2 \sigma^2}\Bigg)
\end{align*}
\normalsize{}

\noindent where, in agreement with our expectations, no angular information is preserved. The covariance is a function of radial distance and vertical separation alone.

For the more useful cases of $n = 2$ and $n = 3$ we obtain respectively:

\scriptsize{}
\begin{align*}
k_{\rho \rho'}|_{n=2} &= \sum_{i} \sum_{i'} R_i^2 R_{i'}^2 \exp \Bigg[\frac{z_i z_{i'}}{\sigma^2}\Bigg] \exp \Bigg[- \frac{r_i^2 + r_{i'}^2}{2 \sigma^2}\Bigg] I_0 \Bigg( \frac{r_i r_{i'}}{\sigma^2}\Bigg) \nonumber \\
&+ 2 \sum_{\substack{i \\ j \neq i}} \sum_{\substack{i' \\ j' \neq i'}} R_i R_{i'} R_j R_{j'} \exp \Bigg[\frac{z_i z_{i'} + z_j z_{j'}}{2 \sigma^2}\Bigg] \exp \Bigg[- \frac{r_i^2 + r_{i'}^2 + r_j^2 + r_{j'}^2}{4 \sigma^2}\Bigg] \nonumber \\
& \quad \quad \times I_0 \Bigg(\frac{1}{2 \sigma^2} \sqrt{r_i^2 r_{i'}^2 + r_j^2 r_{j'}^2 + 2 r_i r_j r_{i'} r_{j'} \cos (\theta_{ij} - \theta_{i'j'})} \Bigg) \\
\\
k_{\rho \rho'}|_{n=3} &= \sum_{i} \sum_{i'} R_i^3 R_{i'}^3 \exp \Bigg[\frac{3}{2} \frac{z_i z_{i'}}{\sigma^2}\Bigg] \exp \Bigg[- \frac{3}{4} \frac{r_i^2 + r_{i'}^2}{\sigma^2}\Bigg] I_0 \Bigg( \frac{3}{2} \frac{r_i r_{i'}}{\sigma^2}\Bigg) \nonumber \\
&+ 3 \sum_{\substack{i \\ j \neq i}} \sum_{\substack{i' \\ j' \neq i'}} R_i^2 R_{i'}^2 R_j R_{j'} \exp \Bigg[\frac{2 z_i z_{i'} + z_j z_{j'}}{2 \sigma^2}\Bigg] \exp \Bigg[- \frac{2 r_i^2 + 2 r_{i'}^2 + r_j^2 + r_{j'}^2}{4 \sigma^2}\Bigg] \nonumber \\
& \quad \quad \times I_0 \Bigg(\frac{1}{2 \sigma^2} \sqrt{4 r_i^2 r_{i'}^2 + r_j^2 r_{j'}^2 + 4 r_i r_j r_{i'} r_{j'} \cos (\theta_{ij} - \theta_{i'j'})} \Bigg) \nonumber \\
&+ 6 \sum_{\substack{i \\ j \neq i \\ k \neq i, j}} \sum_{\substack{i' \\ j' \neq i' \\ k' \neq i', j'}} R_i R_{i'} R_j R_{j'} R_k R_{k'} \exp \Bigg[\frac{z_i z_{i'} + z_j z_{j'} + z_k z_{k'}}{2 \sigma^2}\Bigg] \nonumber \\
& \quad \quad \times \exp \Bigg[- \frac{r_i^2 + r_{i'}^2 + r_j^2 + r_{j'}^2 + r_k^2 + r_{k'}^2}{4 \sigma^2}\Bigg] \nonumber \\
& \quad \quad \times I_0 \Bigg(\frac{1}{2 \sigma^2} \sqrt{\begin{array}{l} r_i^2 r_{i'}^2 + r_j^2 r_{j'}^2 + r_k^2 r_{k'}^2 + \\ 2 r_i r_j r_{i'} r_{j'} \cos (\theta_{ij} - \theta_{i'j'}) + \\ 2 r_j r_k r_{j'} r_{k'} \cos (\theta_{jk} - \theta_{j'k'}) + \\ 2 r_k r_i r_{k'} r_{i'} \cos (\theta_{ki} - \theta_{k'i'}) \end{array}} \Bigg)
\end{align*}
\normalsize{}

\noindent and we can immediately recognise that for the case of $n = 2$ angles projected onto the plane perpendicular to the bond axis for each of the bonds are coupled. For the case of $n = 3$ the coupling is between three angles.

Finally, if the bonds in question are connecting atoms of the same species the covariance function needs to be invariant to reflections about the $x$-$y$ plane. Consequently, we need to sum over the possible reflections while swapping the direction of the $z$ axis. In the above expression all terms dependent on the $z_i$ coordinate are separated into a single exponential. Hence, the summation over the possible reflections of bond $\rho'$ can be achieved by rewriting:

\clearpage 

\begin{equation}
\exp \Bigg[\frac{z_i z_{i'} + z_j z_{j'} + \dots}{2 \sigma^2}\Bigg] \xrightarrow{\text{reflection}} \cosh \Bigg(\frac{z_i z_{i'} + z_j z_{j'} + \dots}{2 \sigma^2}\Bigg) .
\end{equation}

\noindent This is because the summation over the two mirror images is equivalent to:

\begin{equation}
e^x + e^{-x} = 2 \cosh \left(x\right) .
\end{equation}

\section{Implementation Considerations}
\label{chapter:bond-based_soap-gap_potential:section:implementation_considerations}

In order to simplify the implementation of the bond-based SOAP-GAP potential we can rewrite the expression for $k_{\rho \rho'}$ derived in the previous section in an alternative form. If one defines elements of the matrix $\gamma_{ij}$ as:

\begin{equation}
\gamma_{ij} = r_i r_j (\cos \theta_{ij} + i \sin \theta_{ij}) = r_i r_j e^{i \theta_{ij}} ,
\end{equation}

\noindent we can exploit the property:

\begin{align}
\Re (\gamma_{ij} \gamma_{i'j'}^*) = r_i r_j r_{i'} r_{j'} (\cos \theta_{ij} \cos \theta_{i'j'} + \sin \theta_{ij} \sin \theta_{i'j'}) ,
\end{align}

\noindent and consequently we obtain the expressions for $k_{\rho \rho'}$ which for the simple cases of $n = 1$, $n = 2$ and $n = 3$ reduce to:

\scriptsize{}
\begin{align*}
k_{\rho \rho'}|_{n=1} &= \sum_{i} \sum_{i'} R_i R_{i'} \exp \Bigg[\frac{z_i z_{i'}}{2 \sigma^2}\Bigg] \exp \Bigg[- \frac{|\gamma_{ii}| + |\gamma_{i'i'}|}{4 \sigma^2}\Bigg] I_0 \Bigg( \frac{1}{2 \sigma^2} \sqrt{\Re(\gamma_{ii} \gamma_{i'i'}^*)} \Bigg) \nonumber \\
\\
k_{\rho \rho'}|_{n=2} &= \sum_{i} \sum_{i'} R_i^2 R_{i'}^2 \exp \Bigg[\frac{z_i z_{i'}}{\sigma^2}\Bigg] \exp \Bigg[- \frac{|\gamma_{ii}| + |\gamma_{i'i'}|}{2 \sigma^2}\Bigg] I_0 \Bigg( \frac{1}{\sigma^2} \sqrt{\Re(\gamma_{ii} \gamma_{i'i'}^*)} \Bigg) \nonumber \\
&+ 2 \sum_{\substack{i \\ j \neq i}} \sum_{\substack{i' \\ j' \neq i'}} R_i R_{i'} R_j R_{j'} \exp \Bigg[\frac{z_i z_{i'} + z_j z_{j'}}{2 \sigma^2}\Bigg] \exp \Bigg[- \frac{|\gamma_{ii}| + |\gamma_{i'i'}| + |\gamma_{jj}| + |\gamma_{j'j'}|}{4 \sigma^2}\Bigg] \nonumber \\
& \quad \quad \times I_0 \Bigg(\frac{1}{2 \sigma^2} \sqrt{\Re(\gamma_{ii} \gamma_{i'i'}^* + \gamma_{jj} \gamma_{j'j'}^* + 2 \gamma_{ij} \gamma_{i'j'}^*)} \Bigg) \nonumber \\
\end{align*}
\normalsize{}

\scriptsize{}
\begin{align*}
k_{\rho \rho'} |_{n=3} &= \sum_{i} \sum_{i'} R_i^3 R_{i'}^3 \exp \Bigg[\frac{3}{2} \frac{z_i z_{i'}}{\sigma^2}\Bigg] \exp \Bigg[- \frac{3}{4} \frac{|\gamma_{ii}| + |\gamma_{i'i'}|}{\sigma^2}\Bigg] I_0 \Bigg( \frac{3}{2} \frac{1}{\sigma^2} \sqrt{\Re(\gamma_{ii} \gamma_{i'i'}^*)} \Bigg) \nonumber \\
&+ 3 \sum_{\substack{i \\ j \neq i}} \sum_{\substack{i' \\ j' \neq i'}} R_i^2 R_{i'}^2 R_j R_{j'} \exp \Bigg[\frac{2 z_i z_{i'} + z_j z_{j'}}{2 \sigma^2}\Bigg] \exp \Bigg[- \frac{2 |\gamma_{ii}| + 2 |\gamma_{i'i'}| + |\gamma_{jj}| + |\gamma_{j'j'}|}{4 \sigma^2}\Bigg] \nonumber \\
& \quad \quad \times I_0 \Bigg(\frac{1}{2 \sigma^2} \sqrt{\Re(4 \gamma_{ii} \gamma_{i'i'}^* + \gamma_{jj} \gamma_{j'j'}^* + 4 \gamma_{ij} \gamma_{i'j'})} \Bigg) \nonumber \\
&+ 6 \sum_{\substack{i \\ j \neq i \\ k \neq i, j}} \sum_{\substack{i' \\ j' \neq i' \\ k' \neq i', j'}} R_i R_{i'} R_j R_{j'} R_k R_{k'} \exp \Bigg[\frac{z_i z_{i'} + z_j z_{j'} + z_k z_{k'}}{2 \sigma^2}\Bigg] \nonumber \\
& \quad \quad \times \exp \Bigg[- \frac{|\gamma_{ii}| + |\gamma_{i'i'}| + |\gamma_{jj}| + |\gamma_{j'j'}| + |\gamma_{kk}| + |\gamma_{k'k'}|}{4 \sigma^2}\Bigg] \nonumber \\
& \quad \quad \times I_0 \Bigg(\frac{1}{2 \sigma^2} \sqrt{\Re(\gamma_{ii} \gamma_{i'i'}^* + \gamma_{jj} \gamma_{j'j'}^* + \gamma_{kk} \gamma_{k'k'}^* + 2 \gamma_{ij} \gamma_{i'j'}^* + 2 \gamma_{jk} \gamma_{j'k'}^* + 2 \gamma_{ki} \gamma_{k'i'}^*)} \Bigg)
\end{align*}
\normalsize{}

\noindent where the expression for any $n > 3$ is a simple extension using the terms of the multinomial theorem.

We should also recognise that the elements of the of the matrix $\boldsymbol{\gamma}$ correspond to the symmetric matrix $\boldsymbol{\Sigma}$ described in section \ref{chapter:gaussian_approximation_potential:section:description_of_atomic_environments} (and introduced in \cite{weyl1950theory}), with the only difference being that $\boldsymbol{\gamma}$ obeys cylindrical symmetry and $\boldsymbol{\Sigma}$ obeys spherical symmetry. Consequently, we can think of the functional derived for our covariance function $k_{\rho \rho'}$ expressed above as way of introducing permutational invariance, since:

\begin{equation}
k_{\rho \rho'} = k(\{\boldsymbol{\Sigma}, \mathbf{z}\}, \{\boldsymbol{\Sigma}', \mathbf{z}'\}) .
\end{equation}

In order to evaluate the above expression for $k_{\rho \rho'}$ we find that a modified Bessel function of the first kind $I_\nu$ can be computed iteratively (as outlined in \cite{abramowitz1964handbook}):

\begin{equation}
I_\nu(z) = \left( \frac{1}{2} z \right)^\nu \sum_{k=0}^\infty \frac{\left( \frac{1}{4} z^2 \right)^k}{k! \Gamma(\nu + k + 1)} ,
\end{equation}

\noindent which for the special case of $\nu = 0$ simplifies to:

\begin{equation}
I_0 (z) = \sum_{k=0}^\infty \frac{\left( \frac{1}{4} z^2 \right)^k}{(k!)^2} .
\end{equation}

\noindent Furthermore, whenever the argument of $I_0$ is large, in order to ensure numerical stability (for large values of argument when $\gamma_{ij} \gg \sigma$ the negative exponential term approaches zero whereas the modified Bessel function term approaches $\infty$) we use an asymptotic expansion of $I_0$ (more details in \cite{abramowitz1964handbook}):

\begin{equation}
I_0 (z) = \frac{\exp (z)}{\sqrt{2 \pi z}} \left( 1 + \frac{1}{8z} + \frac{9}{2! (8z)^2} + \frac{9 \times 25}{3! (8z)^3} + \dots \right) .
\end{equation}

\noindent Finally, to evaluate the derivatives of the covariance function $k_{\rho \rho'}$ we use a derivative identity for the modified Bessel functions and it becomes simply a matter of applying chain rule sufficient number of times.

We finish this chapter with a brief analysis of the computational complexity associated with the bond-based SOAP-GAP covariance function. It is clear that the number of unique terms in our sum is dictated by the multinomial theorem and depends on the degree of the polynomial $n$, the number of atoms $N$ in the atomic environment of bond $\rho$ and the number of atoms $N'$ in the atomic environment of bond $\rho'$. Consequently, for an arbitrary $n$ the number of terms $\#_{N N'}$ is given by:

\begin{equation}
\#_{N N'} (n) = \frac{(n + NN' - 1)!}{n! (NN' - 1)!} ,
\end{equation}

\noindent and we can see that for the simple cases of $n = 1$, $n = 2$ and $n = 3$:

\begin{align}
\#_{N N'}|_{n = 1} &= NN' \nonumber \\
\#_{N N'}|_{n = 2} &= \frac{1}{2} NN' (NN' + 1) \nonumber \\
\#_{N N'}|_{n = 3} &= \frac{1}{6} NN' (NN' + 1) (NN' + 2) .
\end{align}

\noindent As was the case with the atomic SOAP kernel with no expansion in a radial basis, this is an increasingly intensive task in situations where the bond is surrounded by a large number of neighbours. Hence, in spite of the fact that the bond-based SOAP-GAP covariance function offers an improved accuracy (since no expansion in a spherical harmonics basis is necessary), we find that it emerges as a solution only in systems that are either less-densely packed (in terms of nearest neighbours) or where the bond environment is completely determined by a small number of neighbouring atoms.

\cleardoublepage

\chapter{Conclusions and Further Work}
\label{chapter:conclusions_and_further_work}

Throughout this thesis I explored how the Gaussian Approximation Potential scheme for generating interatomic potentials can be applied to atomistic studies of tungsten --- a bcc transition metal selected as a ``testing ground'' for the development of ``GAP for metals'' methodology. Since the plasticity behaviour of metals is largely controlled by the properties of dislocations and their interactions with other lattice defects, our investigation focused primarily on developing a method that is capable of describing the energetics of these defects with an accuracy approaching that of explicitly quantum-mechanical models. Consequently, the outcome of the research carried out during my doctoral studies can be summarised as follows:

\begin{enumerate}
\item Development of protocol for training GAP potentials for an accurate description of lattice defects in bcc transition metals:
\suspend{enumerate}

In our study we systematically improved our training dataset in order to identify what training data contributes to an accurate representation of specific properties in the resulting potential. Consequently, we find that to reproduce elasticity behaviour, the training data should include primitive lattice cells sampled using a Monte Carlo approach in the lattice space. In order to reproduce the vibrational behaviour, the training data should include large cubic simulation cells sampled using Molecular Dynamics at appropriate temperature, etc.

\resume{enumerate}
\item Development of bispectrum-GAP interatomic potential for tungsten:
\suspend{enumerate}

We find that although the GAP potential based on the bispectrum descriptor of the atomic environment is successful in reproducing the energetics of the lattice defects that are included in the training data set explicitly, its predictive power is significantly limited within the extrapolation regime. Our investigation reveals that this is caused by the ``noisiness'' in the representation of the atomic density function that is used for training of the potential energy surface and it can be attributed to the fact that atoms are represented by Dirac delta functions with a truncated representation in bispectrum space.

\resume{enumerate}
\item Development of SOAP-GAP interatomic potential for tungsten:
\suspend{enumerate}

We improve on the bispectrum-GAP potential by applying the Smooth Overlap of Atomic Positions kernel to the GAP methodology. This uses Gaussian functions to represent atomic density. We demonstrate that this significantly improves the accuracy of the resulting potential, which is capable of reproducing our benchmark data with accuracy approaching that of the DFT model. We confirm that the SOAP kernel allows us to systematically improve the accuracy of the GAP potential by improving the quality of representation of the atomic density function and we also demonstrate that training GAP potentials without a core potential improves the accuracy of the forces and energies.

\resume{enumerate}
\item \begin{sloppypar}Simulation of the mobility of tungsten $\frac{1}{2} \langle 111 \rangle$ screw dislocation and dislocation-vacancy interactions:\end{sloppypar} 
\suspend{enumerate}

We use the SOAP-GAP interatomic potential to calculate the Peierls barrier and the dislocation-vacancy interaction map for tungsten in an isolated dislocation system of $>$100,000 atoms and we verify our results against DFT model in a dislocation dipole system of 135 atoms. We demonstrate that the transition of a $\frac{1}{2} \langle 111 \rangle$ screw dislocation is not mediated by a meta-stable state and we find that our description of the dislocation provides accuracy approaching that of the DFT model.

\resume{enumerate}
\item Development of Smooth Overlap of Atomic Positions methodology for bond-based GAP potentials:
\end{enumerate}

In the last section of my thesis, I derive a method of calculating a rotationally, permutationally (and reflection) invariant covariance function between bond-environments, where the atomic density is expressed in terms of Gaussian functions. Unlike the Smooth Overlap of Atomic Positions kernel for the interatomic potential, we obtain an analytical expression for the value of the covariance function that does not rely on expansion in a spherical harmonics basis and consequently always offers a fully converged result, where the accuracy can be tuned by computing terms of higher order than the power spectrum or bispectrum by coupling multiple angles.

\vspace{1.5cm}

\clearpage

Although we demonstrate our results for tungsten exclusively, the available literature on the bcc transition metals suggests that group V and VI elements share many of their physical properties and therefore we anticipate that our methodology should be equally applicable to these elements (more details in \cite{PhysRevB.54.6941}-\cite{PhysRevB.70.104113}). Consequently, we see multiple avenues for extending the work presented in this thesis in the future:

\begin{enumerate}
\item Application to other metallic systems:
\suspend{enumerate}

While we anticipate that SOAP-GAP potentials for group V and VI elements could be developed relatively easily, we believe that the next step in the development of GAP methodology is to simulate multi-component systems, such as alloys, in order to model interactions of dislocations with impurities. Another challenging system that is of particular interest due to its engineering applications, is iron --- also a bcc metal, but which has significantly more complex properties than tungsten due to multiple allotropic forms and complicated magnetic behaviour

\resume{enumerate}
\item Application to other lattice defects:
\end{enumerate}

We believe that our existing tungsten potential can be extended to include an accurate description of other lattice defects, such as grain boundaries, dislocation jogs and kinks. Since the GAP potential guarantees linear scaling of computational cost, these defects could be simulated in systems containing more than 100,000 atoms (as already demonstrated in this work), in order to predict properties that influence plasticity behaviour and crack propagation, or to compute some of the properties that are involved in controlling the onset of the brittle-to-ductile transition in tungsten, which we anticipate to be a cooperative phenomenon of many dislocations.

\cleardoublepage

%
%

\phantomsection
\addcontentsline{toc}{chapter}{Bibliography}
\begin{sloppypar}
\printbibliography
\end{sloppypar}

\cleardoublepage

%
%

\appendix
\chapter[Tungsten Energy-Volume Phase Diagram]{\texorpdfstring{Tungsten Energy-Volume \\ Phase Diagram}{Tungsten Energy-Volume Phase Diagram}}
\label{appendix:tungsten_energy-volume_phase_diagram}

\vspace{-0.15cm} 

We compute tungsten energy-volume curves for the four most common cubic crystal structures using both the Finnis-Sinclair interatomic potential and DFT model. The results are shown in figure \ref{figure:energy_volume_phases} below.

\begin{figure}[H]
\begin{center}
\resizebox{12cm}{!}{\footnotesize{}\input{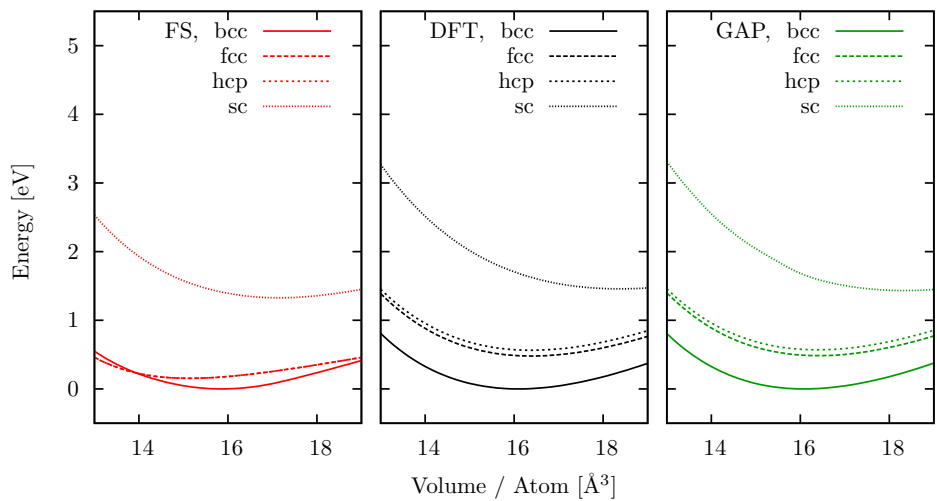}\normalsize{}}
\vspace{0.75cm}
\caption[Energy-volume curves of tungsten phases.]{Energy-volume curves of tungsten for the four most common cubic crystal structures.}
\label{figure:energy_volume_phases}
\end{center}
\end{figure}

\noindent We observe that the Finnis-Sinclair interatomic potential significantly underestimates the cohesive energy of the face-centred cubic (fcc) and simple cubic (sc) phases. It also cannot distinguish between the face-centred cubic and hexagonal close-packed (hcp) structures.

In order to investigate the suitability of the GAP methodology for representation of multiple crystal phases we generate a GAP potential from bcc tungsten training data (with no lattice defects) and the above energy-volume curves. The energy-volume phase diagram computed using the resulting potential is included in figure \ref{figure:energy_volume_phases} and we find it to be in an excellent agreement with the DFT model.

\cleardoublepage

\chapter{Tungsten Di- and Tri-Vacancies}
\label{appendix:tungsten_di-_and_tri-vacancies}

We use the same methodology to calculate formation energies of di- and tri-vacancies as we did for the mono-vacancy outlined in section \ref{chapter:bulk_properties_and_lattice_defects_in_tungsten:section:vacancy}. A schematic representation of the simulation cells comparing that of a mono-vacancy to the systems of di- and tri-vacancies is shown in figure \ref{figure:vacancies} below.

\begin{figure}[H]
\begin{center}
\vspace{0.25cm}
\resizebox{12cm}{!}{\footnotesize{}\input{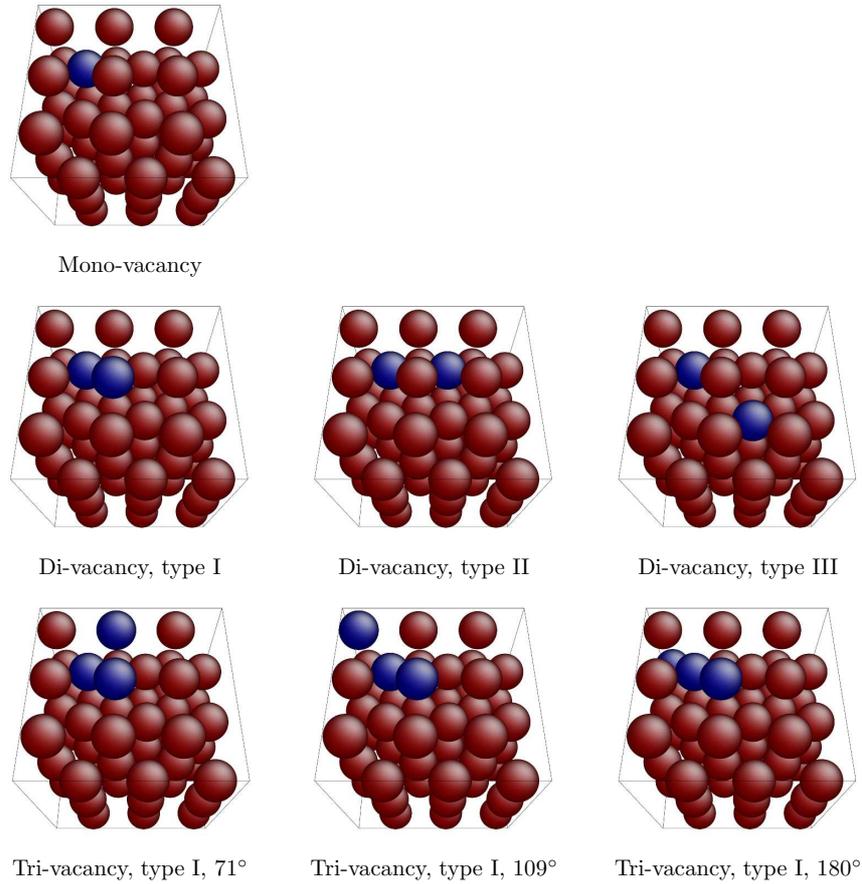}\normalsize{}}
\vspace{0.5cm}
\caption[Mono-, di- and tri-vacancy simulation cells.]{Representation of mono-, di- and tri-vacancy simulation cells. The atoms coloured blue are removed and vacancy introduced instead.}
\label{figure:vacancies}
\end{center}
\end{figure}

We carry out a preliminary investigation of the suitability of the GAP methodology for description of systems of di- and tri-vacancies. We compute formation energies of di- and tri-vacancies using DFT in a 53 atom simulation cell. We then generate a GAP potential from bcc tungsten training data (with no lattice defects) and relaxation trajectories of the di- and tri-vacancies. We verify the formation energies computed using DFT model with the Finnis-Sinclair interatomic potential and the resulting GAP potential. This is done by recomputing the energies of the DFT-minimised structures. The results are shown in table \ref{table:vacancies} below:

\begin{table}[H]
\vspace{0.5cm}
\begin{center}
\begin{tabular}{ l c c c c }
& & FS & DFT & GAP \\
\midrule
Di-vacancy, type I & [eV] & $7.02$ & $12.56$ & $12.19$ \\
\midrule
Di-vacancy, type II & [eV] & $7.02$ & $12.96$ & $12.93$ \\
\midrule
Di-vacancy, type III & [eV] & $7.54$ & $12.71$ & $12.72$ \\
\midrule
Tri-vacancy, type I, 71$^{\circ}$ & [eV] & $9.80$ & $15.78$ & $15.87$ \\
\midrule
Tri-vacancy, type I, 109$^{\circ}$ & [eV] & $10.32$ & $15.89$ & $15.95$ \\
\midrule
Tri-vacancy, type I, 180$^{\circ}$ & [eV] & $10.20$ & $15.70$ & $15.77$ \\
\midrule
\end{tabular}
\caption[Formation energies of di- and tri-vacancies.]{Formation energies of di- and tri-vacancies computed using DFT model and verified using FS and GAP potentials.}
\label{table:vacancies}
\end{center}
\end{table}

\noindent The results of this preliminary investigation suggest that the GAP potential can describe the energetics of the di- and tri-vacancies --- it is in a good agreement with the DFT model.

\clearpage

%
%

\newpage
\thispagestyle{empty}
\mbox{}

\end{document}